%% file: main.tex
\pdfoutput=1
\documentclass[12pt,a4paper]{article}

\usepackage{ifthen}
\newboolean{pdflatex}
\setboolean{pdflatex}{true}

\newboolean{articletitles}
\setboolean{articletitles}{true}
\newboolean{uprightparticles}
\setboolean{uprightparticles}{false}
\newcommand{\comment}[1]{}

\def\paperauthors{LHCb collaboration}
\def\paperasciititle{Angular analysis of the rare decay BsToPhimm}
\def\papertitle{Angular analysis of the rare decay $B_s^0\to\phi\mu^+\mu^-$}
\def\paperkeywords{{High Energy Physics}, {LHCb}}
\def\papercopyright{\the\year\ CERN for the benefit of the LHCb collaboration}
\def\paperlicence{CC BY 4.0 licence}
\def\paperlicenceurl{https://creativecommons.org/licenses/by/4.0/}

\input{preamble}
\usepackage{longtable}
\begin{document}

\renewcommand{\thefootnote}{\fnsymbol{footnote}}
\setcounter{footnote}{1}

\input{title-LHCb-PAPER}

\renewcommand{\thefootnote}{\arabic{footnote}}
\setcounter{footnote}{0}

\cleardoublepage

\pagestyle{plain}
\setcounter{page}{1}
\pagenumbering{arabic}


\section{Introduction}
\label{sec:Introduction}
Transitions of a \bquark quark to an \squark quark and a pair of oppositely charged leptons are forbidden at tree level in the Standard Model (SM) and only proceed via higher-order electroweak (loop) diagrams.
These transitions constitute powerful probes for New Physics (NP) contributions beyond the SM that can appear in competing diagrams and significantly affect branching fractions and angular distributions of \bsll\ decays.
Recent studies of \bsll\ decays have observed tensions with SM predictions in measurements of branching fractions~\cite{LHCb-PAPER-2013-017,LHCb-PAPER-2014-006,LHCb-PAPER-2015-009,LHCb-PAPER-2015-023,LHCb-PAPER-2016-012, LHCb-PAPER-2021-014}, angular observables~\cite{LHCb-PAPER-2015-051,LHCb-PAPER-2020-002,LHCb-PAPER-2020-041,Aaboud:2018krd,Khachatryan:2015isa,Sirunyan:2017dhj,Wehle:2016yoi} and tests of lepton universality~\cite{LHCb-PAPER-2014-024,LHCb-PAPER-2017-013,LHCb-PAPER-2019-009,LHCb-PAPER-2019-040,LHCb-PAPER-2021-004,Wehle:2016yoi,Lees:2012tva,Abdesselam:2019lab,Abdesselam:2019wac}.
One of the most significant deviations from SM expectations is observed in the determination of the branching fraction of \BsToPhimm decays~\cite{LHCb-PAPER-2021-014}.\footnote{The inclusion of the charge-conjugated mode is implied throughout this paper unless otherwise stated, and the shorthand $\phi$ refers to the $\phi(1020)$ meson in the following.}
The measured branching fraction is found to be 3.6 standard deviations ($\sigma$) below a precise SM prediction~\cite{Altmannshofer:2014rta,Straub:2015ica,Straub:2018kue,Horgan:2013pva,Horgan:2015vla} in the squared dimuon mass (\qsq) region $1.1<\qsq<6.0\gevgevcccc$.
Angular analyses of \bsll\ decays provide information complementary to branching fraction measurements, allowing to probe the operator structure of potential NP contributions.
An angular analysis of the decay \BsToPhimm~\cite{LHCb-PAPER-2015-023}, using proton-proton ($pp$) collision data corresponding to $3\invfb$  recorded by the \lhcb experiment during 2011--2012, found the angular distributions to be compatible with SM predictions.

This paper presents an updated angular analysis of \BsToPhimm\ decays, where the $\phi$ meson is reconstructed in the $\Kp\Km$ final state, using $pp$ collisions recorded by the \lhcb experiment corresponding to a total integrated luminosity of $8.4\invfb$.
The data were collected at centre-of-mass energies of $7\tev$ (2011), $8\tev$ (2012) and $13\tev$ (2016--2018) during the LHC Run~1 and Run~2, respectively.  For the purposes of this analysis, the data are split according to the 2011--2012 ($3\invfb$), 2016 ($1.7\invfb$) and 2017--2018 ($3.7\invfb$) data-taking periods.
The higher $b\bar{b}$ production cross-section in Run~2\cite{LHCB-PAPER-2016-031,LHCb-PAPER-2017-037} yields an approximate four-fold increase in the total number of produced \Bs\ mesons compared to the Run~1 data.
The criteria used to select candidates in this analysis are identical to those of Ref.~\cite{LHCb-PAPER-2021-014}, with an adapted $q^{2}$ binning scheme.

Neglecting the natural width of the $\phi$ meson, the $\decay{\Bs}{\phi(\to\Kp\Km)\mumu}$ decay rate depends on \qsq, three decay angles, \thetal, \thetak, and $\phi$, and the decay time of the \Bs meson~\cite{Descotes-Genon:2015hea}.
The angle \thetal (\thetak) is defined as the angle of the \mun (\Km) with respect to the direction of flight of the \Bs meson in the \mumu (\Kp\Km) centre-of-mass frame, and $\phi$ as the angle between the \mumu and the \Kp\Km planes in the \Bs meson centre-of-mass frame.
As the decay flavour of the \Bs\ meson cannot be determined from the flavour-symmetric final state, the same angular definition is used for both \Bs\ and \Bsb\ decays.

The untagged \CP-averaged angular decay rate, $\Gamma+\overline{\Gamma}$, is measured integrated over the \Bs decay time and is given for a specific \qsq region by
{
\small
\begin{align}
\frac{1}{{\deriv}(\Gamma+\overline{\Gamma})/{\deriv}\qsq}\frac{\deriv^3(\Gamma+\overline{\Gamma})}{\deriv\cos\thetal\, \deriv\cos\thetak\, \deriv\phi} &=
    \frac{9}{32\pi}
    \Big[
        \tfrac{3}{4}(1-\FL) \sin^2\thetak(1+\tfrac{1}{3}\cos 2\thetal) \nonumber\\
        & +  {\FL} \cos^2\thetak(1-\cos 2\thetal)\nonumber
          +  {S_3} \sin^2\thetak\sin^2\thetal \cos 2\phi\\&
        +  {S_4} \sin 2\thetak\sin 2\thetal \cos \phi\nonumber
          +  {A_5} \sin 2\thetak\sin \thetal \cos \phi
        \\&
        +  \tfrac{4}{3}\AFBCP \sin^2\thetak\cos\thetal\nonumber
          +  {S_7} \sin 2\thetak \sin\thetal \sin\phi \\ &
        +  {A_8} \sin 2\thetak \sin 2\thetal \sin\phi
          +  {A_9} \sin^2\thetak \sin^2\thetal \sin 2\phi
    \Big]\,,\label{eq:untagged}
\end{align}
}
where the angular observables \FL and $S_{3,4,7}$ are \CP averages, and \AFBCP and $A_{5,8,9}$ are \CP asymmetries~\cite{Altmannshofer:2008dz,Bobeth:2008ij}.
The presence of \CP asymmetries in Eq.~\ref{eq:untagged} is due to the need to use identical angular definitions for the \Bs and \Bsb modes~\cite{Descotes-Genon:2015hea, Bobeth:2008ij}.
Of particular interest are the $T$-odd \CP\ asymmetries $A_8$ and $A_9$, which are predicted to be close to zero in the SM, but can be large in the presence of NP contributions~\cite{Bobeth:2008ij}.
As the decay flavour of the \Bs meson is unknown, the \CP-averaged observable $S_5$ ($P_5^\prime$), which has received a lot of attention in the study of \BdToKstmm decays~\cite{LHCb-PAPER-2020-002}, cannot be accessed by this analysis.

\section{Detector and simulation}
\label{sec:Detector}
The \lhcb detector~\cite{LHCb-DP-2008-001,LHCb-DP-2014-002} is a single-arm forward spectrometer covering the \mbox{pseudorapidity} range $2<\eta <5$, designed for the study of particles containing \bquark or \cquark quarks. The detector includes a high-precision tracking system consisting of a silicon-strip vertex detector surrounding the $pp$ interaction region~\cite{LHCb-DP-2014-001}, a large-area silicon-strip detector located upstream of a dipole magnet with a bending power of about $4{\mathrm{\,Tm}}$, and three stations of silicon-strip detectors and straw drift tubes~\cite{LHCb-DP-2013-003,LHCb-DP-2017-001} placed downstream of the magnet.
The tracking system provides a measurement of the momentum, \ptot, of charged particles with a relative uncertainty that varies from 0.5\% at low momentum to 1.0\% at 200\gevc.
The minimum distance of a track to a primary $pp$ collision vertex~(PV), the impact parameter (IP), is measured with a resolution of $(15+29/\pt)\mum$, where \pt is the component of the momentum transverse to the beam, in\,\gevc.
Different types of charged hadrons are distinguished using information from two ring-imaging Cherenkov detectors~\cite{LHCb-DP-2012-003}.
Muons are identified by a system composed of alternating layers of iron and multiwire proportional chambers~\cite{LHCb-DP-2012-002}.

The online event selection is performed by a trigger system~\cite{LHCb-DP-2012-004}.
In this analysis, an initial hardware stage uses information from the muon system to require at least one muon with significant \pt\ in the event.
Events passing the hardware trigger enter the software trigger, where a full event reconstruction is applied.
At this stage, further requirements are placed on the kinematics of the muon candidates and on the topology of the signal candidate.

Simulated samples are used to determine the effect of reconstruction and selection on the angular distributions of the signal candidates, as well as to estimate expected signal yields and contamination from specific background processes.
The $pp$ collisions are simulated using \pythia~\cite{Sjostrand:2007gs,*Sjostrand:2006za}
with a specific \lhcb configuration~\cite{LHCb-PROC-2010-056}. Decays of unstable particles are described by \evtgen~\cite{Lange:2001uf}, in which final-state radiation is generated using \photos~\cite{davidson2015photos}. The interaction of the generated particles with the detector, and its response, are implemented using the \geant toolkit~\cite{Allison:2006ve, *Agostinelli:2002hh} as described in Ref.~\cite{LHCb-PROC-2011-006}. Residual mismodelling of the particle identification performance, the \pt spectrum of the \Bs mesons, the track multiplicity and the efficiency of the hardware trigger is corrected using high-yield control samples from data.

\section{Selection of signal candidates}
\label{sec:Selection}

All tracks in the $\Kp\Km\mumu$ final-state are required to have
significant \chisqip with respect to any PV, where \chisqip\ is defined as the difference in the vertex-fit \chisq of a given PV reconstructed with and without the track being considered.
The final-state particles are further required to be well identified using particle identification information.

The \Bs\ decay vertex, determined by fitting the four final-state tracks, is required to be of good quality and to be significantly displaced from any PV in the event.
The angle between the vector connecting the associated PV with the \Bs\ decay vertex and the momentum of the \Bs candidate ($\theta_{\rm DIRA}$) is required to be small.
The associated PV is defined as that which fits best to the flight direction of the \Bs candidate.

Candidates are accepted if the invariant reconstructed $\Kp\Km\mumu$ mass is in the range $5270 < m(\Kp\Km\mumu) < 5700 \mevcc$ and the invariant mass of the $\Kp\Km$ system is within $12\mevcc$ of the known $\phi$ mass~\cite{PDG2020}.
Candidates are further required to have a \qsq\ value in the range $0.1<\qsq<18.9\gevgevcccc$.

The resonant \mbox{$\decay{\Bs}{\phi(\to\mumu)\phi}$}, \mbox{$\decay{\Bs}{\jpsi(\to\mumu)\phi}$} and \mbox{$\decay{\Bs}{\psitwos(\to\mumu)\phi}$} decays dominate the experimental \qsq spectrum in the \qsq regions of \mbox{$0.98 < \qsq <1.1\gevgevcccc$}, \mbox{$8 < \qsq <11\gevgevcccc$} and \mbox{$12.5 < \qsq <15\gevgevcccc$}, respectively.
These \qsq regions are therefore excluded from the signal selection but \BsToJPsiPhi candidates in the \mbox{$8 < \qsq <11\gevgevcccc$} region are retained as a control mode to develop selection criteria, validate fit behaviour and derive corrections to the simulation.

Background originating from a random combination of tracks (combinatorial background) is reduced using a boosted decision tree (BDT)~\cite{Breiman} classifier trained with the AdaBoost algorithm~\cite{AdaBoost} as implemented in the TMVA package~\cite{Hocker:2007ht,*TMVA4}.
The BDT classifier is trained on data and its performance verified using standard cross-validation techniques~\cite{Blum:1999:BHB:307400.307439}.
The upper mass sideband, defined as $m(\Kp\Km\mumu) >5567\mevcc$, is used as a proxy for the combinatorial background and enriched in background by relaxing the requirement on the invariant mass of the $\Kp\Km$ system from $12\mevcc$ to $50\mevcc$ around the known $\phi$ mass.
As a signal proxy, a sample of \BsToJPsiPhi candidates from data in a $50\mevcc$ mass range around the known \Bs mass~\cite{PDG2020} is used, for which background contributions have been statistically subtracted~\cite{Pivk:2004ty} using the invariant reconstructed \Kp\Km\mup\mun mass as the separating variable.

The classifier combines the transverse momentum of the \Bs, the angle $\theta_{\rm DIRA}$, the fit quality of the \Bs vertex~(vertex-fit $\chi^2$), its displacement from the associated PV and the \chisqip and particle identification information of all final-state tracks.
The selection criterion on the BDT output is chosen according to the figure of merit $N_{\rm sig}/\sqrt{N_{\rm sig}+N_{\rm bkg}}$, where $N_{\rm sig}$~($N_{\rm bkg}$) is the expected number of signal (background) events in the signal region. With respect to the previously described selection criteria, this requirement results in a signal efficiency of 96\% and a background rejection of 96\%, where the latter considers only contributions from combinatorial background.

Decays of $b$ hadrons where one or more of the final-state particles have been misidentified constitute another important background source, referred to as peaking backgrounds. Contributions include decays of the form $\decay{\Bs}{\jpsi\phi}$, $\decay{\Bs}{\psitwos\phi}$, \mbox{$\decay{\Bd}{\jpsi\Kstarz}$} and $\decay{\Bd}{\psitwos\Kstarz}$, where a hadron is misidentified as a muon and vice versa. Once misidentified, these decays can contaminate the signal \qsq regions. To further suppress these contributions, more stringent particle identification requirements are placed on candidates where the invariant mass of the $\mupm \Kmp$ system under the dimuon mass hypothesis is close to the known \jpsi or \psitwos mass~\cite{PDG2020}.

Other sources of peaking background include \LbTopKmm decays, where the proton is misidentified as a kaon, and \decay{\Bd}{\Kstarz(\to\Kp\pim)\mumu} decays, where the pion is misidentified as a kaon. The \LbTopKmm decay is additionally suppressed by applying more stringent particle identification criteria if the invariant mass of a candidate under the relevant misidentification hypothesis is close to the known \Lb mass~\cite{PDG2020}.
No single source of peaking background is found to contribute more than 0.5\% of the total signal yield after all selection criteria are applied. Peaking background contributions are therefore neglected in the fit model and a systematic uncertainty is assigned to account for potential residual background pollution.

\begin{figure}
    \centering
    \includegraphics[width=.45\textwidth]{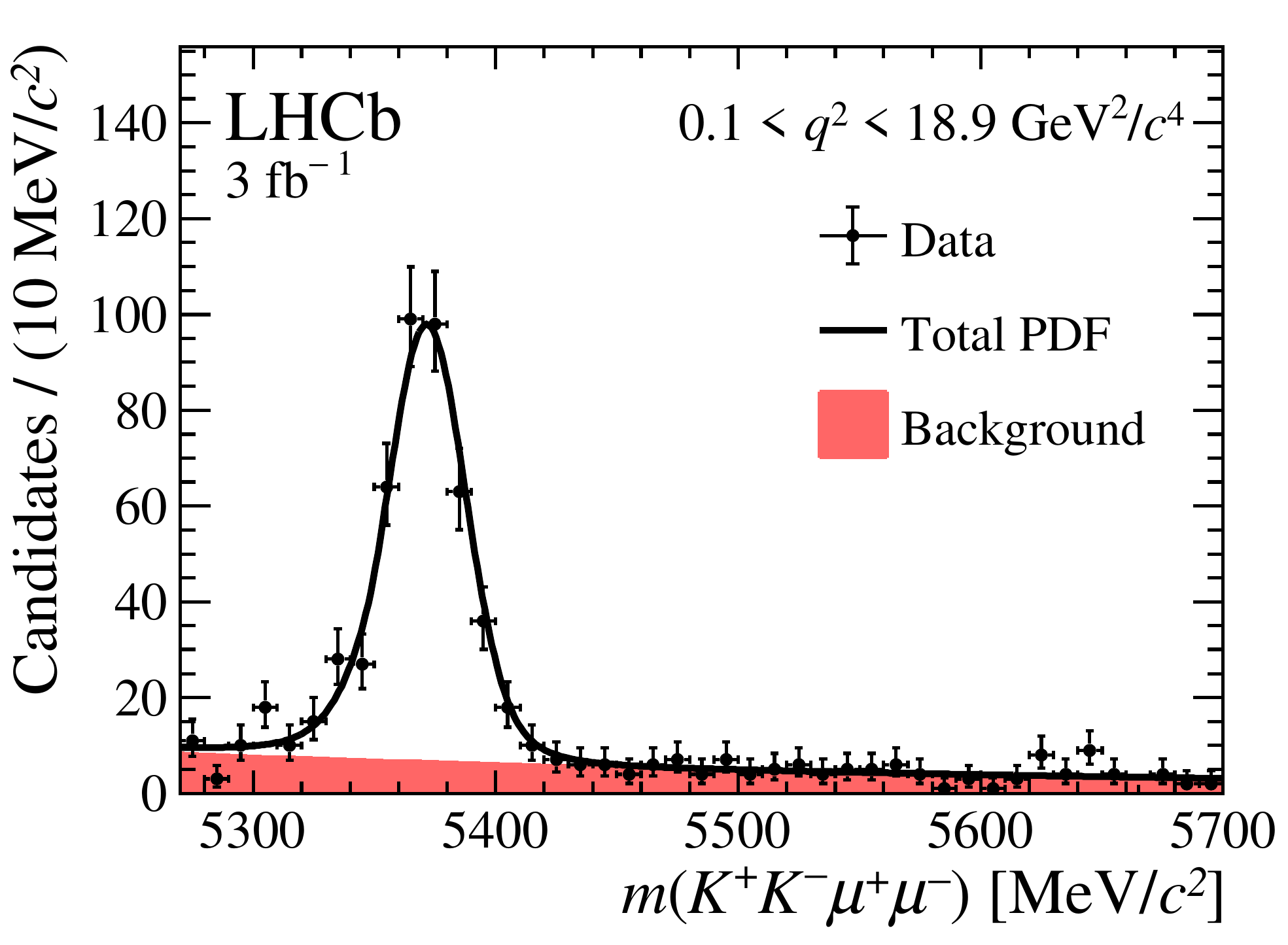}
    \includegraphics[width=.45\textwidth]{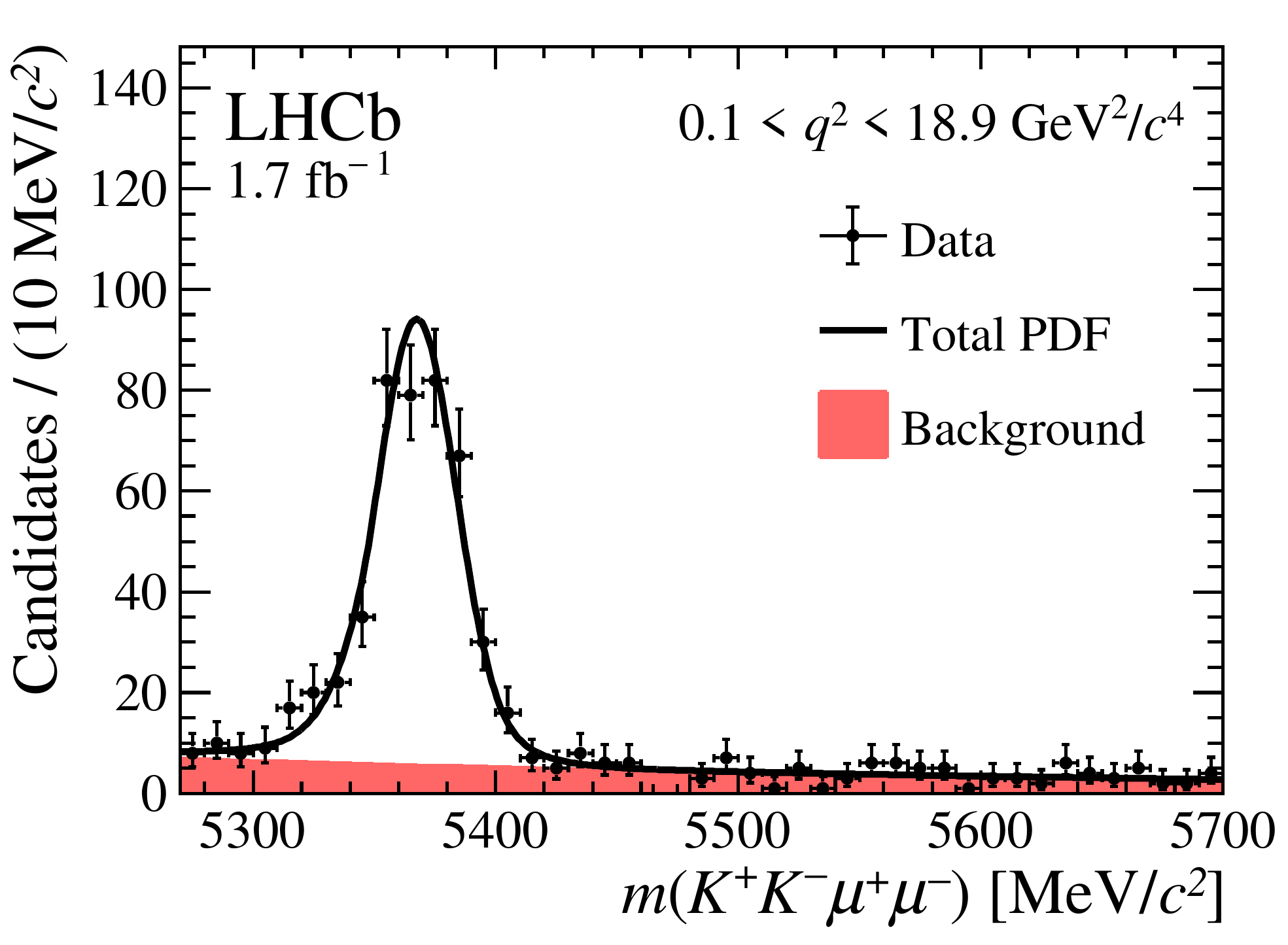}
    \includegraphics[width=.45\textwidth]{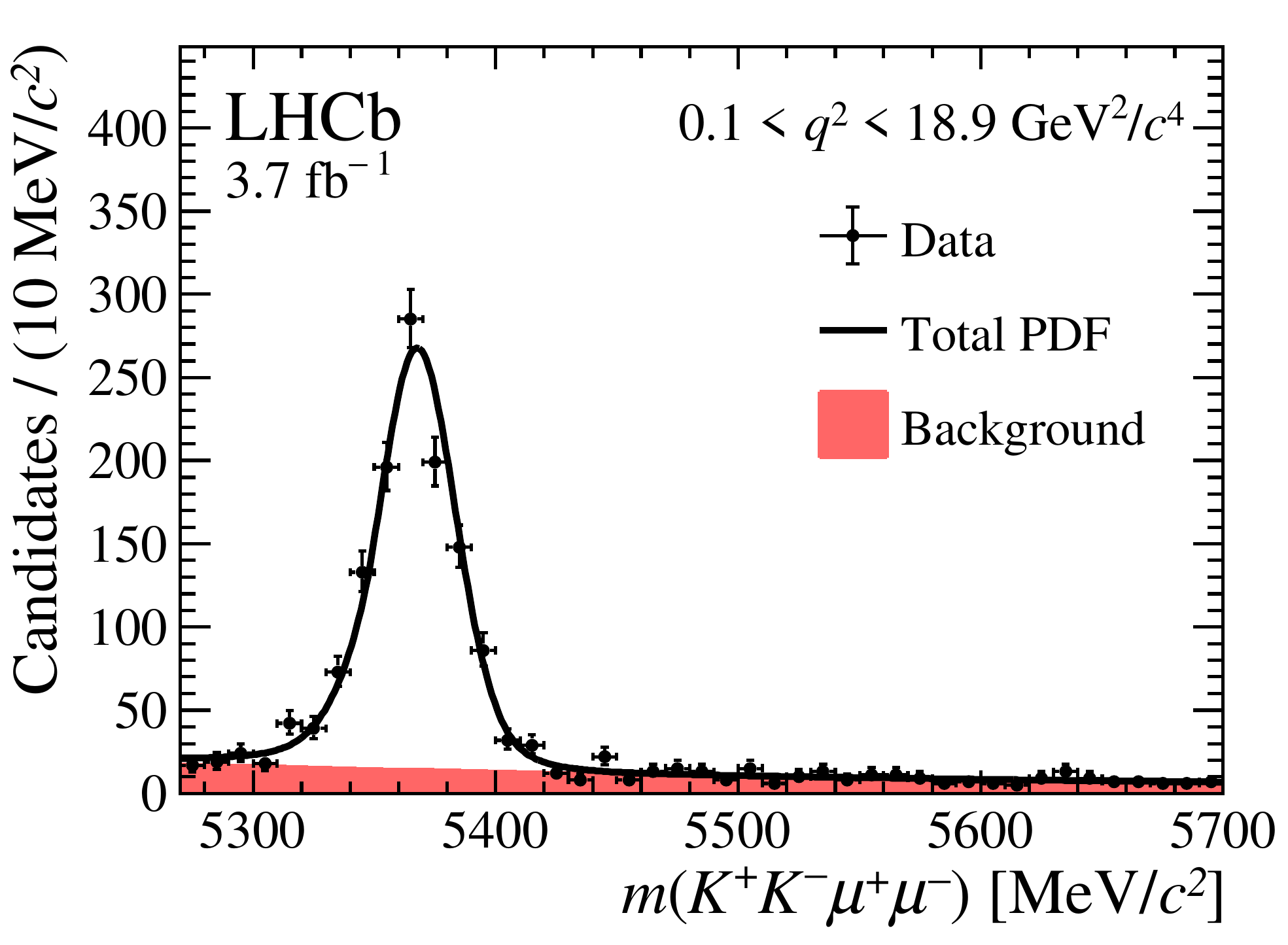}
    \caption{\label{fig:results_mass_only} The $m(\Kp\Km\mumu)$ distribution for \BsToPhimm candidates integrated over the $0.1 < \qsq <0.98\gevgevcccc$, $1.1 < \qsq <8\gevgevcccc$, $11.0 < \qsq < 12.5\gevgevcccc$ and \mbox{$15.0 < \qsq <18.9\gevgevcccc$} regions for the data-taking periods 2011-2012 (top left), 2016 (top right), and 2017--2018 (bottom). The data are overlaid with the PDF used to describe the $m(\Kp\Km\mumu)$ spectrum, fitted separately for each data set.
    }
\end{figure}

Figure~\ref{fig:results_mass_only} shows the $m(\Kp\Km\mumu)$ distribution for all candidates passing the selection, integrated over the $0.1 < \qsq <18.9\gevgevcccc$ region for the separate data sets, excluding the \qsq regions contaminated by the resonant $\decay{\Bs}{\phi(\to\mumu)\phi}$, $\decay{\Bs}{\jpsi(\to\mumu)\phi}$ and $\decay{\Bs}{\psitwos(\to\mumu)\phi}$ decays.
The data are overlaid with the fitted probability density function (PDF) described in Sec.~\ref{sec:Angular}.
Signal yields of $408\pm23$, $402\pm23$ and $1120\pm40$ are found for the 2011--2012, 2016 and 2017--2018 data sets, where the uncertainties are statistical only.

\section{Angular analysis}
\label{sec:Angular}
The angular observables are determined using an unbinned maximum likelihood fit to the invariant \Kp\Km\mumu mass distribution and the three decay angles, \thetal, \thetak, and $\phi$. In the \qsq region below 12.5\gevgevcccc, the fit is performed separately in narrow \qsq regions of around 2\gevgevcccc width and in an additional wide \qsq region defined as $[1.1,6.0]\gevgevcccc$. Above 15\gevgevcccc, a single wide region is used, defined as $[15.0,18.9]\gevgevcccc$.
A finer binning scheme compared to Ref.~\cite{LHCb-PAPER-2015-023} is chosen to maximise sensitivity to potential short-distance NP contributions whilst ensuring stable fit behaviour.

The $m(\Kp\Km\mumu)$ distribution is included in the fit to improve the separation power between signal and background. The signal component is modelled in $m(\Kp\Km\mumu)$ by a sum of two Gaussian functions with a common mean and power-law tails towards the upper or lower mass side~\cite{Skwarnicki:1986xj}.
The parameters describing the power-law tails are determined using simulated \BsToJPsiPhi  events. The parameters describing the widths and the mean of the Gaussian functions are fixed in the signal mode to the values from a fit to \BsToJPsiPhi candidates in data.
An additional \qsq-dependent scaling factor is determined from simulation and applied to the widths of the Gaussian distributions to account for the \qsq dependence of the $m(\Kp\Km\mumu)$ invariant-mass resolution. The angular distribution for the signal candidates is parameterised using Eq.~\ref{eq:untagged}.
The combinatorial background in the $m(\Kp\Km\mumu)$ distribution is described using an exponential function and in the angular distributions using a product of first-order Chebyshev polynomials.
The factorisation of the background angular distributions is validated using data candidates selected in the upper mass sideband.
The fraction of \decay{\Bs}{\Kp\Km\mumu} decays with the \Kp\Km system in an S-wave configuration, $F_{\rm S}$, is expected to be at the level of 1--2\%~\cite{LHCb-PAPER-2012-040,LHCb-PAPER-2014-059, LHCb-PAPER-2019-013}. These contributions are therefore not modelled in the fit and a systematic uncertainty is assigned to account for this choice.

The selection and reconstruction can distort the angular and \qsq distributions observed in data. These effects are described by an angular acceptance, $\epsilon(\ctl,\ctk,\phi,\qsq)$.
The acceptance is parameterised using a product of Legendre polynomials $P_i$ of order $i$ according to
\begin{equation}
    \epsilon(\ctl,\ctk,\phi,\qsq)=\sum_{k,l,m,n} c_{klmn} P_k(\ctl)P_l(\ctk)P_m(\phi)P_n(\qsq)\,,
\end{equation}
where the coefficients $c_{klmn}$ are determined on a large sample of simulated \BsToPhimm events by exploiting the orthogonality of the Legendre polynomials.
Given that the acceptance is parameterised in terms of the key degrees of freedom used in the decay description (\ie\ the three angles and \qsq) there is minimal dependency on the model used to simulate the events.
The orders used to model the efficiency are $k \leq 4$, $l \leq 2$ ($l \leq 4$ in Run~2), $m \leq 6$ and $n\leq 7$ in \ctl, \ctk, $\phi$ and \qsq, respectively. Where different sets of acceptance orders give a similar description of the acceptance function, the set of lowest orders is chosen.
Given the flavour-symmetric final state, only even orders are considered in the description of the three decay angles.
The choice of orders used to describe the angular acceptance is assessed as a source of systematic uncertainty.

In the narrow \qsq regions, the PDF describing the signal candidates is constructed from the product of the acceptance function evaluated at the median of the \qsq region and the signal fit model.
For the wide \qsq\ regions, the acceptance is taken into account by weighting each event by the inverse efficiency.
The shape of the angular acceptance is found to vary according to the data-taking conditions. The different trigger thresholds during the 2016 and 2017--2018 data-taking periods require the Run~2 data to be further separated.
The angular acceptance is therefore derived separately for the 2011--2012, 2016 and 2017--2018 data sets.
The data are split according to these periods and a simultaneous fit is performed. In the fit, the angular observables and angular background parameters are shared across the three data sets. The sharing of the angular background parameters improves the fit behaviour and the resulting small bias due to this choice is added as a systematic uncertainty.
All other nuisance parameters are determined independently for each data set.
In order to avoid experimenter's bias, all decisions regarding the fit strategy and candidate selection were made before the results were examined.

Pseudoexperiments, generated using the results of the best fit to data, are used to
assess the bias and coverage of the simultaneous fit. The majority of observables have a bias of less than 10\% of the statistical uncertainty. The observables $S_3$ and $S_4$ in the \qsq region $[4.0, 6.0] \gevgevcccc$ and \AFBCP in the \qsq region $[0.1, 0.98] \gevgevcccc$ exhibit a fit bias at the level of 15\% of the statistical uncertainty.
An additional systematic uncertainty equal to the size of the fit bias is assigned for all observables and the statistical uncertainty is corrected to account for any under- or over-coverage, which is at the level of 14\% or less.

The angular acceptance corrections for each data set are validated using both fits to \BsToJPsiPhi candidates and fits to simulated \BsToPhimm candidates, where the latter are generated according to a physics model using inputs from Ref.~\cite{Ball:2004rg}.
The angular observables extracted from fits to \BsToJPsiPhi candidates are in good agreement with previous measurements~\cite{LHCb-PAPER-2014-059, LHCb-PAPER-2019-013}. In addition, the angular observables extracted from fits to simulated events are in good agreement with the values used in their generation.

\section{Results}

\begin{figure}
    \centering
    \includegraphics[width=.45\textwidth]{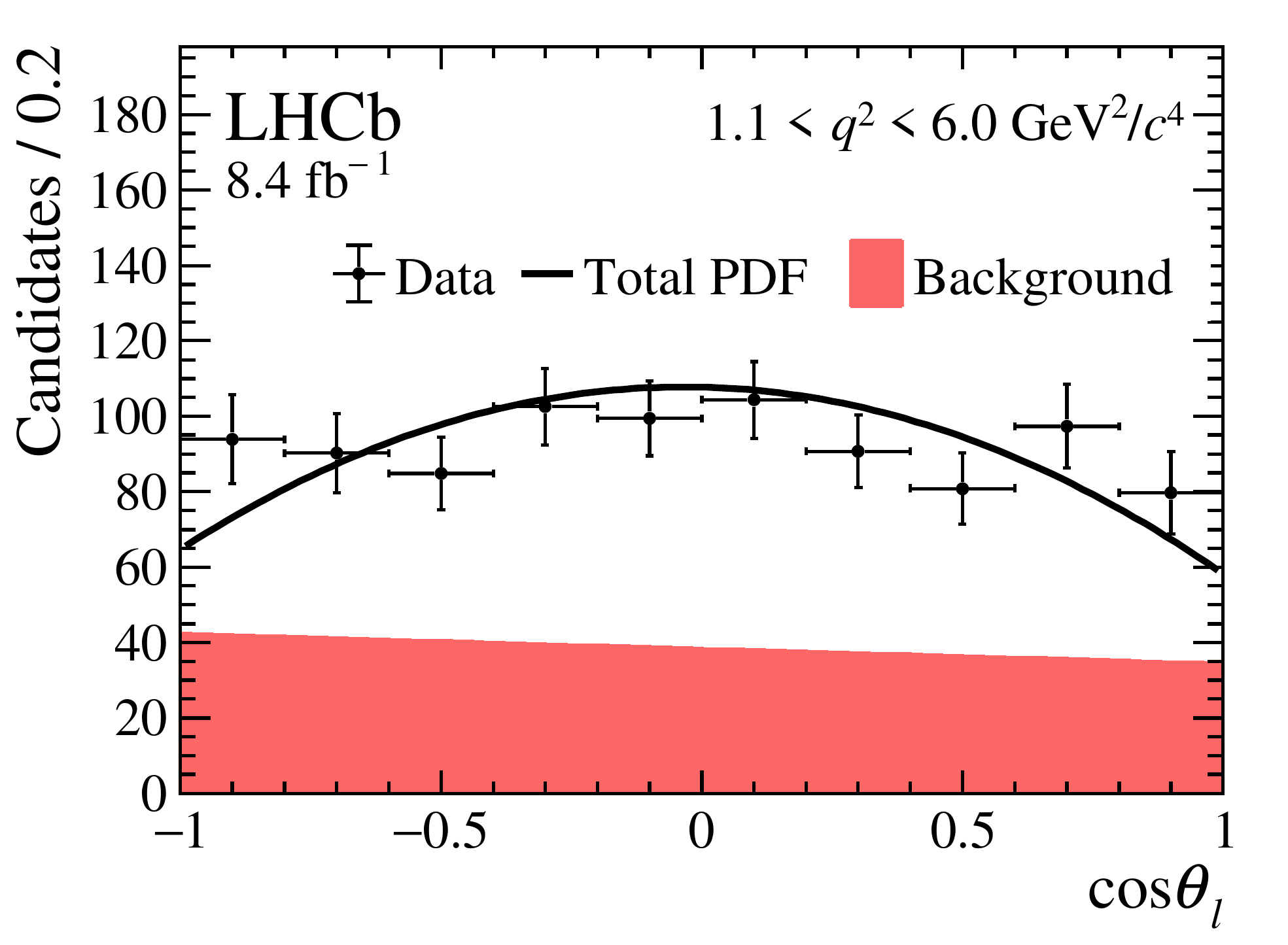}
    \includegraphics[width=.45\textwidth]{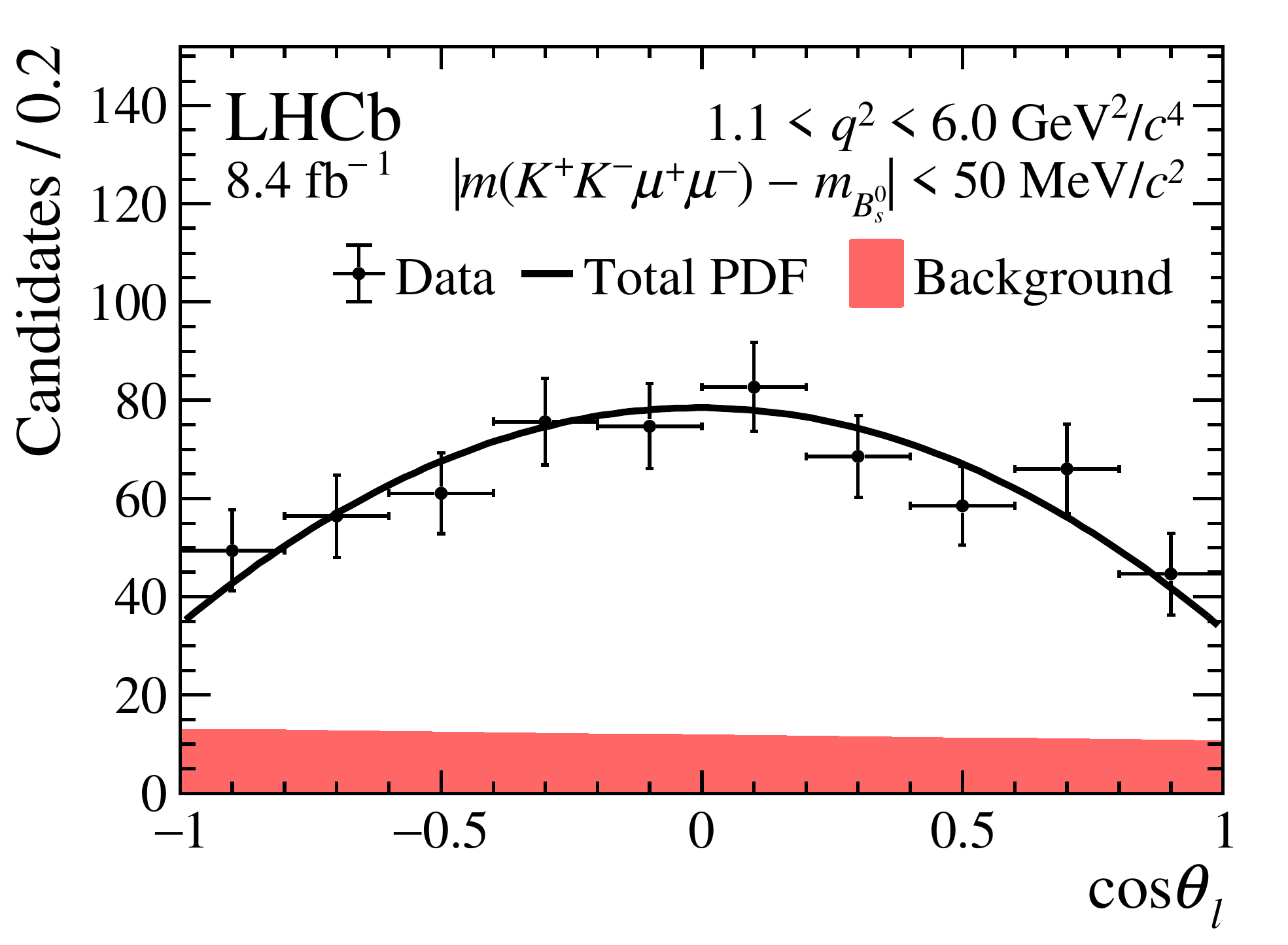}\\
    \includegraphics[width=.45\textwidth]{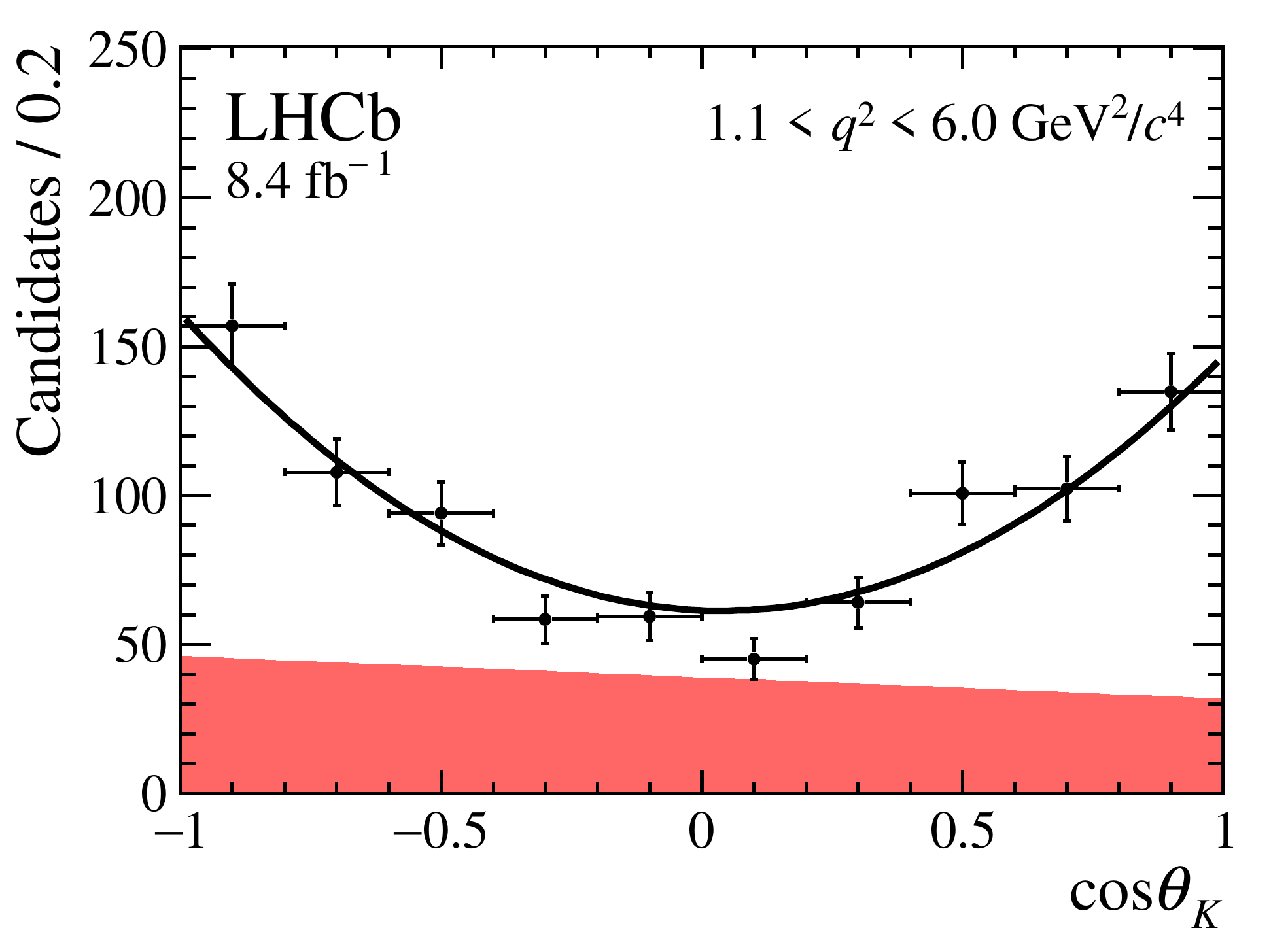}
    \includegraphics[width=.45\textwidth]{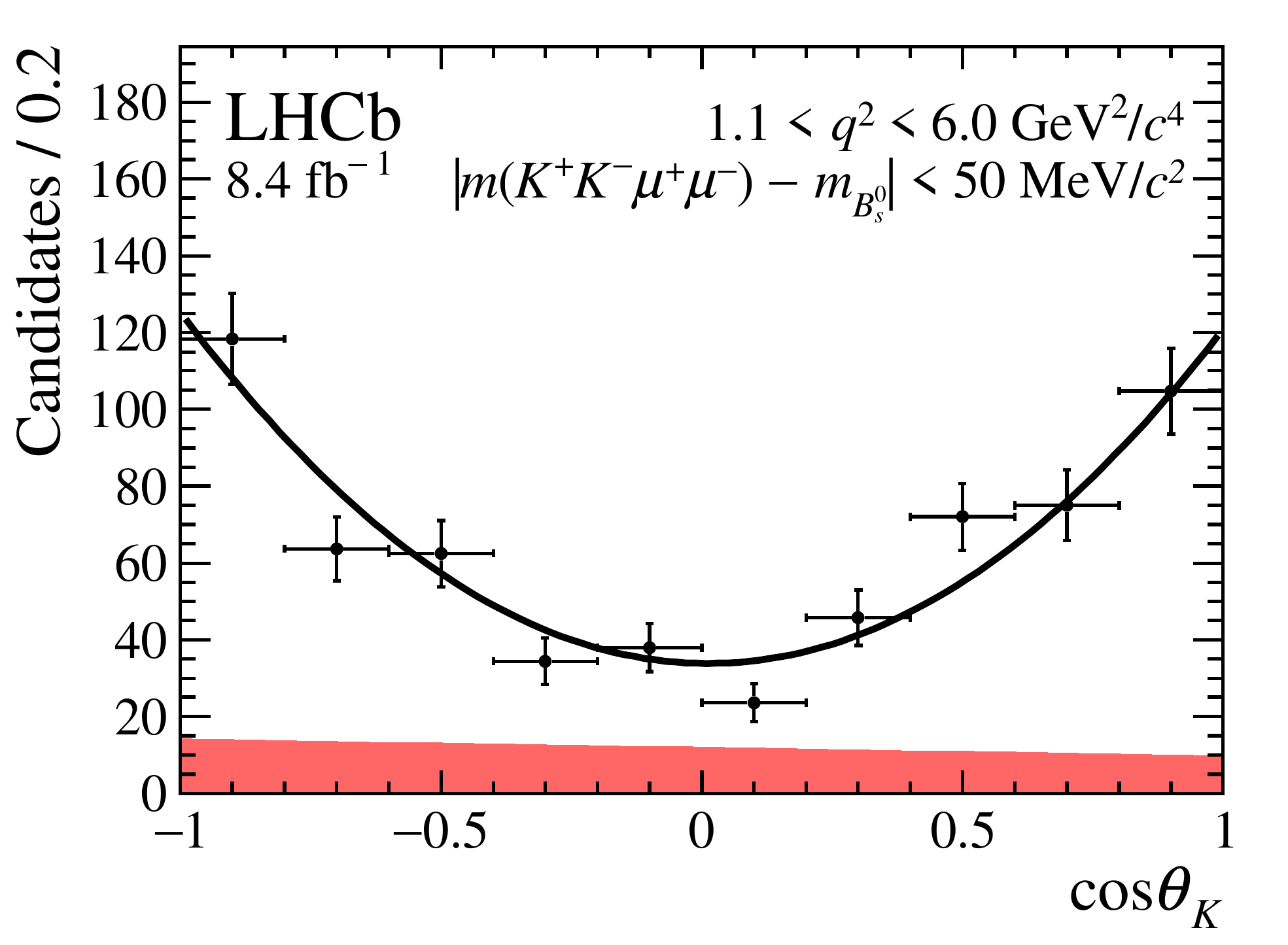}\\
    \includegraphics[width=.45\textwidth]{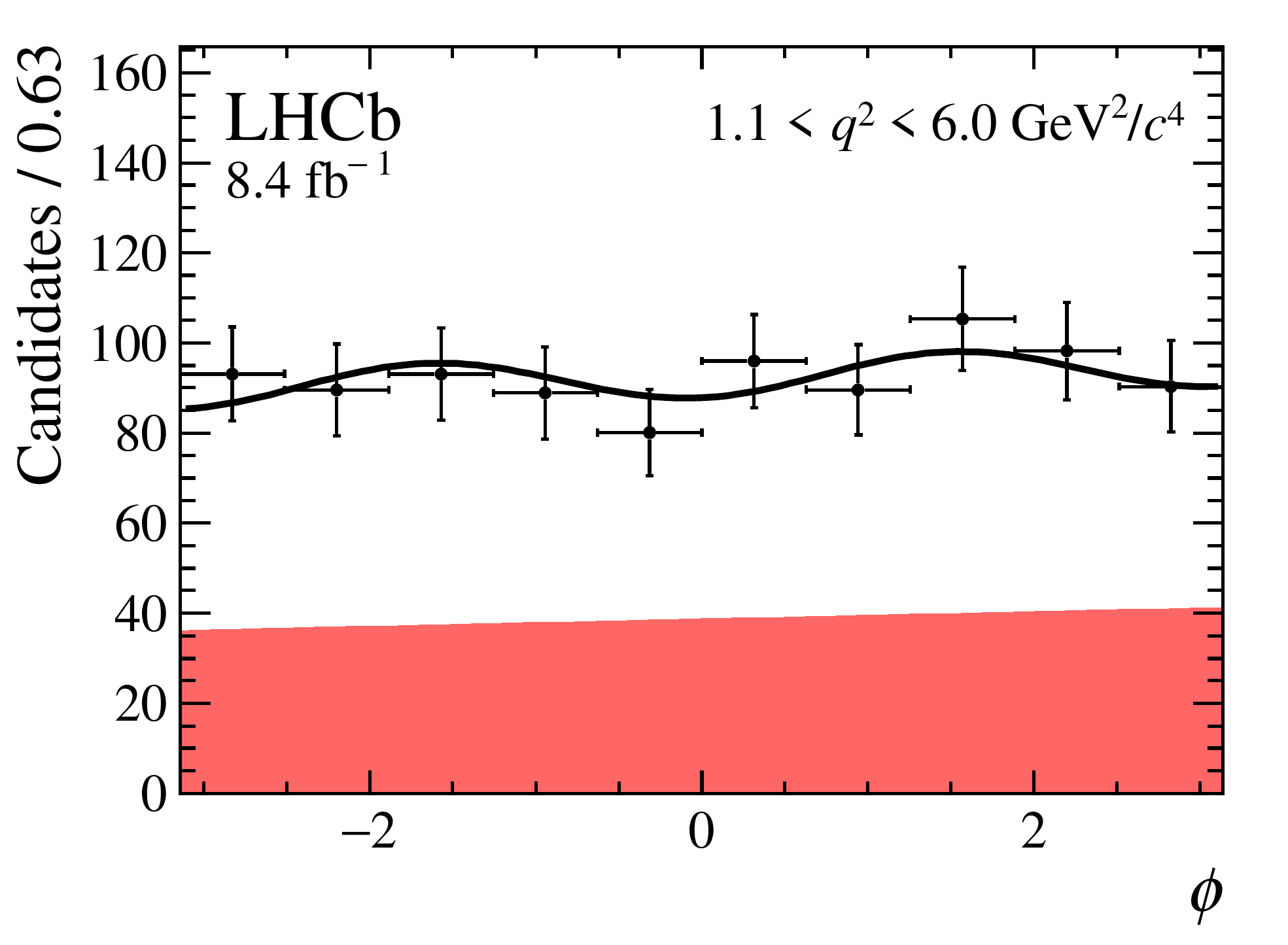}
    \includegraphics[width=.45\textwidth]{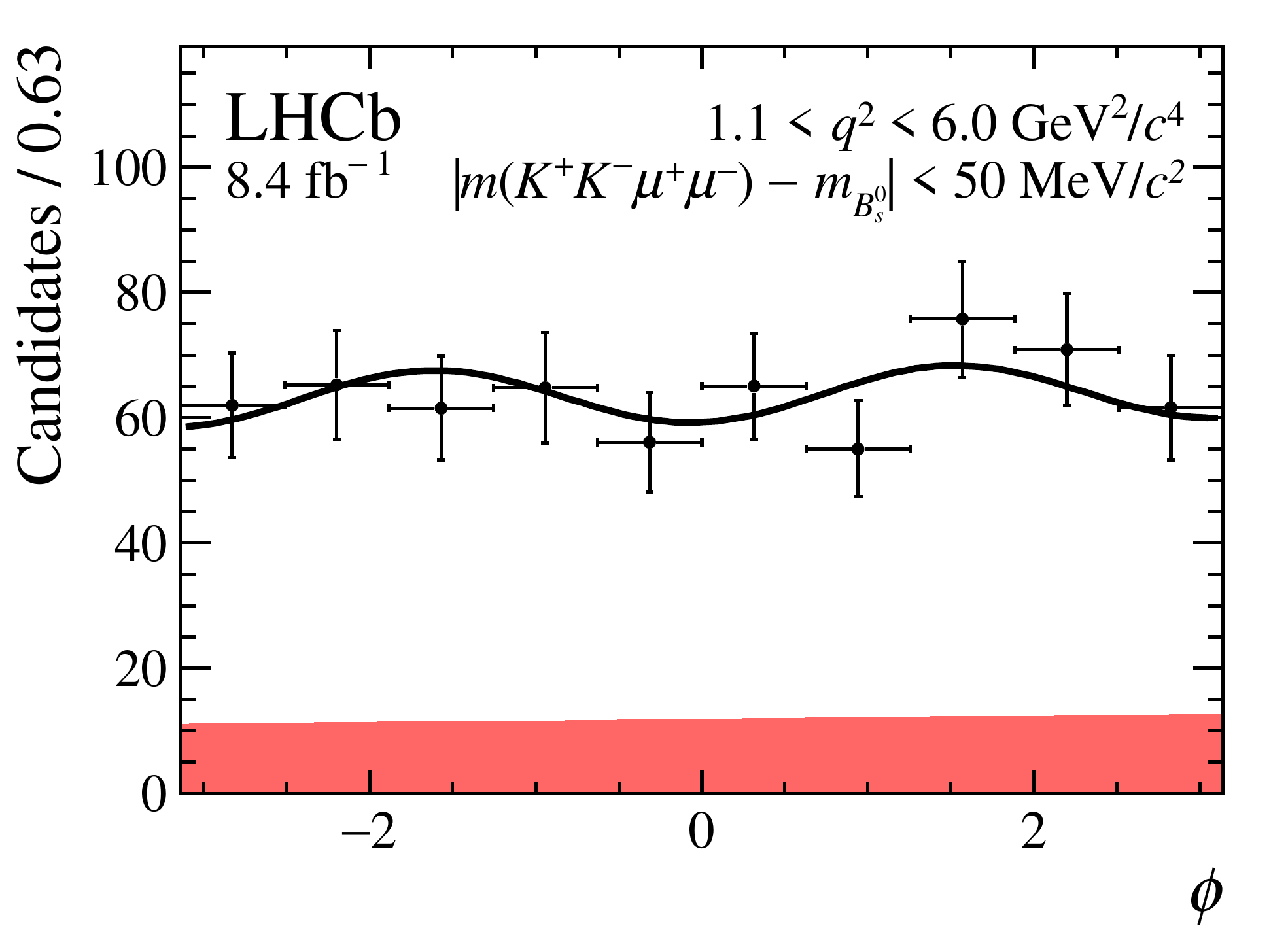}
    \caption{\label{fig:angular_proj} Angular projections in the
    region $1.1 < \qsq <6.0\gevgevcccc$
    for the combined 2011--2012, 2016 and 2017--2018 data sets. The data are overlaid with the projection of the combined PDF.
     The red shaded area indicates the background component and the solid black line the total PDF.
    The angular projections are given for candidates in (left) the entire mass region used to determine the observables in this paper and (right) the signal mass window $\pm50\mevcc$ around the known \Bs mass.
    }
\end{figure}

The angular distributions for the combined 2011--2012, 2016 and 2017--2018 data set are shown in Fig.~\ref{fig:angular_proj} for all candidates in the $[1.1,6]\gevgevcccc$ \qsq region and for candidates within $\pm 50\mevcc$ of the known \Bs mass. The data are overlaid with the projection of the fitted PDF, combined across the data sets.
The fit projections for all \qsq regions and individual data sets are provided
in Appendix~\ref{sec:fit-proj}.

\begin{figure}
    \centering
    \includegraphics[width=.45\textwidth]{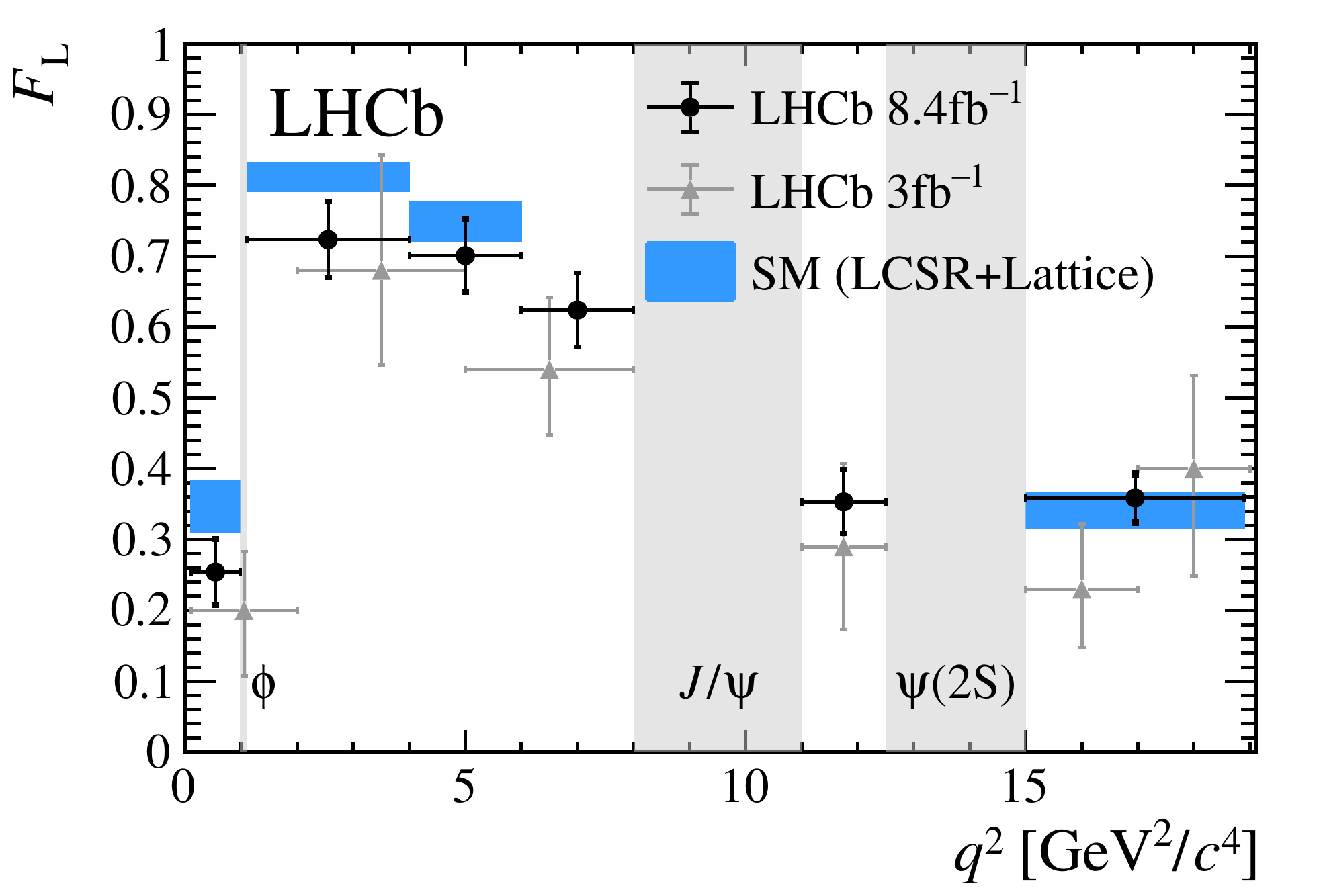}~
    \includegraphics[width=.45\textwidth]{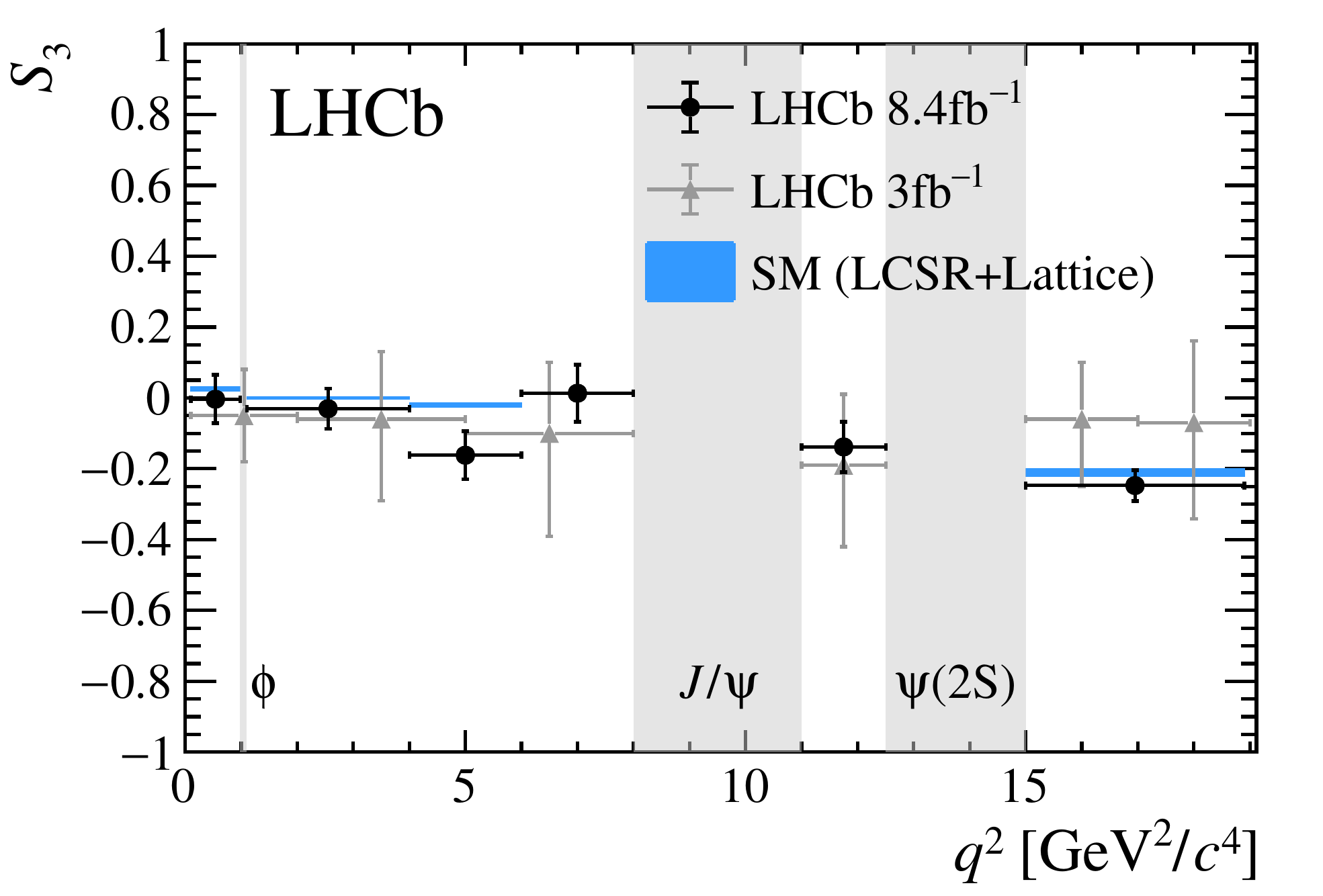}\\
    \includegraphics[width=.45\textwidth]{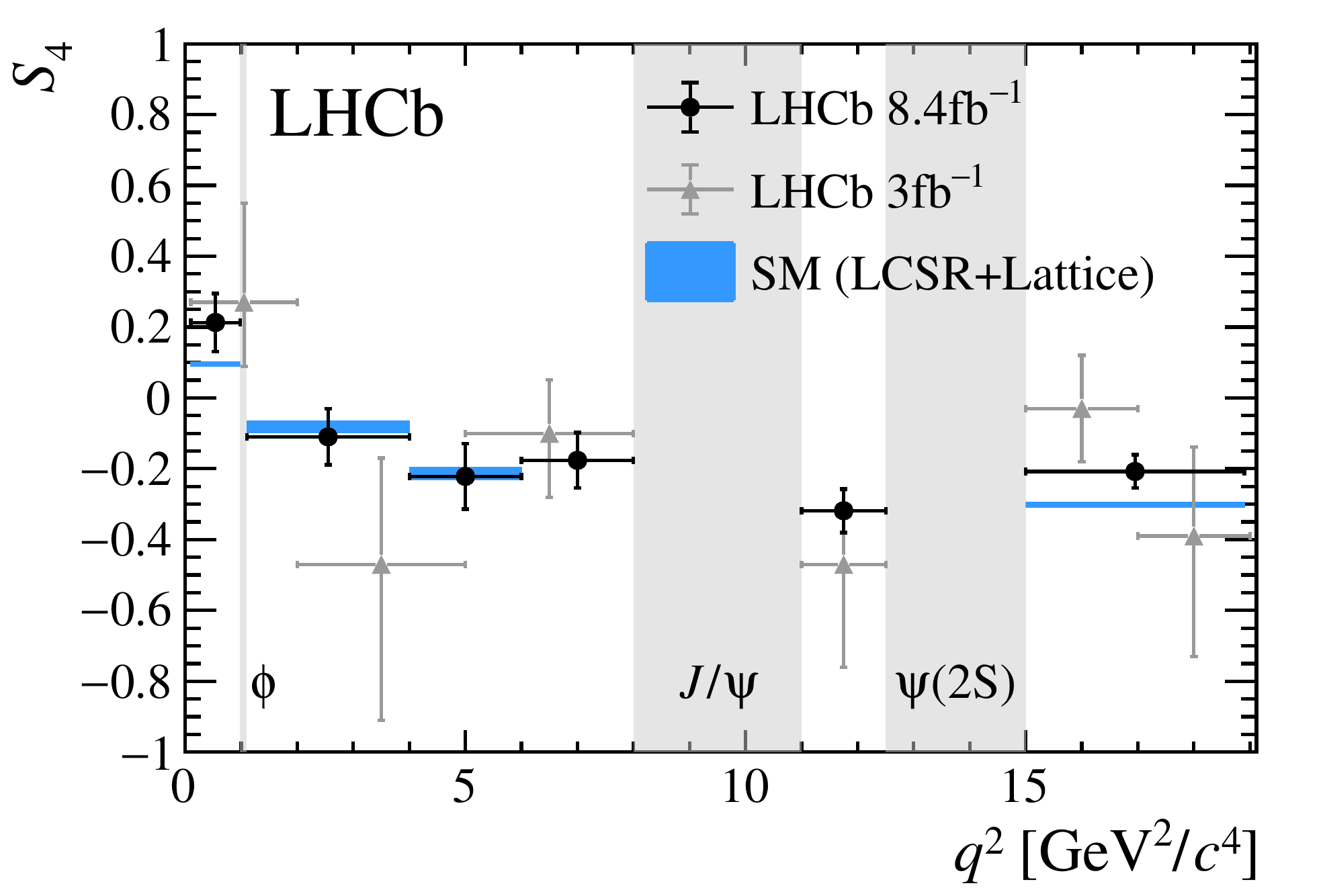}~
    \includegraphics[width=.45\textwidth]{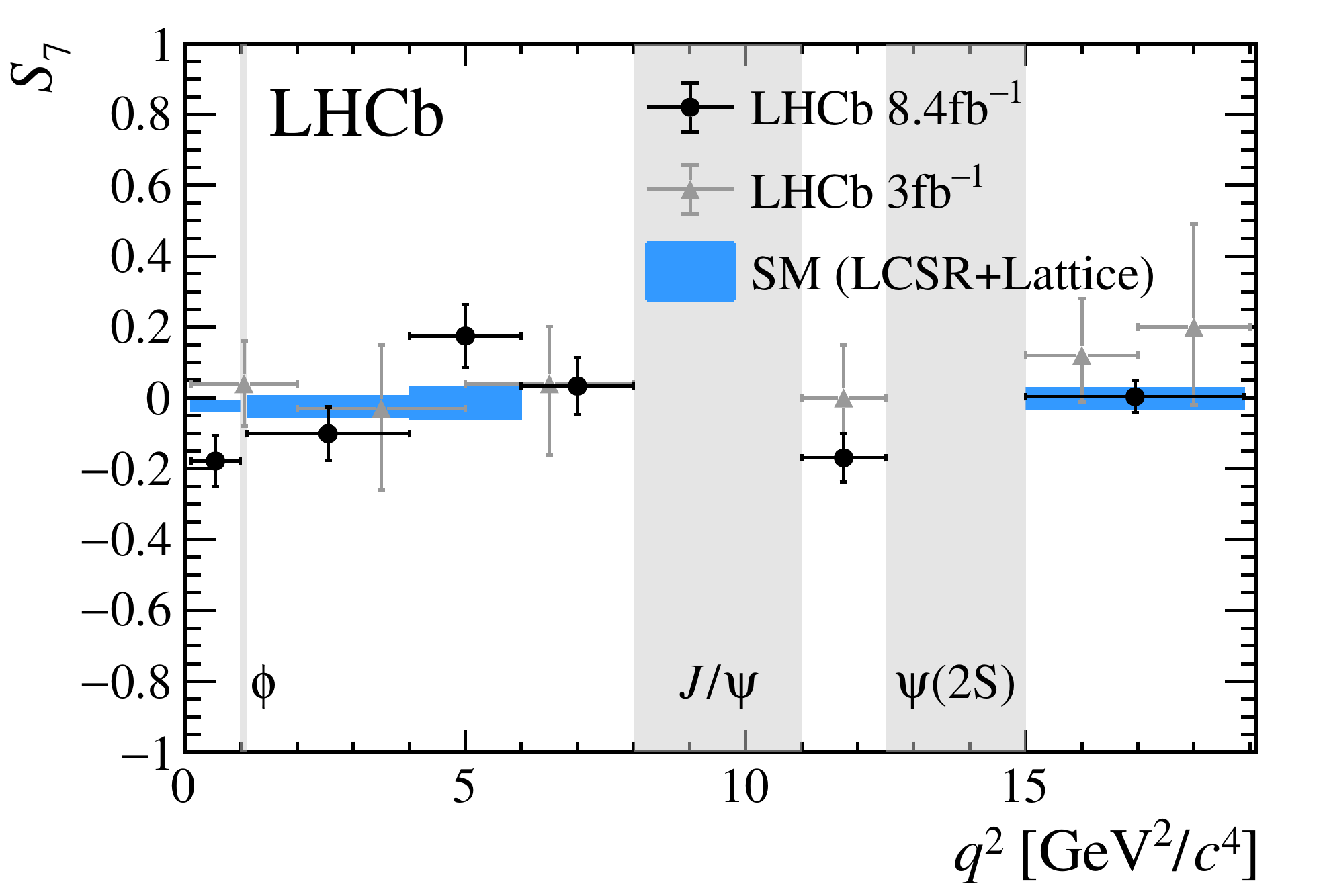}\\
    \includegraphics[width=.45\textwidth]{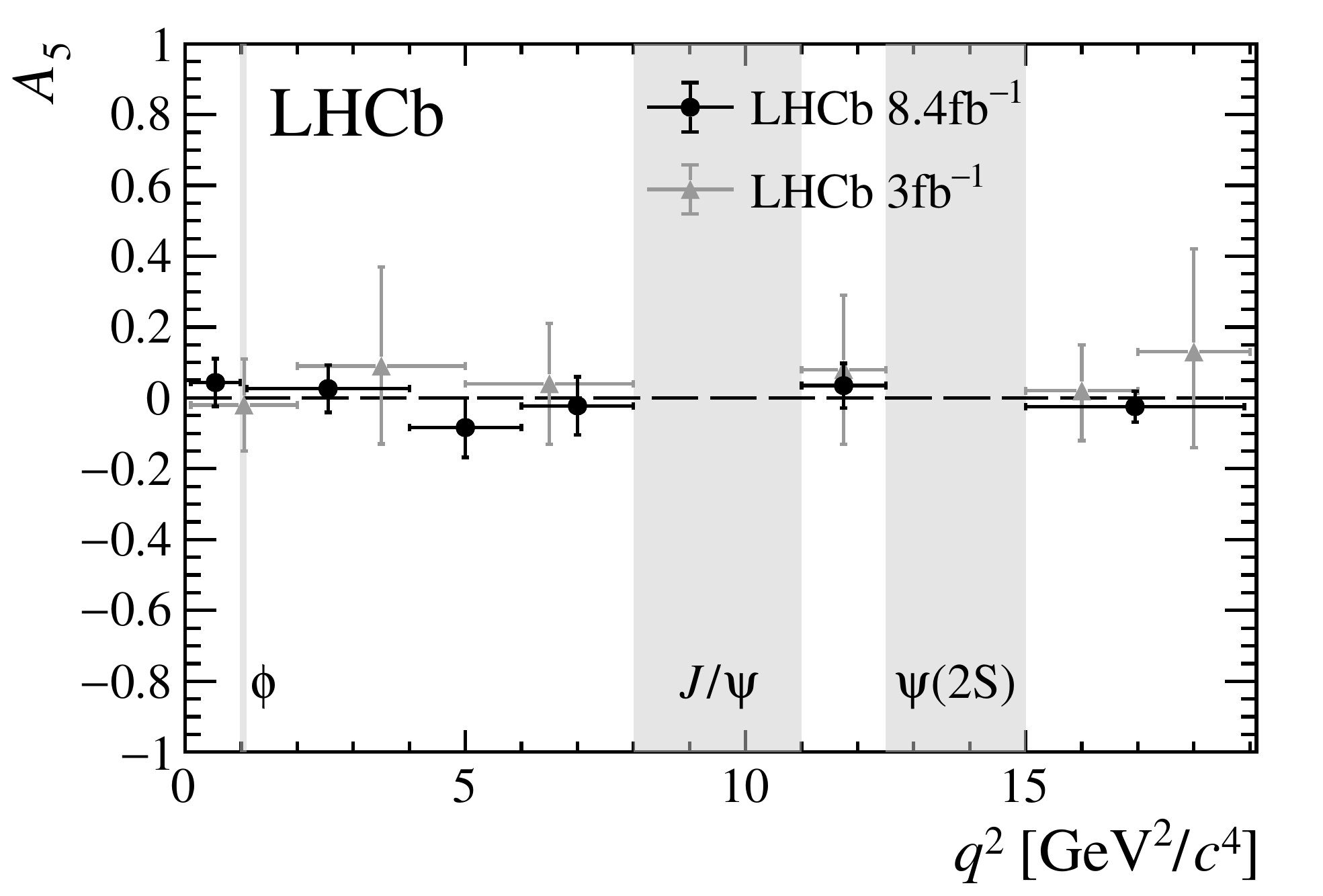}~
    \includegraphics[width=.45\textwidth]{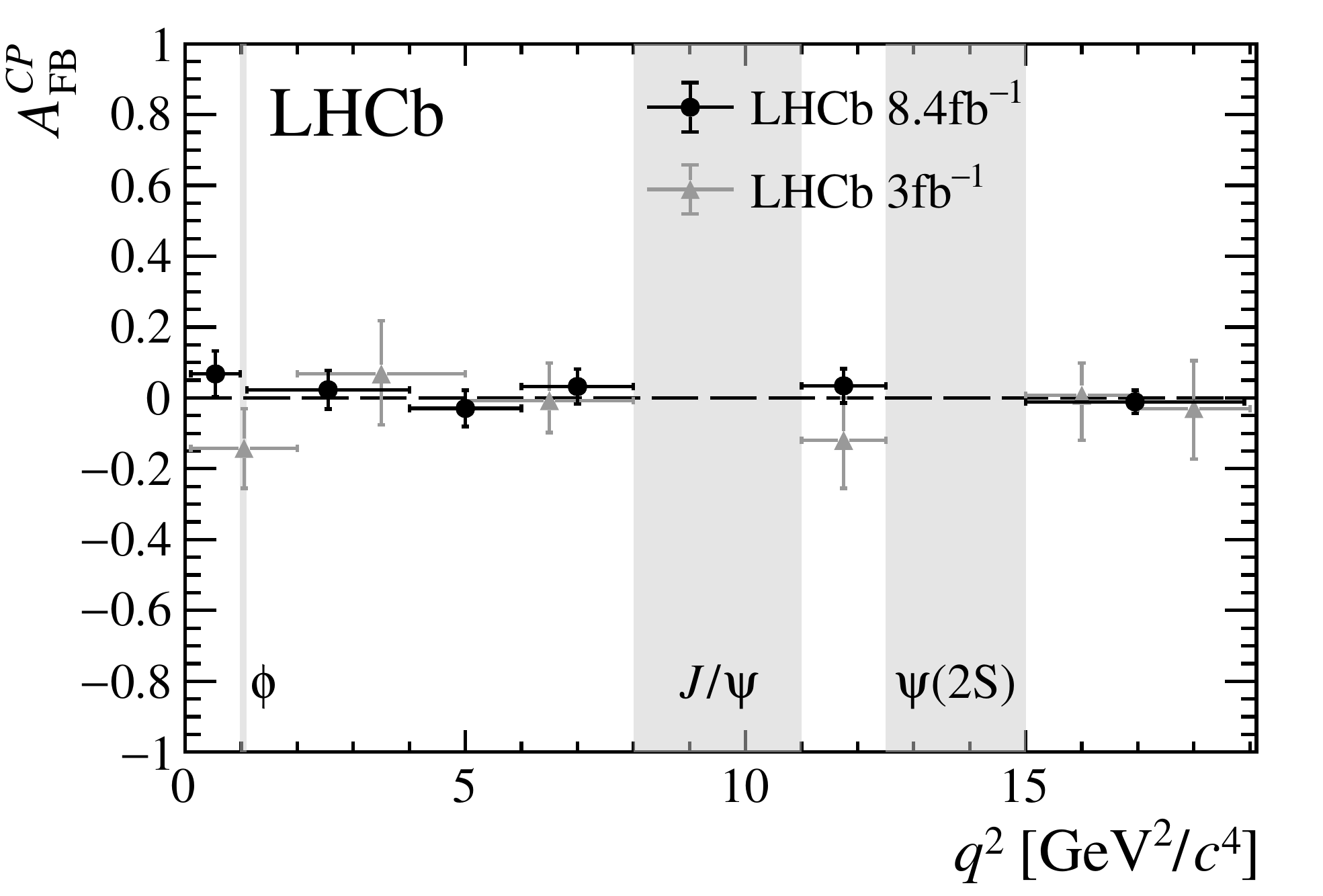}\\
    \includegraphics[width=.45\textwidth]{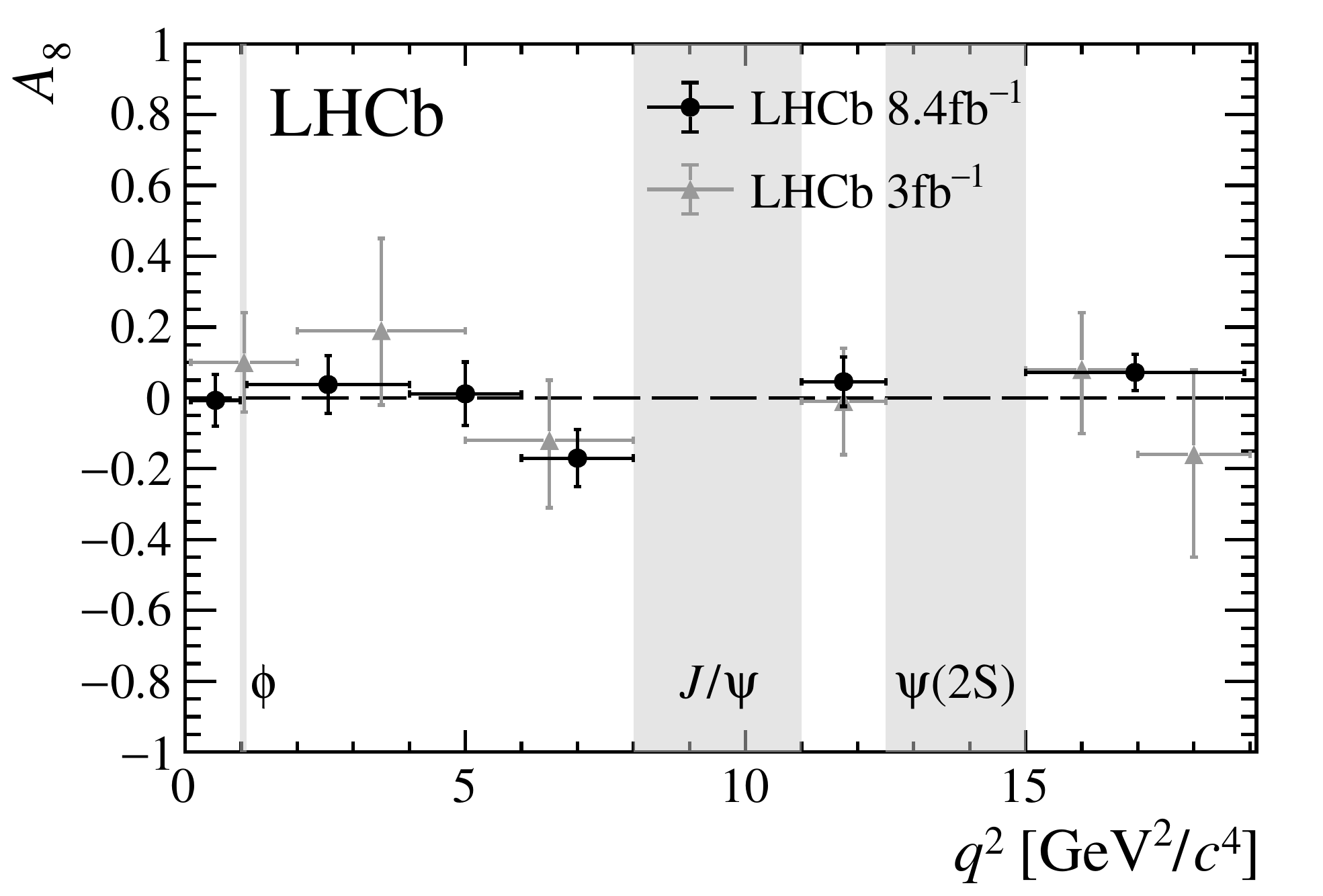}~
    \includegraphics[width=.45\textwidth]{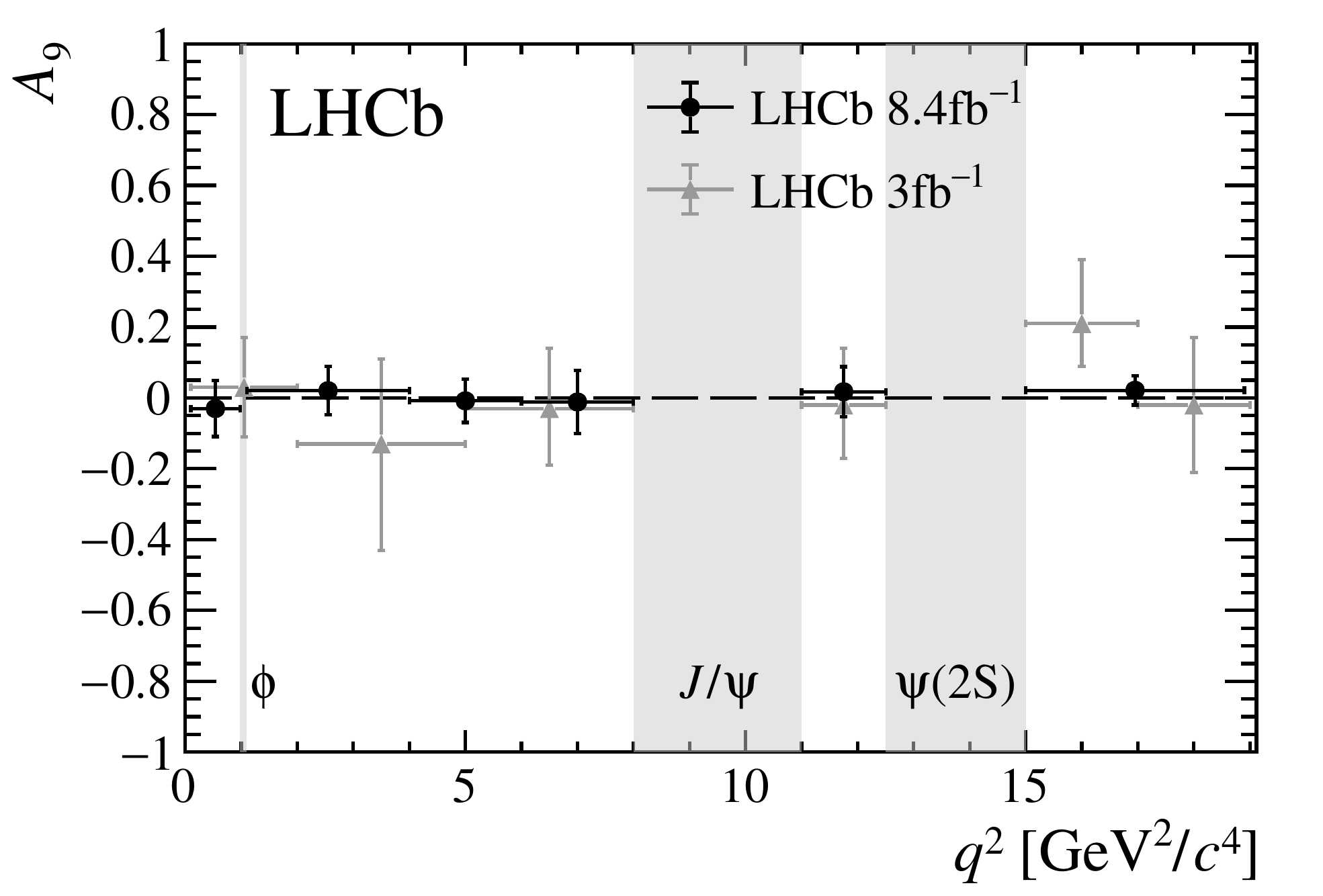}
    \caption{\label{fig:angular_obs} \CP-averaged angular observables \FL and $S_{3,4,7}$ and \CP-asymmetries \AFBCP and $A_{5,8,9}$ shown by black crosses, overlaid with the SM prediction~\cite{Straub:2015ica,Straub:2018kue,Horgan:2013pva,Horgan:2015vla} as blue boxes, where available. The grey crosses indicate the results from Ref.~\cite{LHCb-PAPER-2015-023}. The grey bands indicate the regions of the charmonium resonances and the \decay{\Bs}{\phi\phi} region. }
\end{figure}

The numerical results for the angular observables are given in Table~\ref{tab:results}, including systematic uncertainties as discussed in Sec.~\ref{sec:systematics}.
The linear-correlation matrices for the angular observables are provided in Tables~\ref{tab:correlation1}, \ref{tab:correlation2} and \ref{tab:correlation3} in Appendix~\ref{app:correlations}.
A graphical comparison of the results with the SM predictions~\cite{Straub:2015ica,Straub:2018kue,Horgan:2013pva,Horgan:2015vla} is shown in Fig.~\ref{fig:angular_obs}.
Overall, the results are in good agreement with the SM predictions, with the \CP\ asymmetries compatible with zero as expected in the SM.
For the \CP\ averages, a mild tension in \FL\ is observed at low \qsq, where the data are found to lie below the SM prediction.

To determine the compatibility of the angular observables with the SM,
the \texttt{flavio}~software package~\cite{Straub:2018kue} is used.
The Wilson coefficient representing the real part of the $bs\mu\mu$ vector coupling, ${\cal R}e({\cal C}_9)$, is varied in a fit of the \CP-averaged angular observables \FL, $S_3$, $S_4$ and $S_7$ in the \qsq\ regions $[0.1,0.98]$, $[1.1,4.0]$, $[4.0,6.0]$ and $[15,18.9]\gevgevcccc$. The
$[6.0,8.0]$ and $[11.0,12.5]\gevgevcccc$ regions are excluded from the fit as they are particularly sensitive to long-distance effects from charmonium resonances, which cannot currently be calculated from first principles in the SM~\cite{Altmannshofer:2008dz}. The asymmetries are excluded as they offer little sensitivity to ${\cal R}e({\cal C}_9)$.
The best fit value is given by $\Delta {\cal R}e({\cal C}_9)=-1.3\,^{+0.7}_{-0.6}$ and is preferred over the SM hypothesis ($\Delta {\cal R}e({\cal C}_{9}) = 0$) at the level of $1.9\,\sigma$.

\begin{table}
    \begin{center}
    \caption{\label{tab:results} \CP averages \FL and $S_{3,4,7}$ and \CP asymmetries \AFBCP and $A_{5,8,9}$ obtained from the maximum likelihood fit. The first uncertainty is statistical and the second is the total systematic uncertainty, as described in Sec.~\ref{sec:systematics}.}
    \renewcommand*{\arraystretch}{1.1}
    \resizebox{.99\textwidth}{!}{%
    \begin{tabular}{rrrrr}\hline\noalign{\smallskip}%
        $q^2$ [\gevgevcccc] &\multicolumn{1}{c}{\FL}&\multicolumn{1}{c}{$S_{3}$}&\multicolumn{1}{c}{$S_{4}$}&\multicolumn{1}{c}{$S_{7}$}\\\noalign{\smallskip}\hline\hline
        $[0.1,0.98]$&$0.254\pm0.045\pm0.017$&$-0.004\pm0.068\pm0.014$&$0.213\pm0.082\pm0.005$&$-0.178\pm0.072\pm0.001$\\
        $[1.1,4.0]$&$0.723\pm0.053\pm0.015$&$-0.030\pm0.057\pm0.004$&$-0.110\pm0.079\pm0.002$&$-0.101\pm0.075\pm0.001$\\
        $[4.0,6.0]$&$0.701\pm0.050\pm0.016$&$-0.162\pm0.067\pm0.012$&$-0.222\pm0.092\pm0.010$&$0.175\pm0.089\pm0.003$\\
        $[6.0,8.0]$&$0.624\pm0.051\pm0.012$&$0.013\pm0.080\pm0.009$&$-0.176\pm0.078\pm0.006$&$0.033\pm0.081\pm0.002$\\
        $[11.0,12.5]$&$0.353\pm0.044\pm0.012$&$-0.138\pm0.071\pm0.013$&$-0.319\pm0.061\pm0.008$&$-0.170\pm0.069\pm0.000$\\\hline
        $[1.1,6.0]$&$0.715\pm0.036\pm0.013$&$-0.083\pm0.047\pm0.006$&$-0.155\pm0.058\pm0.004$&$0.020\pm0.059\pm0.001$\\
        $[15.0,18.9]$&$0.359\pm0.031\pm0.019$&$-0.247\pm0.042\pm0.014$&$-0.208\pm0.047\pm0.006$&$0.003\pm0.046\pm0.002$\\
    \hline\end{tabular}%
    }\\[0.5cm]
    \resizebox{\textwidth}{!}{%
    \begin{tabular}{rrrrr}\hline\noalign{\smallskip}
        $q^2$  [\gevgevcccc]&\multicolumn{1}{c}{$A_{5}$}&\multicolumn{1}{c} \AFBCP &\multicolumn{1}{c}{$A_{8}$}&\multicolumn{1}{c}{$A_{9}$}\\\noalign{\smallskip}\hline\hline
        $[0.1,0.98]$&$0.043\pm0.067\pm0.001$&$0.068\pm0.064\pm0.009$&$-0.007\pm0.073\pm0.004$&$-0.030\pm0.079\pm0.001$\\
        $[1.1,4.0]$&$0.026\pm0.067\pm0.002$&$0.023\pm0.054\pm0.001$&$0.038\pm0.082\pm0.001$&$0.020\pm0.068\pm0.001$\\
        $[4.0,6.0]$&$-0.084\pm0.084\pm0.003$&$-0.030\pm0.051\pm0.002$&$0.012\pm0.090\pm0.002$&$-0.008\pm0.061\pm0.001$\\
        $[6.0,8.0]$&$-0.022\pm0.082\pm0.001$&$0.032\pm0.049\pm0.001$&$-0.170\pm0.080\pm0.002$&$-0.012\pm0.090\pm0.001$\\
        $[11.0,12.5]$&$0.035\pm0.063\pm0.001$&$0.034\pm0.048\pm0.001$&$0.046\pm0.070\pm0.002$&$0.017\pm0.071\pm0.001$\\\hline
        $[1.1,6.0]$&$-0.007\pm0.051\pm0.001$&$0.006\pm0.036\pm0.001$&$0.016\pm0.062\pm0.001$&$0.009\pm0.046\pm0.001$\\
        $[15.0,18.9]$&$-0.025\pm0.043\pm0.001$&$-0.011\pm0.033\pm0.001$&$0.072\pm0.051\pm0.002$&$0.021\pm0.042\pm0.001$\\
    \hline\end{tabular}
    }
    \end{center}
\end{table}

\section{Systematic uncertainties}
\label{sec:systematics}

Systematic effects may change the measured angular observables. The size of these effects is determined using high-yield pseudoexperiments, generated using an alternative PDF which encodes the systematic effect under study.
The pseudoexperiments are fitted with both the default and alternative PDFs, and the resulting difference in angular observables is assigned as a systematic uncertainty. Each systematic effect is studied using approximately one hundred million generated events.

The simulated samples used to derive the default acceptance are produced according to the phase space of the three-body \BsToPhimm decay with the lifetime difference between the \Bs and \Bsb system, $\Delta\Gamma_{s}$, set to zero.
A systematic uncertainty is determined by weighting the angular, \qsq and \Bs lifetime distributions of simulated events to an alternative model description, in which the value for $\Delta\Gamma_{s}/\Gamma_{s}$ is taken as 0.17~\cite{PDG2014} and the \BsToPhimm decay is described using a more realistic physics model with form factor calculations taken from  Refs.~\cite{Straub:2015ica,Straub:2018kue,Horgan:2013pva,Horgan:2015vla}.

As the angular observables are measured integrated over the \Bs decay time, neglecting the $\Bs$ decay time dependence of the acceptance induces an additional source of bias.
The size of this effect remains small compared to the statistical uncertainty and is accounted for with a systematic uncertainty.

To assess the systematic uncertainty associated with the description of the angular background distribution, pseudoexperiments are generated using second-order Chebyshev polynomials, the coefficients for which are obtained from a fit to candidates in the upper mass sideband. Similarly, the impact of sharing the angular background parameters across data sets is determined by generating pseudoexperiments using first order Chebyshev polynomials with coefficients derived separately for each data set.

To evaluate the impact of the corrections to the track multiplicity, \Bs \pt spectrum and hardware trigger response in simulated events on the angular observables, the angular acceptance is rederived, each time removing a correction. The largest resulting deviation in a given angular observable is assigned as a systematic uncertainty.
For the particle identification response, the corrections are determined using an alternative model and a new angular acceptance is rederived.

To account for neglected \decay{\Bs}{\Kp\Km\mumu} decays, where the $\Kp\Km$ system is in an S-wave configuration, pseudoexperiments are generated according to the combined P- and S-wave decay rate, where $F_{\rm S}$ is conservatively taken to be 2\%.
The pseudoexperiments are fitted with the default model and the resulting shift in the angular observables is assigned as a systematic uncertainty.
The impact of peaking background contributions is assessed in a similar fashion by injecting additional events drawn from the reconstructed \Bs\ mass and angular distributions of the background in question.

The influence of the choice for the maximum order of the Legendre polynomials used in the acceptance parameterisation is evaluated by rederiving the acceptance using a higher order.
Further sources of systematic uncertainty include the size of the simulated samples used to derive the acceptance, the evaluation of the acceptance at a single point in \qsq for the narrow \qsq regions and the signal mass model, all of which yield negligible contributions to the overall systematic uncertainty.

The systematic uncertainties are summarised in Table~\ref{tab:sourcesyst}.
As the size of a systematic effect can vary strongly depending on the observable and \qsq region in question, only the magnitude of the largest systematic uncertainty across all regions, rounded up to the next multiple of 0.005, is indicated.

\begin{table}
\centering
\caption{\label{tab:sourcesyst}
Sources of the systematic uncertainties associated with the angular observables. As the size of a systematic effect can vary strongly depending on the observable and \qsq region in question, only the magnitude of the largest systematic uncertainty across all regions, rounded up to the next multiple of 0.005, is indicated.
}
    \renewcommand*{\arraystretch}{1.1}
    \begin{tabular}{lccc}\hline\noalign{\smallskip}
        Systematic source & {\centering \FL } & {\centering$S_{3,4,7} $ } & {\centering$A_{5,8,9}, \AFBCP $ }\\
        \noalign{\smallskip}\hline\hline
        Physics model  & $ <0.015 $& $ <0.015 $& $ <0.005 $\\
        Time integration         & $ <0.010 $& $ <0.010 $& $ <0.005 $\\
        Fit bias                 & $ <0.005 $& $ <0.015 $& $ <0.010 $\\
        Angular background model & $ <0.015 $& $ <0.005 $& $ <0.005 $\\
        Simulation corrections   & $ <0.015 $& $ <0.005 $& $ <0.005 $\\
        S-wave and peaking bkg.  & $ <0.010 $& $ <0.010 $& $ <0.005 $\\
        Acceptance order        & $ <0.010 $& $ <0.005 $& $ <0.005 $\\
        Simulation statistics   & $ <0.010 $& $ <0.005 $& $ <0.005 $\\
        Signal mass model       & $ <0.005 $& $ <0.005 $& $ <0.005 $\\
        $q^2$ evaluation point       & $ <0.005 $& $ <0.005 $& $ <0.005 $\\
        \hline
    \end{tabular}
\end{table}

\section{Conclusions}
\label{sec:Conclusion}
This paper presents an angular analysis of the \BsToPhimm decay using $pp$ collisions corresponding to 8.4 \invfb of data recorded by the LHCb experiment during the Run~1 and Run~2 data-taking periods. The angular observables are extracted using an unbinned maximum likelihood fit to the angular distributions of untagged \BsToPhimm decays in regions of the square of the dimuon mass, \qsq. The results in this paper constitute the most precise measurement of the \BsToPhimm angular observables to date, with an approximate two-fold increase in sensitivity compared to the results of Ref.~\cite{LHCb-PAPER-2015-023}, which are superseded by this paper. The results are found to be compatible with SM predictions.

\input{acknowledgements}

\addcontentsline{toc}{section}{References}
\bibliographystyle{LHCb}
\bibliography{main,standard,LHCb-PAPER,LHCb-CONF,LHCb-DP,LHCb-TDR}

\clearpage

\section*{Appendices}
\appendix

\section{Fit projections for the rare decay \texorpdfstring{\BsToPhimm}{BsToPhimm}}
\label{sec:fit-proj}

\begin{figure}[hb]
    \centering
    \includegraphics[width=.45\textwidth]{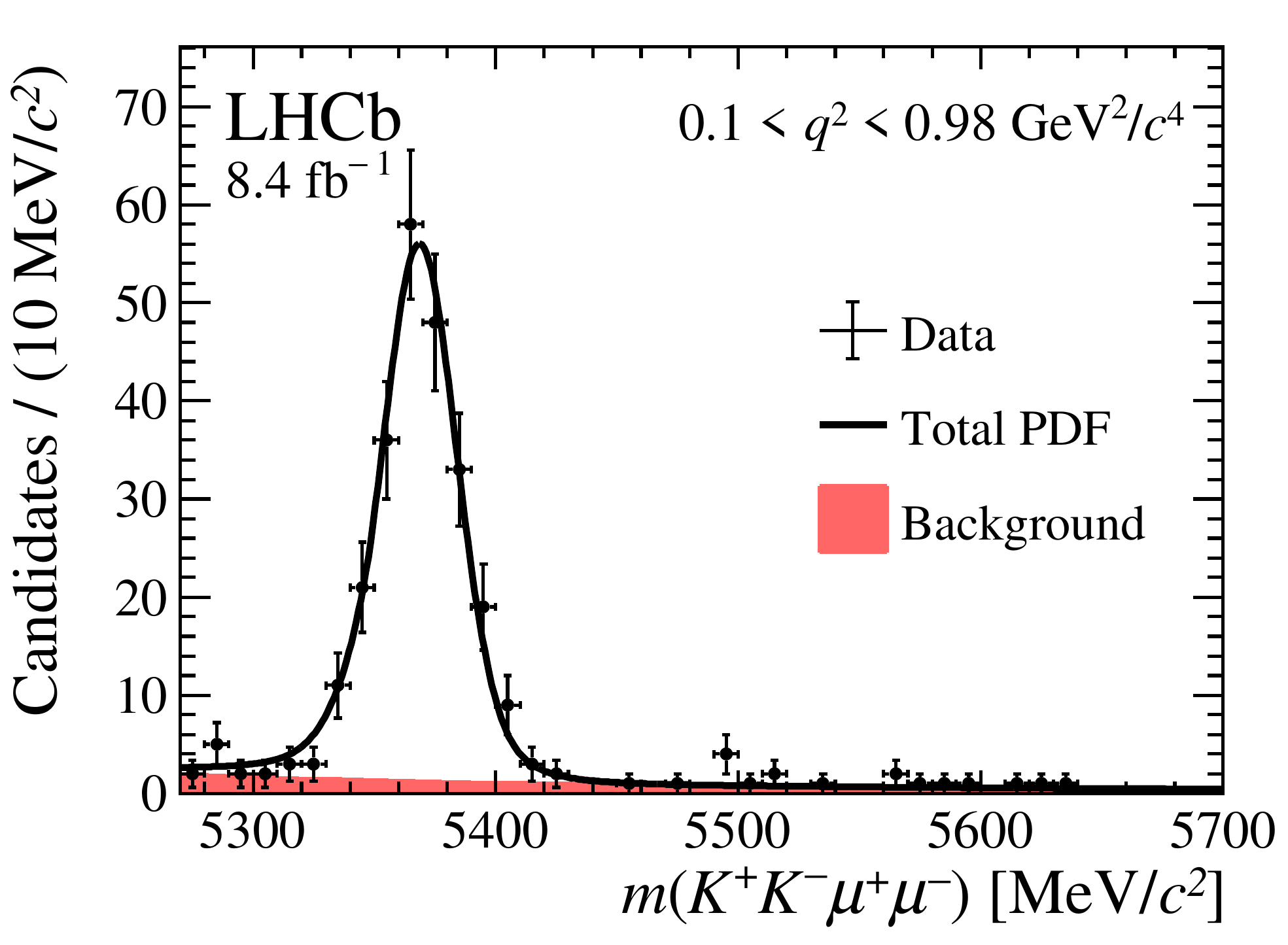}~
    \includegraphics[width=.45\textwidth]{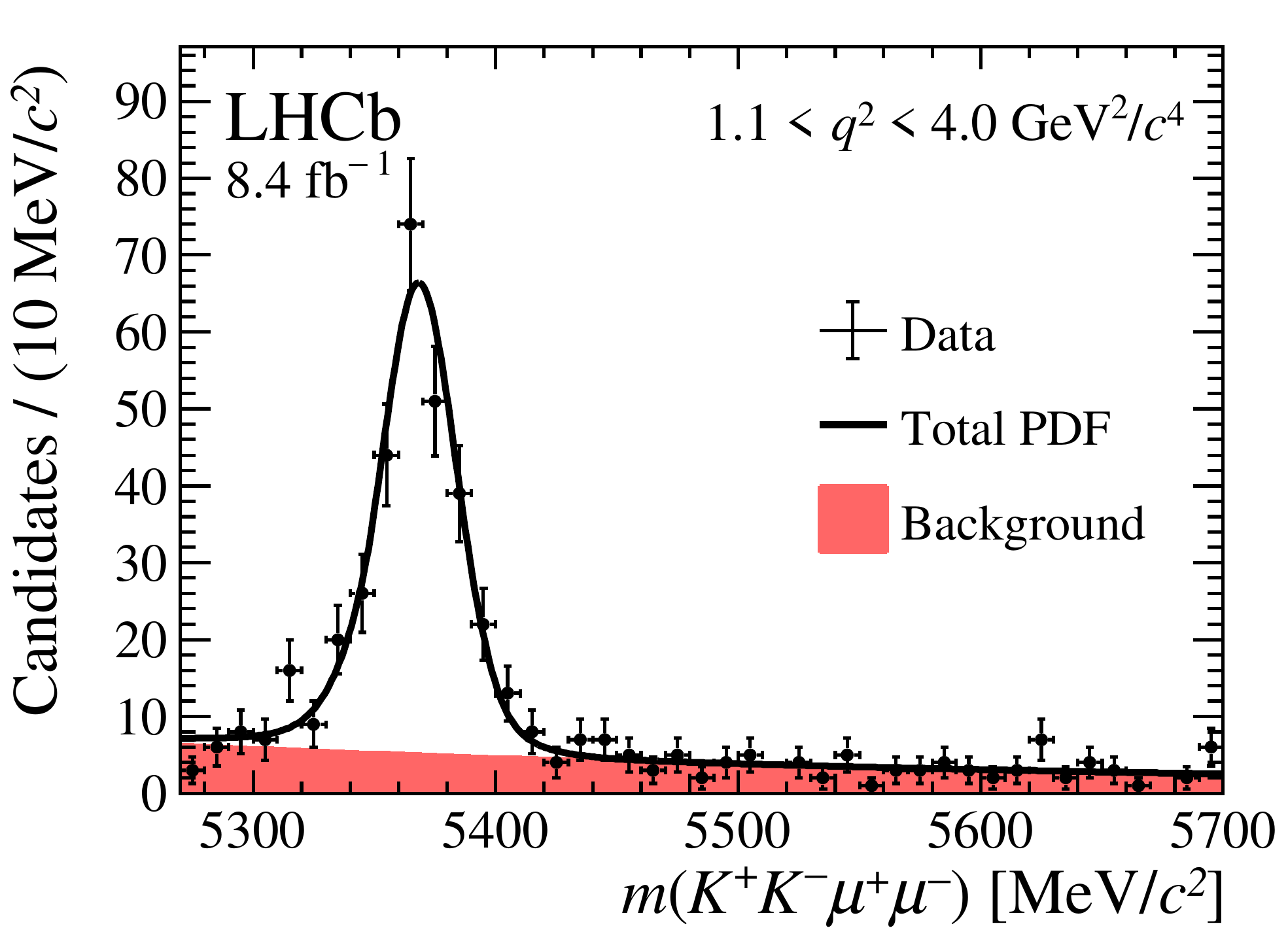}\\
    \includegraphics[width=.45\textwidth]{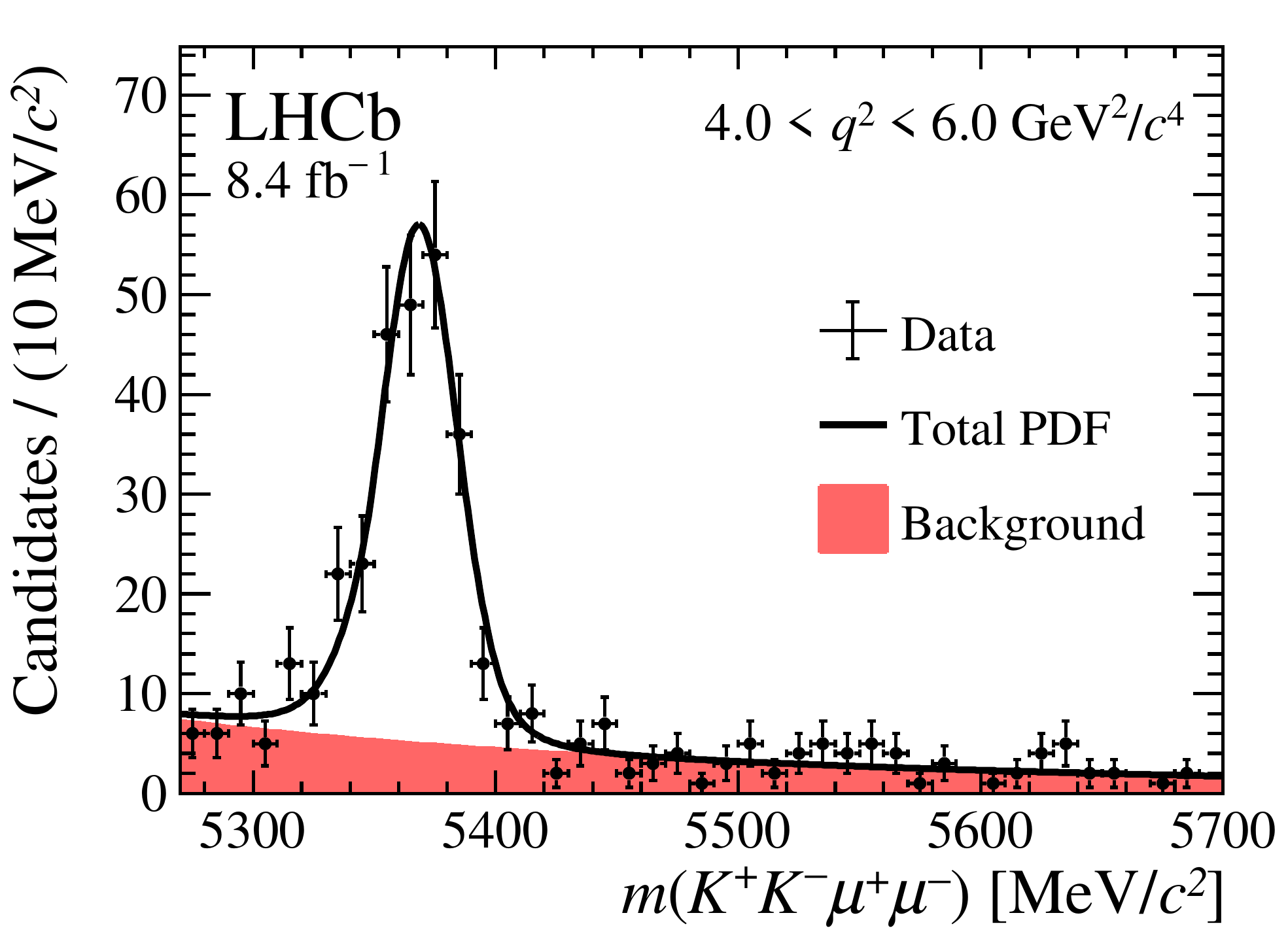}~
    \includegraphics[width=.45\textwidth]{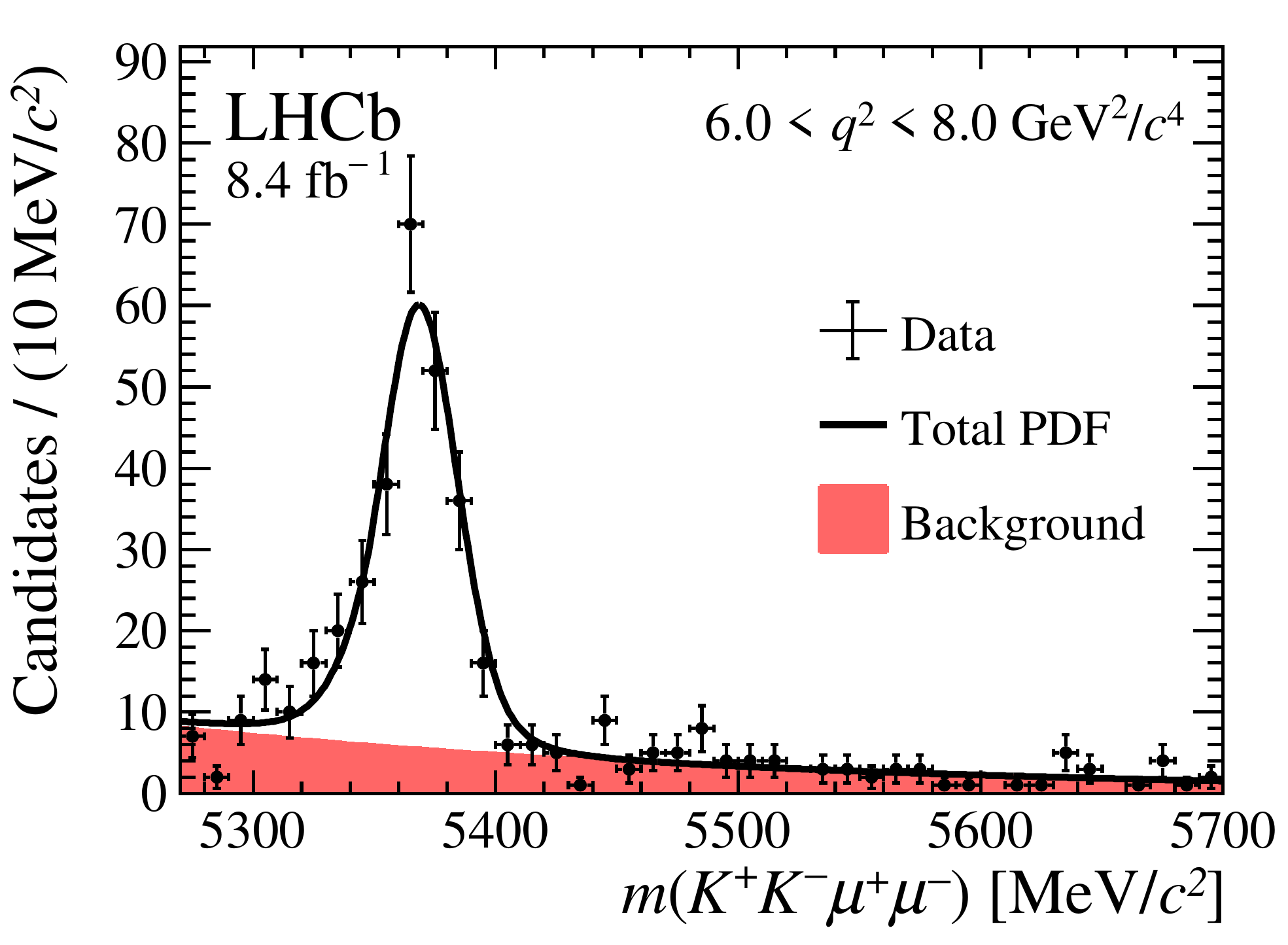}\\
    \includegraphics[width=.45\textwidth]{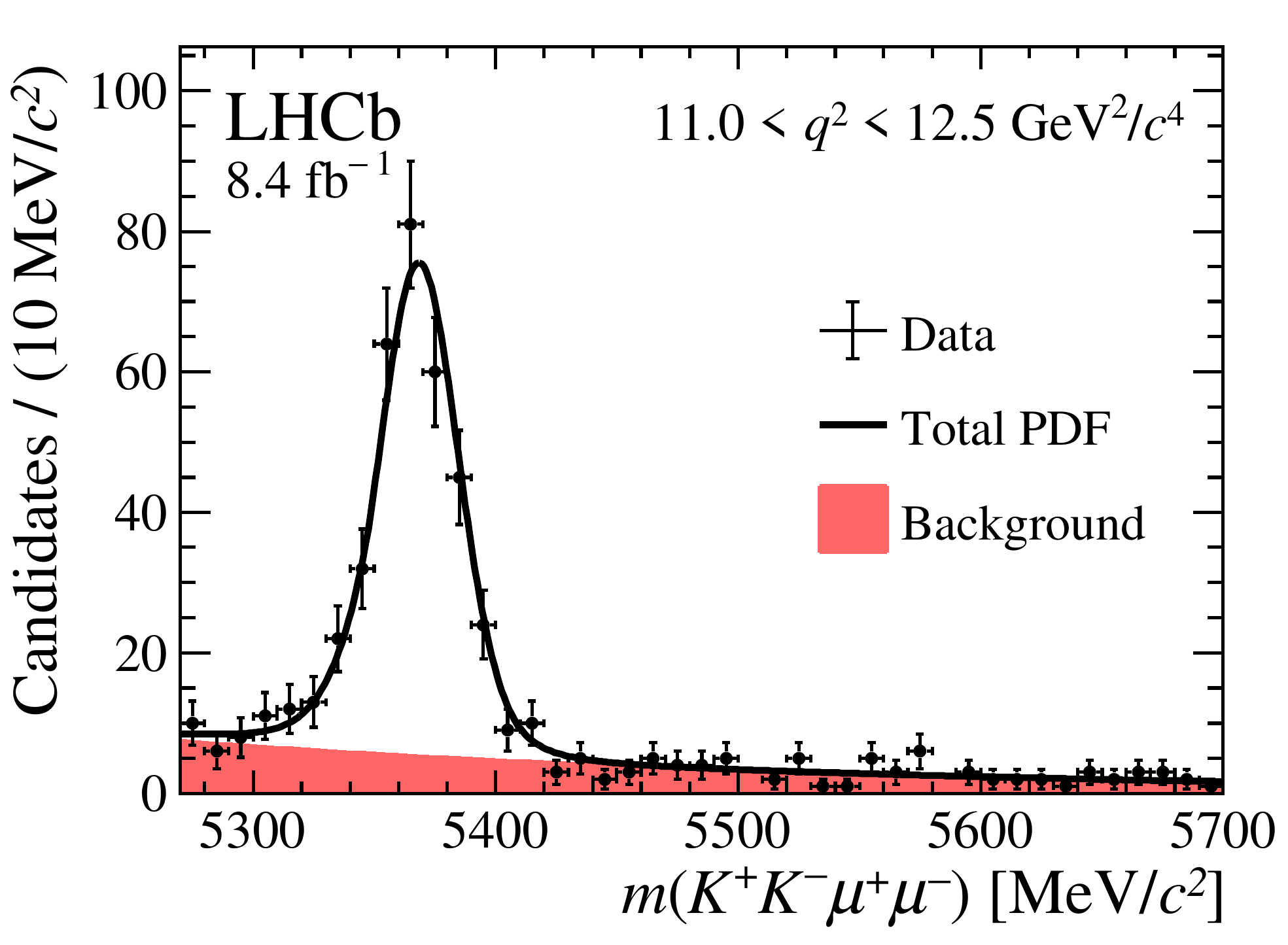}
    \includegraphics[width=.45\textwidth]{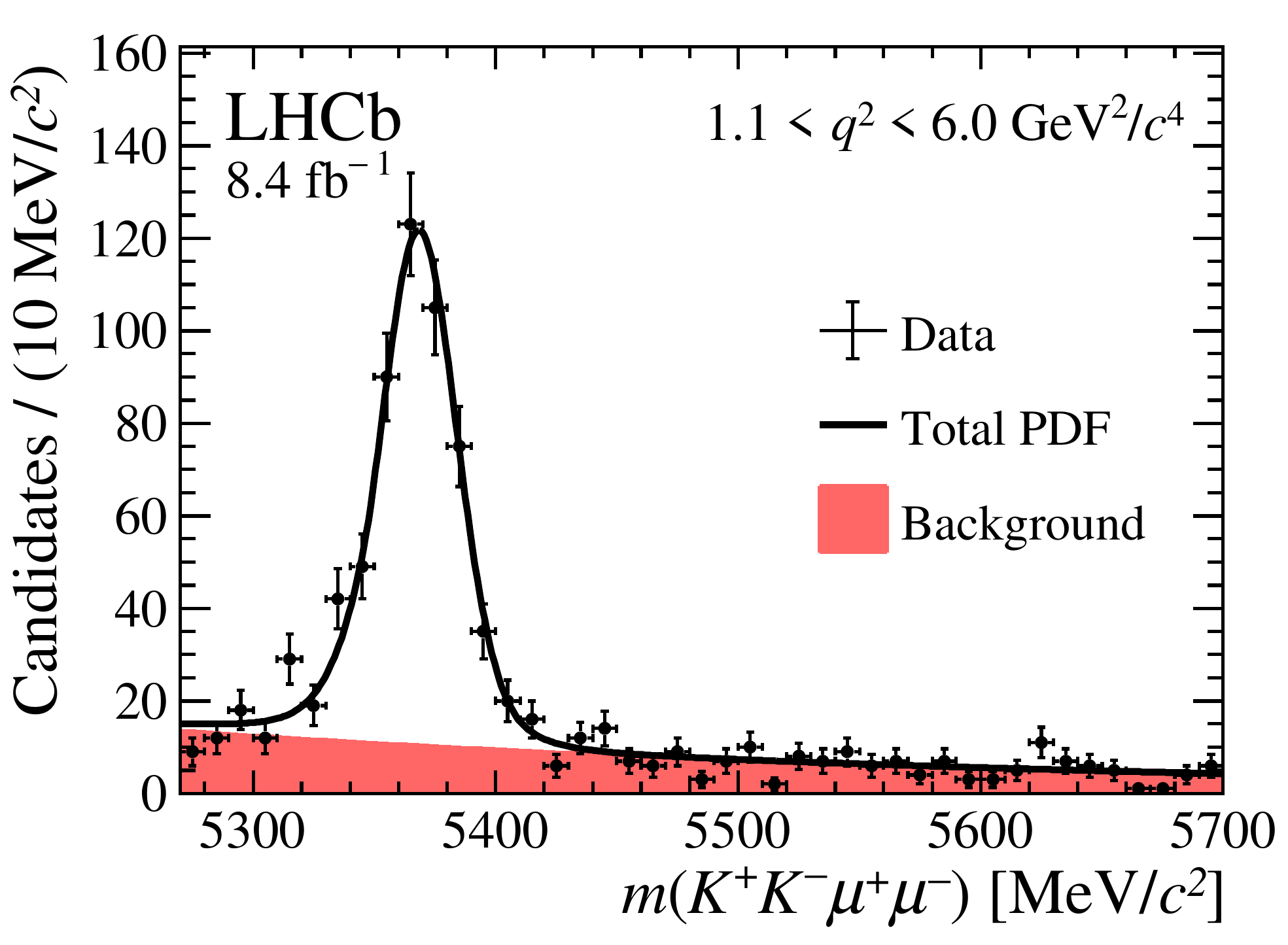}\\
    \includegraphics[width=.45\textwidth]{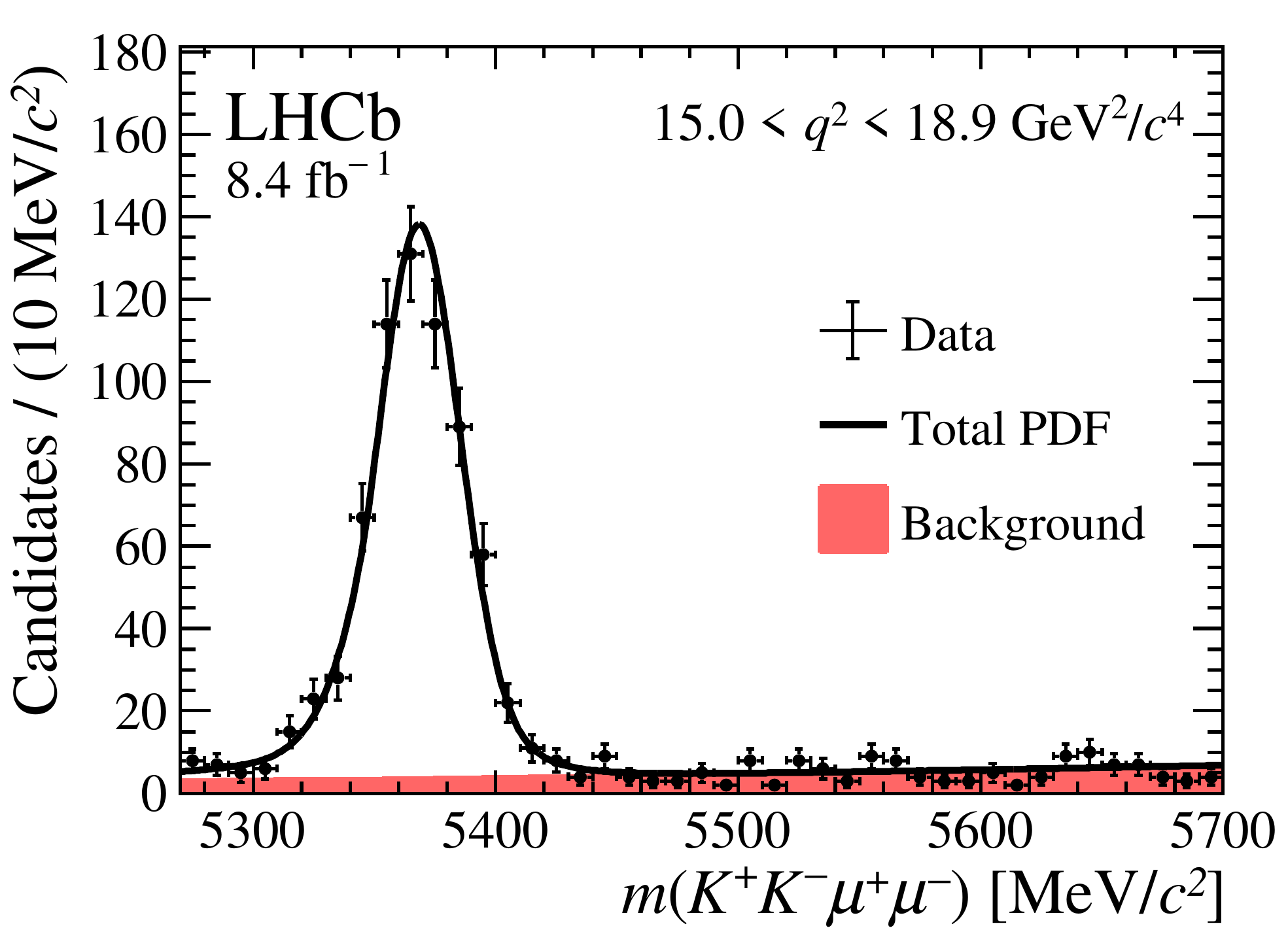}\\
    \caption{\label{fig:results_mass_comb} Mass distributions of \BsToPhimm\ candidates in the different \qsq\ regions for the combined 2011--2012, 2016 and 2017--2018 data sets. The data are overlaid with the projection of the combined PDF. The red shaded area indicates the background component and the solid black line the total PDF. }
\end{figure}

\begin{figure}[hb]
    \centering
    \includegraphics[width=.45\textwidth]{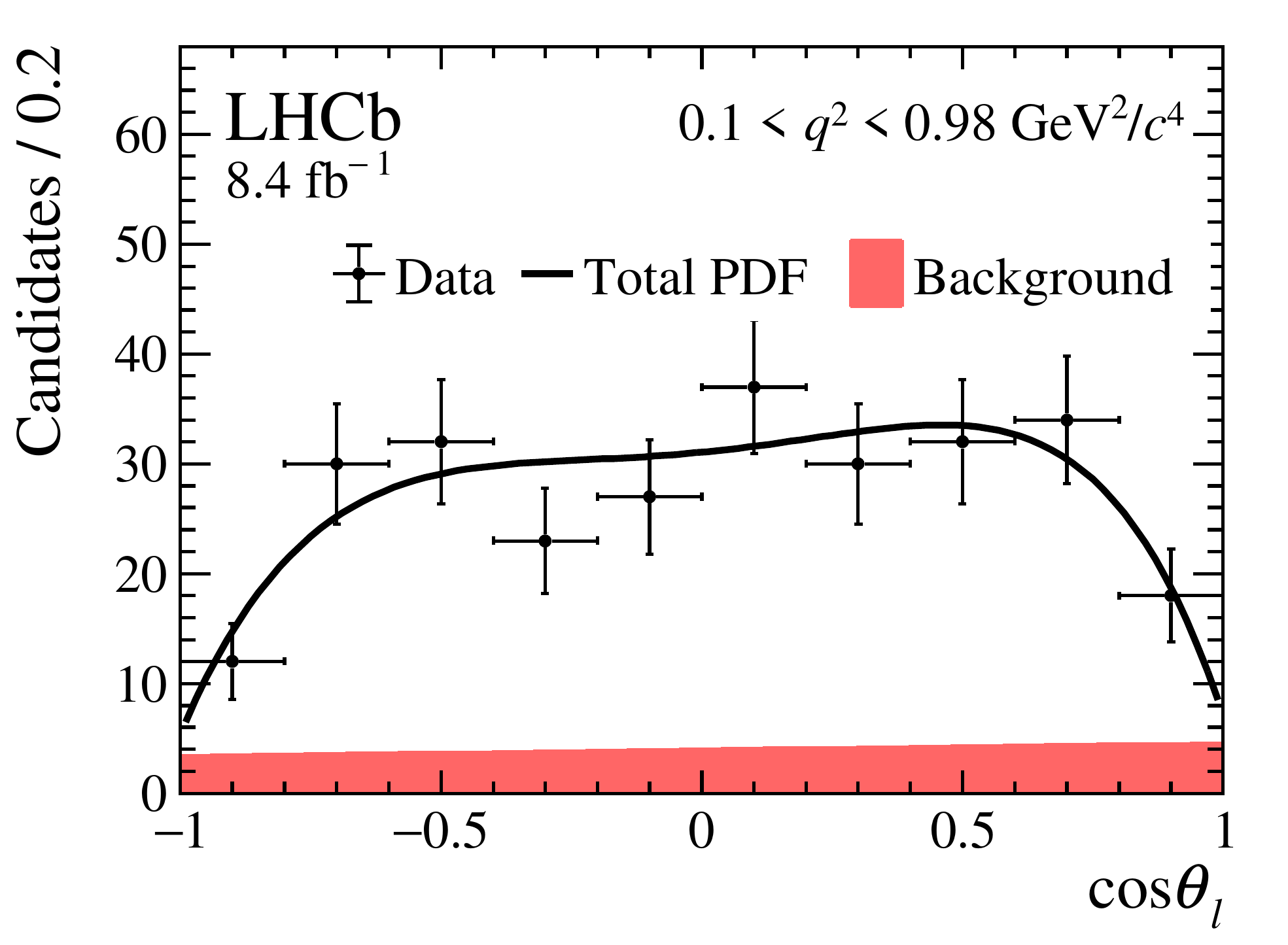}
    \includegraphics[width=.45\textwidth]{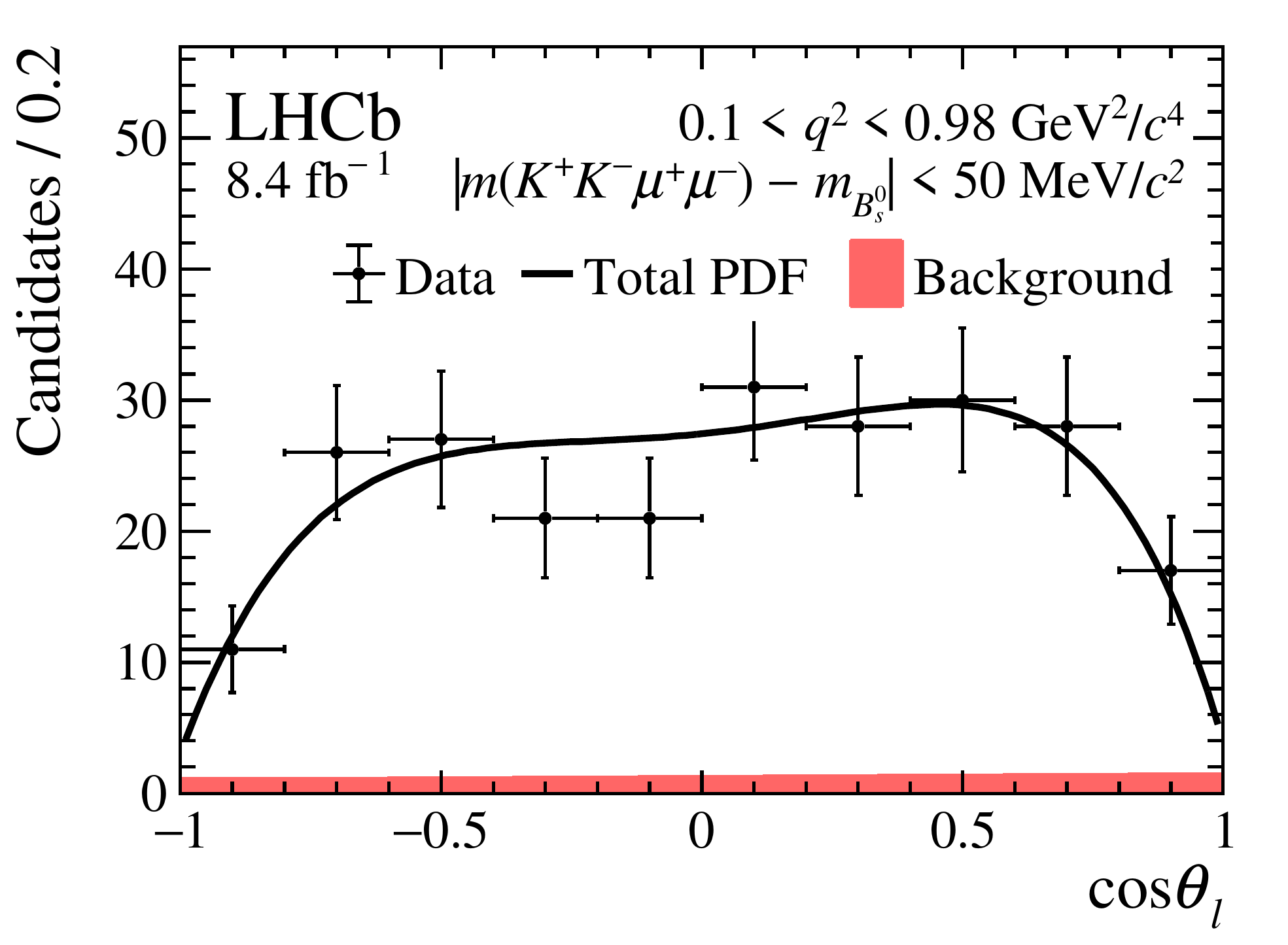}\\
    \includegraphics[width=.45\textwidth]{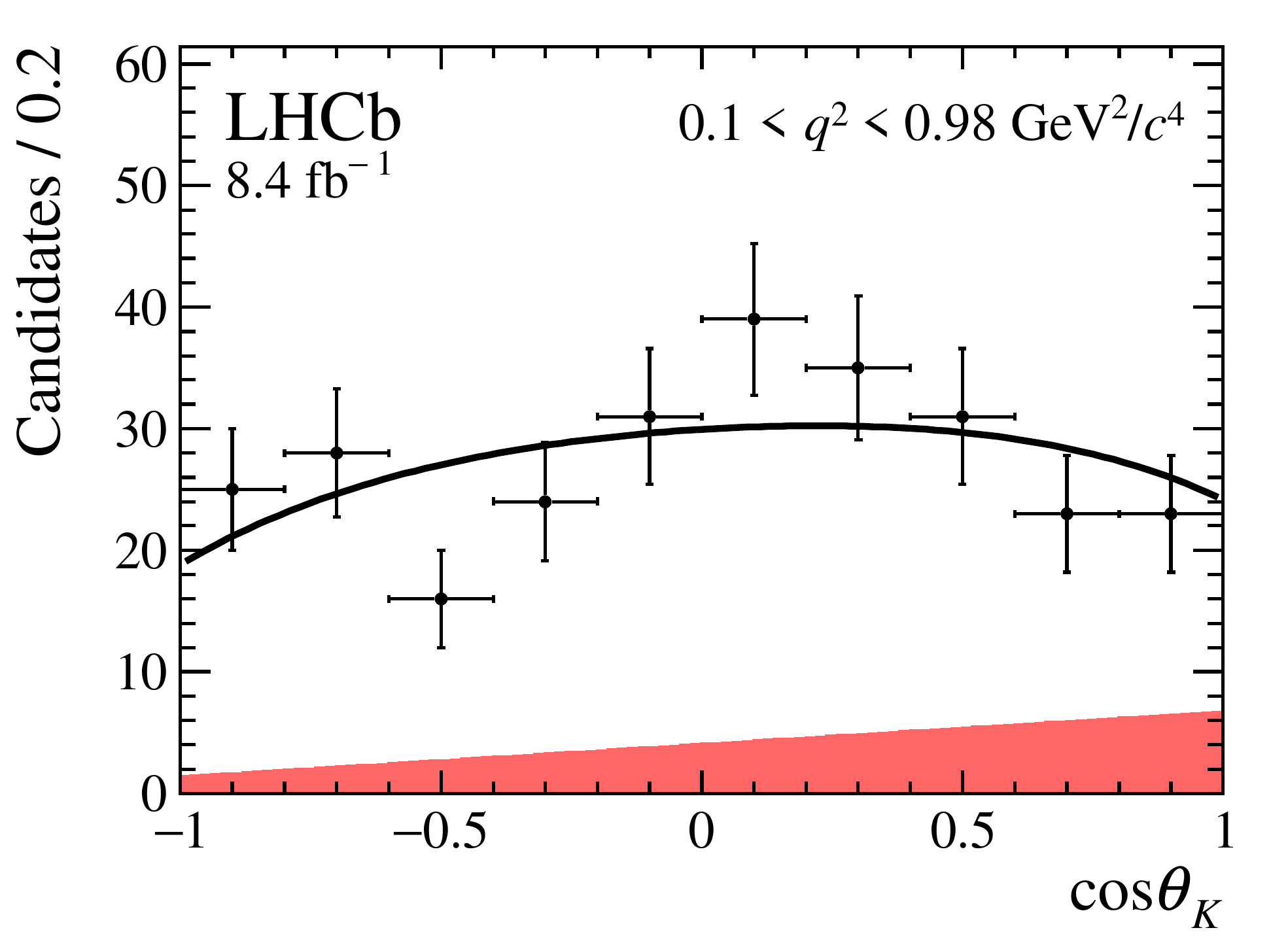}
    \includegraphics[width=.45\textwidth]{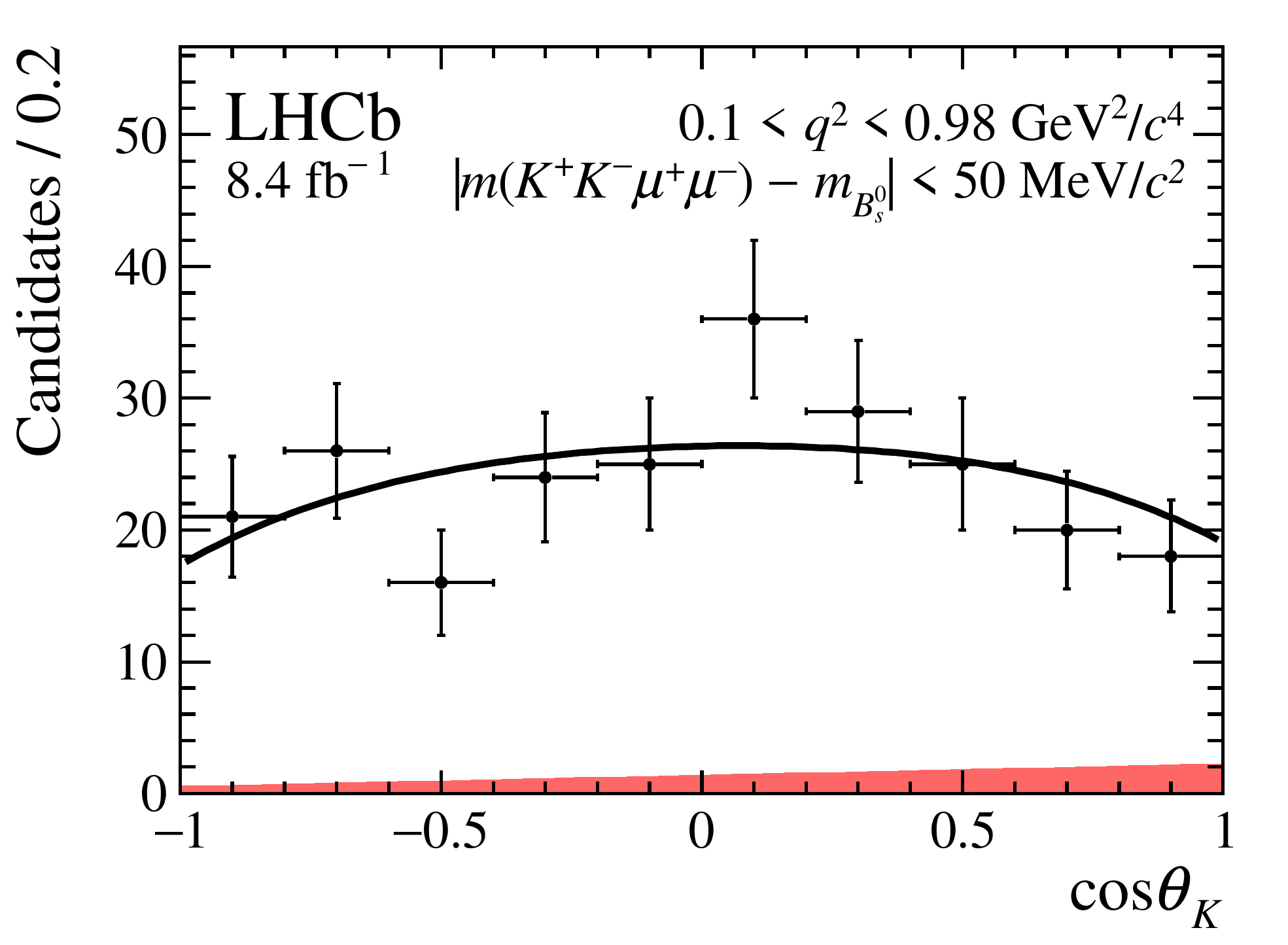}\\
    \includegraphics[width=.45\textwidth]{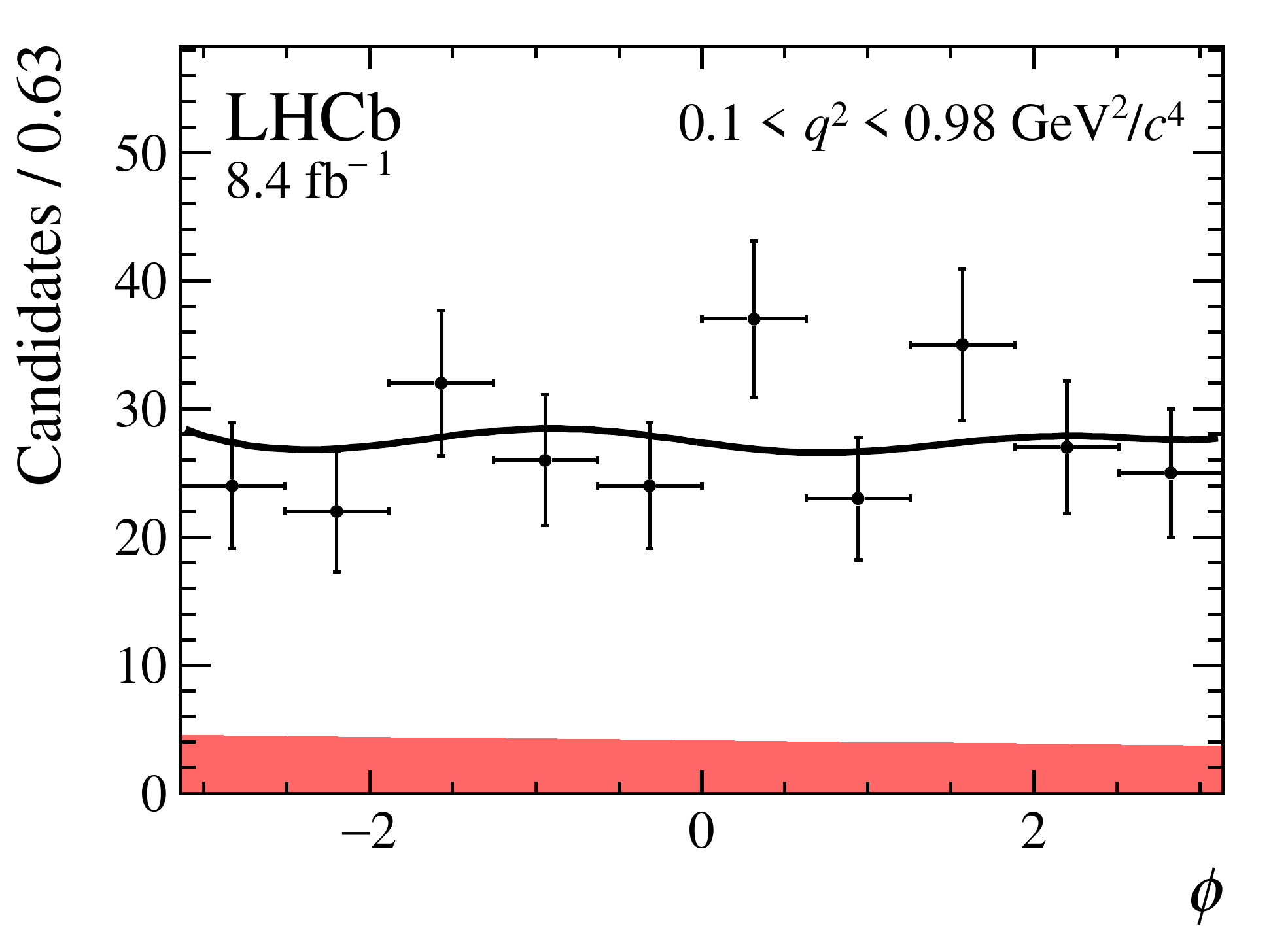}
    \includegraphics[width=.45\textwidth]{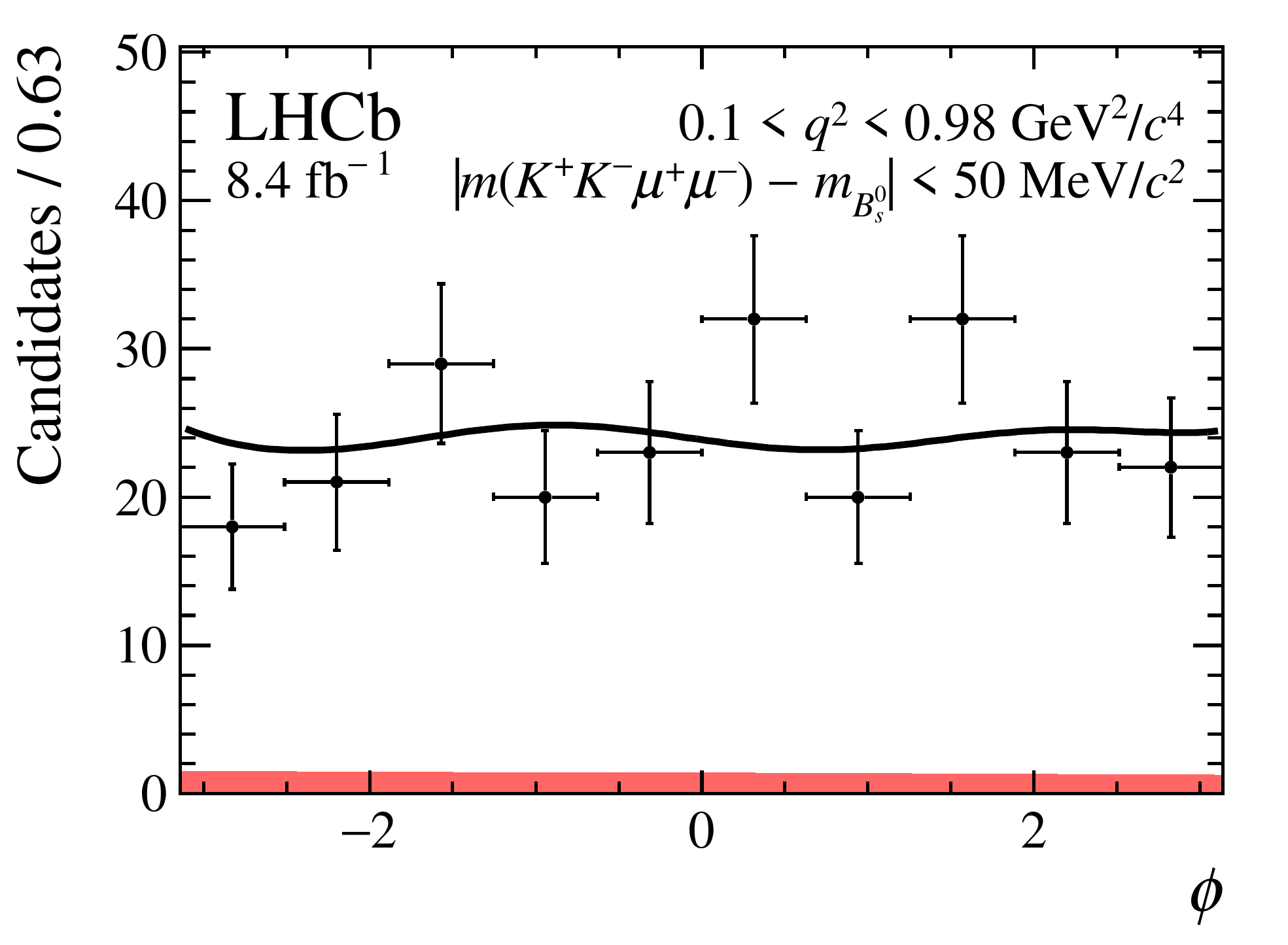}
    \caption{\label{fig:results_bin1_comb} Projections in the region \mbox{$0.1 < \qsq <0.98\gevgevcccc$} for the angular distributions of the combined 2011--2012, 2016 and 2017--2018 data sets. The data are overlaid with the projection of the combined PDF. The red shaded area indicates the background component and the solid black line the total PDF.
    The angular projections are given for candidates in (left) the entire mass region used to determine the observables in this paper and (right) the signal mass window $\pm50\mevcc$ around the known \Bs mass.}
\end{figure}

\begin{figure}[hb]
    \centering
    \includegraphics[width=.45\textwidth]{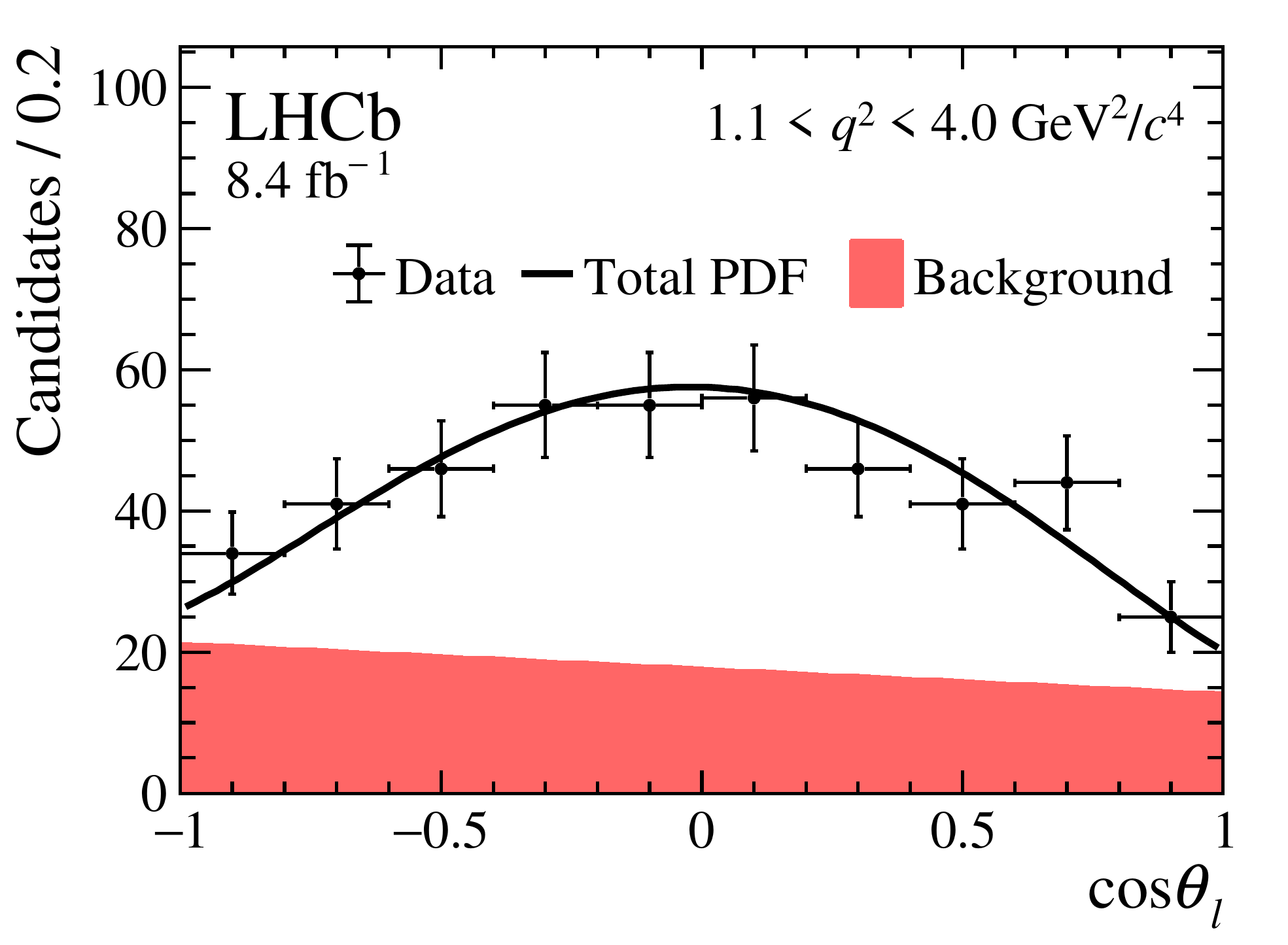}
    \includegraphics[width=.45\textwidth]{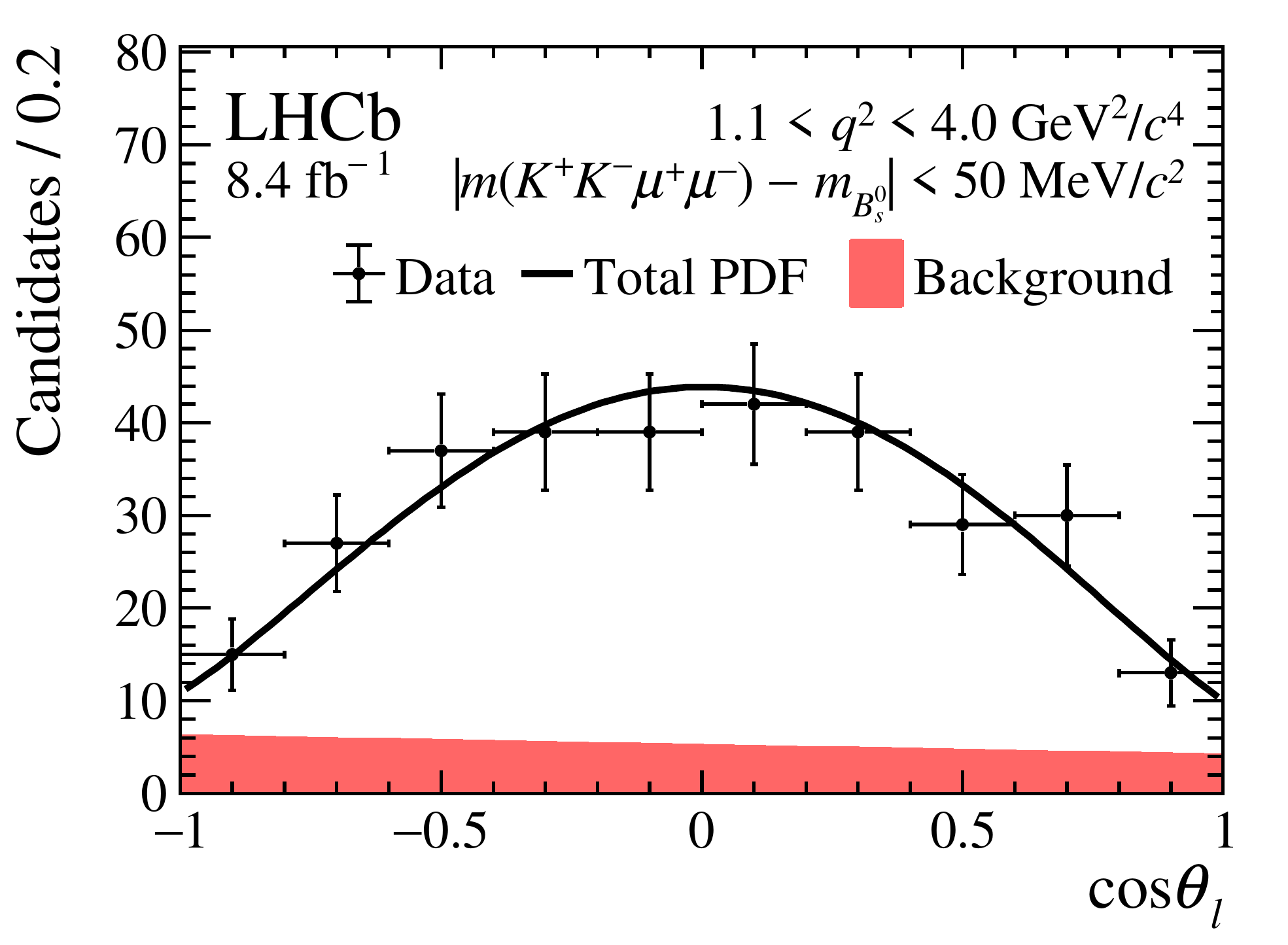}\\
    \includegraphics[width=.45\textwidth]{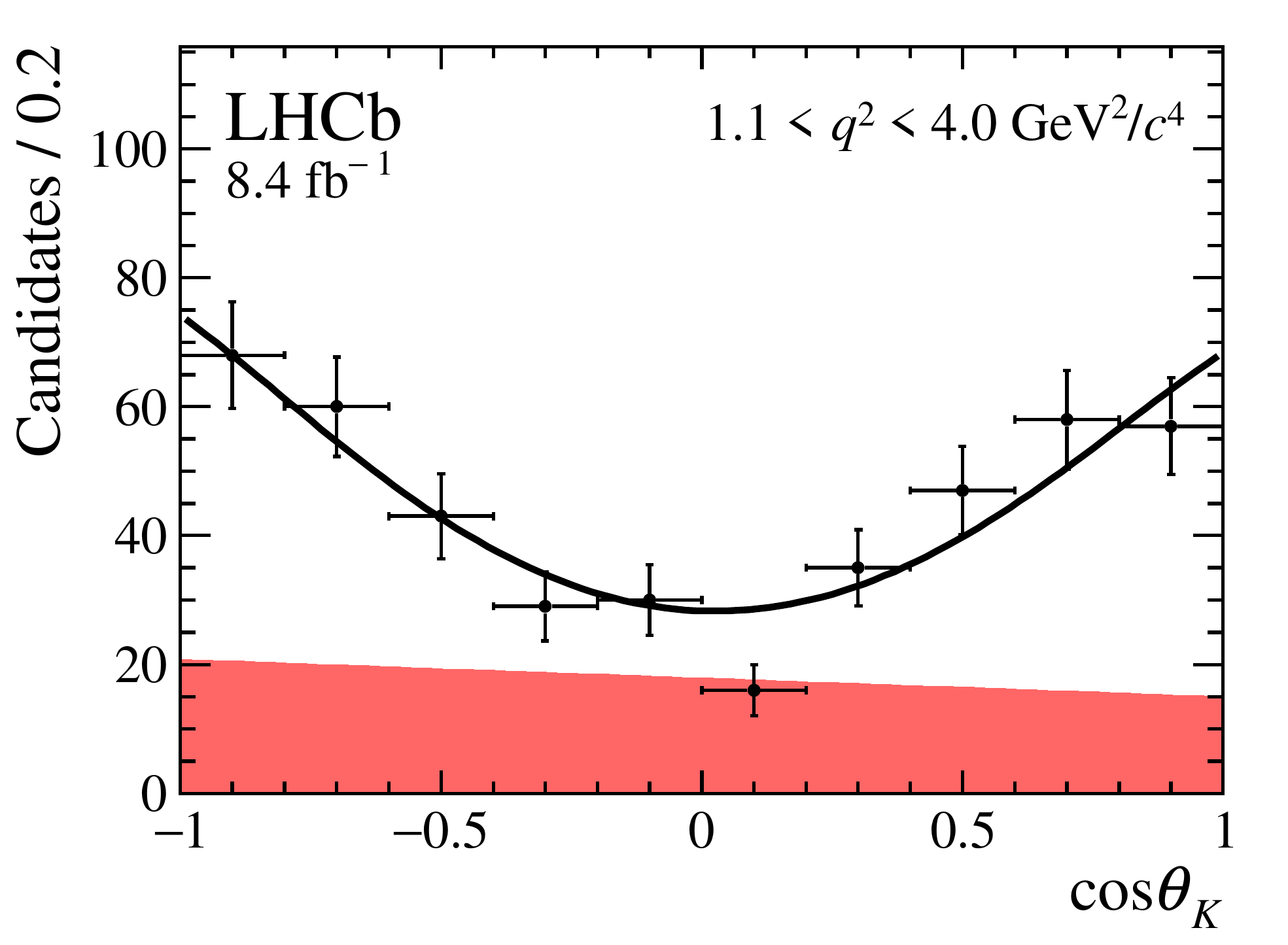}
    \includegraphics[width=.45\textwidth]{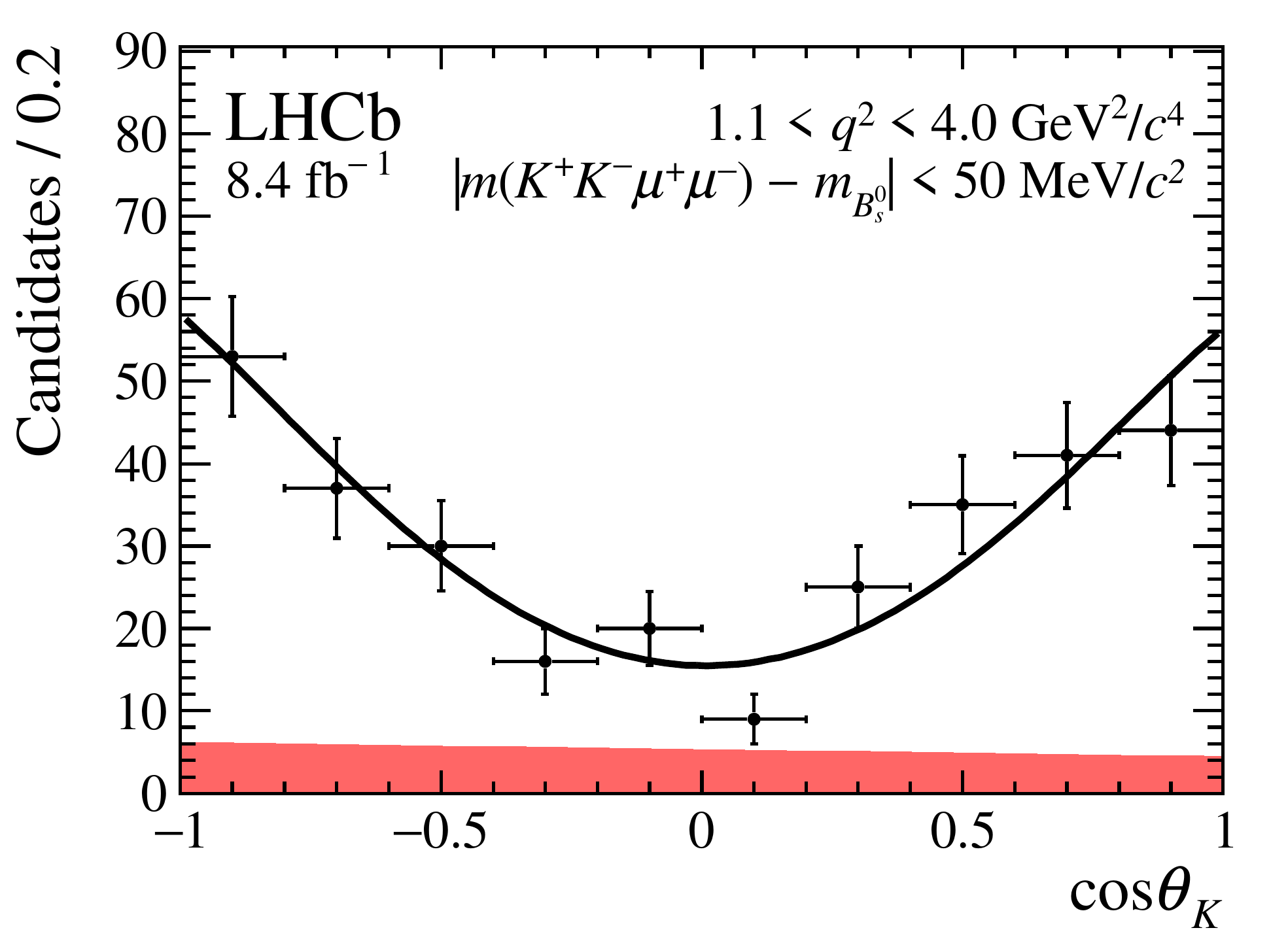}\\
    \includegraphics[width=.45\textwidth]{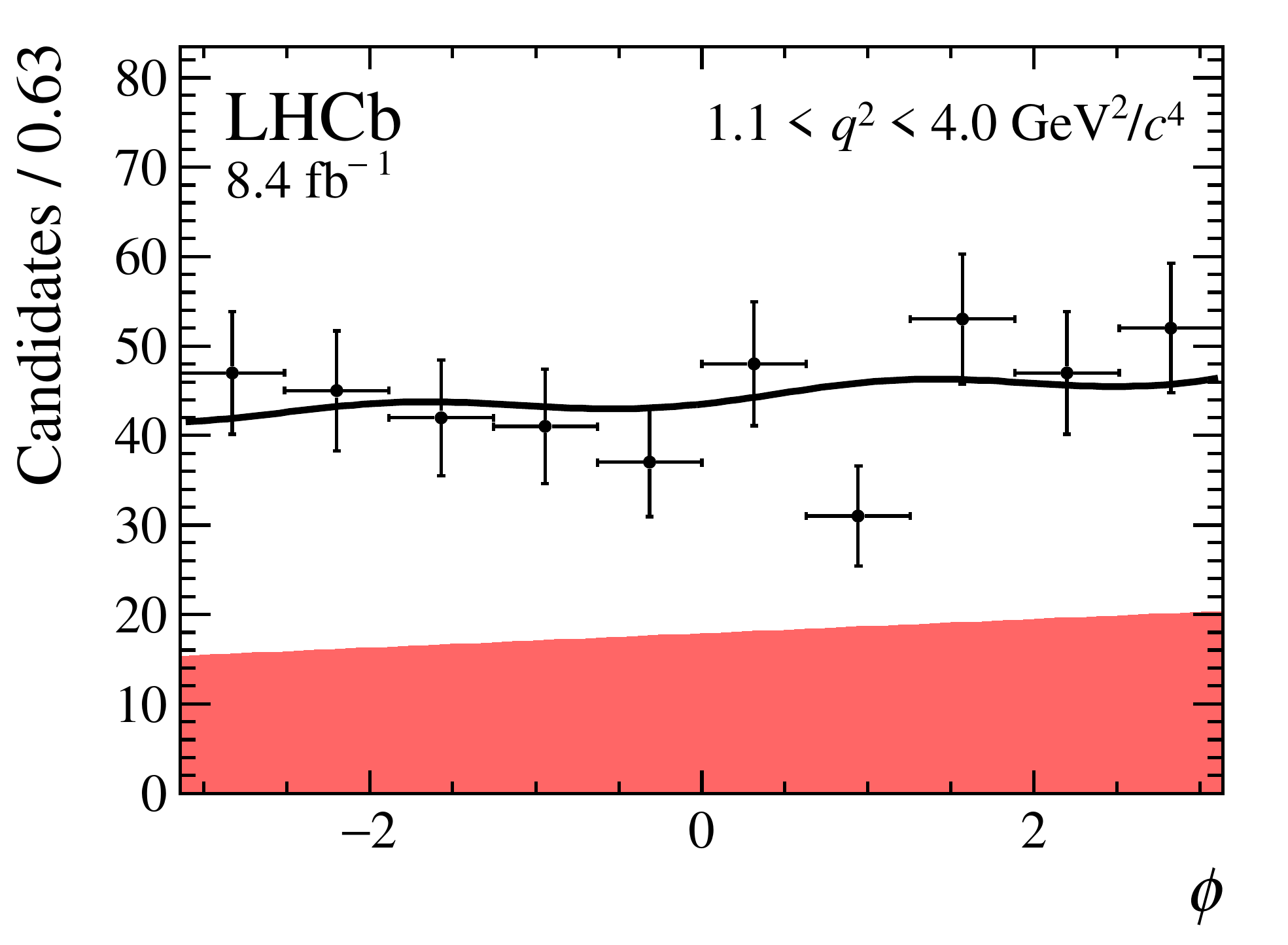}
    \includegraphics[width=.45\textwidth]{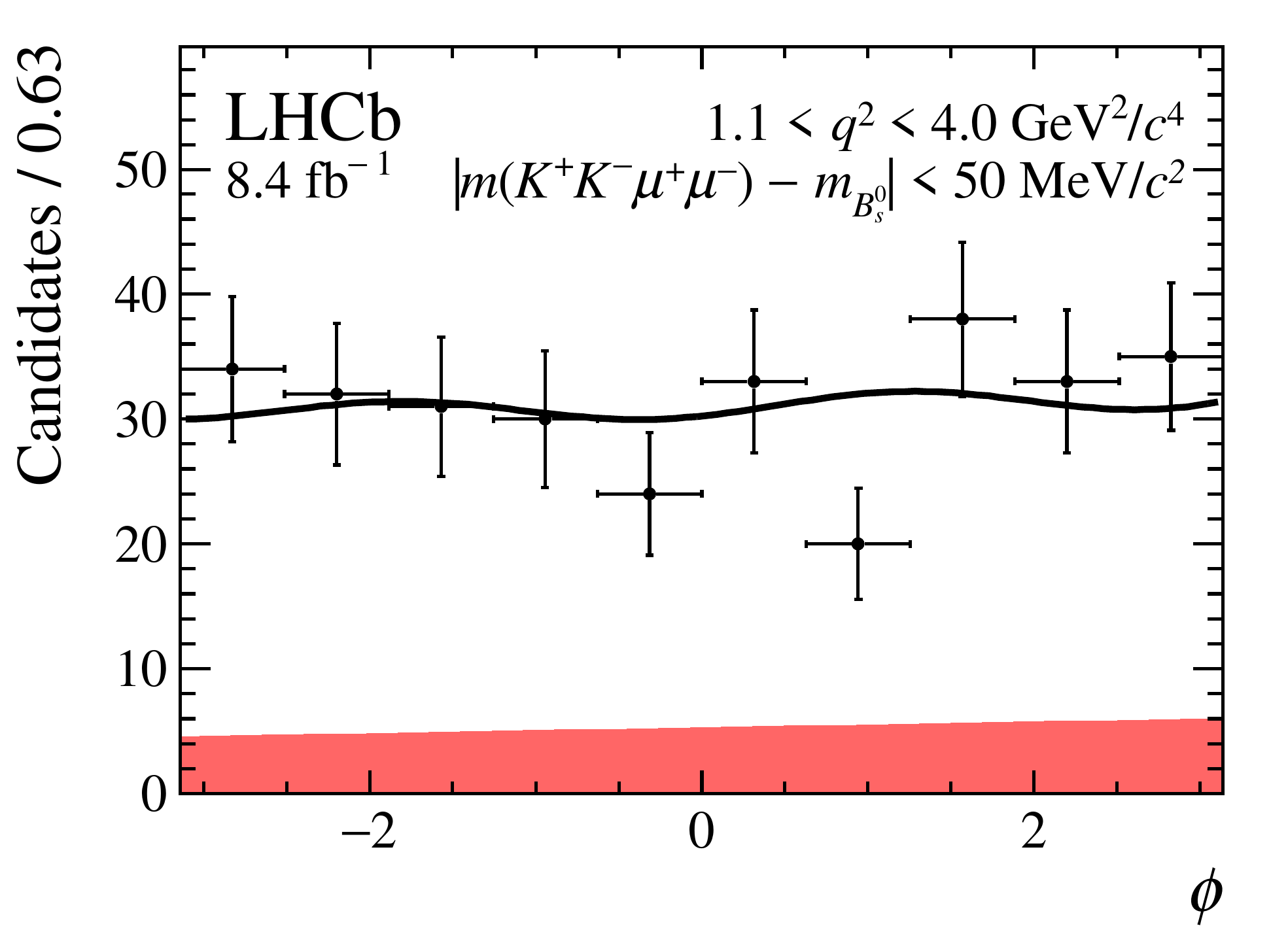}
    \caption{\label{fig:results_bin2_comb} Projections in the region \mbox{$1.1 < \qsq <4.0\gevgevcccc$} for the angular distributions of the combined 2011--2012, 2016 and 2017--2018 data set. The data are overlaid with the projection of the combined PDF. The red shaded area indicates the background component and the solid black line the total PDF.
    The angular projections are given for candidates in (left) the entire mass region used to determine the observables in this paper and (right) the signal mass window $\pm50\mevcc$ around the known \Bs mass.}
\end{figure}

\begin{figure}[hb]
    \centering
    \includegraphics[width=.45\textwidth]{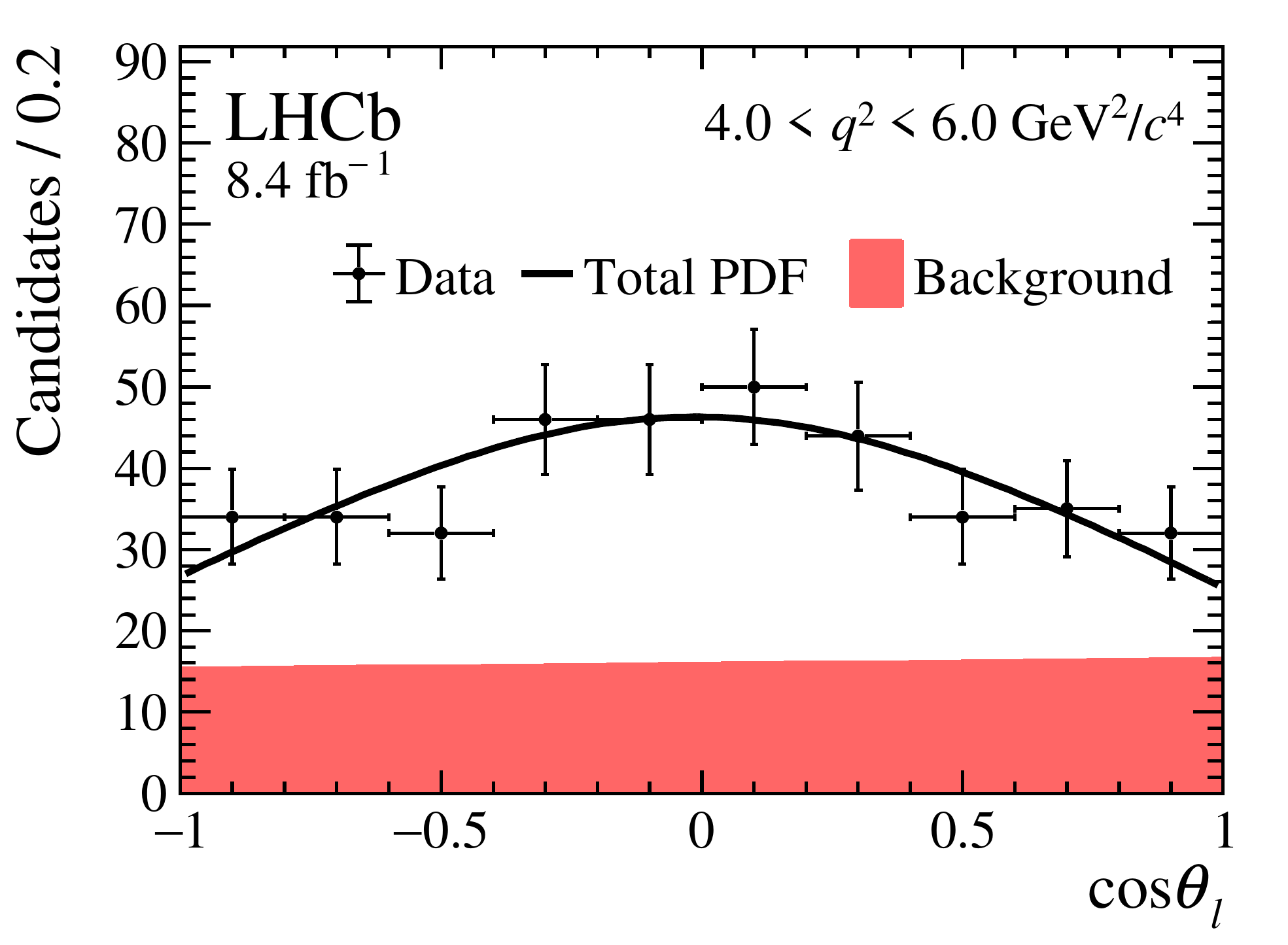}
    \includegraphics[width=.45\textwidth]{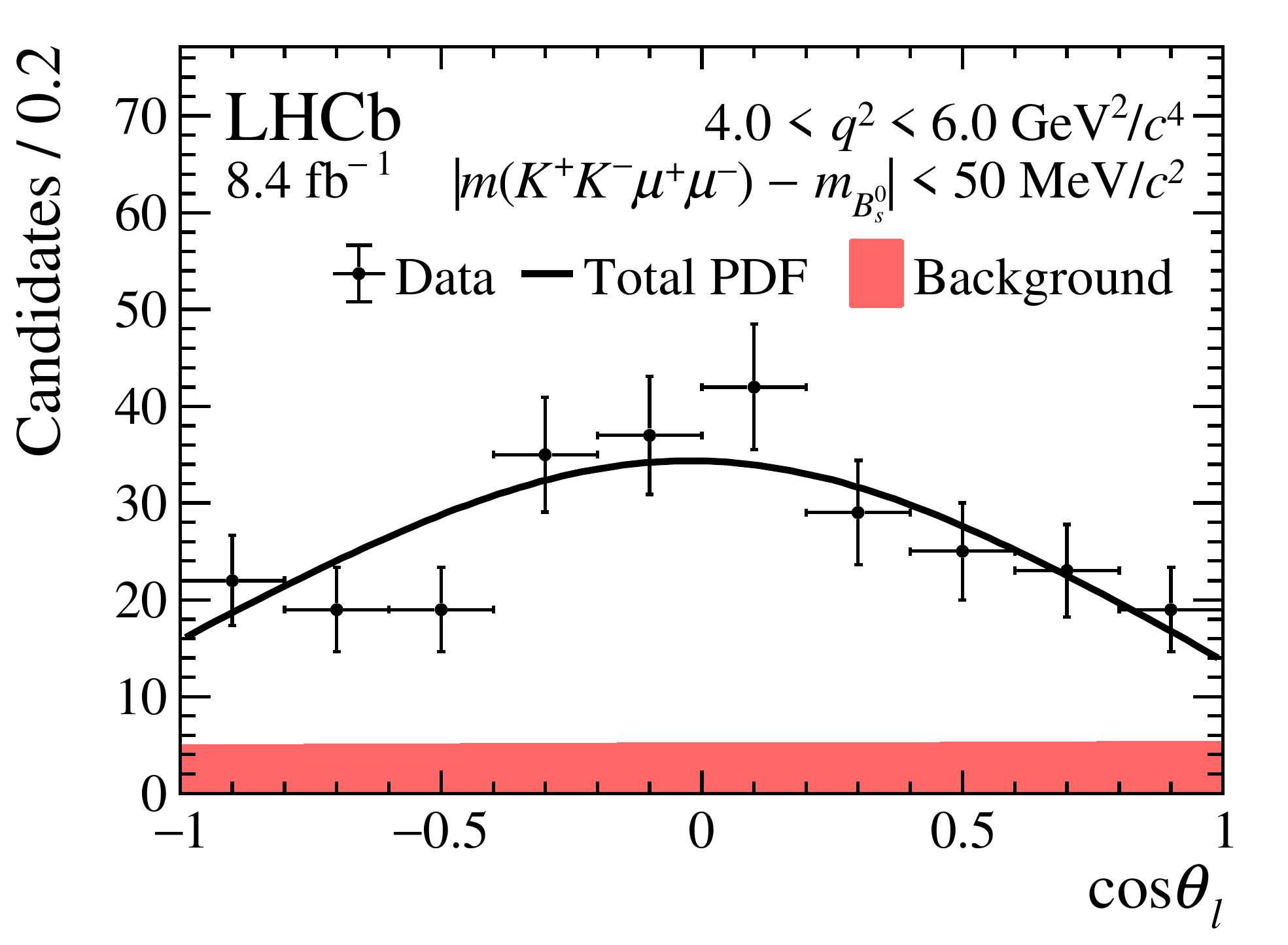}\\
    \includegraphics[width=.45\textwidth]{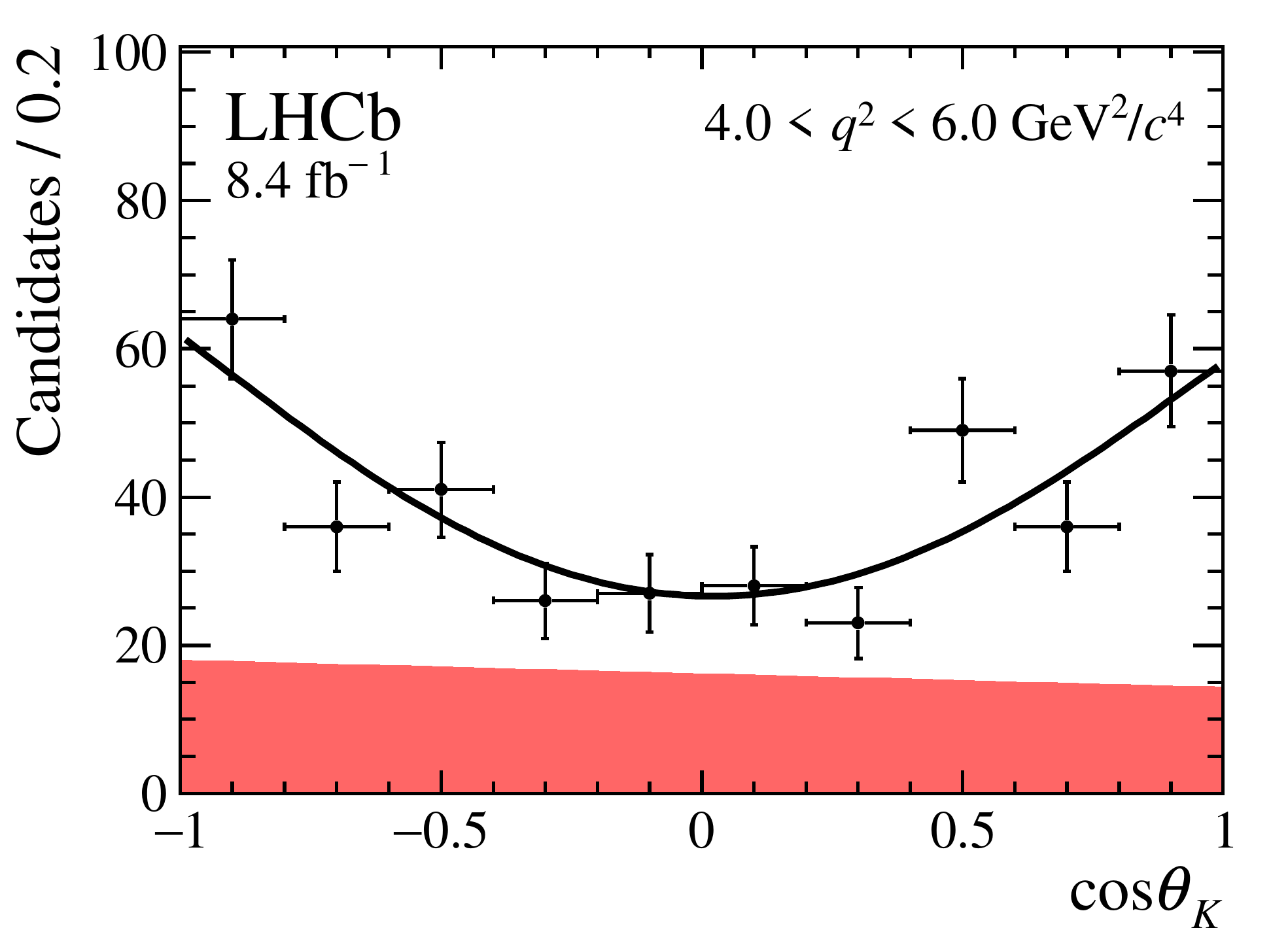}
    \includegraphics[width=.45\textwidth]{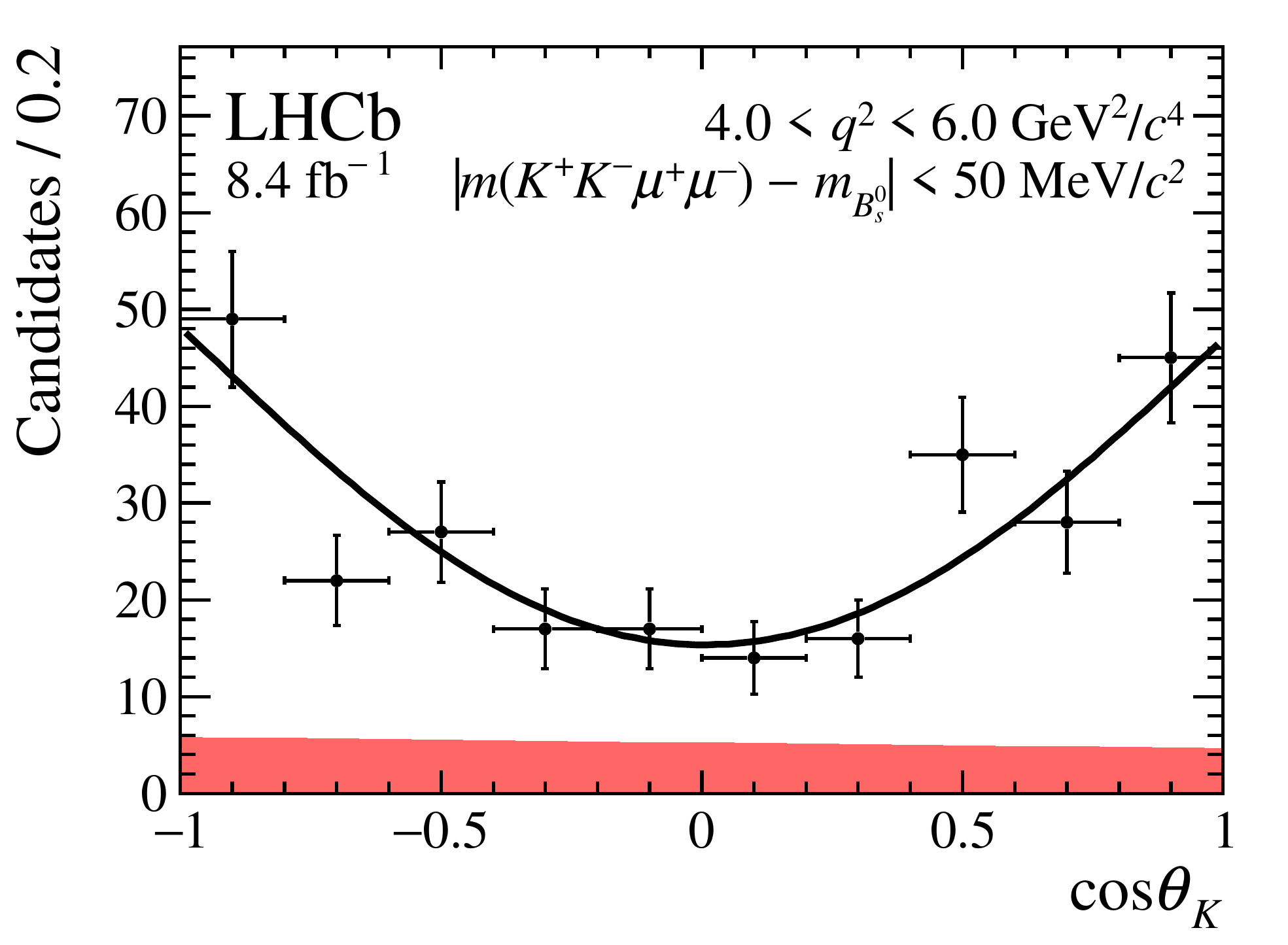}\\
    \includegraphics[width=.45\textwidth]{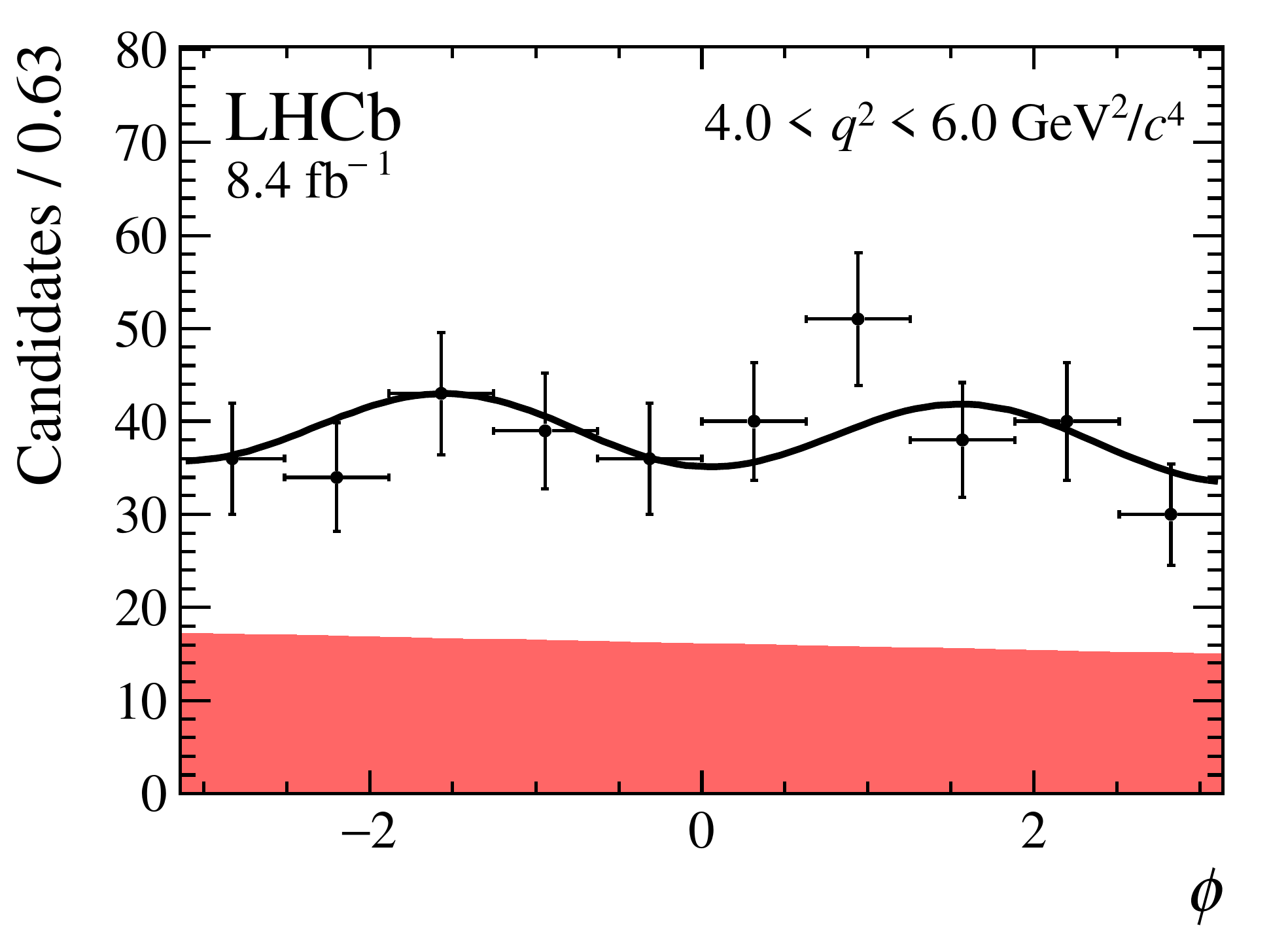}
    \includegraphics[width=.45\textwidth]{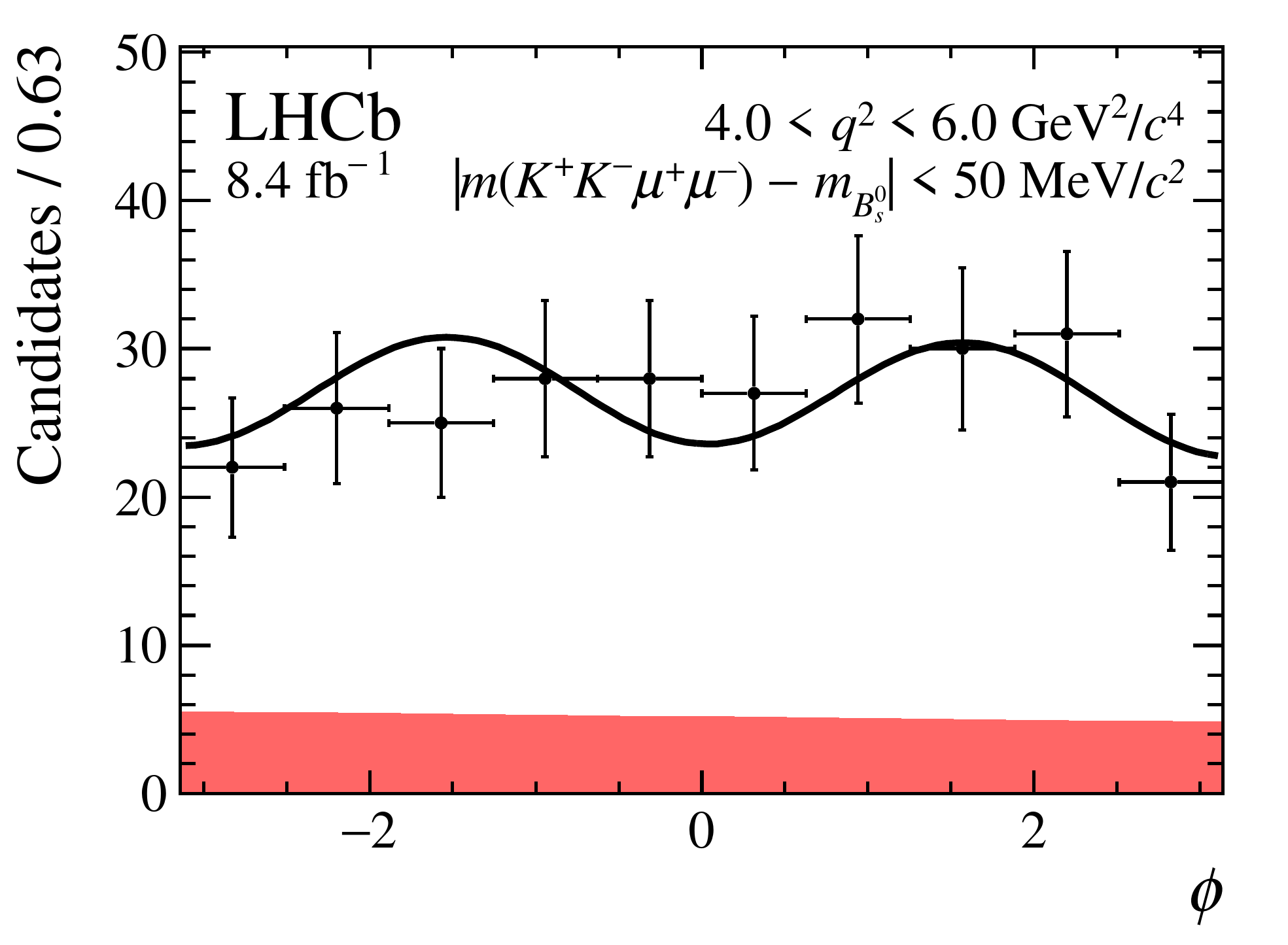}
    \caption{\label{fig:results_bin3_comb} Projections in the region \mbox{$4.0 < \qsq <6.0\gevgevcccc$} for the angular distributions of the combined 2011--2012, 2016 and 2017--2018 data sets. The data are overlaid with the projection of the combined PDF. The red shaded area indicates the background component and the solid black line the total PDF.
    The angular projections are given for candidates in (left) the entire mass region used to determine the observables in this paper and (right) the signal mass window $\pm50\mevcc$ around the known \Bs mass.}
\end{figure}

\begin{figure}[hb]
    \centering
    \includegraphics[width=.45\textwidth]{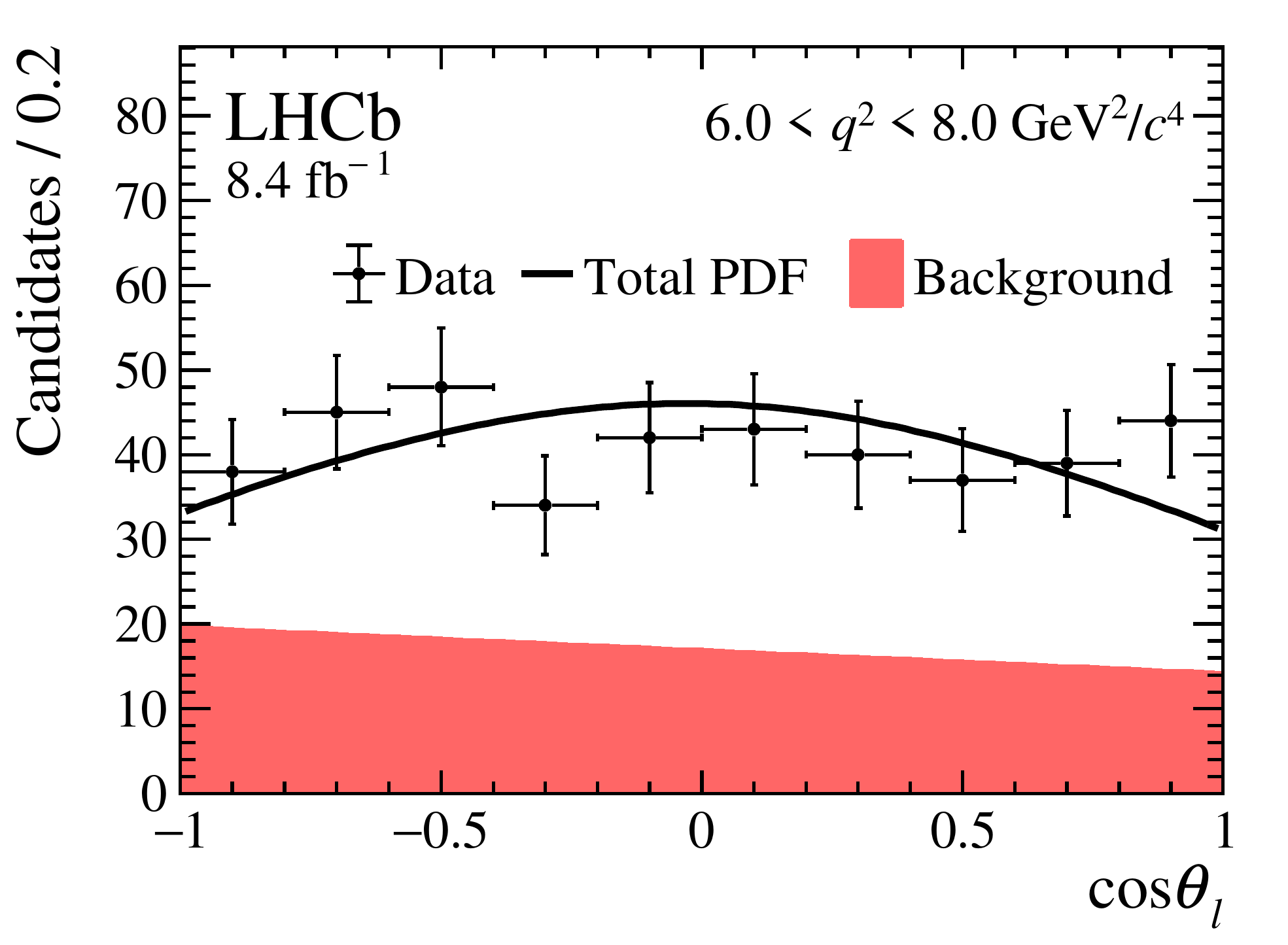}
    \includegraphics[width=.45\textwidth]{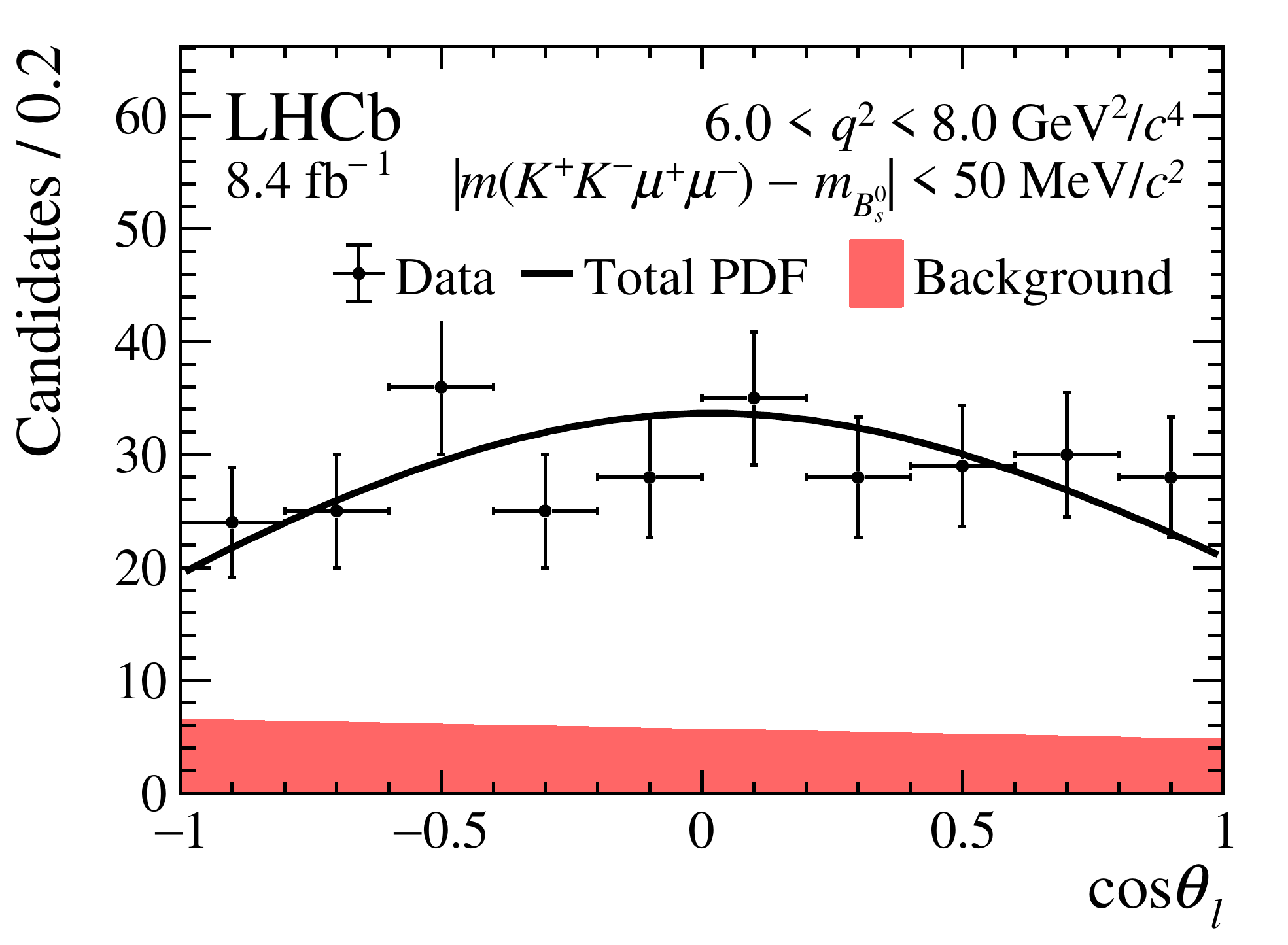}\\
    \includegraphics[width=.45\textwidth]{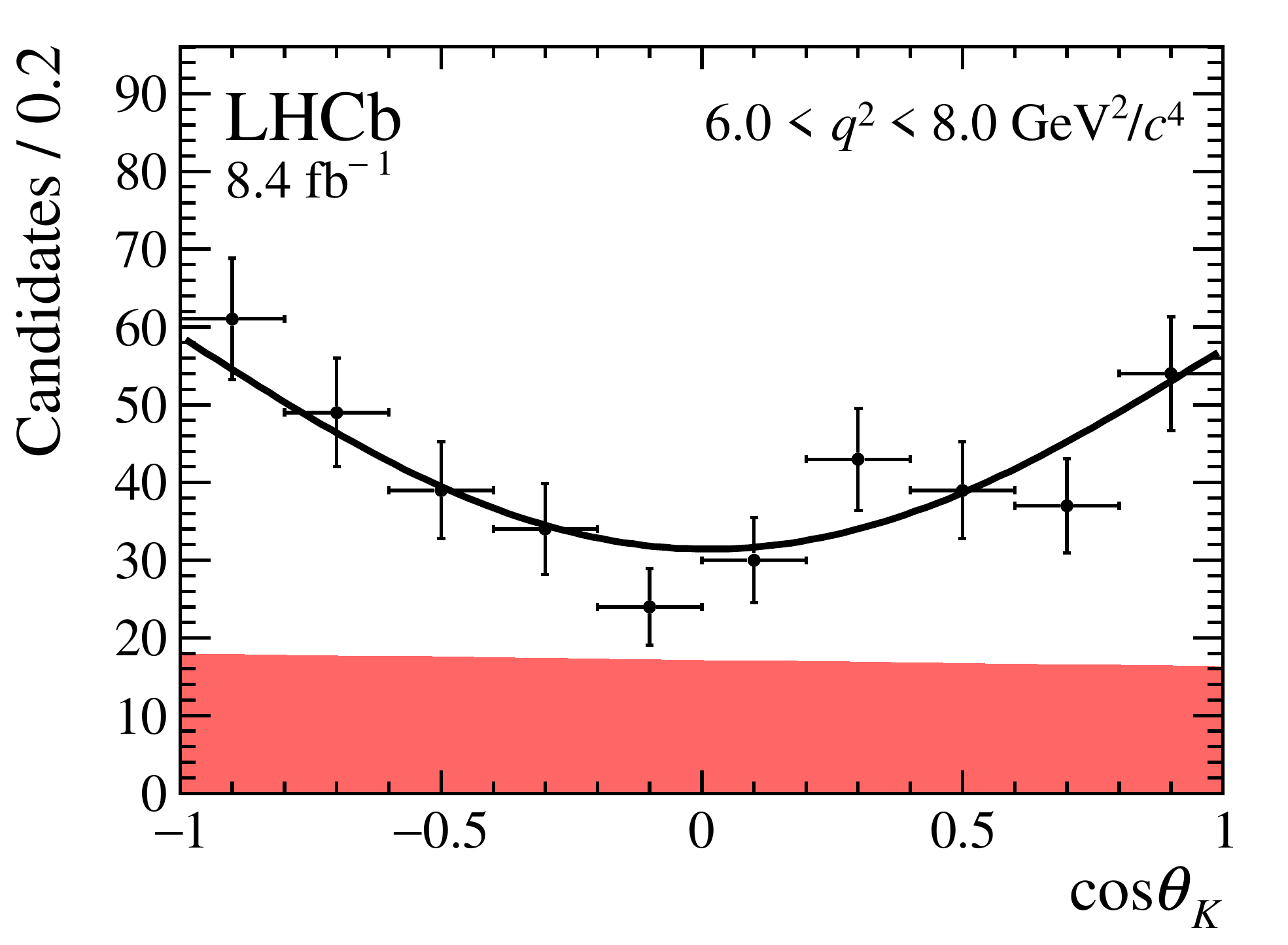}
    \includegraphics[width=.45\textwidth]{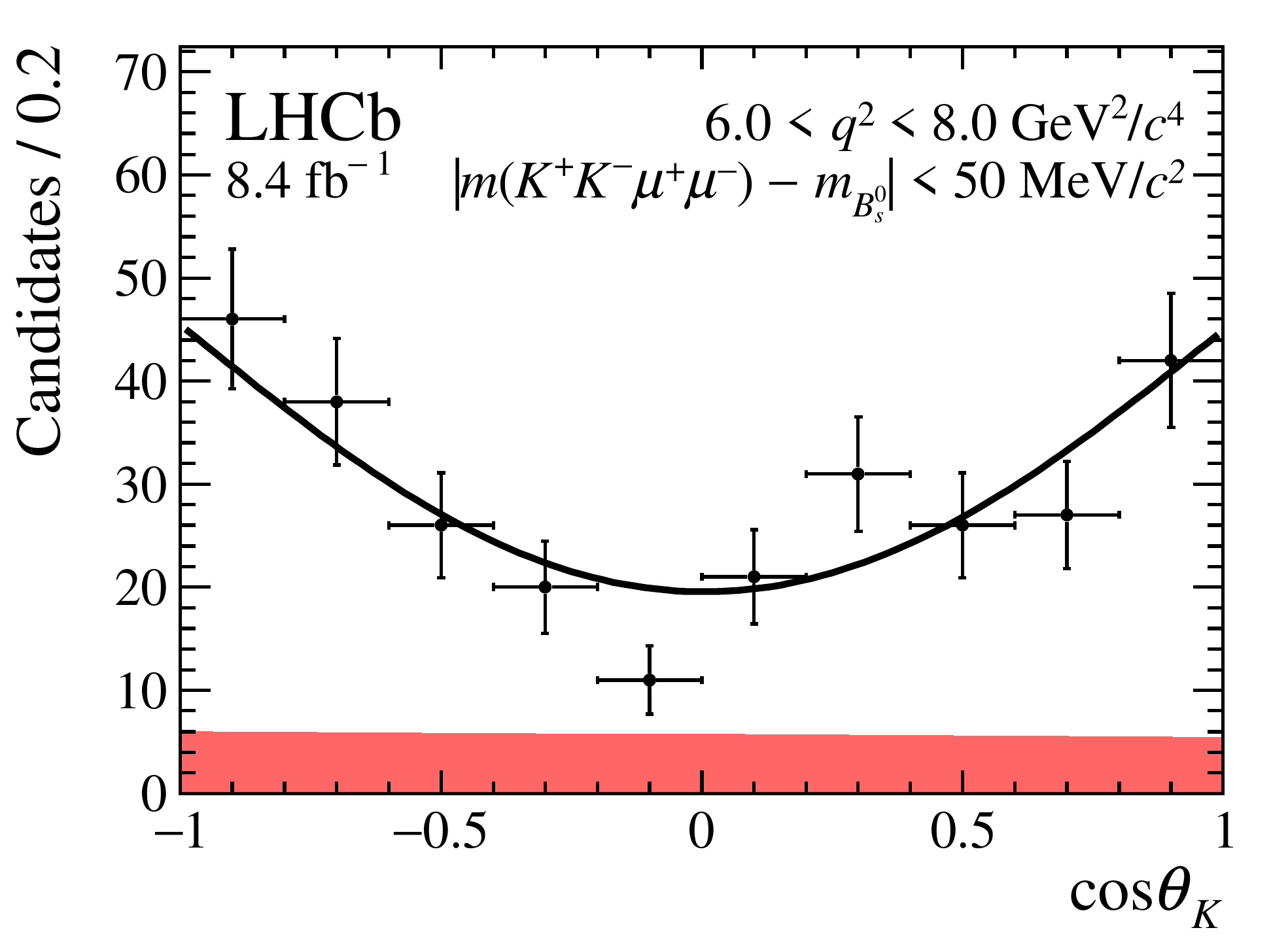}\\
    \includegraphics[width=.45\textwidth]{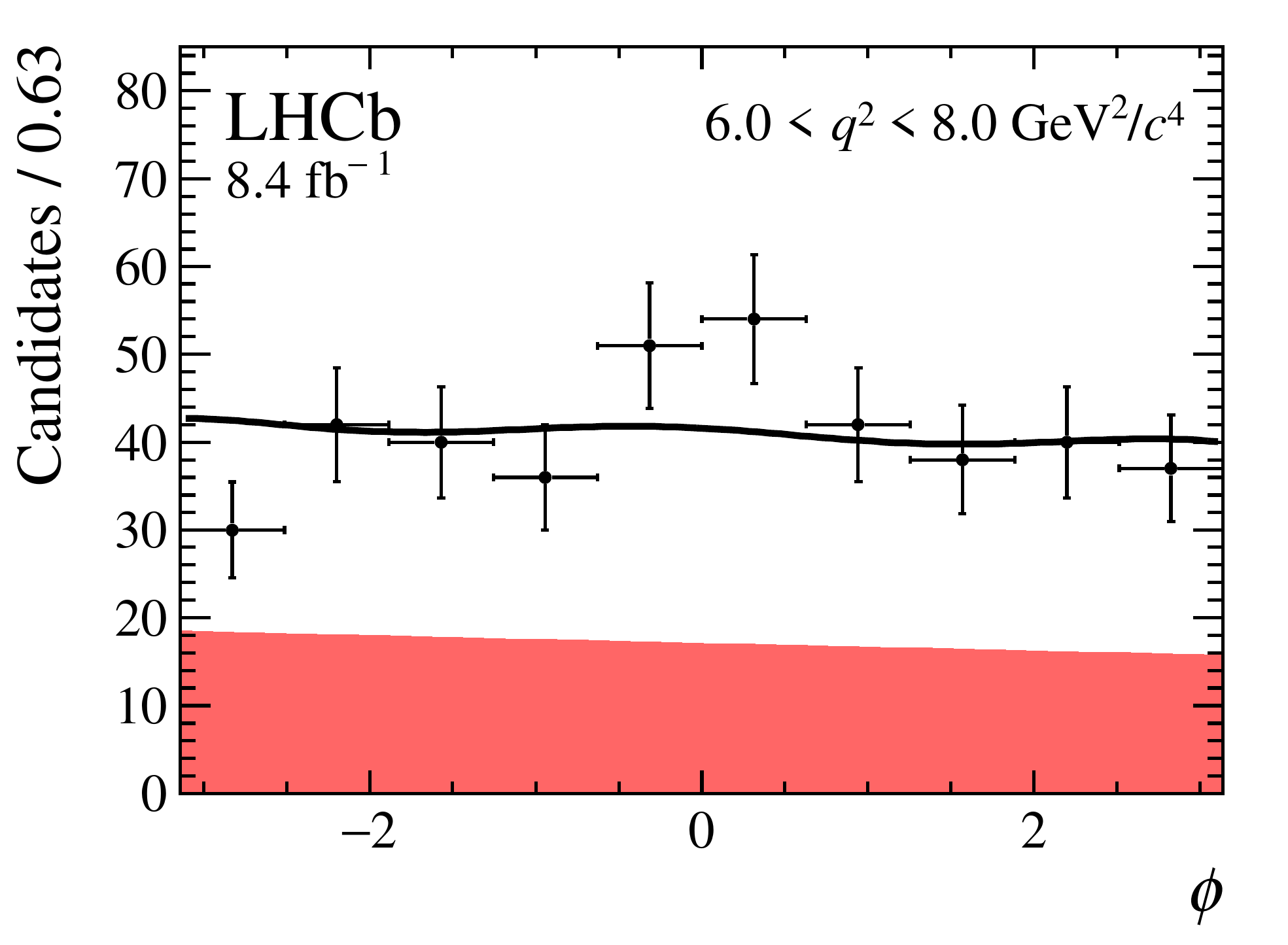}
    \includegraphics[width=.45\textwidth]{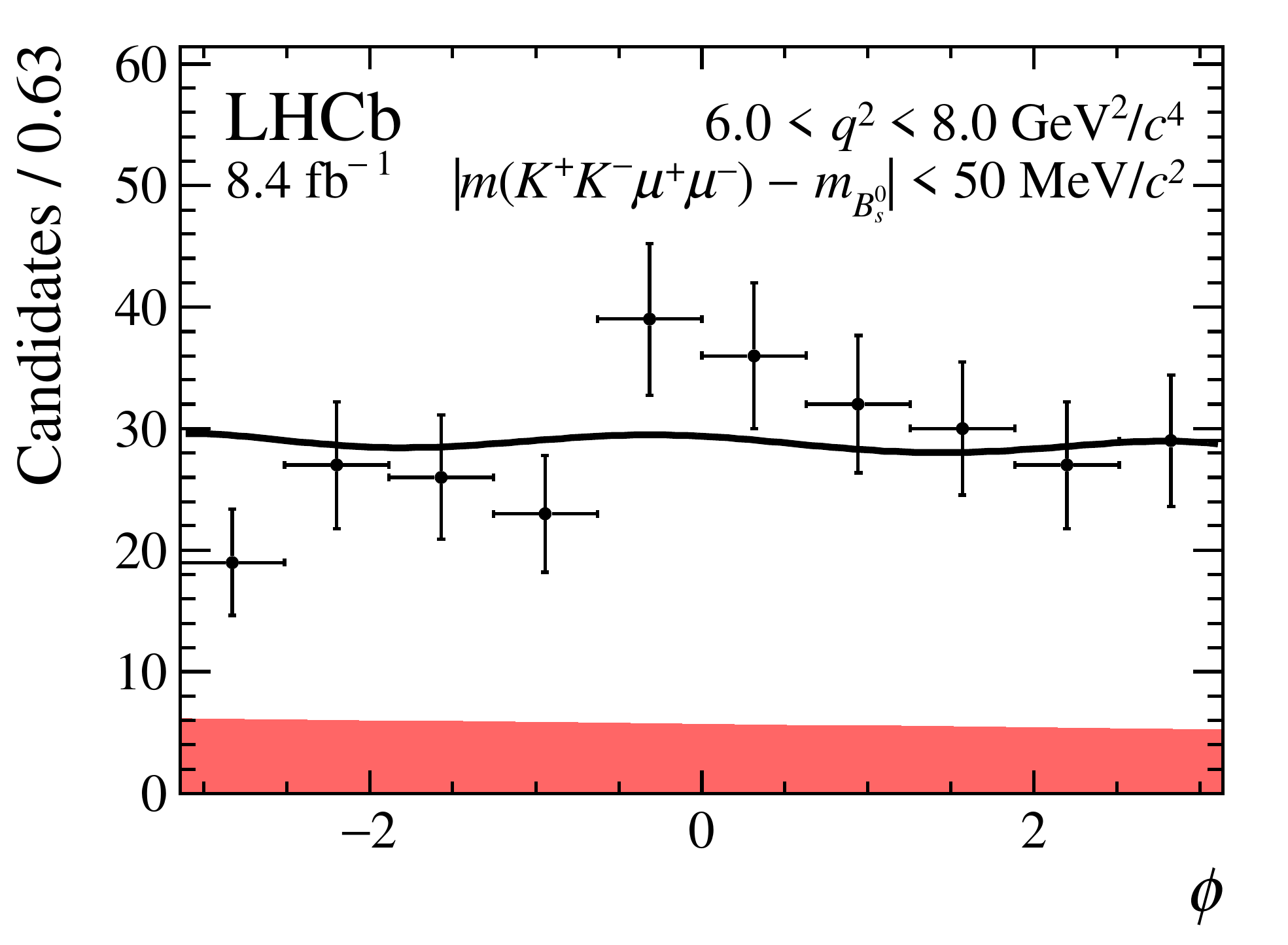}
    \caption{\label{fig:results_bin4_comb} Projections in the region \mbox{$6.0 < \qsq <8.0\gevgevcccc$} for the angular distributions of the combined 2011--2012, 2016 and 2017--2018 data sets. The data are overlaid with the projection of the combined PDF. The red shaded area indicates the background component and the solid black line the total PDF.
    The angular projections are given for candidates in (left) the entire mass region used to determine the observables in this paper and (right) the signal mass window $\pm50\mevcc$ around the known \Bs mass.}
\end{figure}

\begin{figure}[hb]
    \centering
    \includegraphics[width=.45\textwidth]{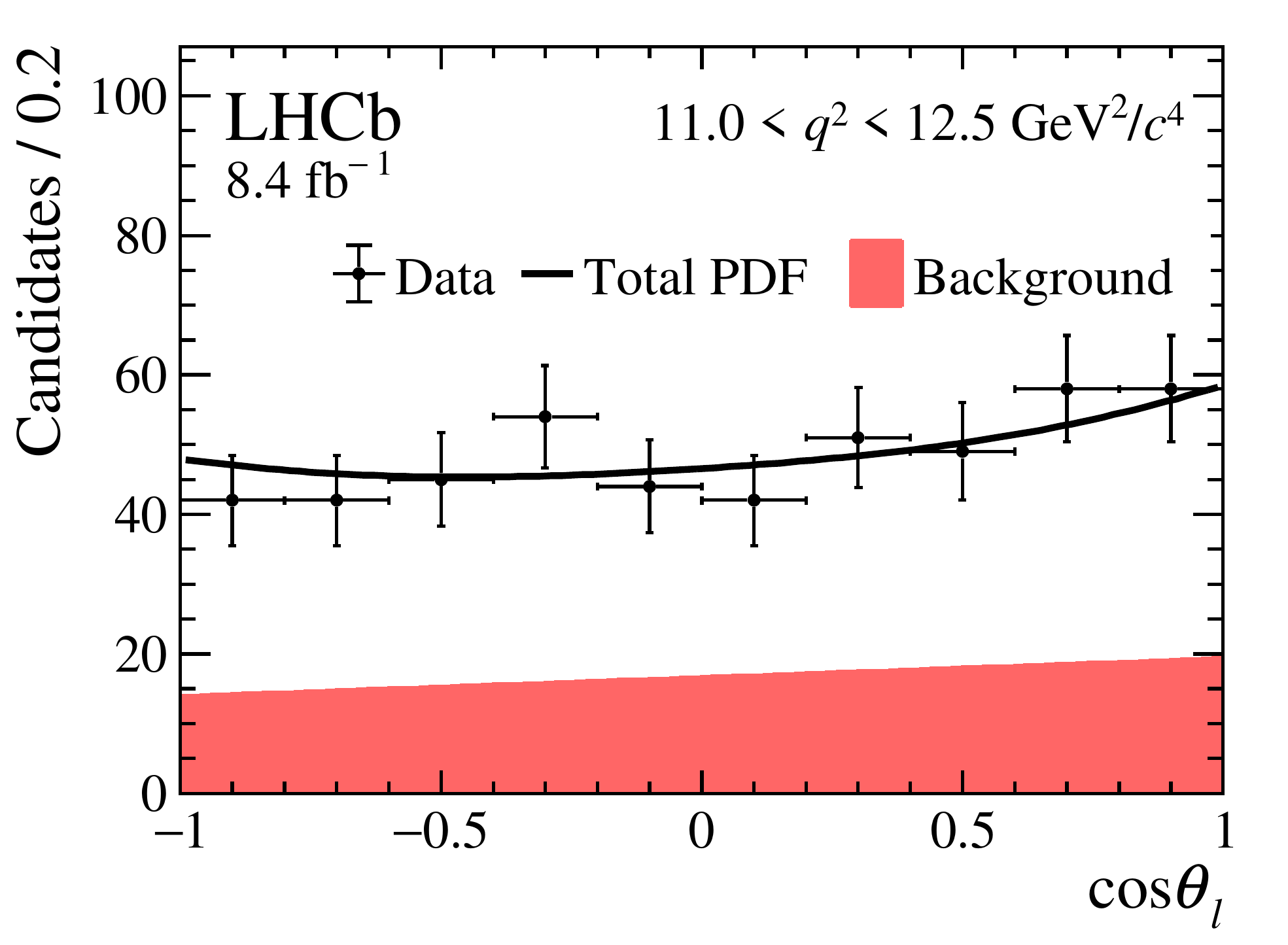}
    \includegraphics[width=.45\textwidth]{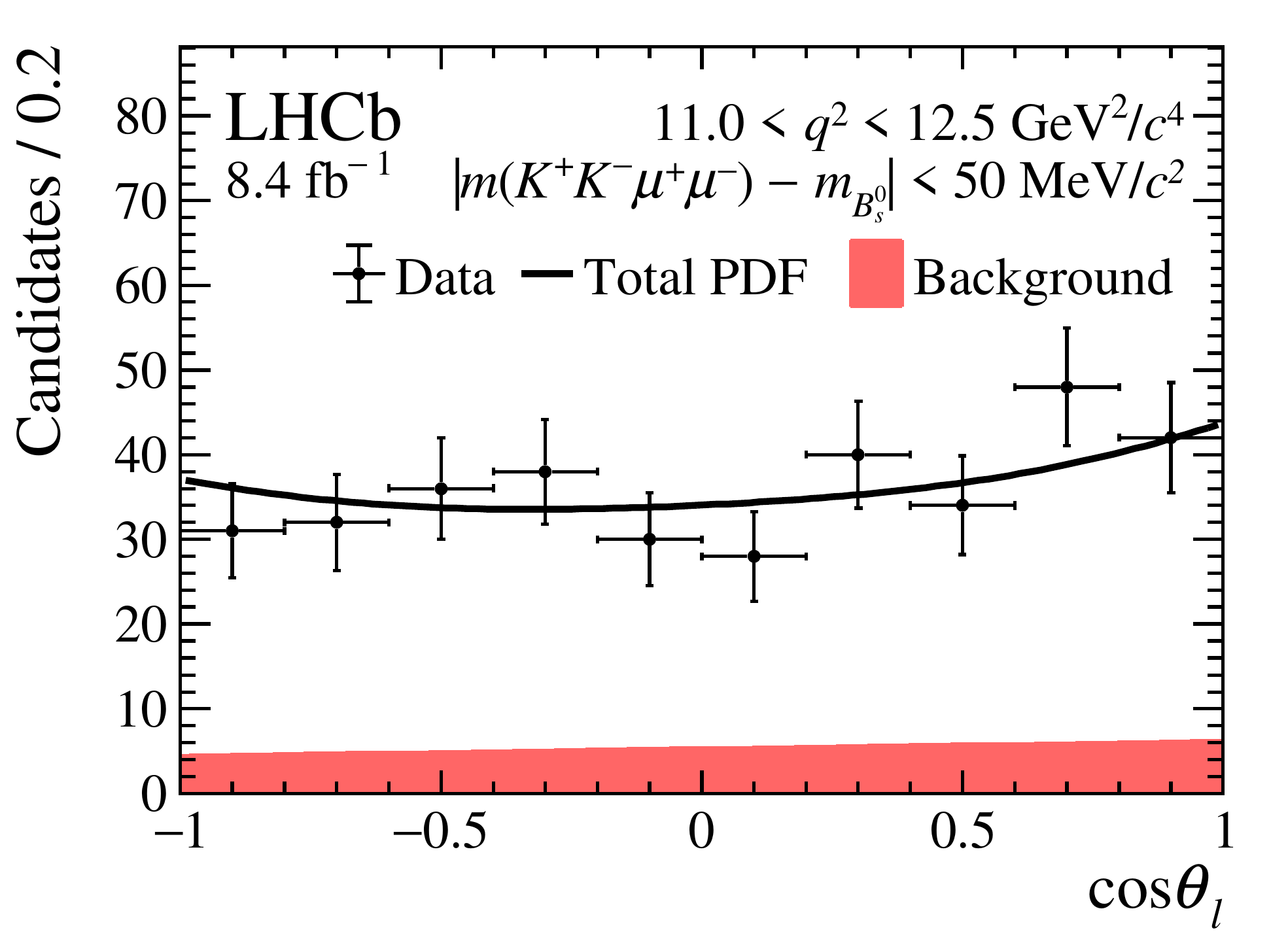}\\
    \includegraphics[width=.45\textwidth]{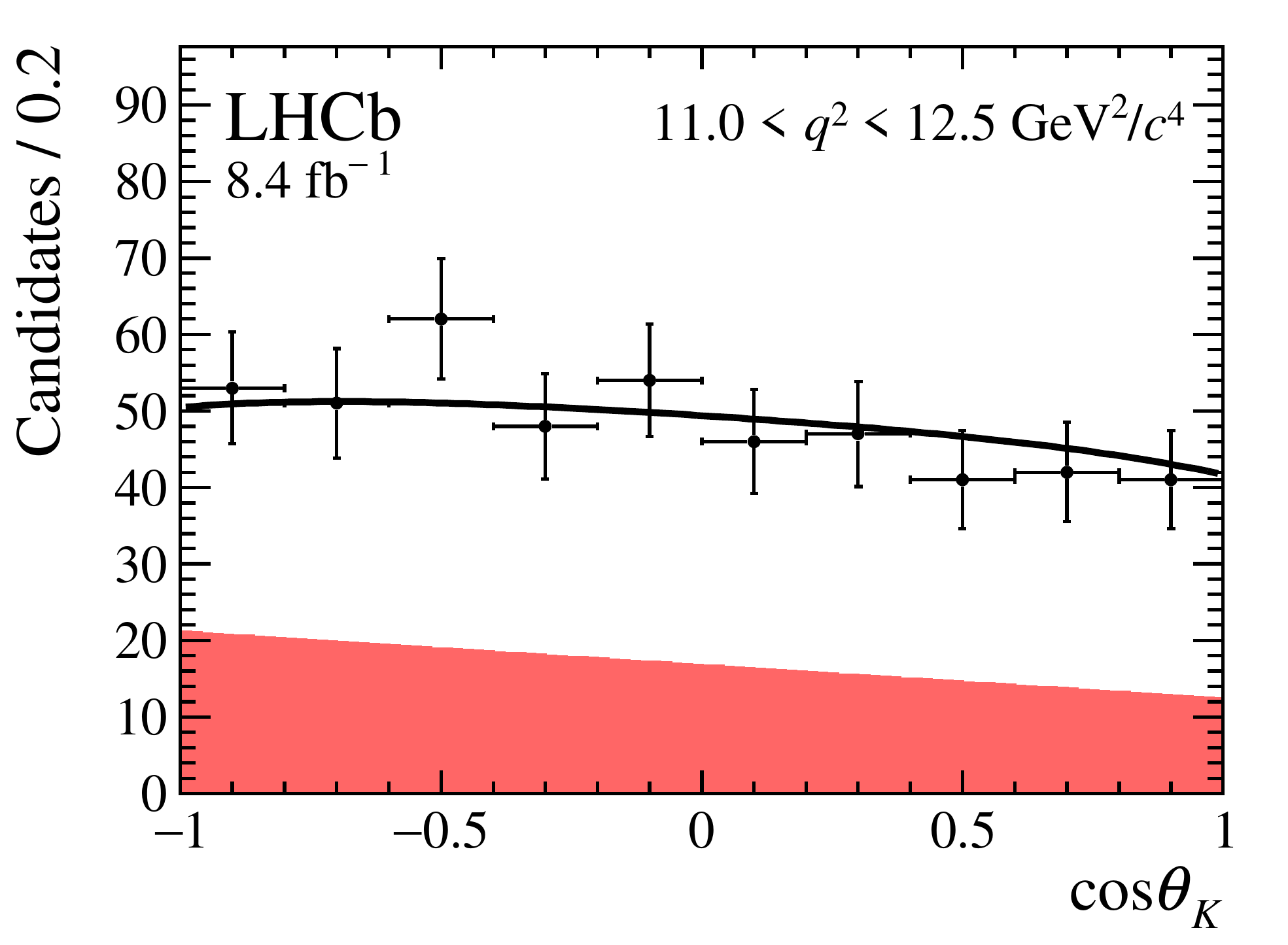}
    \includegraphics[width=.45\textwidth]{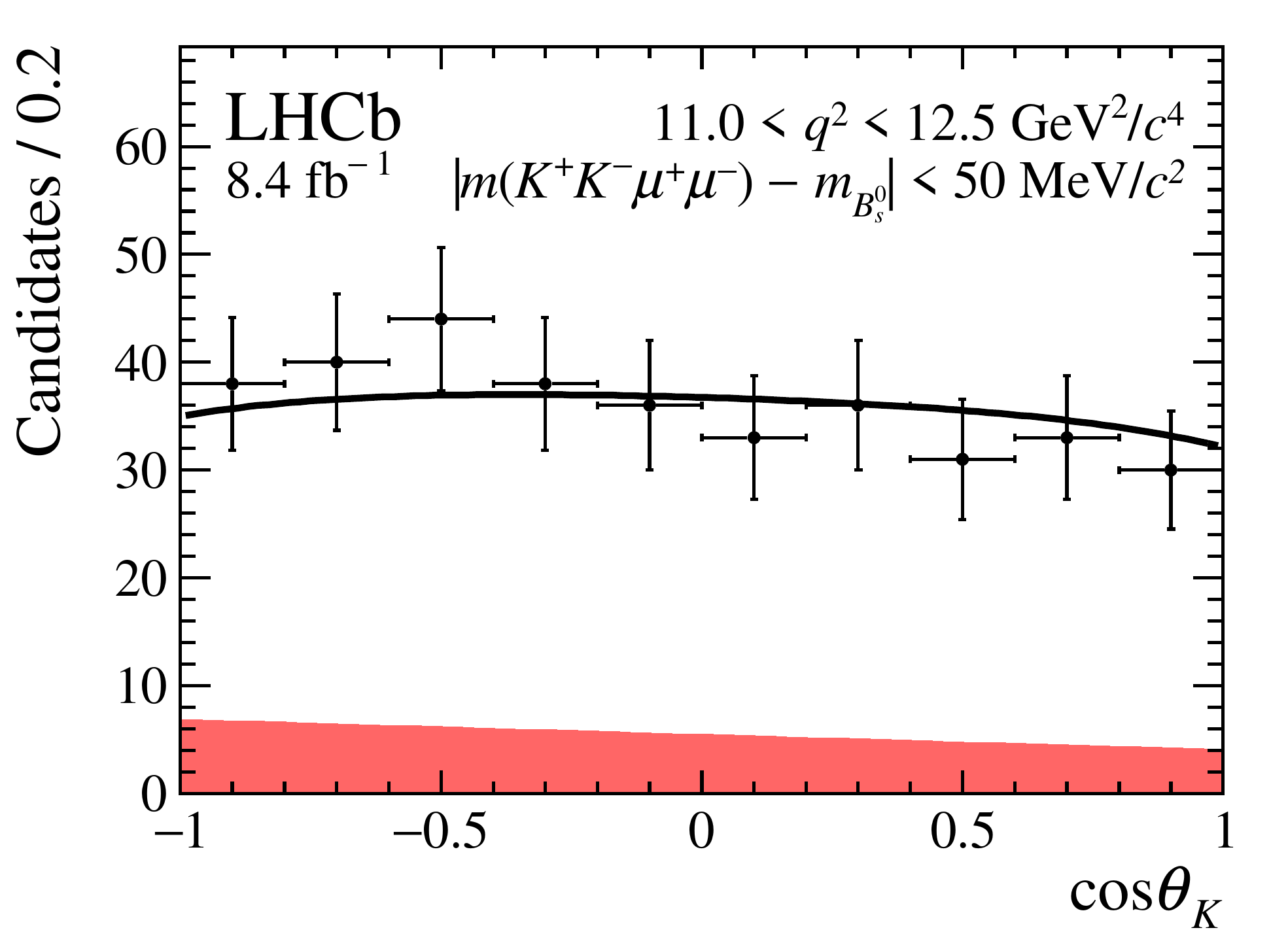}\\
    \includegraphics[width=.45\textwidth]{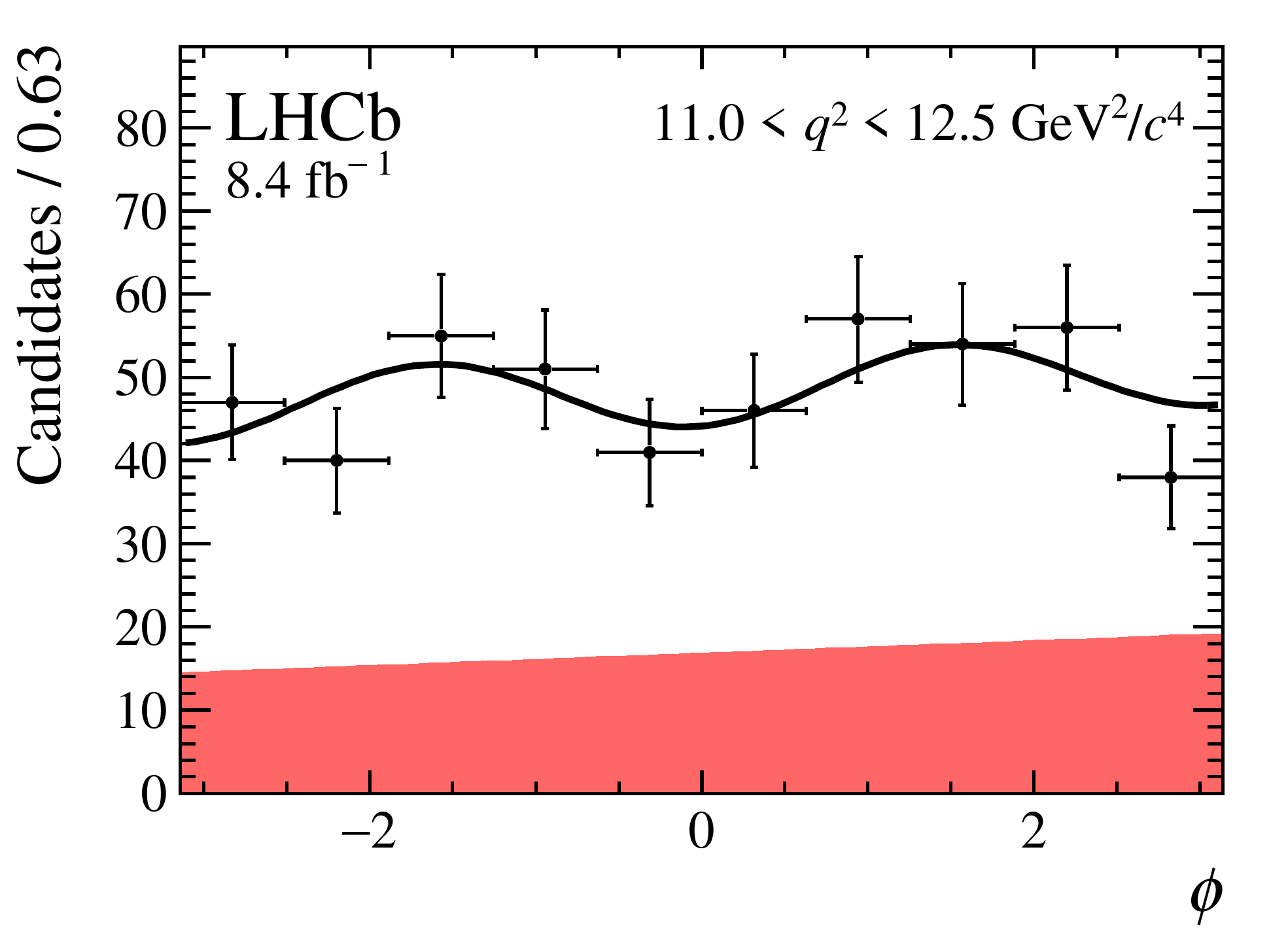}
    \includegraphics[width=.45\textwidth]{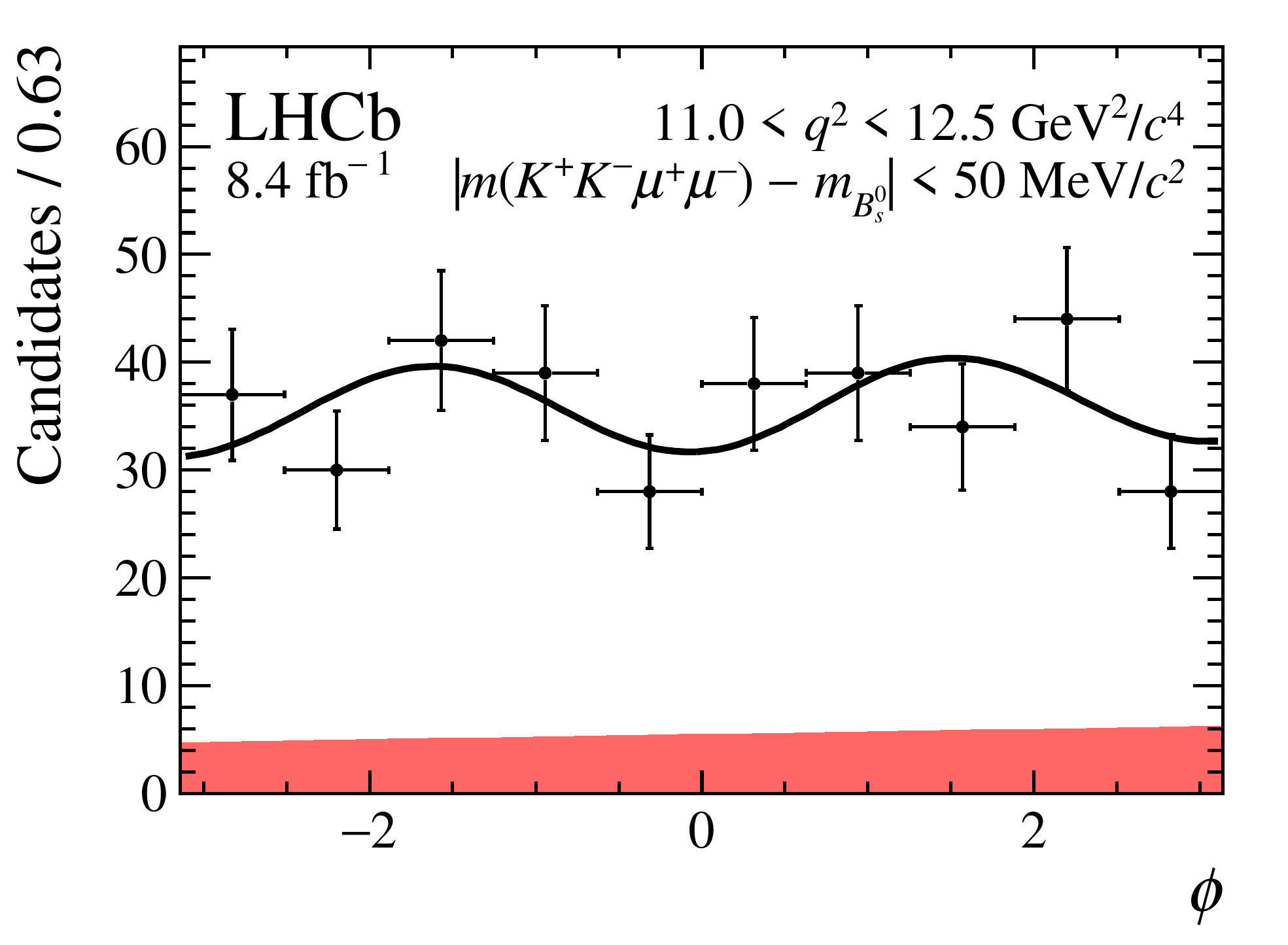}
    \caption{\label{fig:results_bin5_comb} Projections in the region \mbox{$11.0 < \qsq <12.5\gevgevcccc$} for the angular distributions of the combined 2011--2012, 2016 and 2017--2018 data sets. The data are overlaid with the projection of the combined PDF. The red shaded area indicates the background component and the solid black line the total PDF.
    The angular projections are given for candidates in (left) the entire mass region used to determine the observables in this paper and (right) the signal mass window $\pm50\mevcc$ around the known \Bs mass.}
\end{figure}

\begin{figure}[hb]
    \centering
    \includegraphics[width=.45\textwidth]{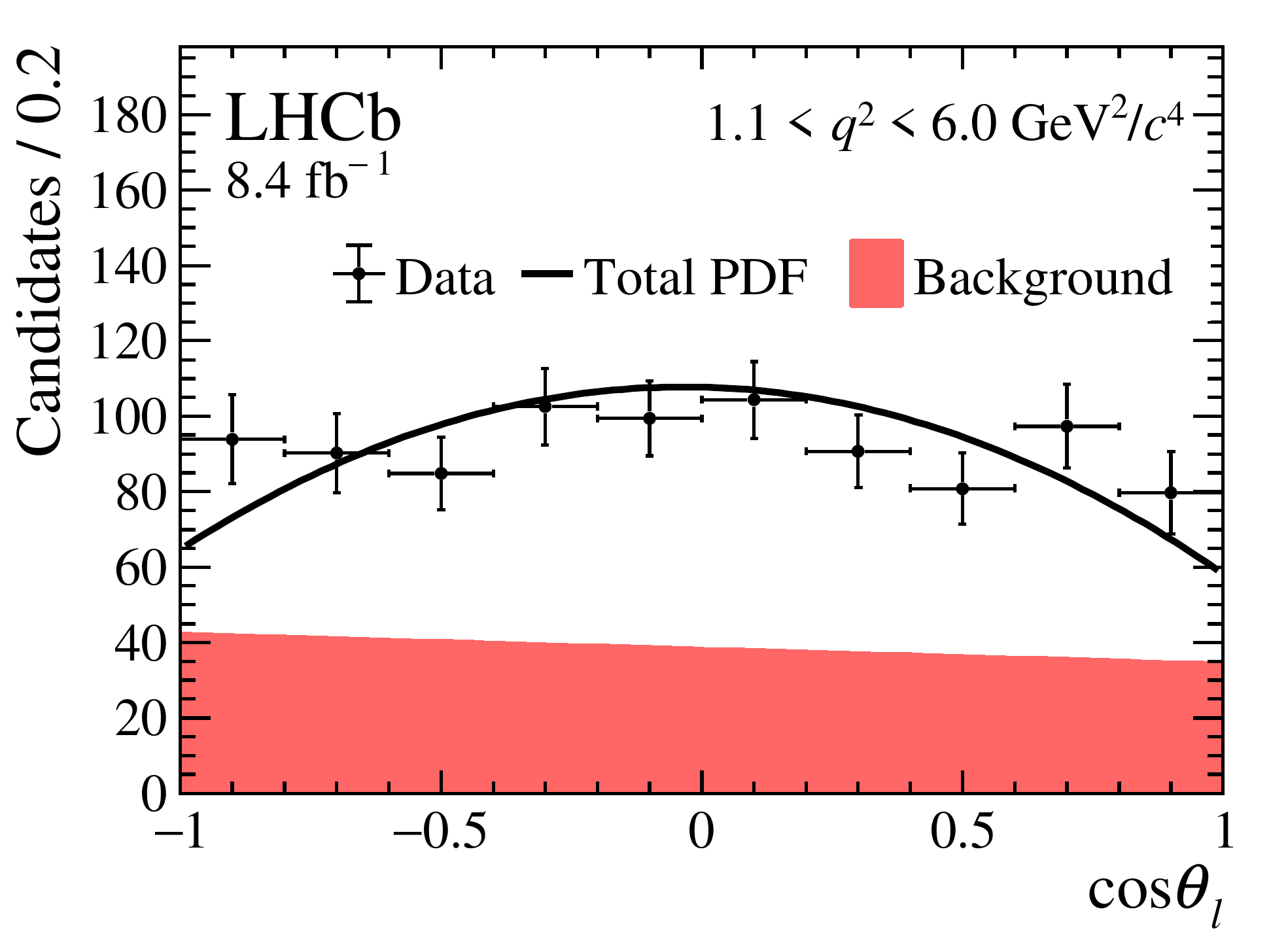}
    \includegraphics[width=.45\textwidth]{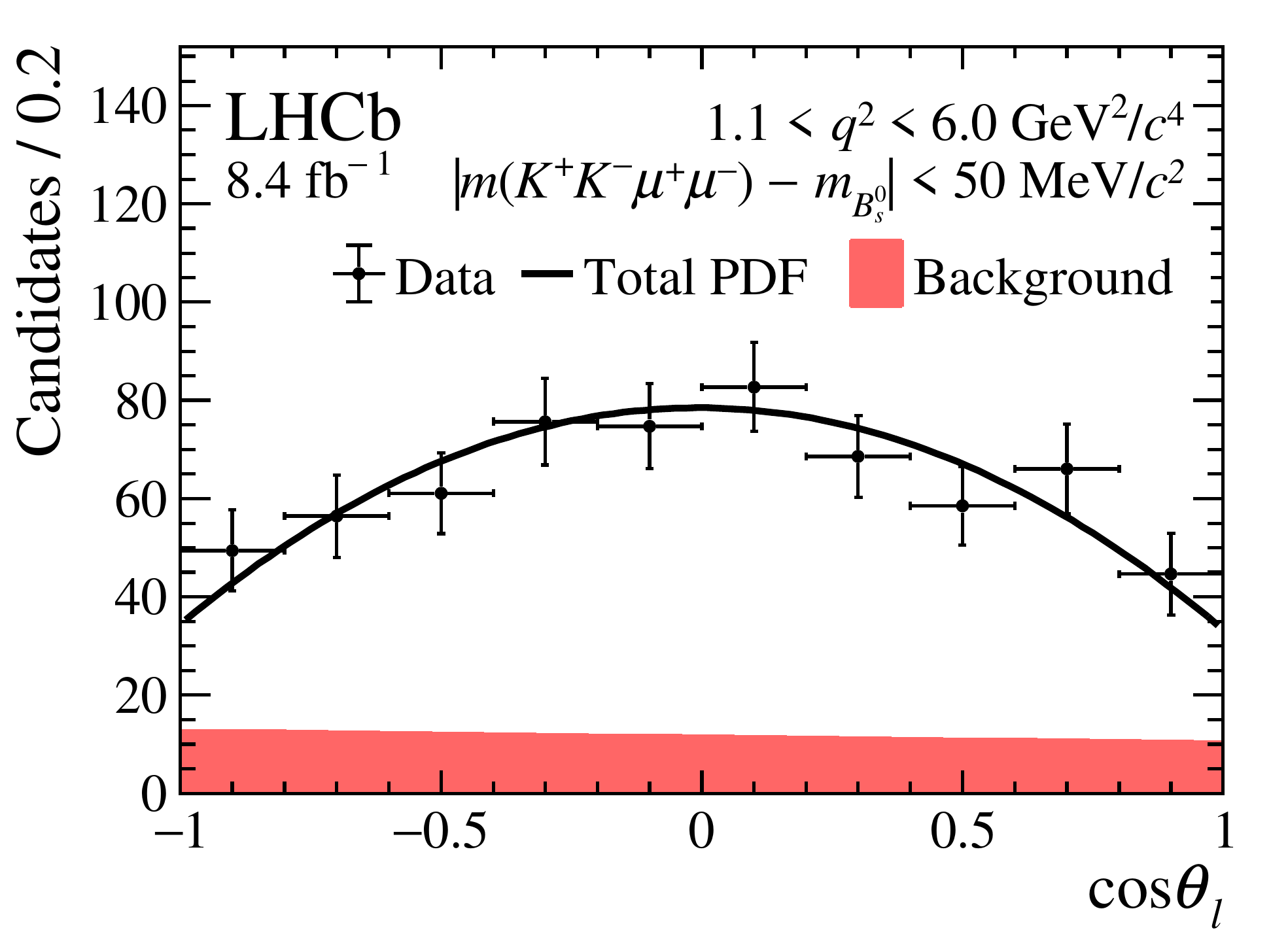}\\
    \includegraphics[width=.45\textwidth]{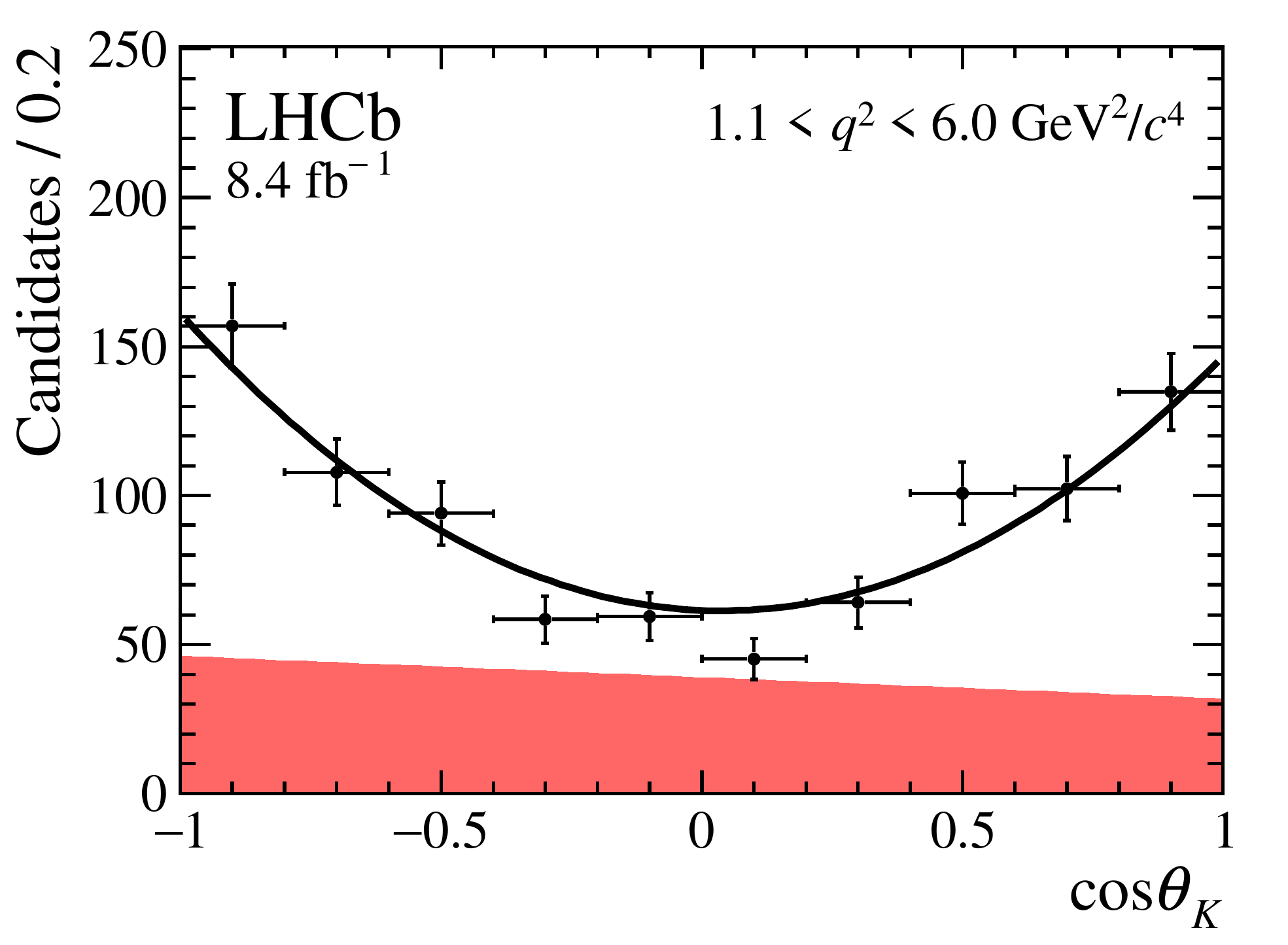}
    \includegraphics[width=.45\textwidth]{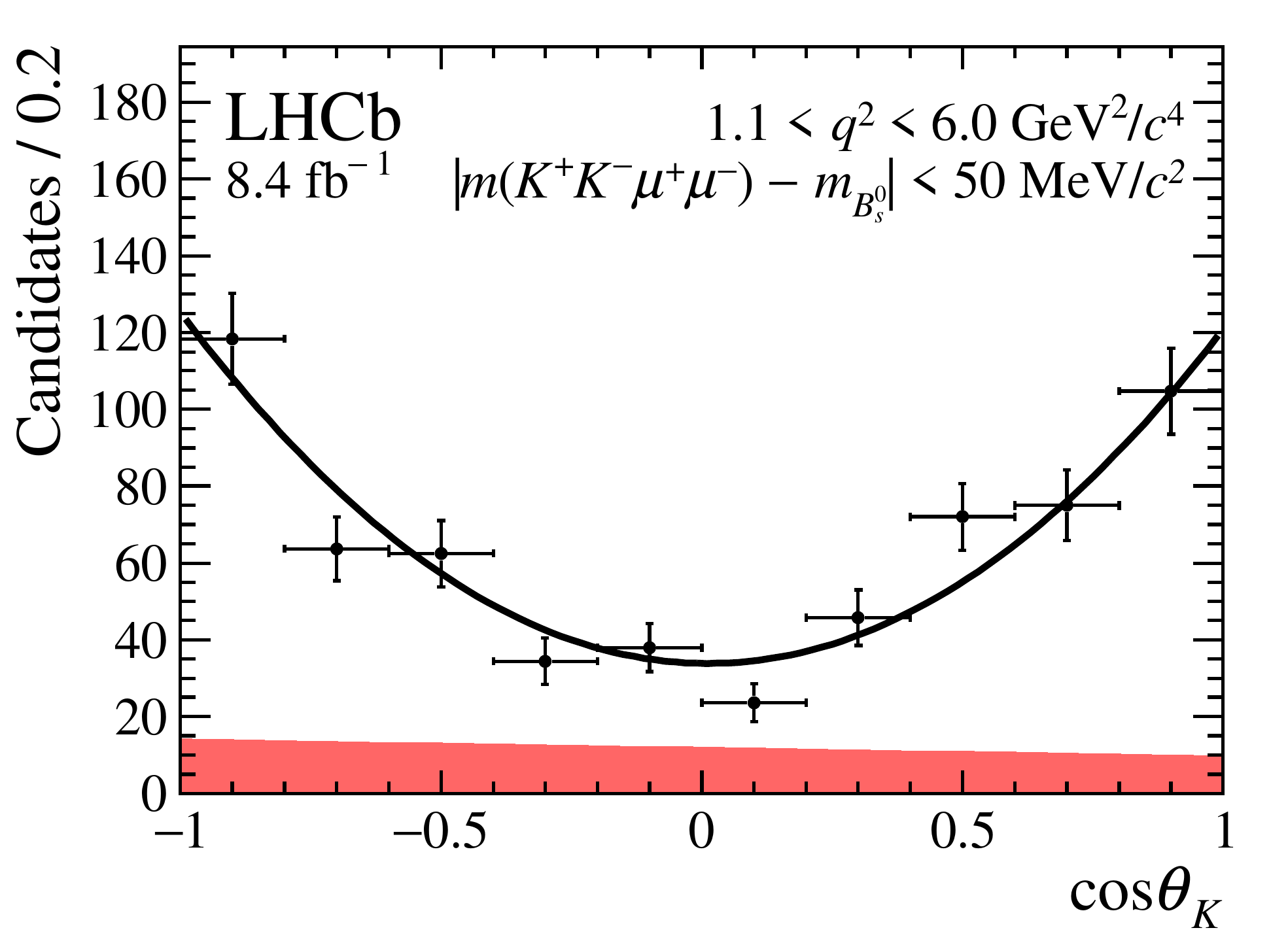}\\
    \includegraphics[width=.45\textwidth]{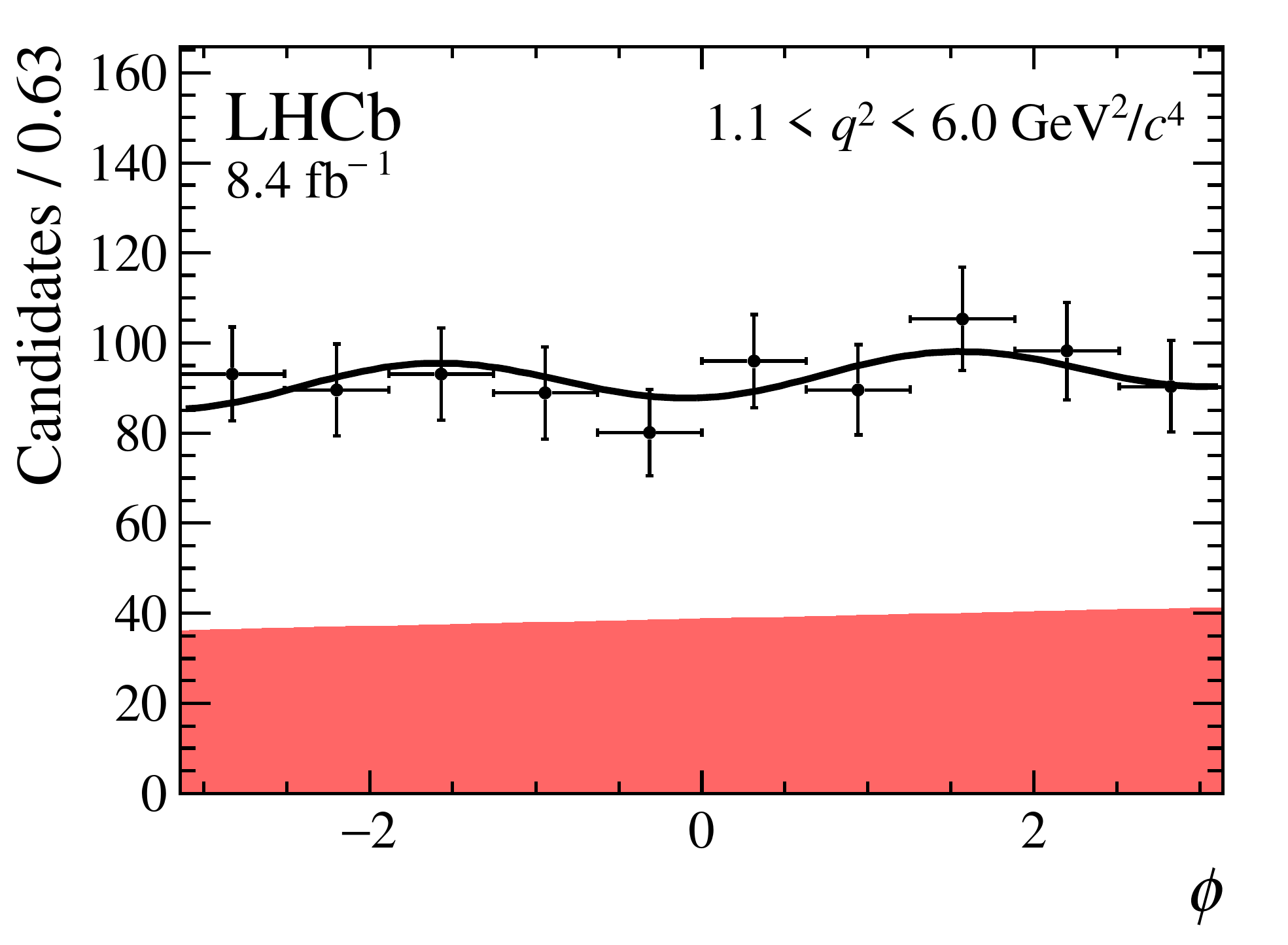}
    \includegraphics[width=.45\textwidth]{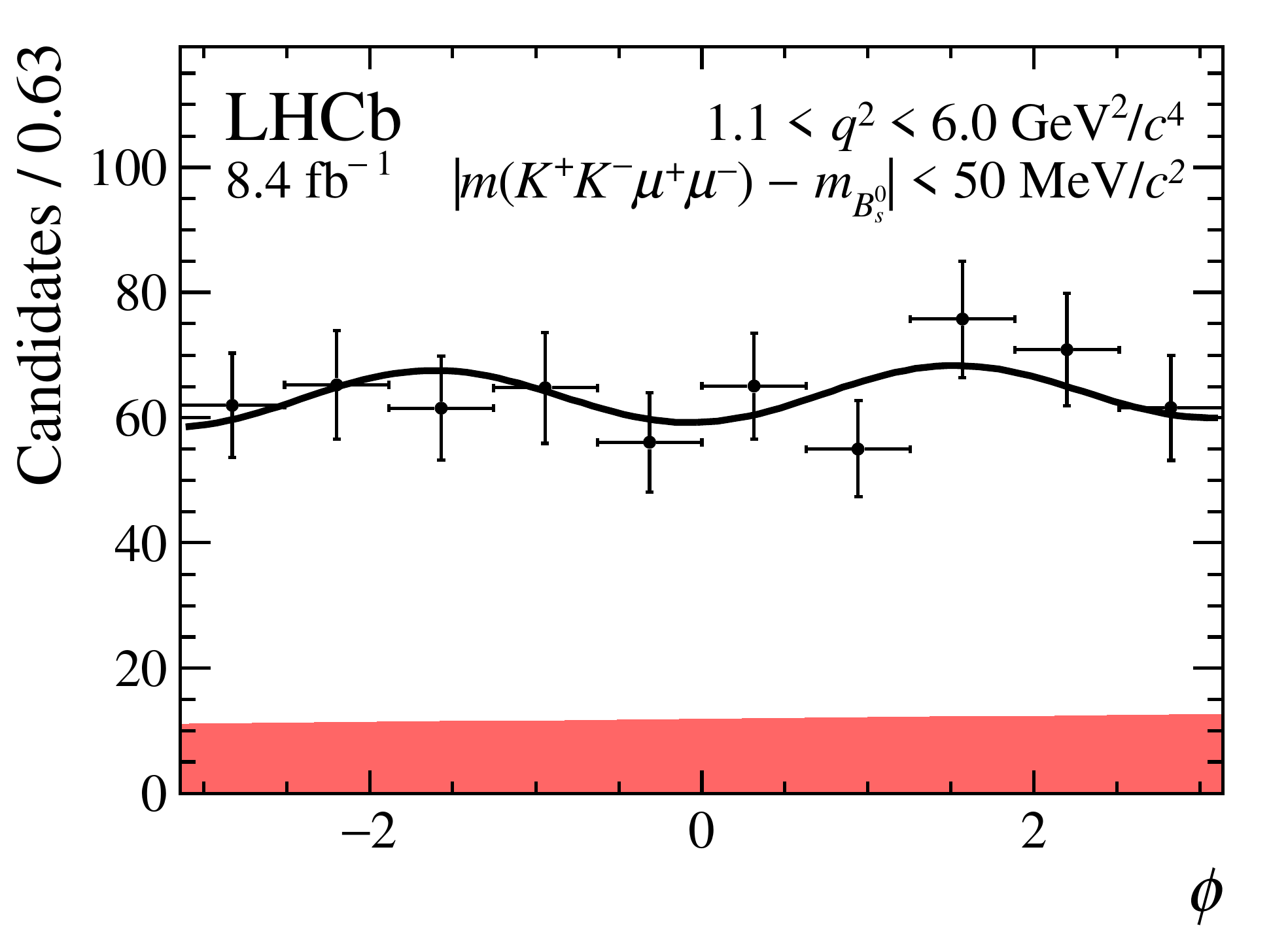}
    \caption{\label{fig:results_bin6_comb} Projections in the region \mbox{$1.1 < \qsq <6.0\gevgevcccc$} for the angular distributions of the combined 2011--2012, 2016 and 2017--2018 data sets. The data are overlaid with the projection of the combined PDF. The red shaded area indicates the background component and the solid black line the total PDF.
    The angular projections are given for candidates in (left) the entire mass region used to determine the observables in this paper and (right) the signal mass window $\pm50\mevcc$ around the known \Bs mass.}
\end{figure}

\begin{figure}[hb]
    \centering
    \includegraphics[width=.45\textwidth]{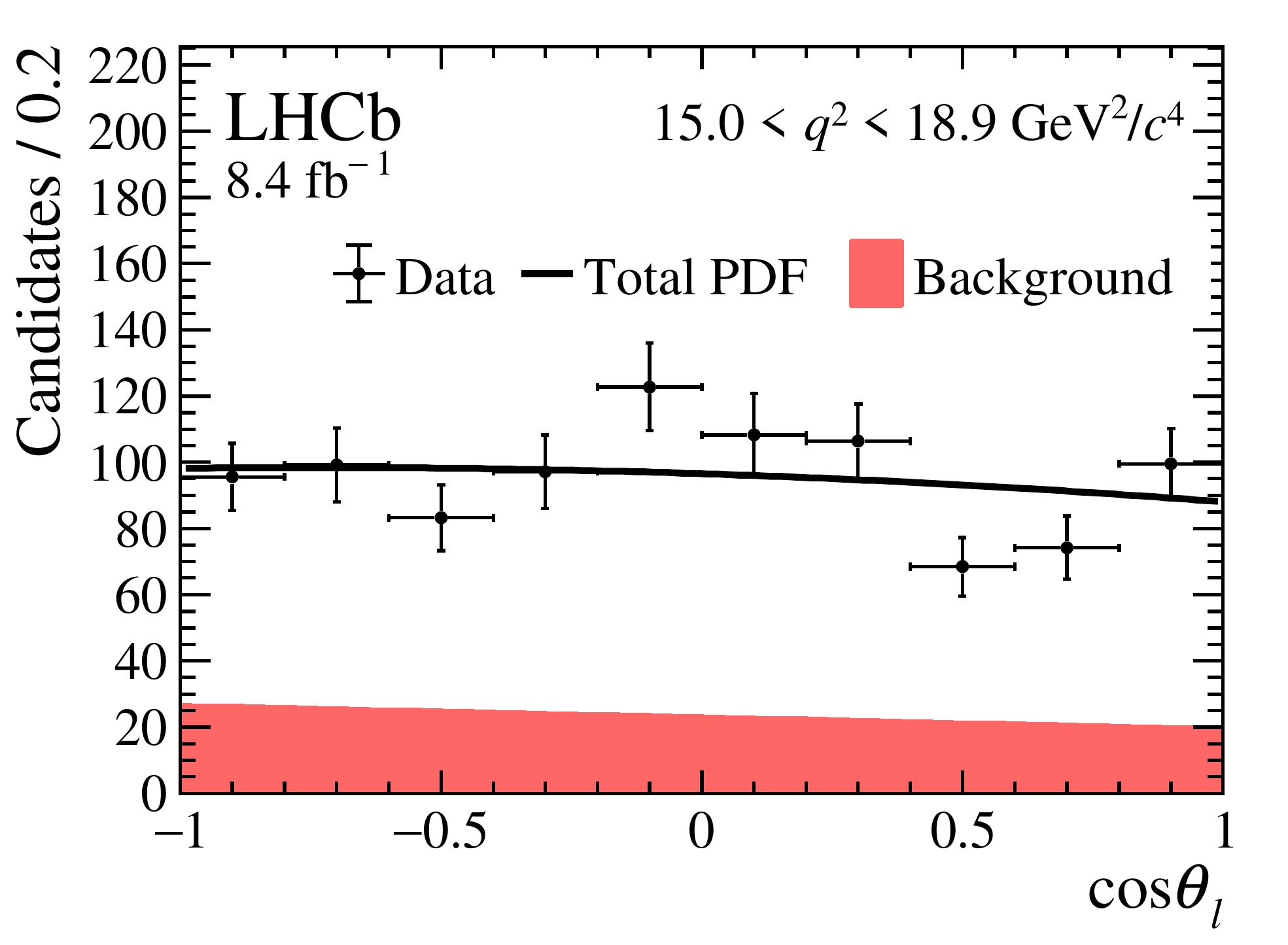}
    \includegraphics[width=.45\textwidth]{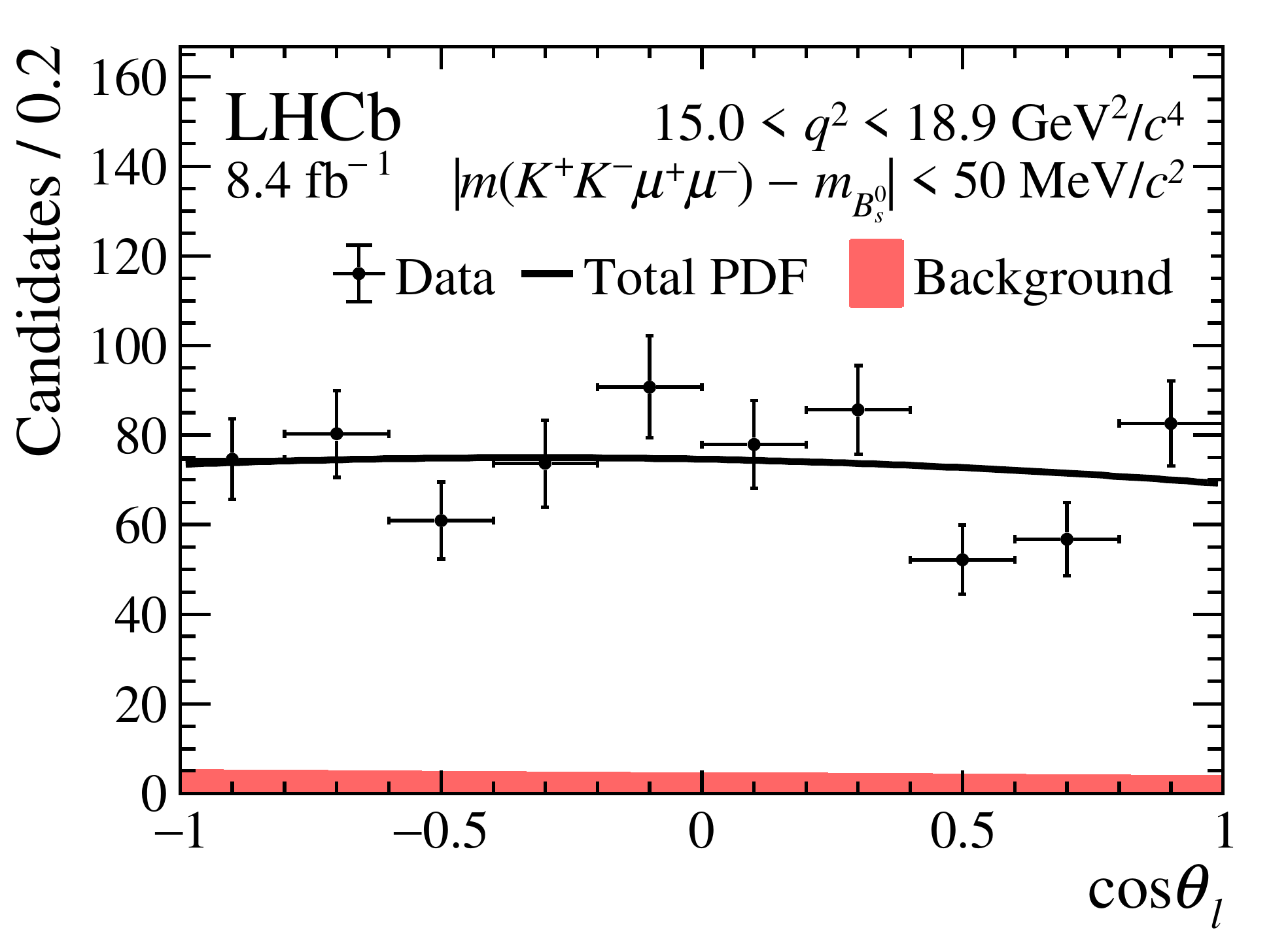}\\
    \includegraphics[width=.45\textwidth]{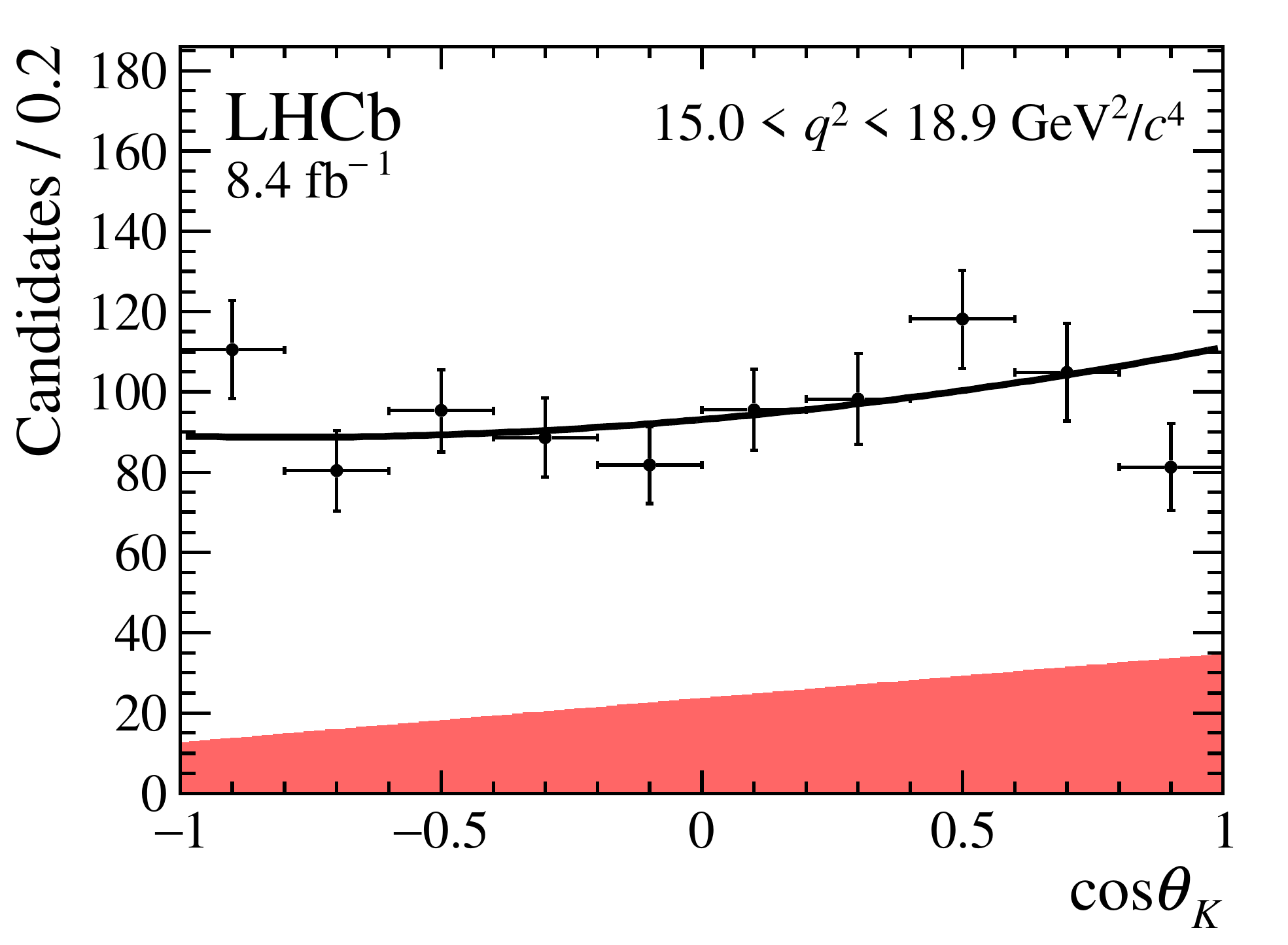}
    \includegraphics[width=.45\textwidth]{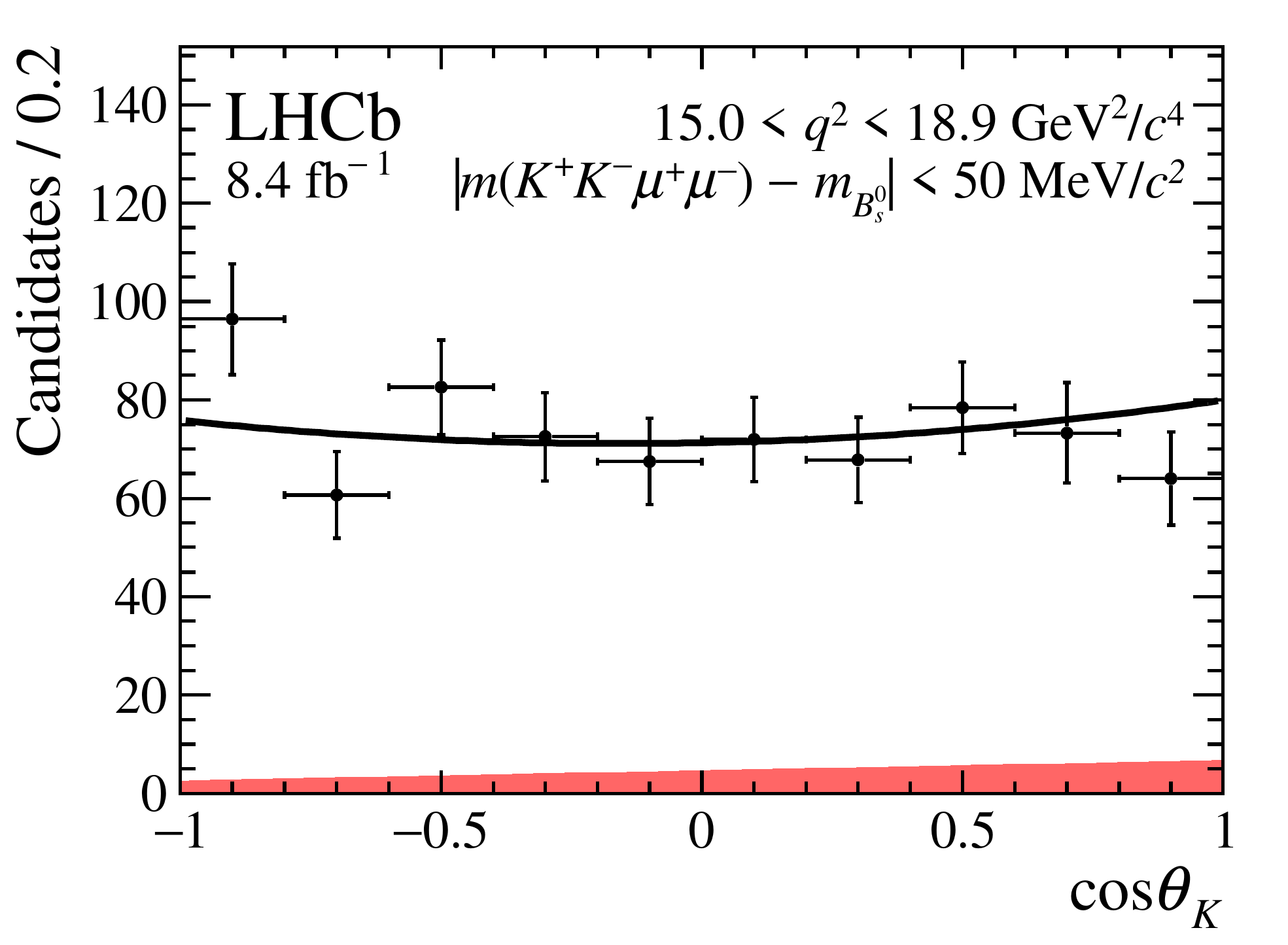}\\
    \includegraphics[width=.45\textwidth]{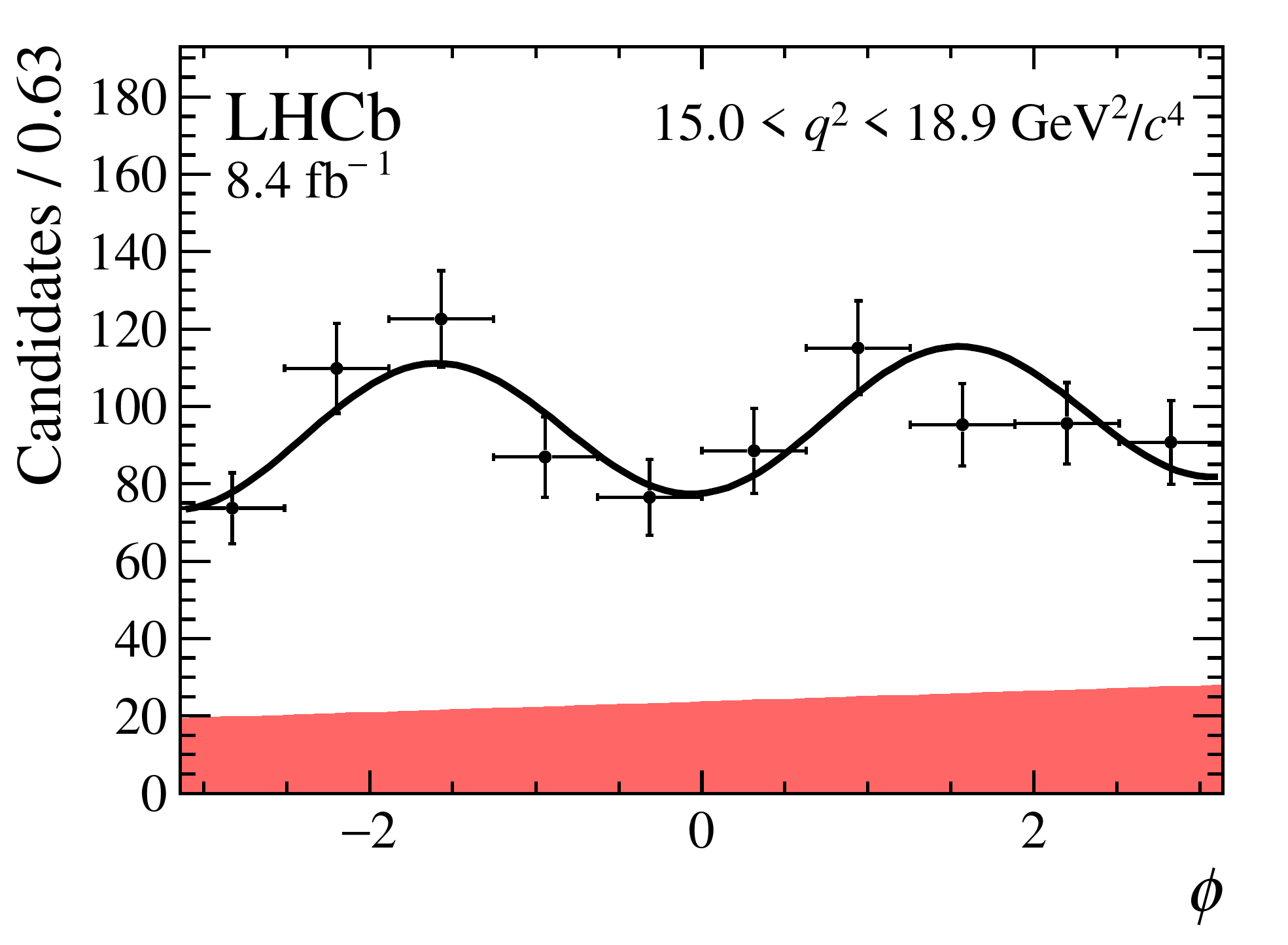}
    \includegraphics[width=.45\textwidth]{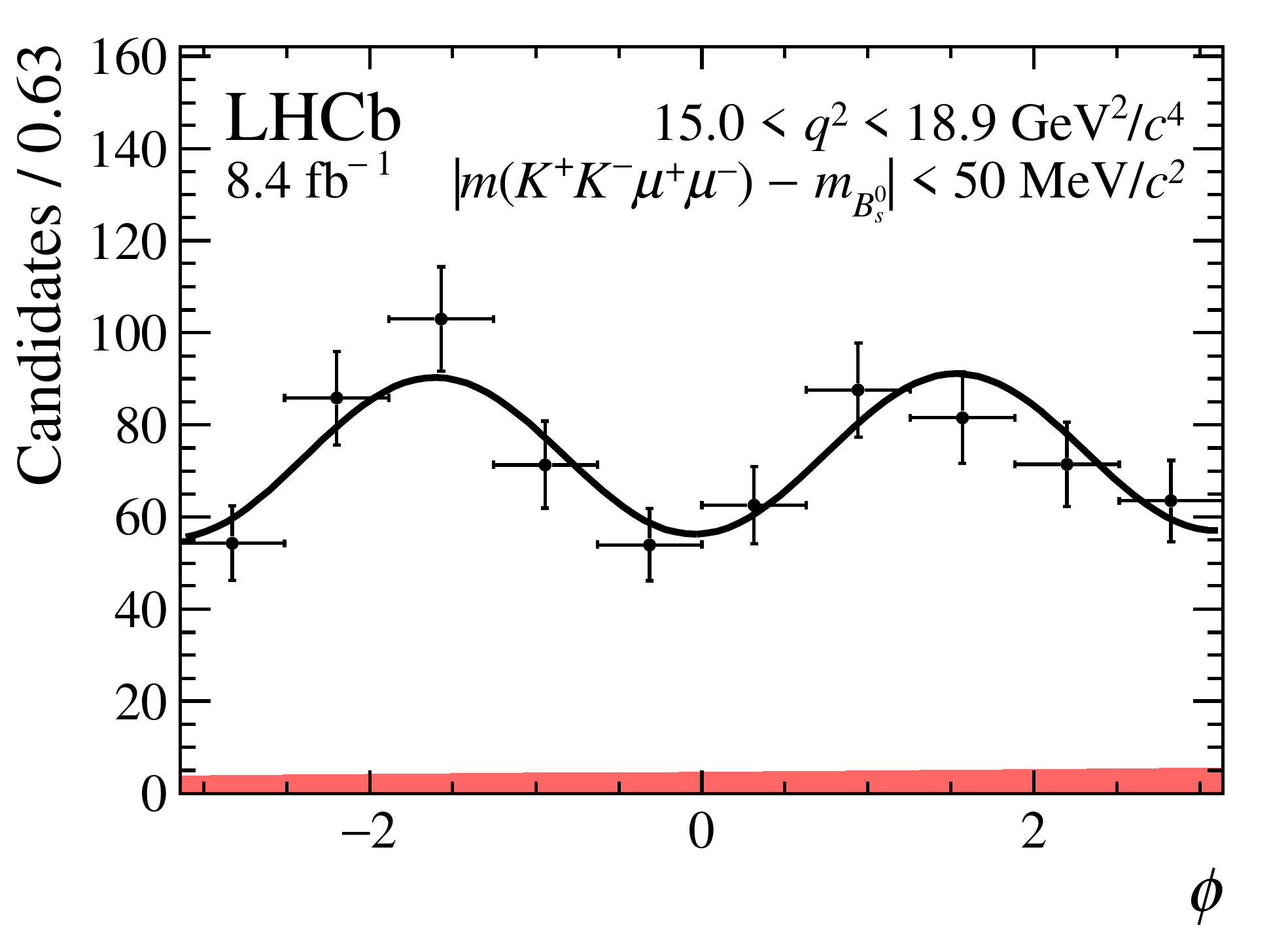}
    \caption{\label{fig:results_bin7_comb} Projections in the region \mbox{$15.0 < \qsq <18.9\gevgevcccc$} for the angular distributions of the combined 2011--2012, 2016 and 2017--2018 data sets. The data are overlaid with the projection of the combined PDF. The red shaded area indicates the background component and the solid black line the total PDF.
    The angular projections are given for candidates in (left) the entire mass region used to determine the observables in this paper and (right) the signal mass window $\pm50\mevcc$ around the known \Bs mass.}
\end{figure}

\begin{figure}[hb]
    \centering
    \includegraphics[width=.4\textwidth]{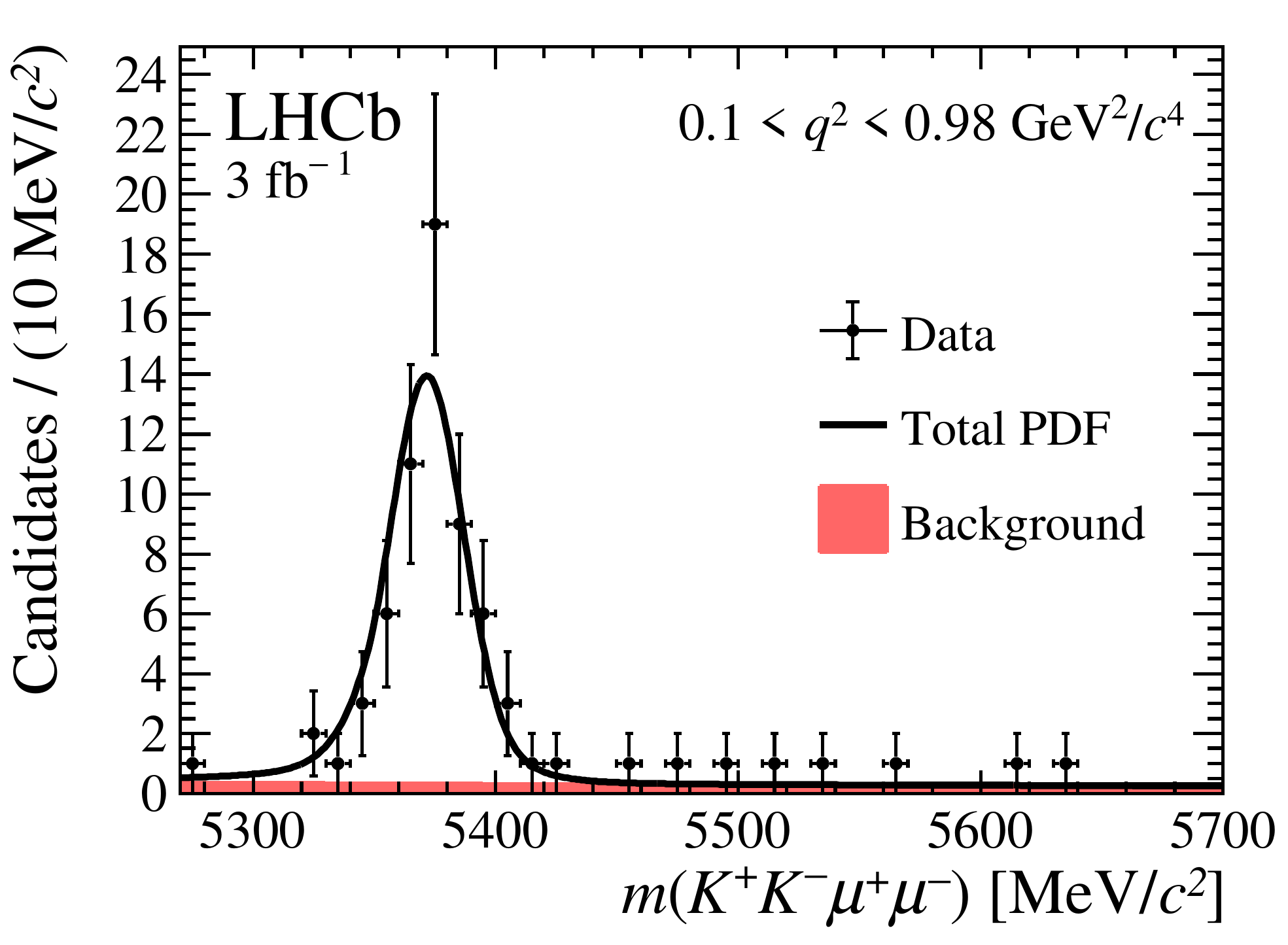}~
    \includegraphics[width=.4\textwidth]{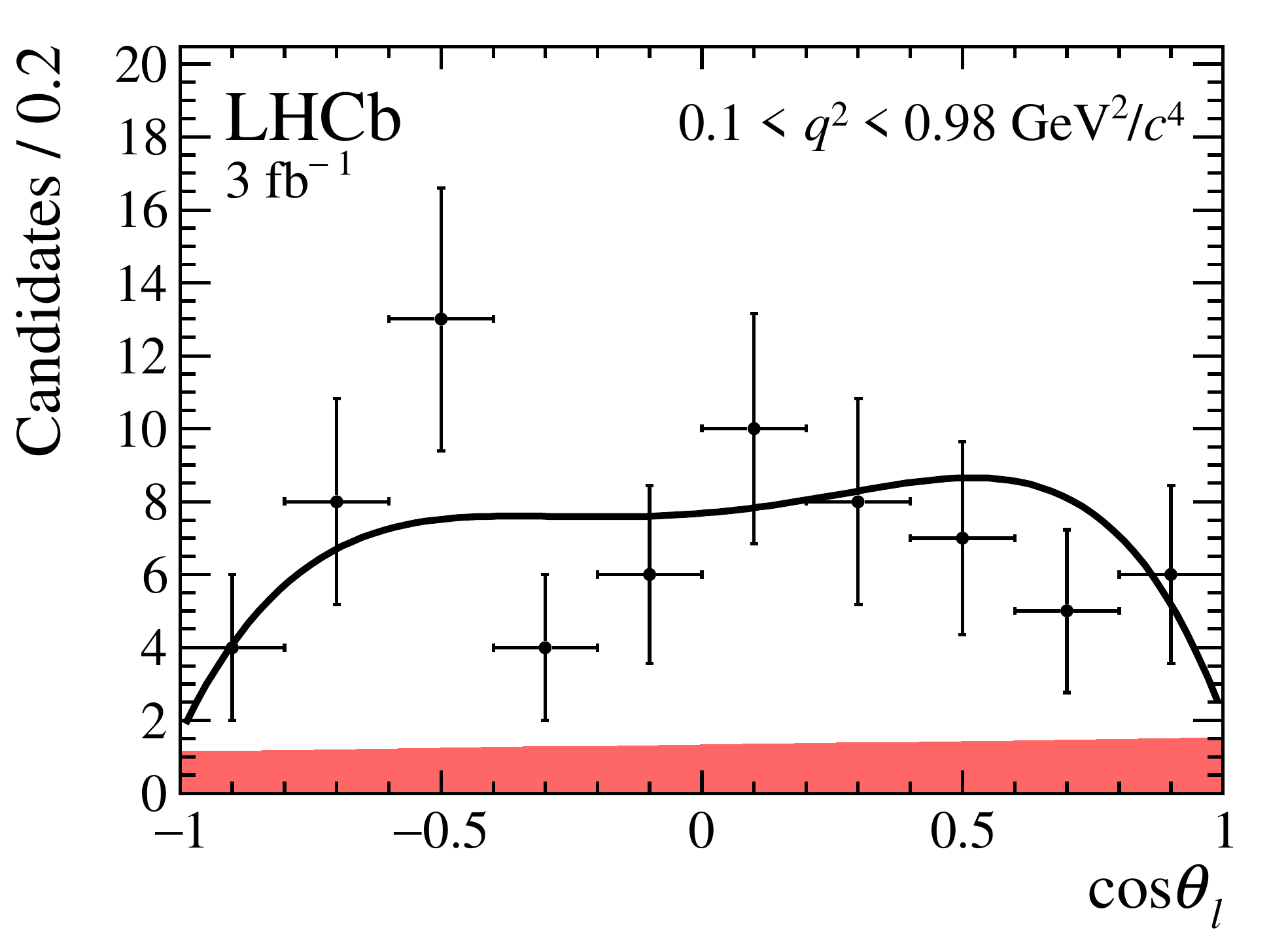}\\
    \includegraphics[width=.4\textwidth]{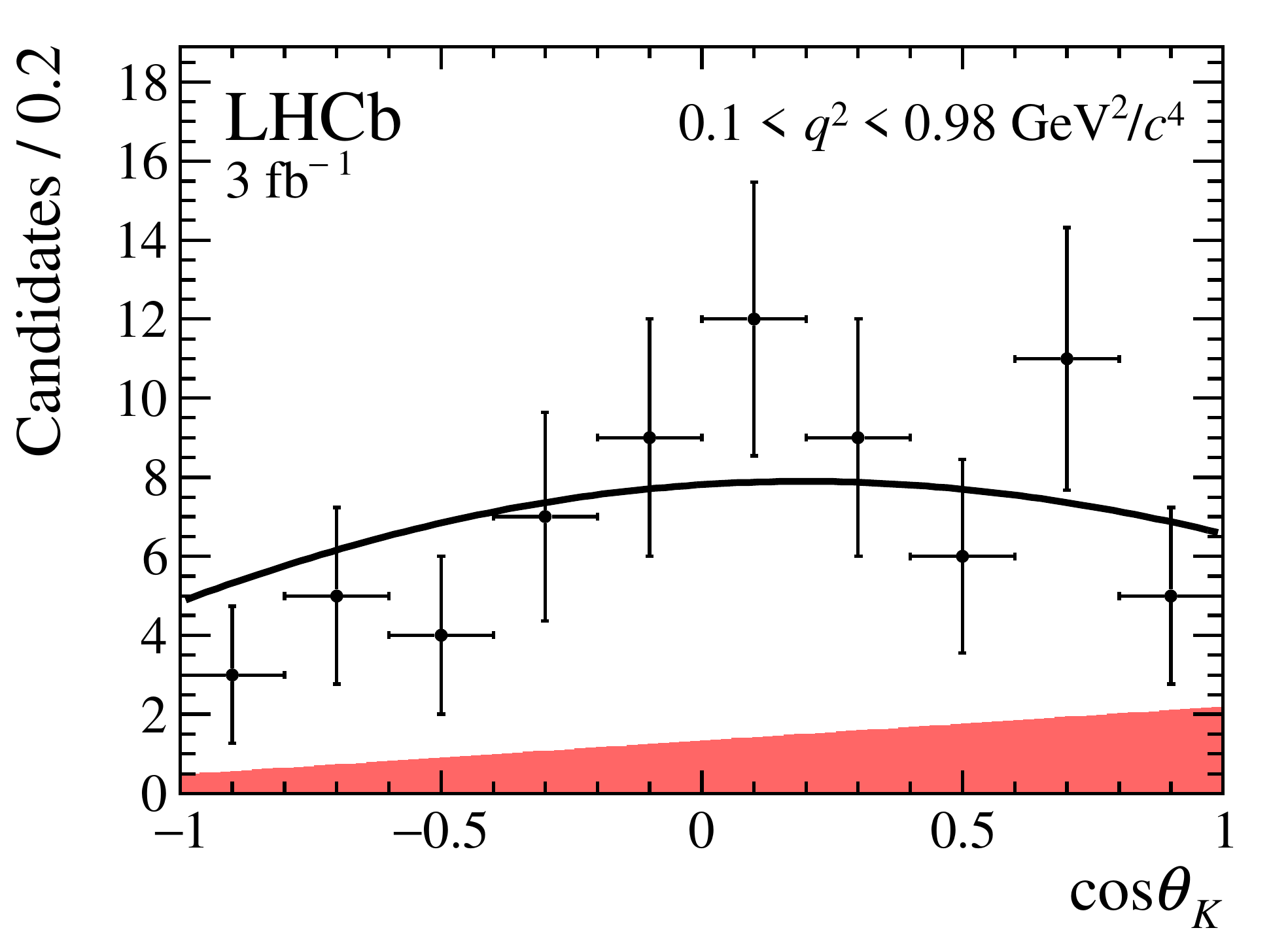}~
    \includegraphics[width=.4\textwidth]{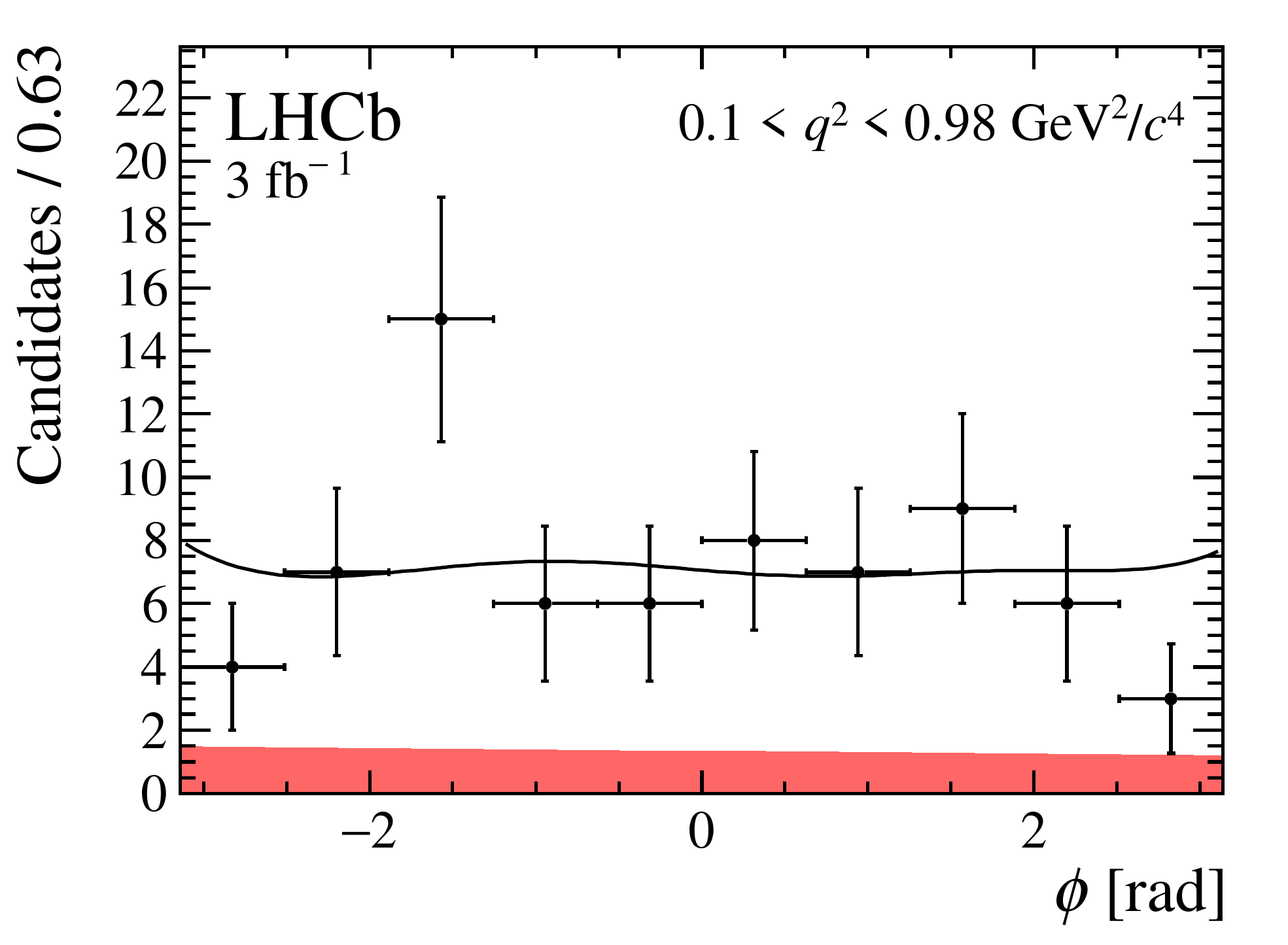}~
    \caption{\label{fig:results_bin1_run1} Mass and angular distributions of \BsToPhimm\ candidates in the region \mbox{$0.1<\qsq<0.98\gevgevcccc$} for data taken in 2011--2012. The data are overlaid with the projections of the fitted PDF. }
\end{figure}

\begin{figure}[hb]
    \centering
    \includegraphics[width=.4\textwidth]{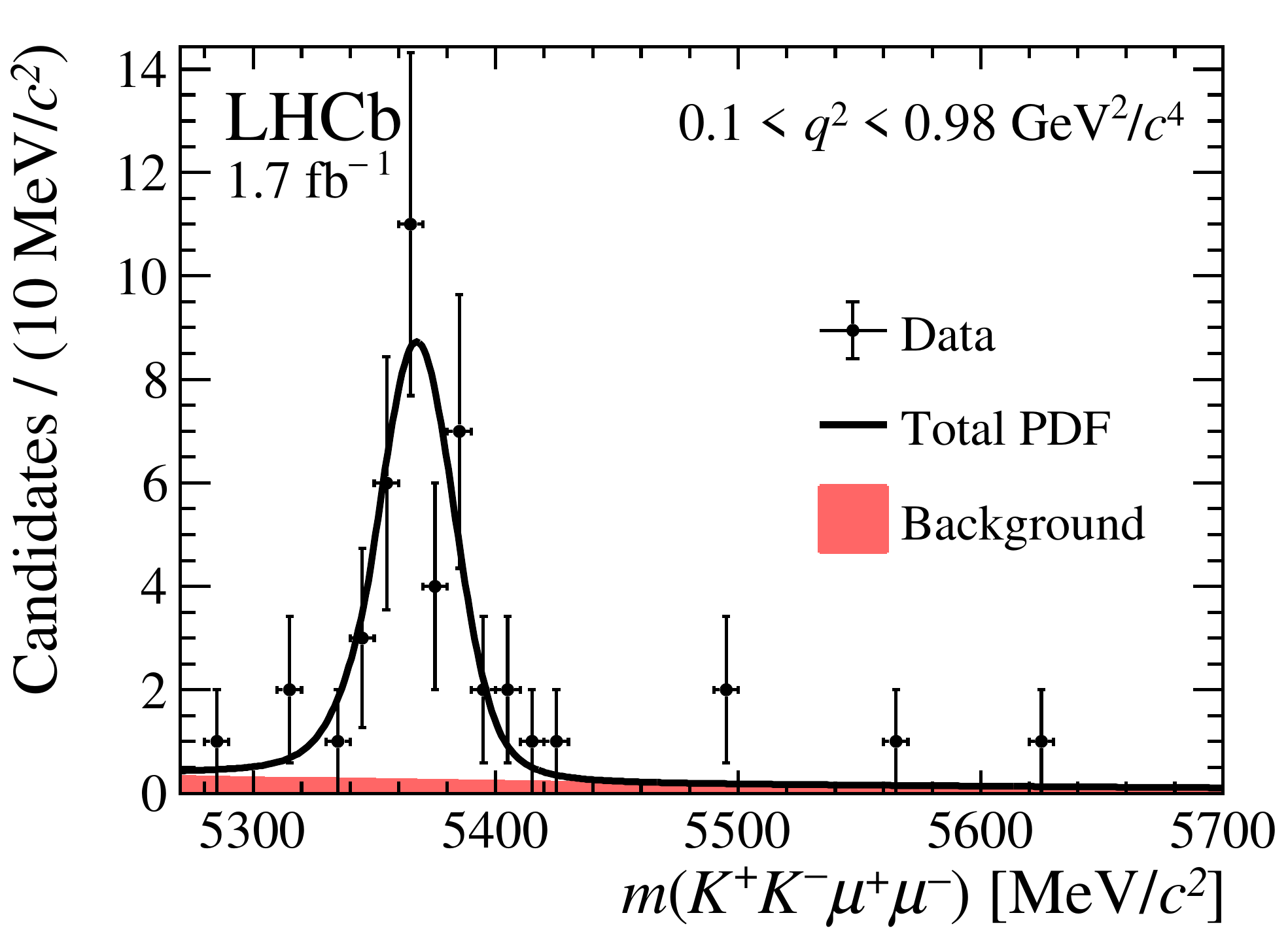}~
    \includegraphics[width=.4\textwidth]{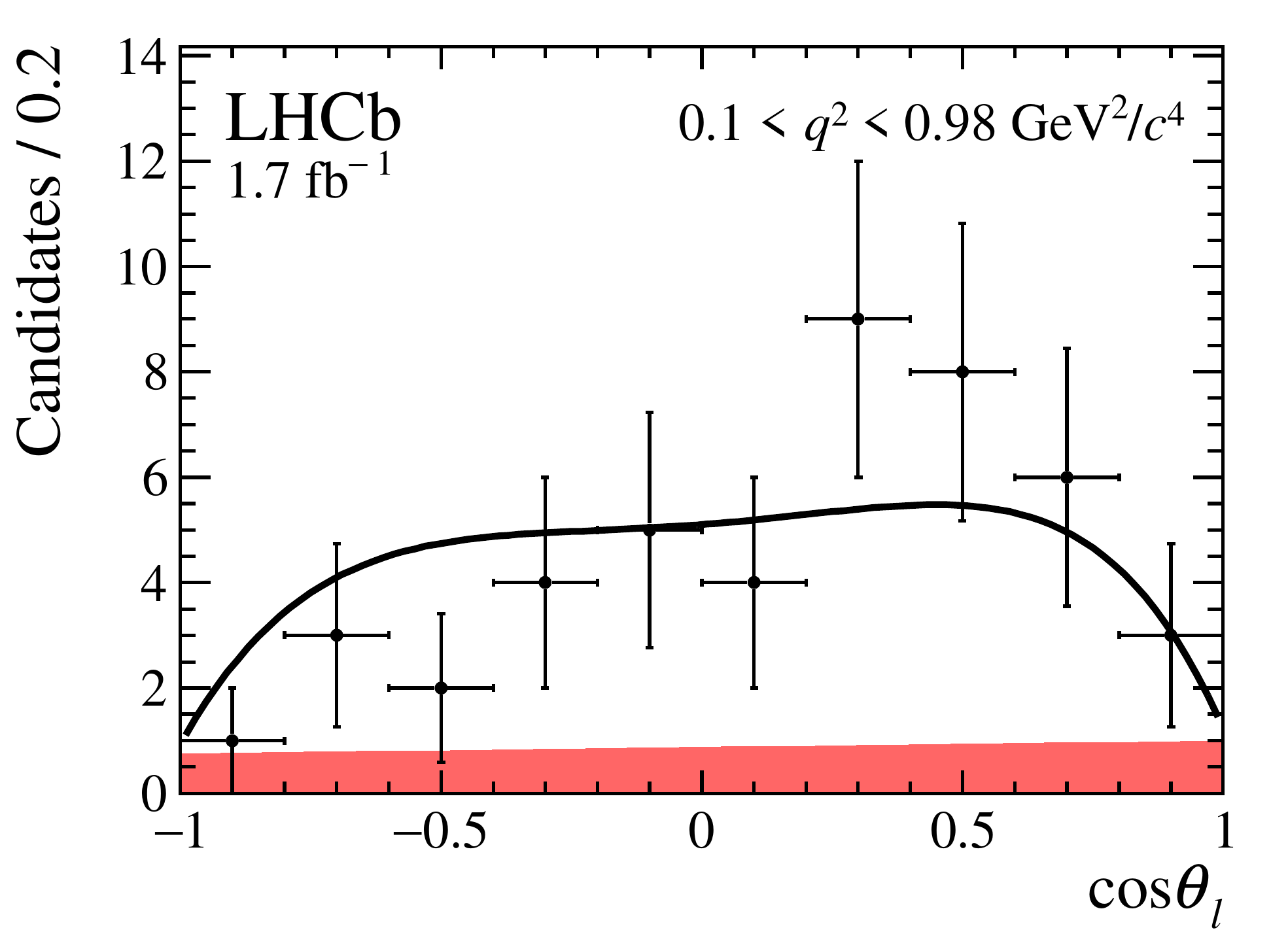}\\
    \includegraphics[width=.4\textwidth]{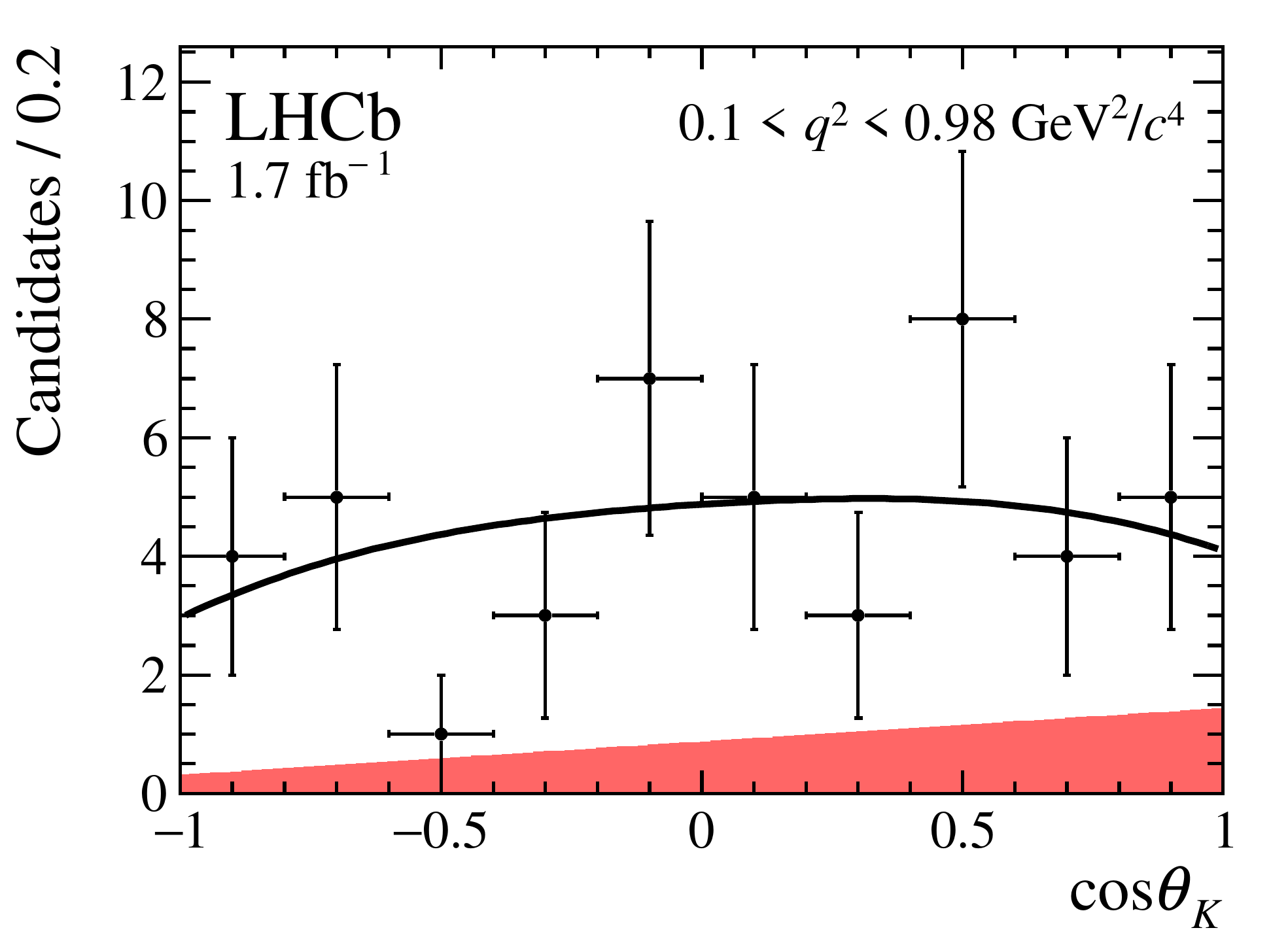}~
    \includegraphics[width=.4\textwidth]{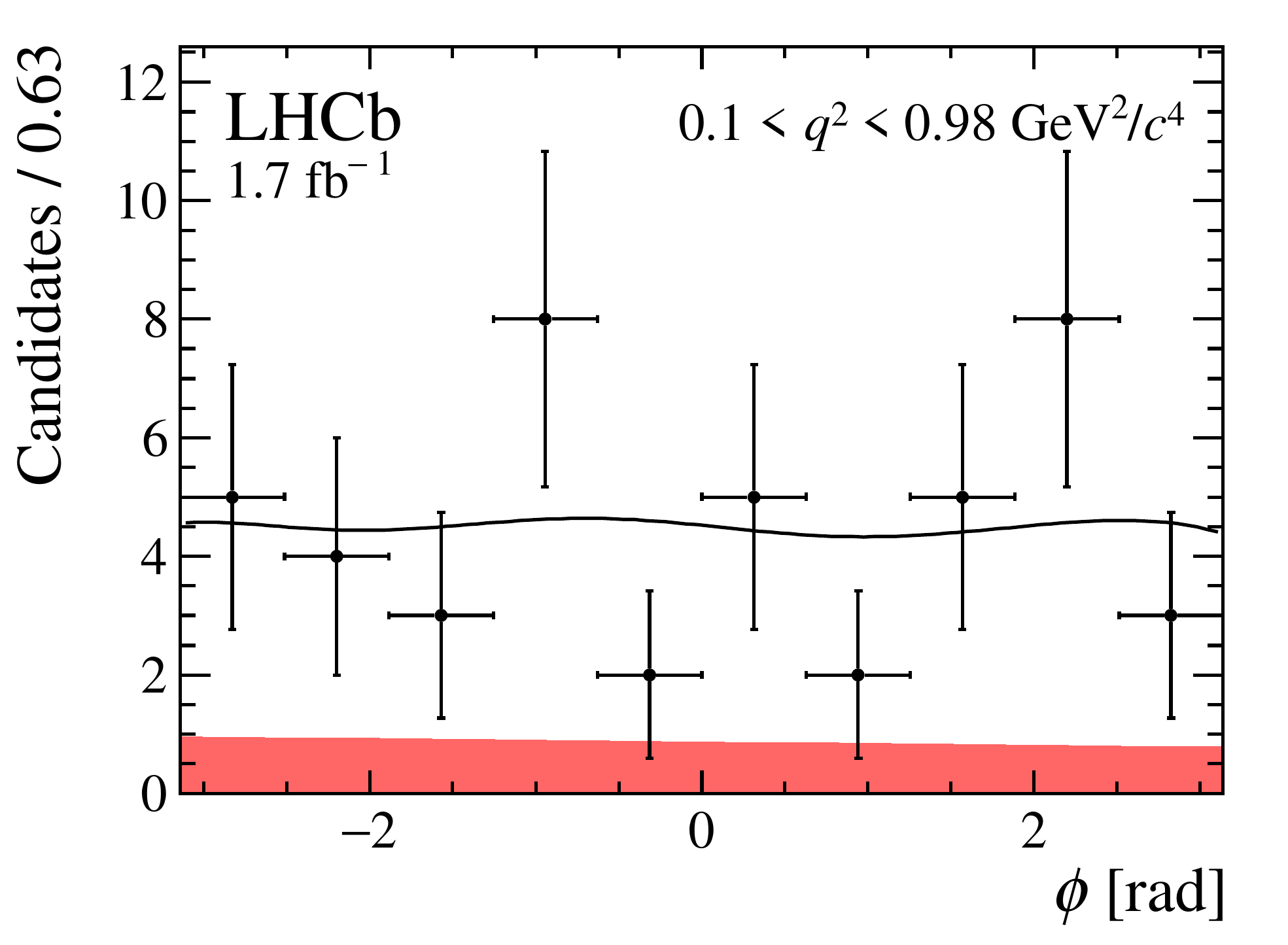}~
    \caption{\label{fig:results_bin1_run2p1} Mass and angular distributions of \BsToPhimm\ candidates in the region \mbox{$0.1<\qsq<0.98\gevgevcccc$} for data taken in 2016. The data are overlaid with the projections of the fitted PDF. }
\end{figure}

\begin{figure}[hb]
    \centering
    \includegraphics[width=.4\textwidth]{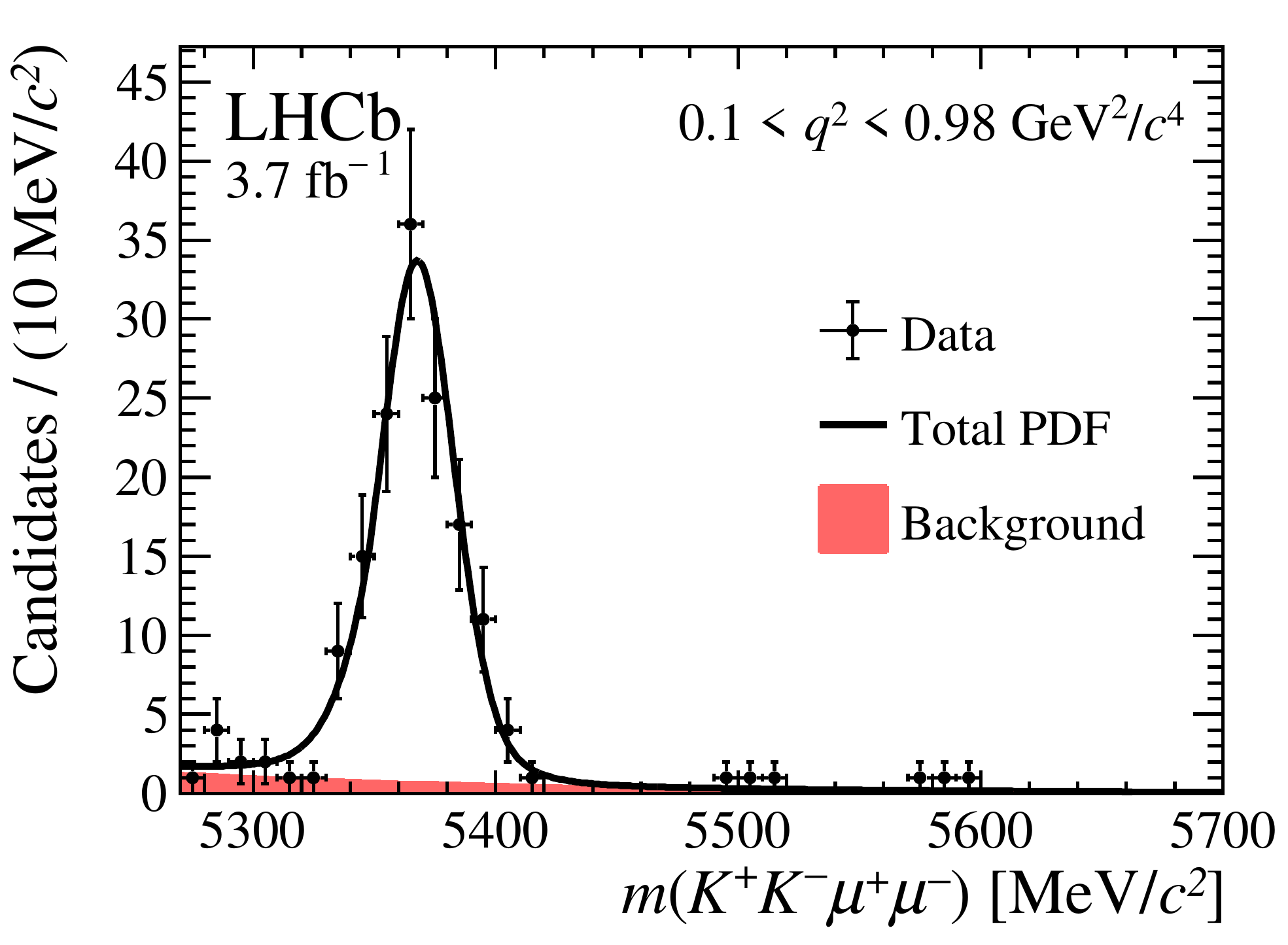}~
    \includegraphics[width=.4\textwidth]{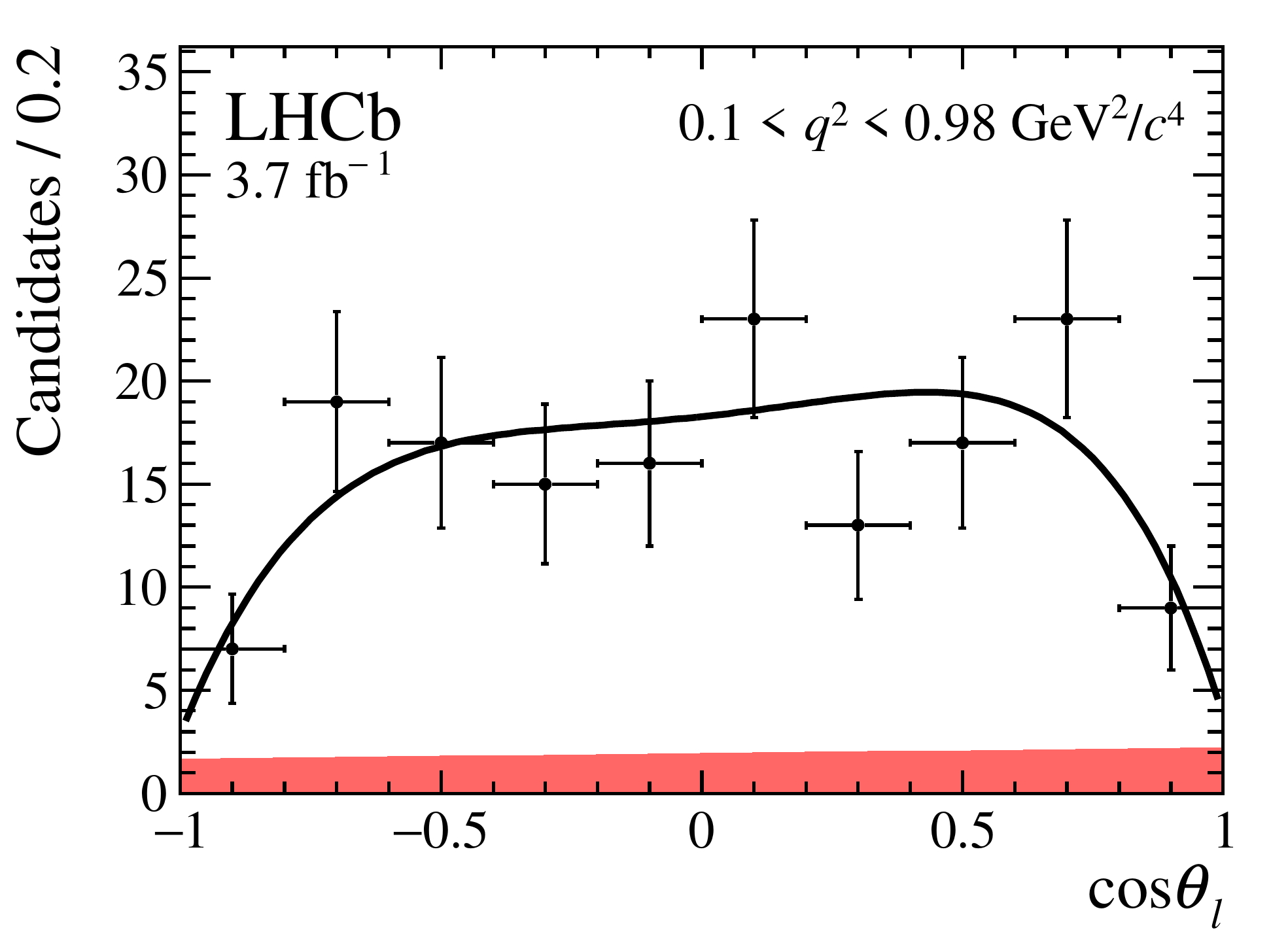}\\
    \includegraphics[width=.4\textwidth]{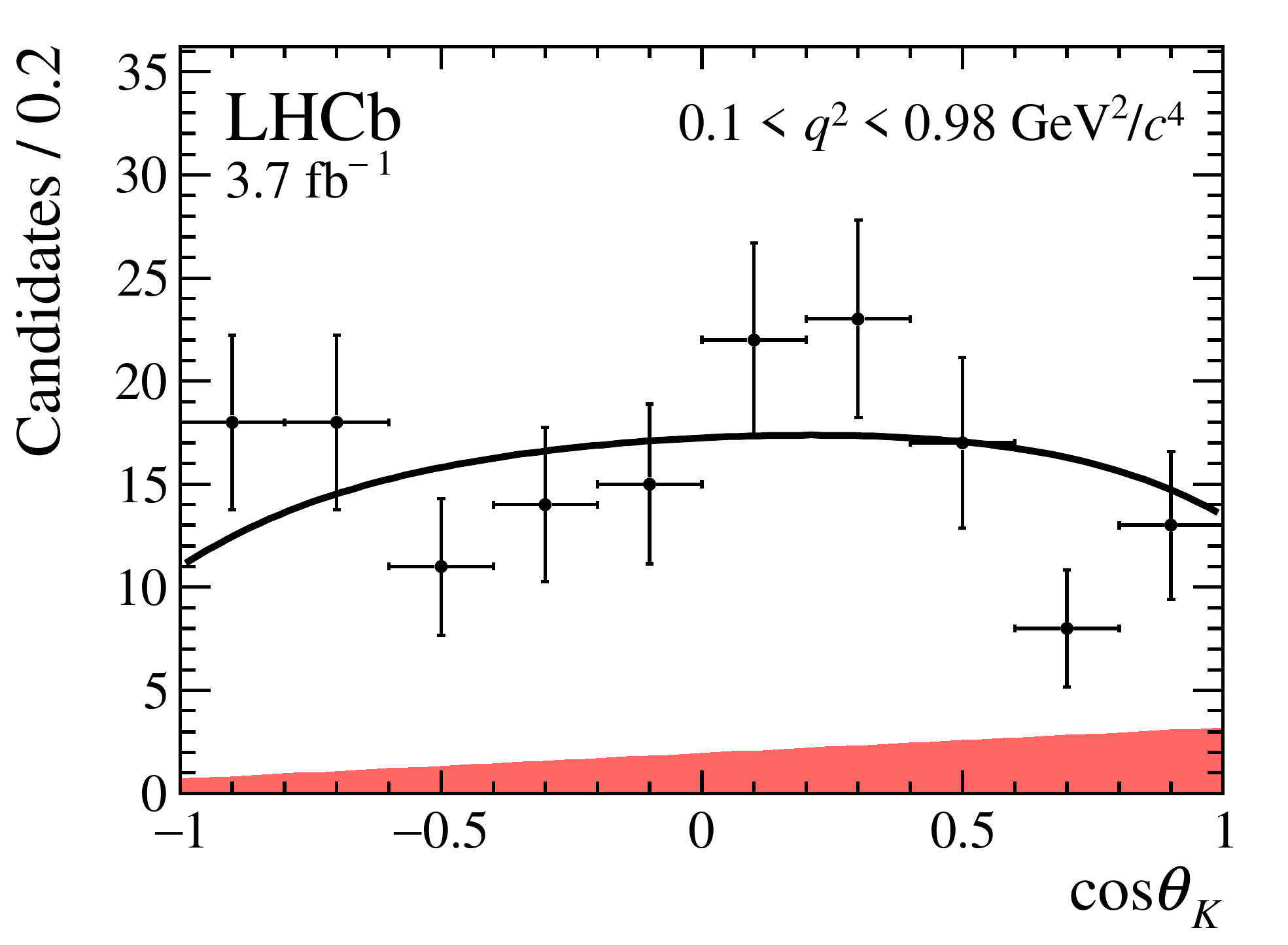}~
    \includegraphics[width=.4\textwidth]{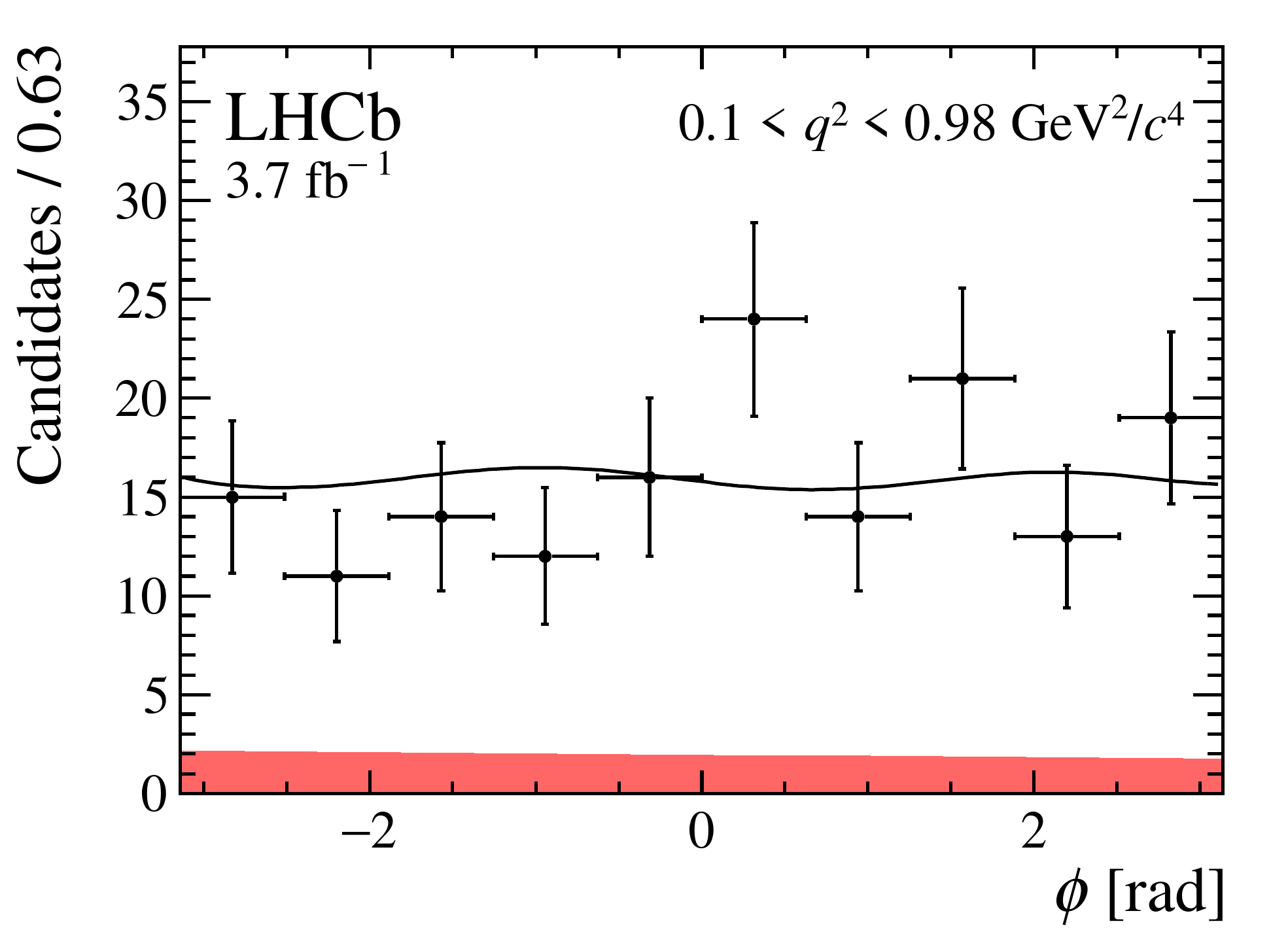}~
    \caption{\label{fig:results_bin1_run2p2} Mass and angular distributions of \BsToPhimm\ candidates in the region \mbox{$0.1<\qsq<0.98\gevgevcccc$} for data taken in 2017--2018. The data are overlaid with the projections of the fitted PDF.}
\end{figure}

\begin{figure}[hb]
    \centering
    \includegraphics[width=.4\textwidth]{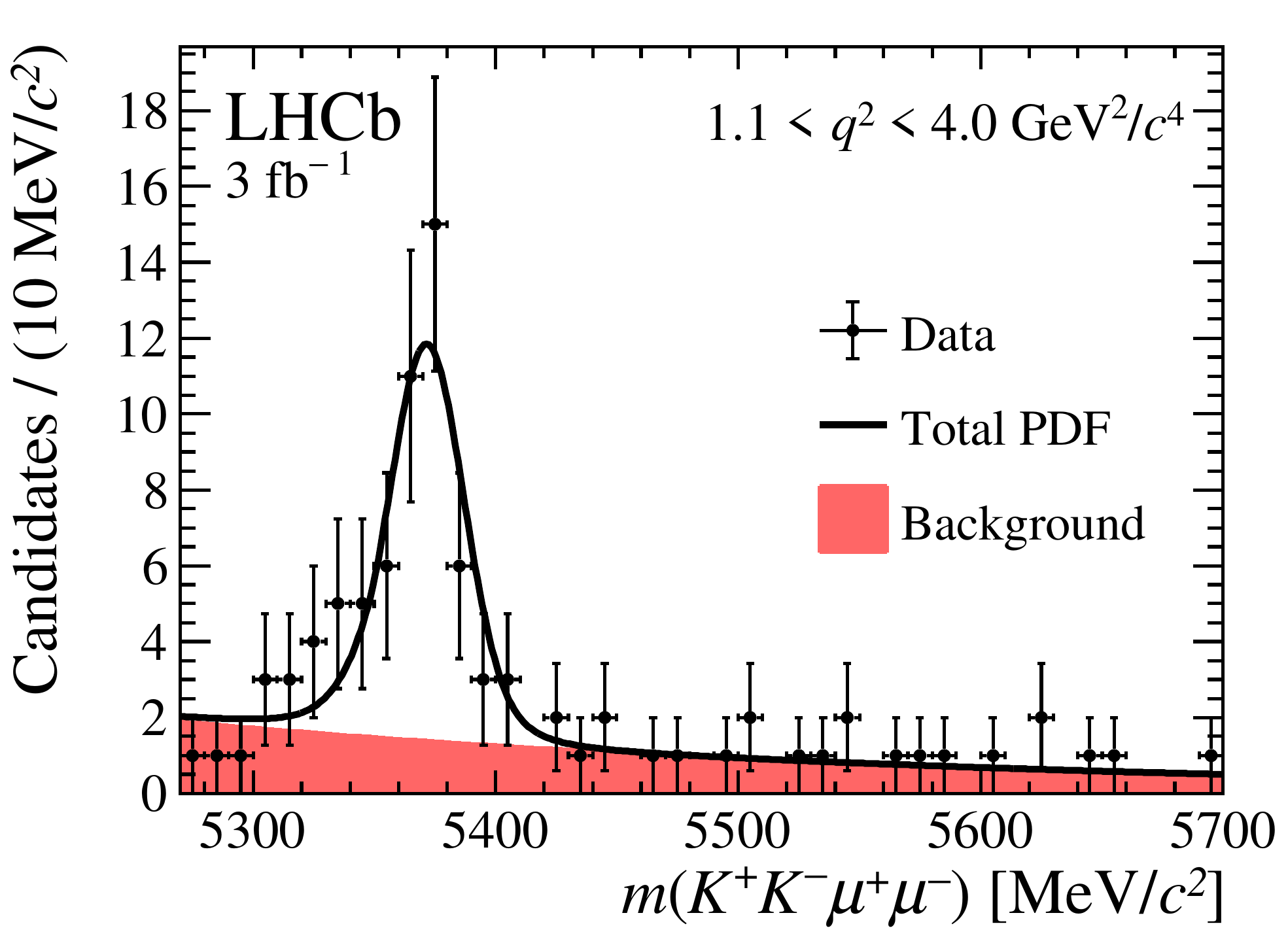}~
    \includegraphics[width=.4\textwidth]{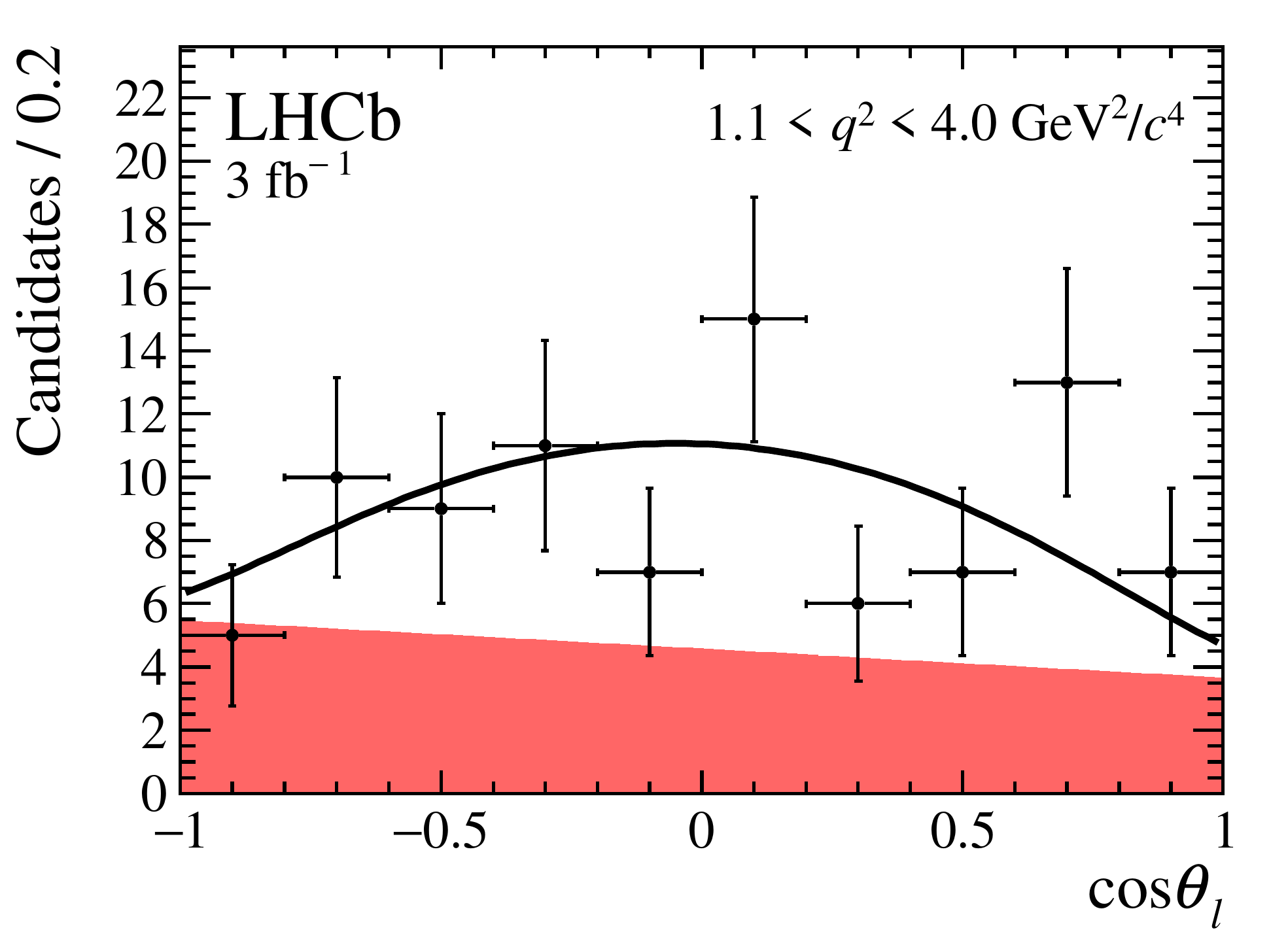}\\
    \includegraphics[width=.4\textwidth]{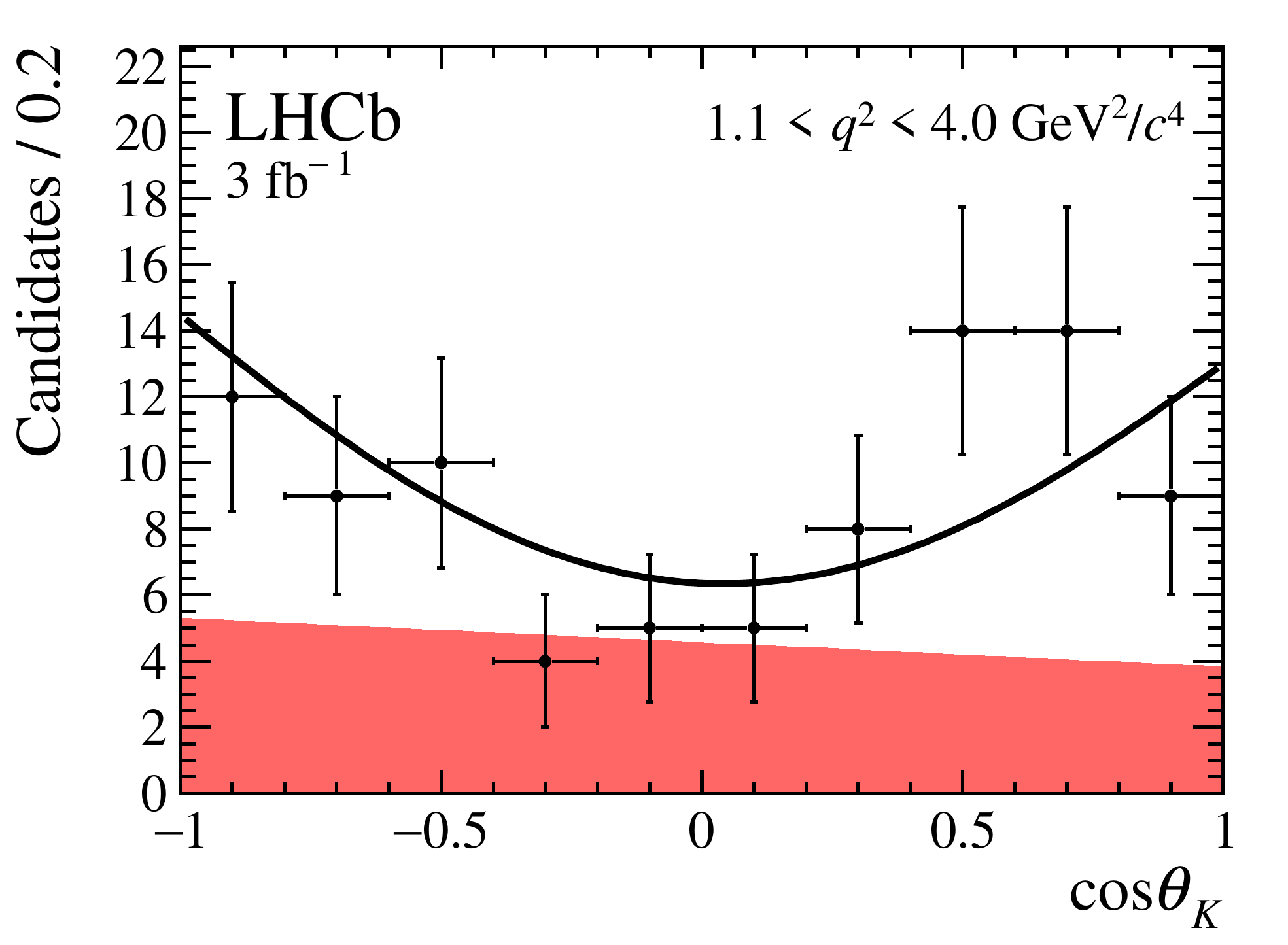}~
    \includegraphics[width=.4\textwidth]{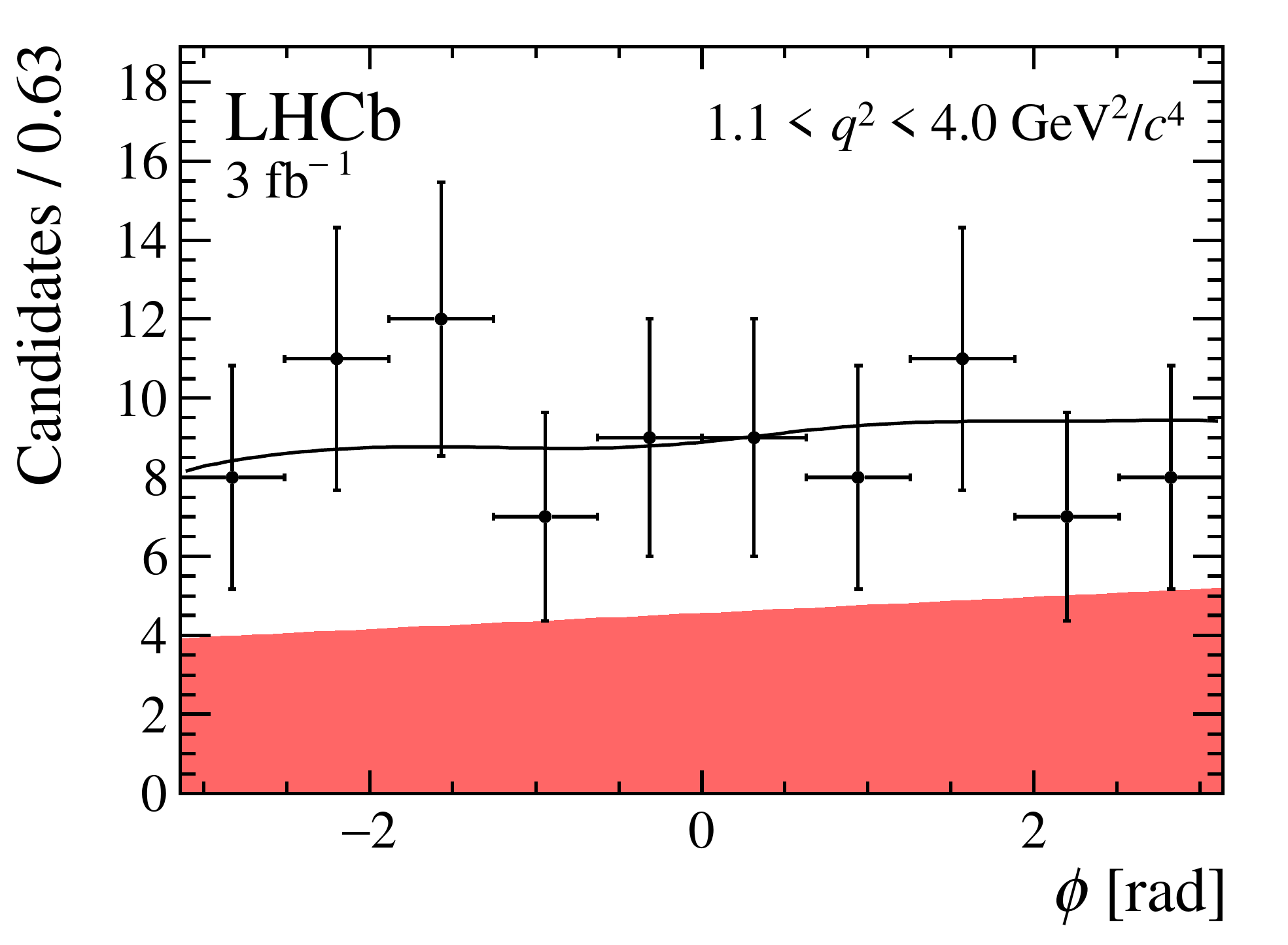}~
    \caption{\label{fig:results_bin2_run1} Mass and angular distributions of \BsToPhimm\ candidates in the region \mbox{$1.1<\qsq<4.0\gevgevcccc$} for data taken in 2011--2012. The data are overlaid with the projections of the fitted PDF.}
\end{figure}

\begin{figure}[hb]
    \centering
    \includegraphics[width=.4\textwidth]{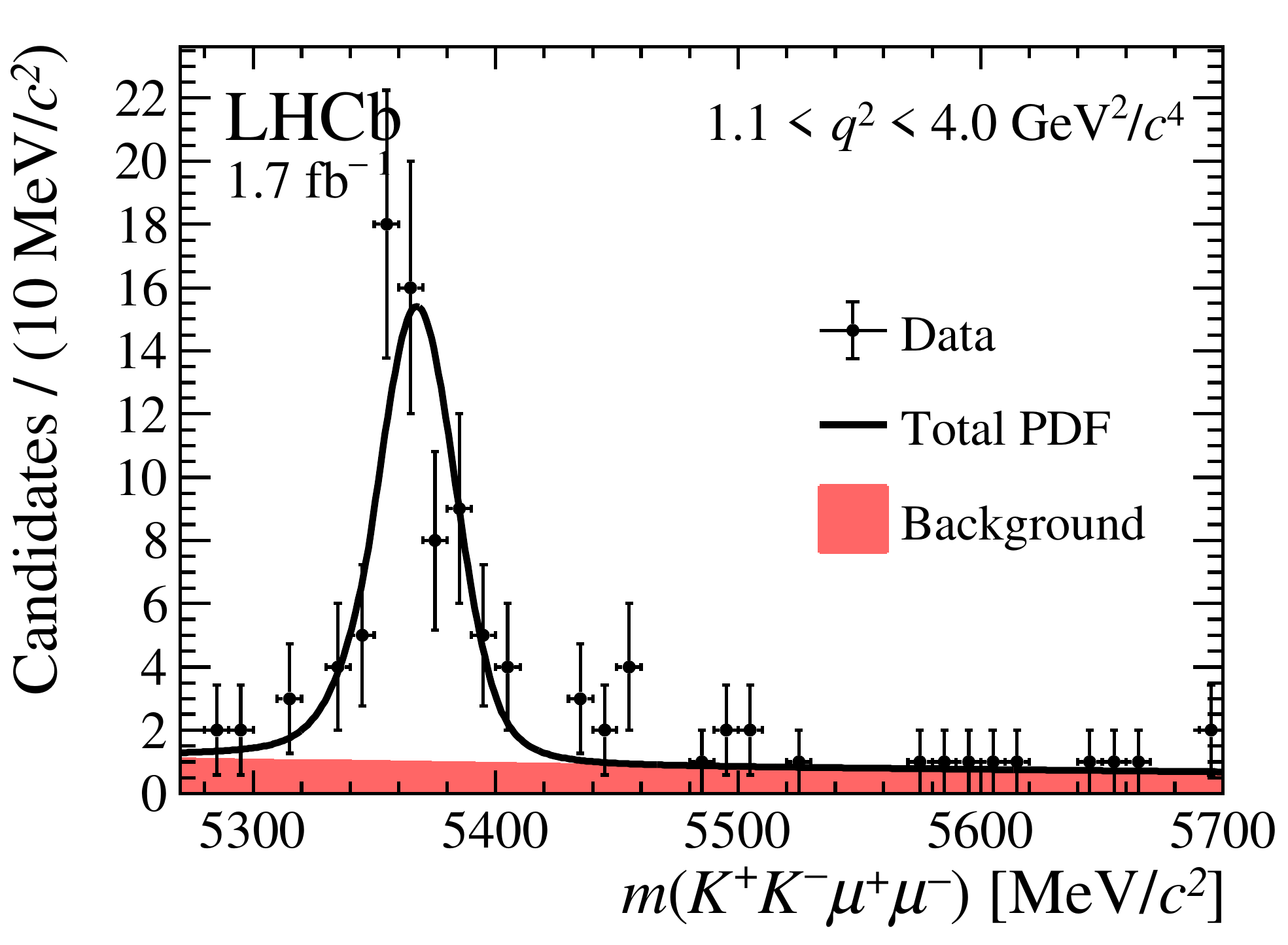}~
    \includegraphics[width=.4\textwidth]{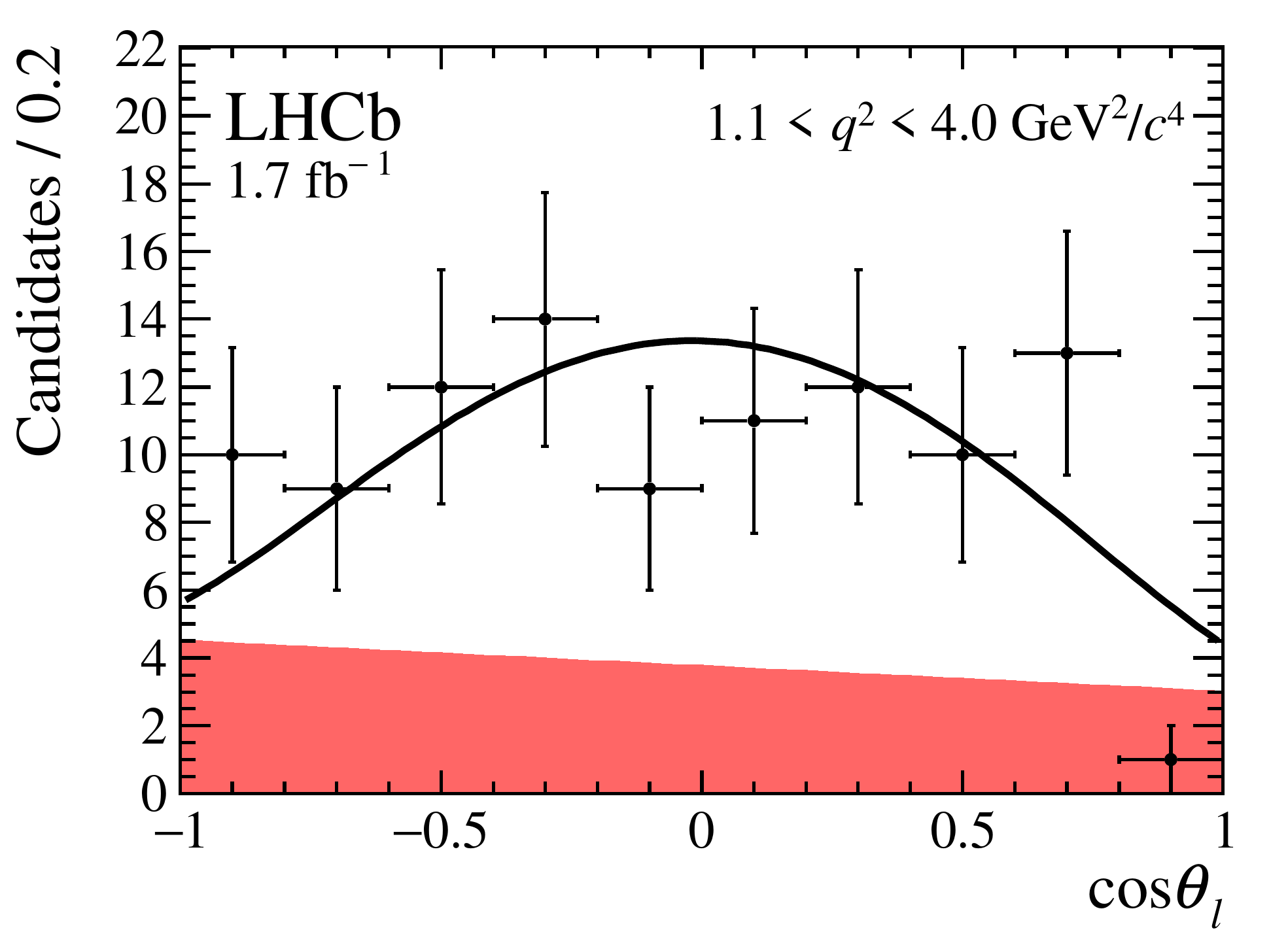}\\
    \includegraphics[width=.4\textwidth]{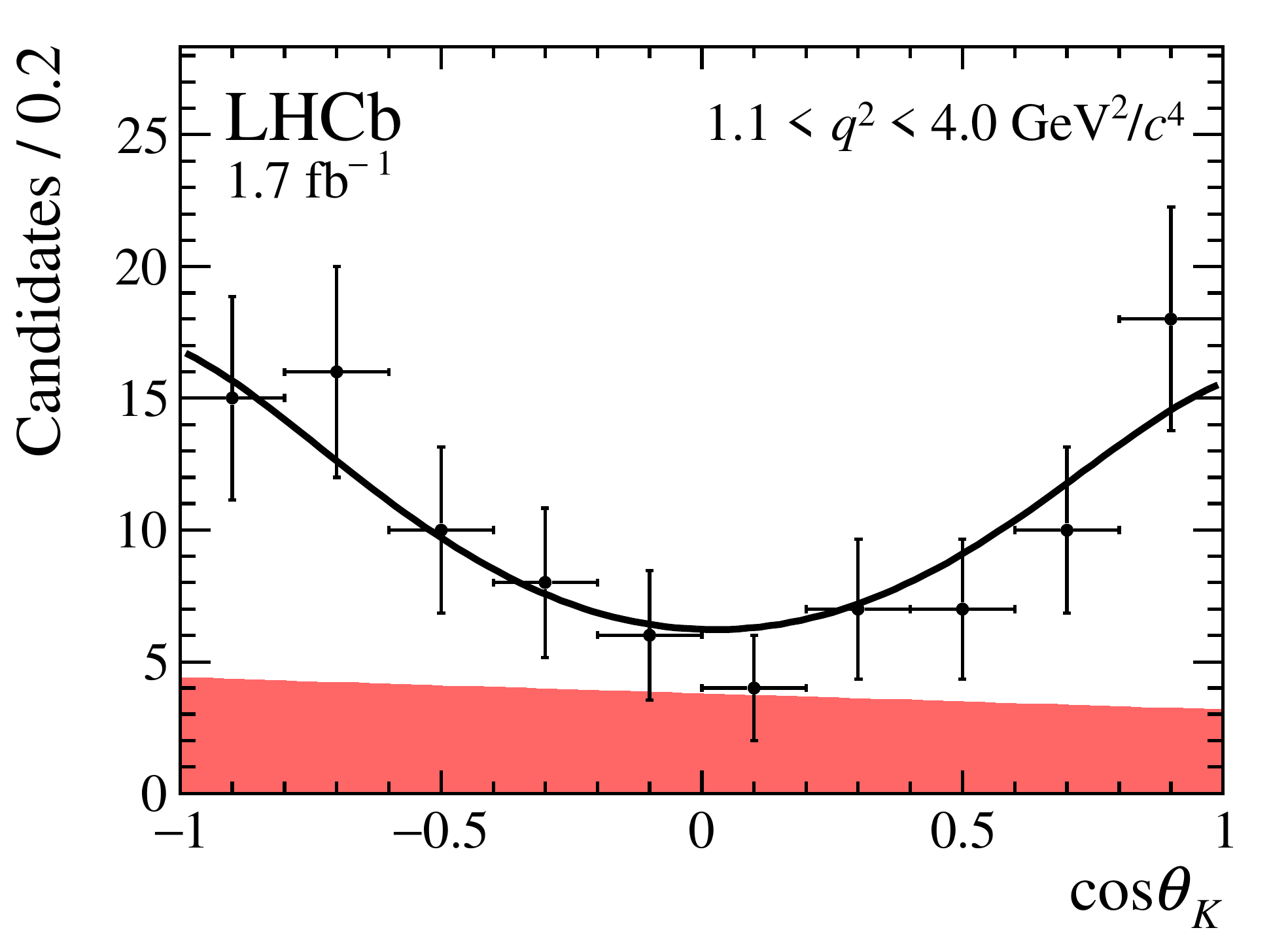}~
    \includegraphics[width=.4\textwidth]{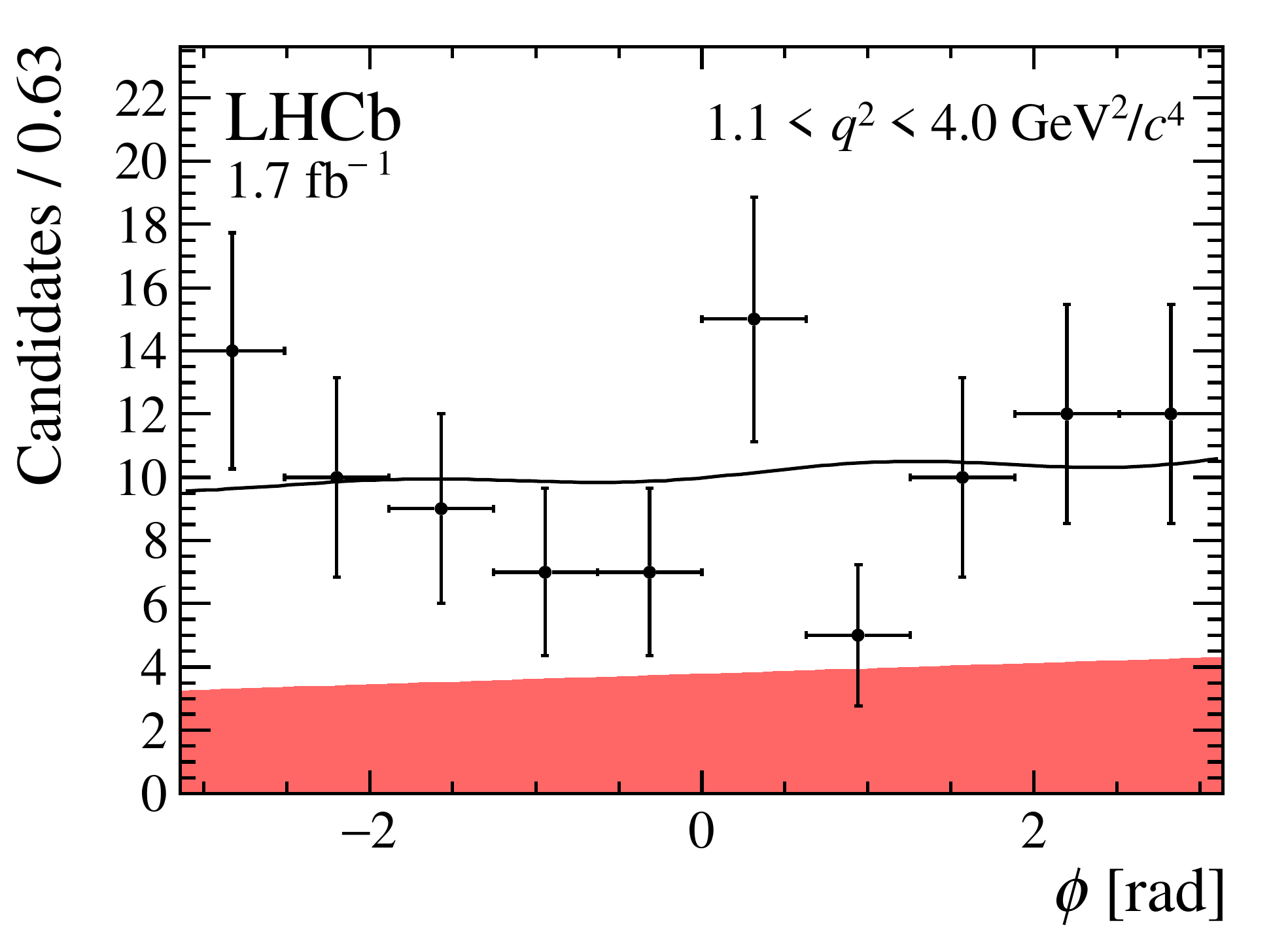}~
    \caption{\label{fig:results_bin2_run2p1} Mass and angular distributions of \BsToPhimm\ candidates in the region \mbox{$1.1<\qsq<4.0\gevgevcccc$} for data taken in 2016. The data are overlaid with the projections of the fitted PDF.}
\end{figure}

\begin{figure}[hb]
    \centering
    \includegraphics[width=.4\textwidth]{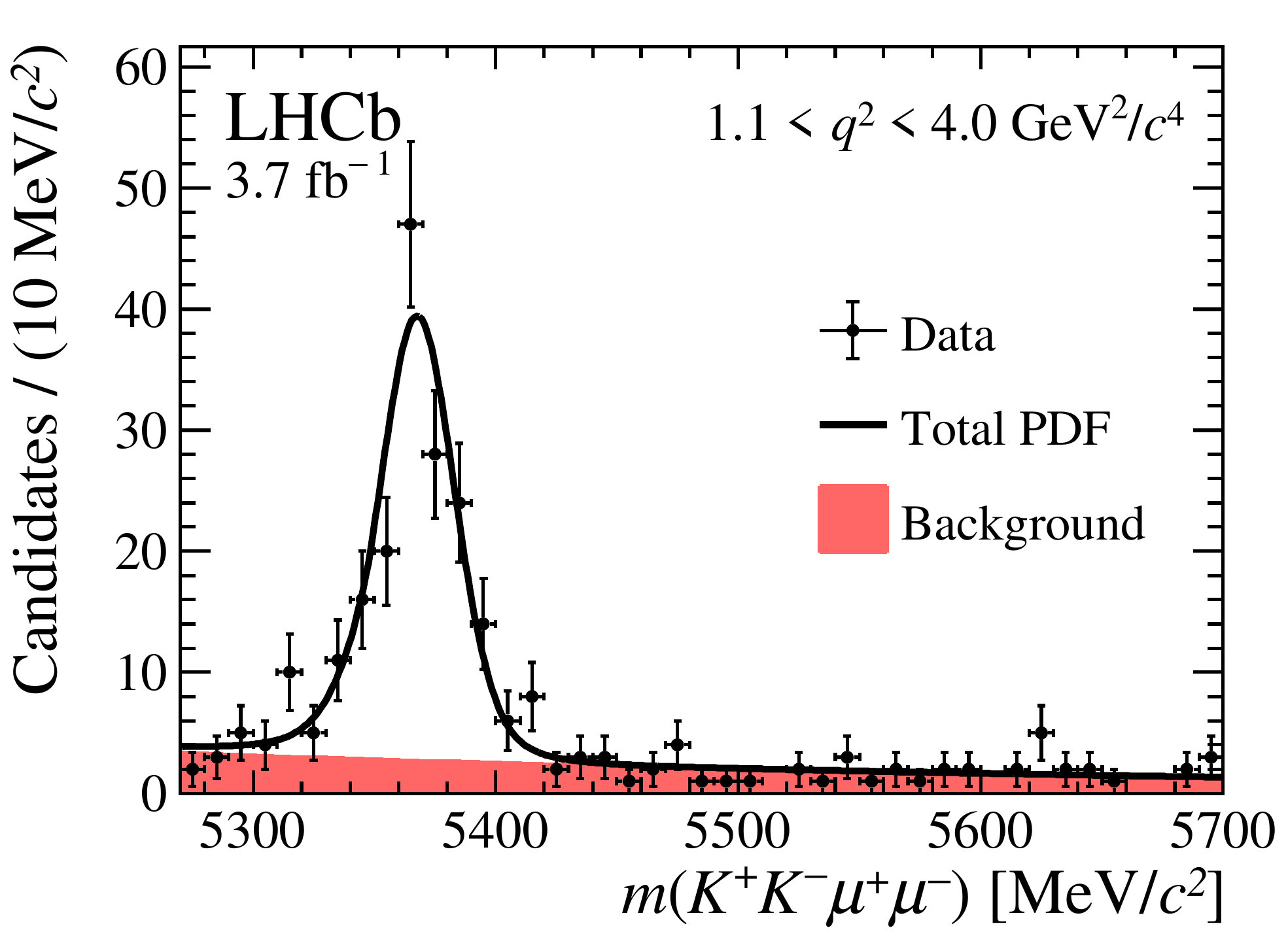}~
    \includegraphics[width=.4\textwidth]{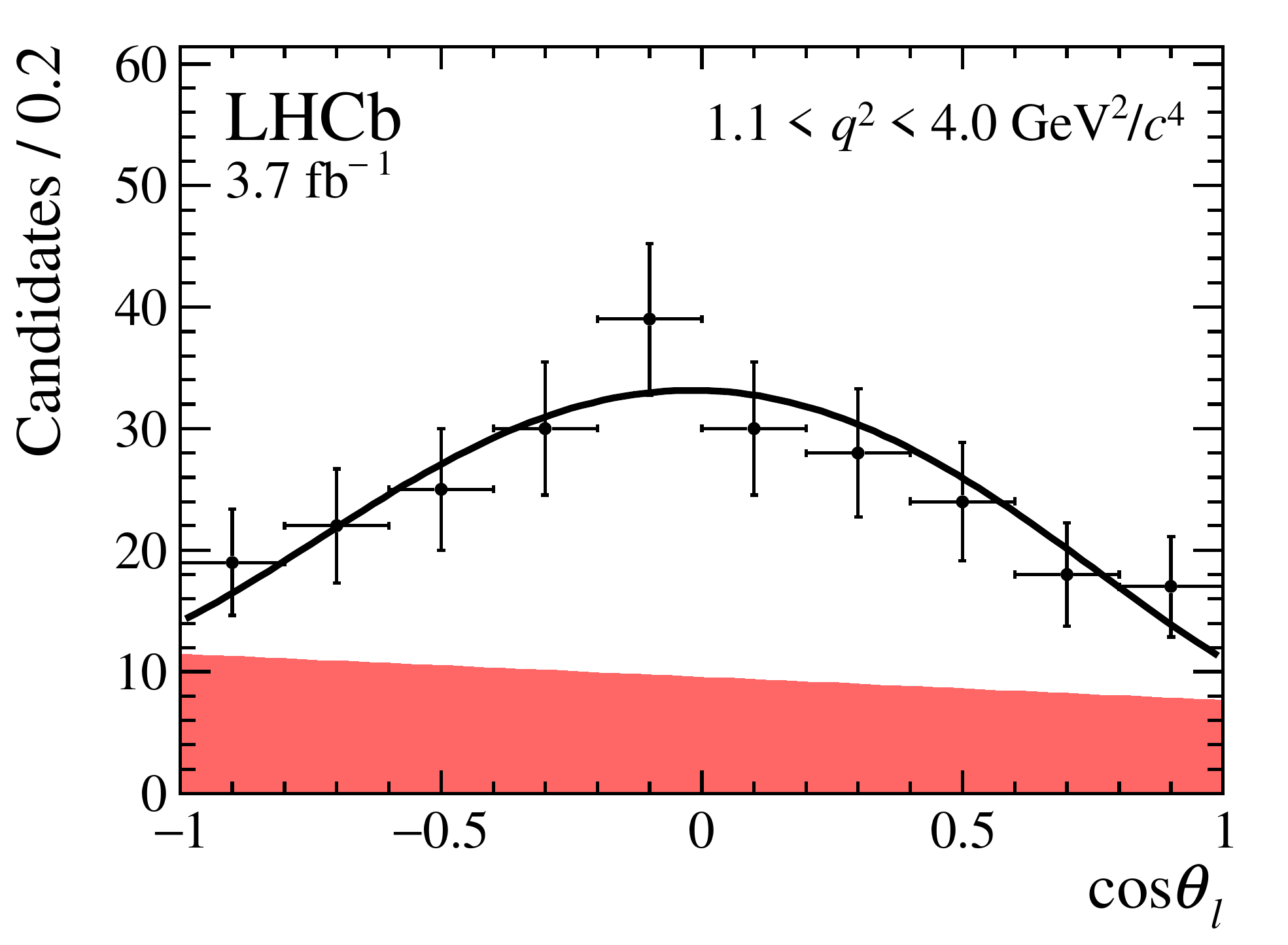}\\
    \includegraphics[width=.4\textwidth]{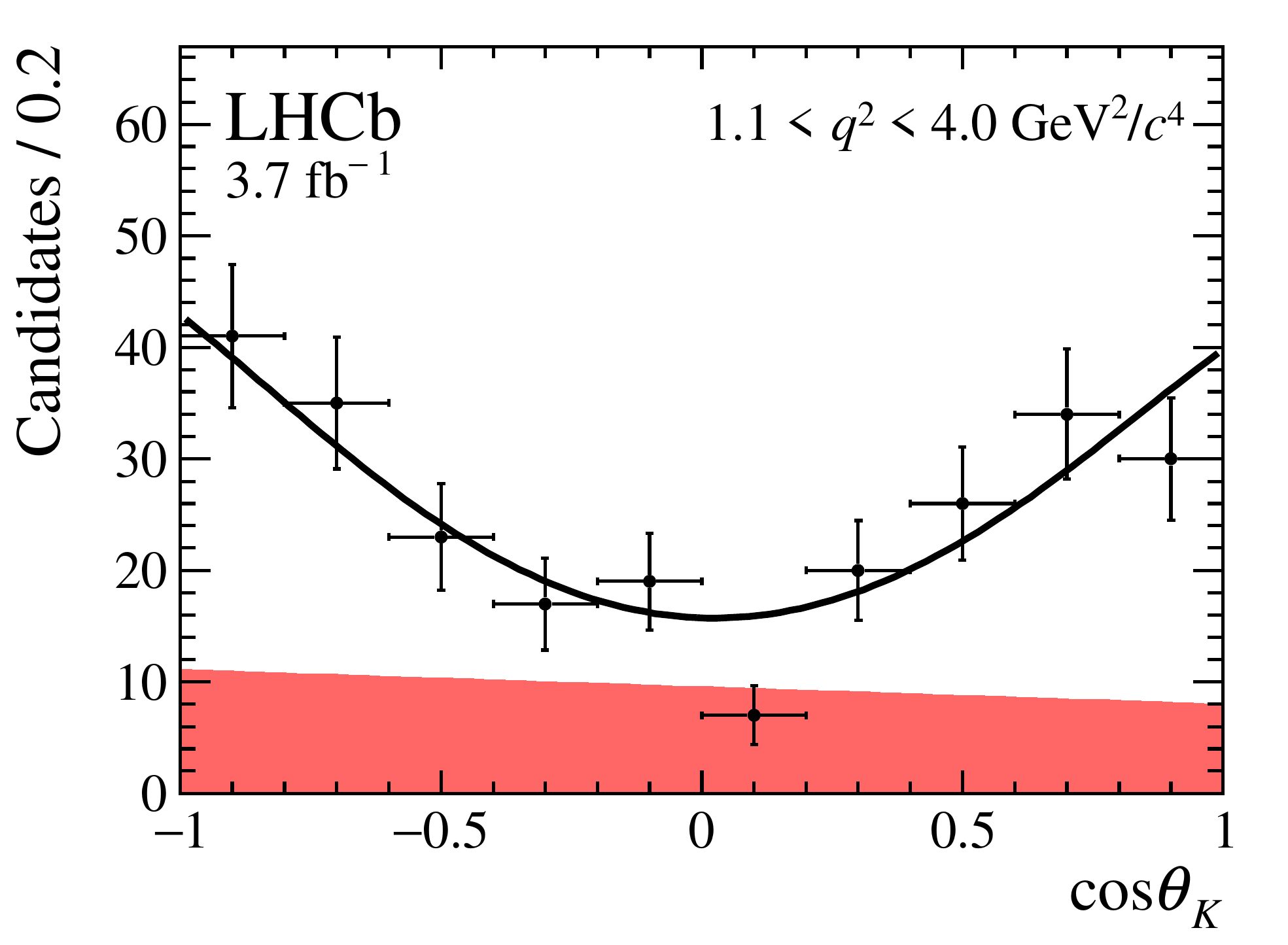}~
    \includegraphics[width=.4\textwidth]{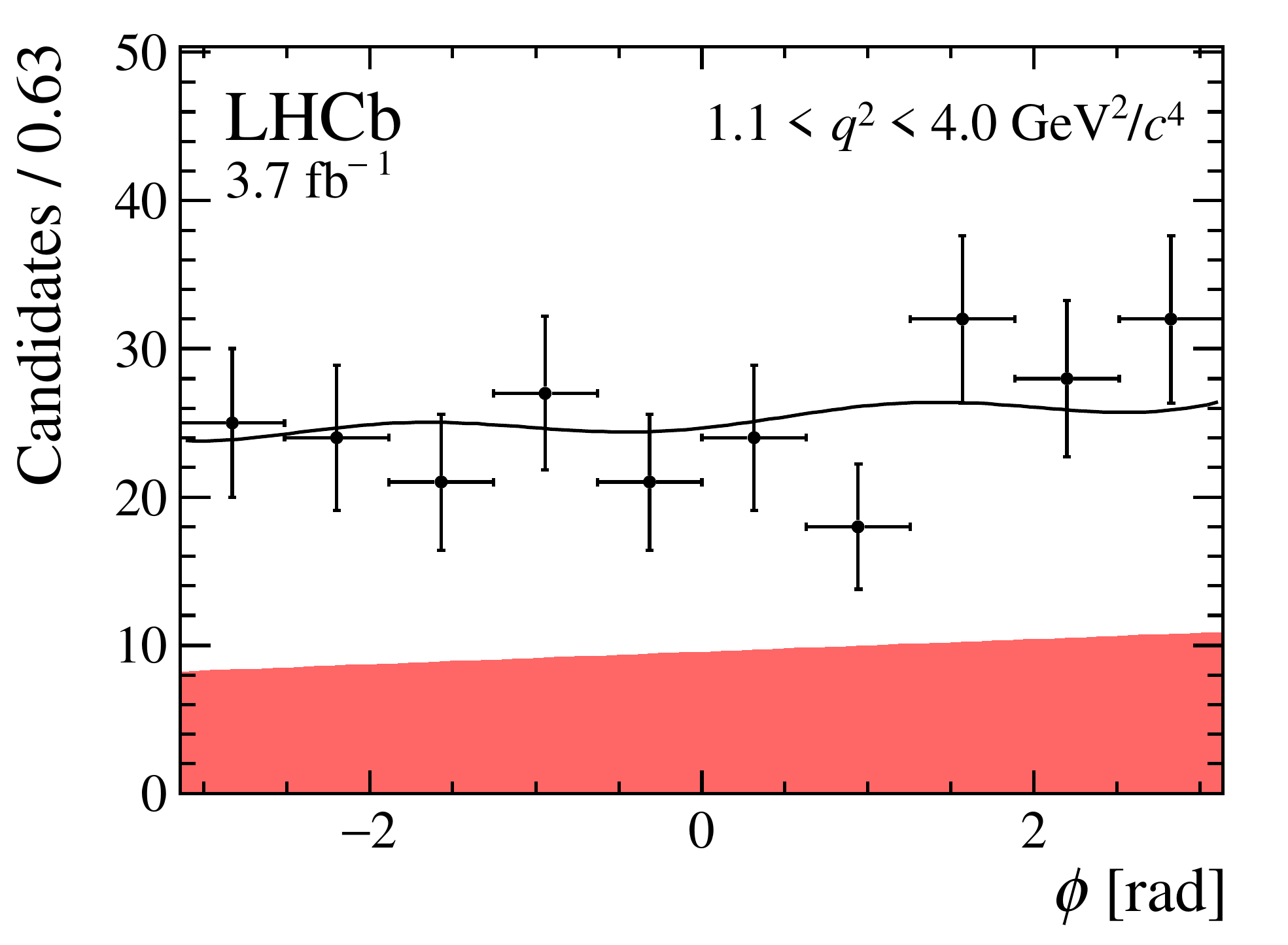}~
    \caption{\label{fig:results_bin2_run2p2} Mass and angular distributions of \BsToPhimm\ candidates in the region \mbox{$1.1<\qsq<4.0\gevgevcccc$} for data taken in 2017--2018. The data are overlaid with the projections of the fitted PDF.}
\end{figure}

\begin{figure}[hb]
    \centering
    \includegraphics[width=.4\textwidth]{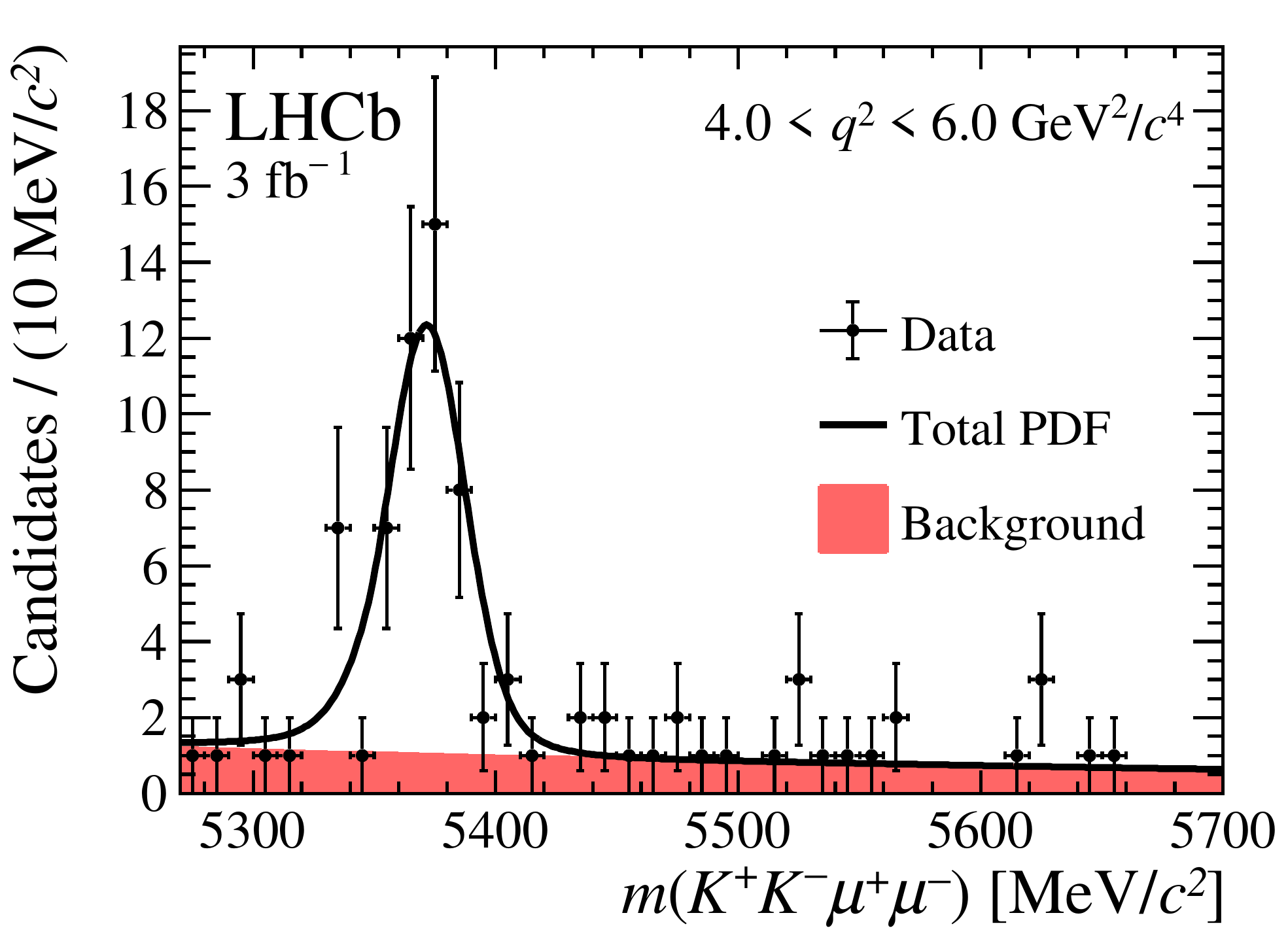}~
    \includegraphics[width=.4\textwidth]{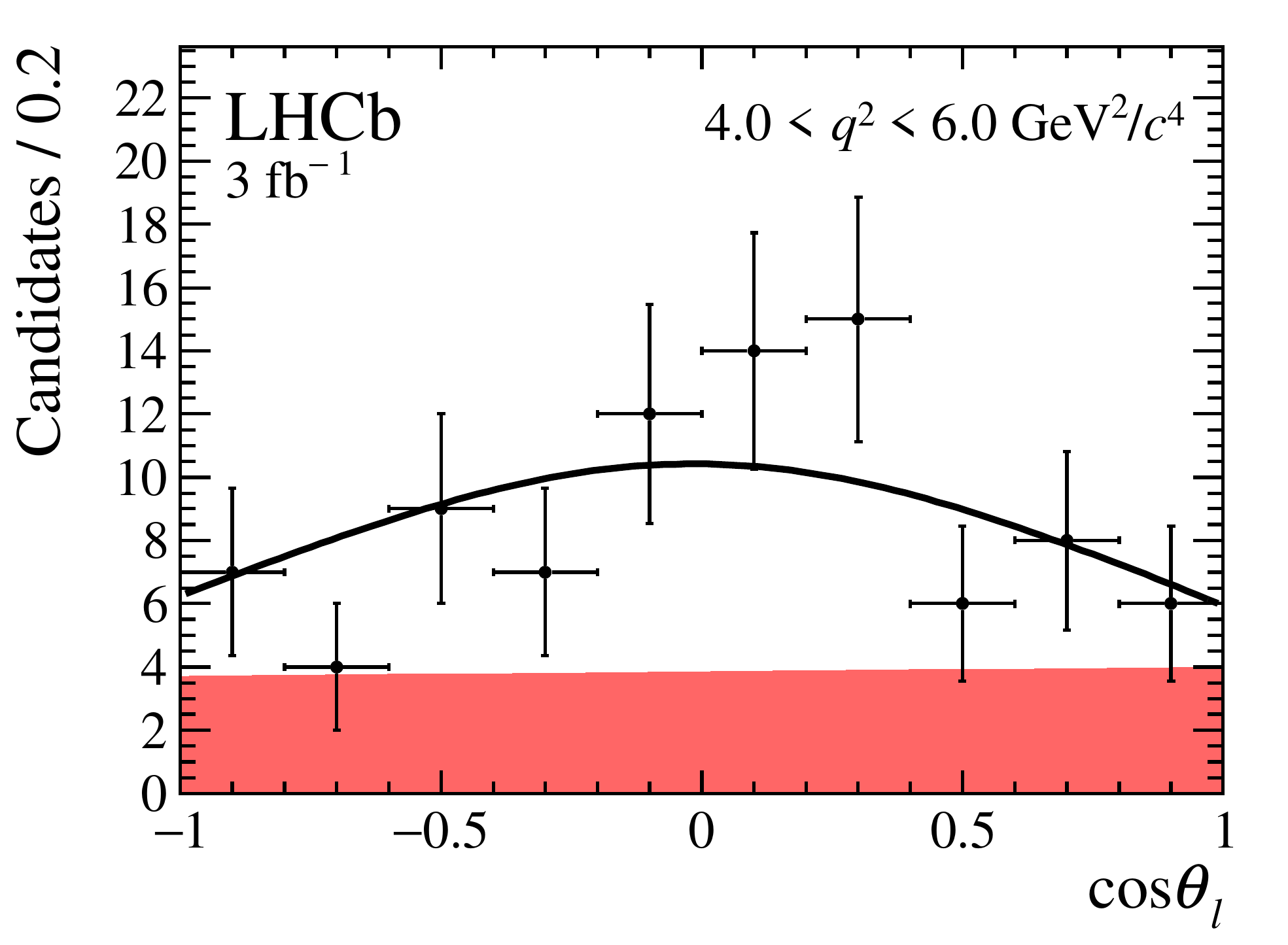}\\
    \includegraphics[width=.4\textwidth]{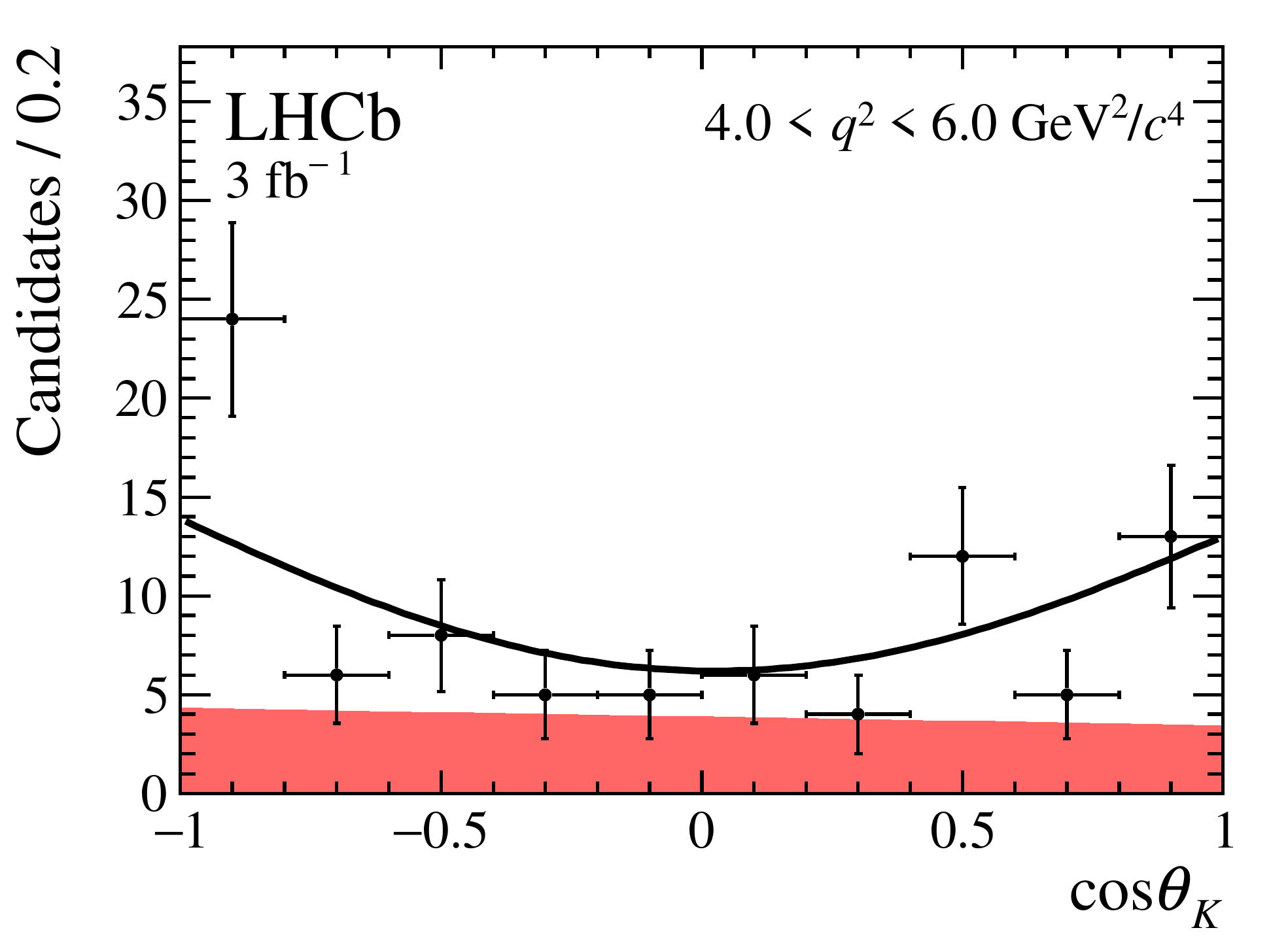}~
    \includegraphics[width=.4\textwidth]{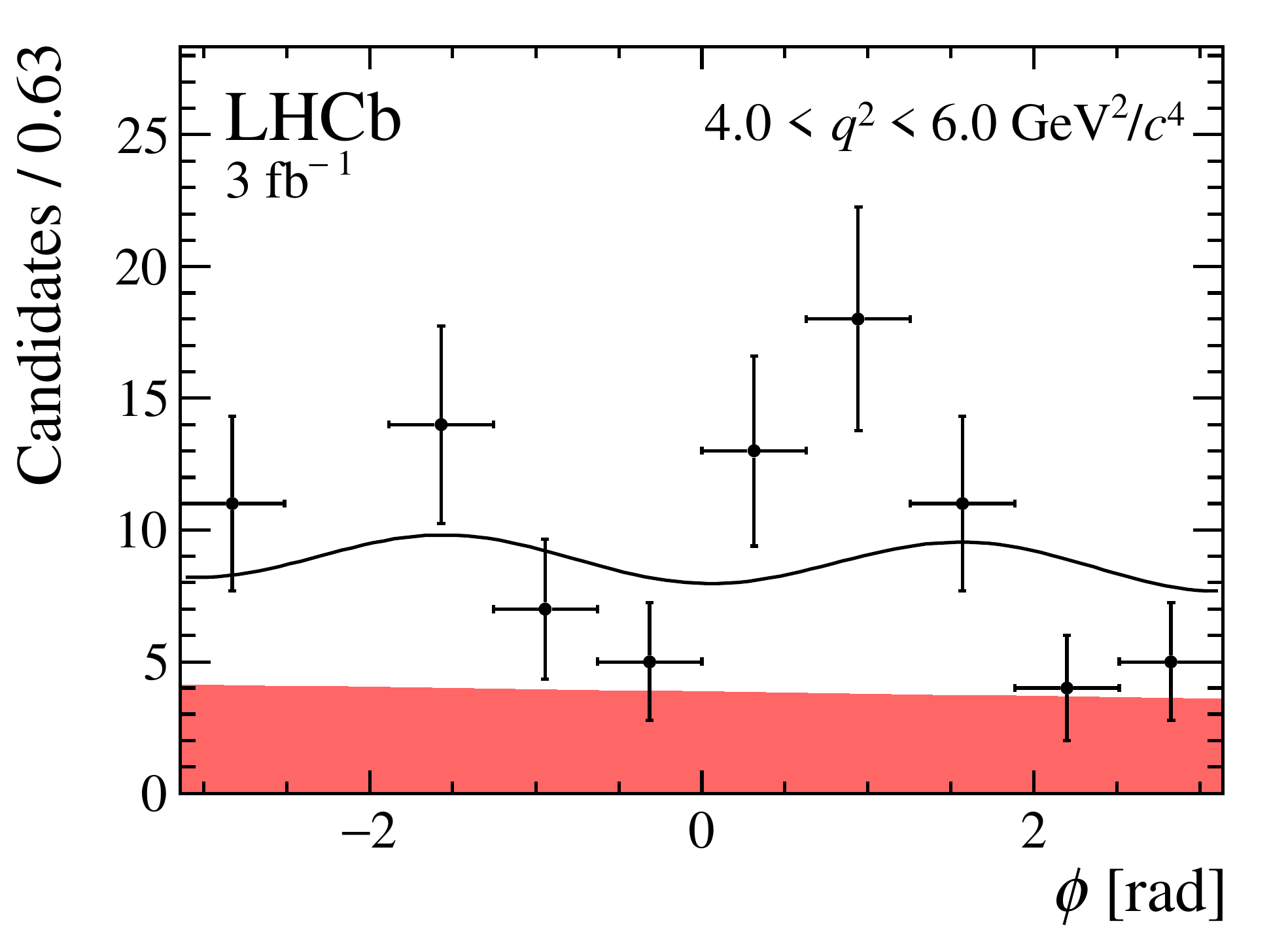}~
    \caption{\label{fig:results_bin3_run1} Mass and angular distributions of \BsToPhimm\ candidates in the region \mbox{$4.0<\qsq<6.0\gevgevcccc$} for data taken in 2011--2012. The data are overlaid with the projections of the fitted PDF.}
\end{figure}

\begin{figure}[hb]
    \centering
    \includegraphics[width=.4\textwidth]{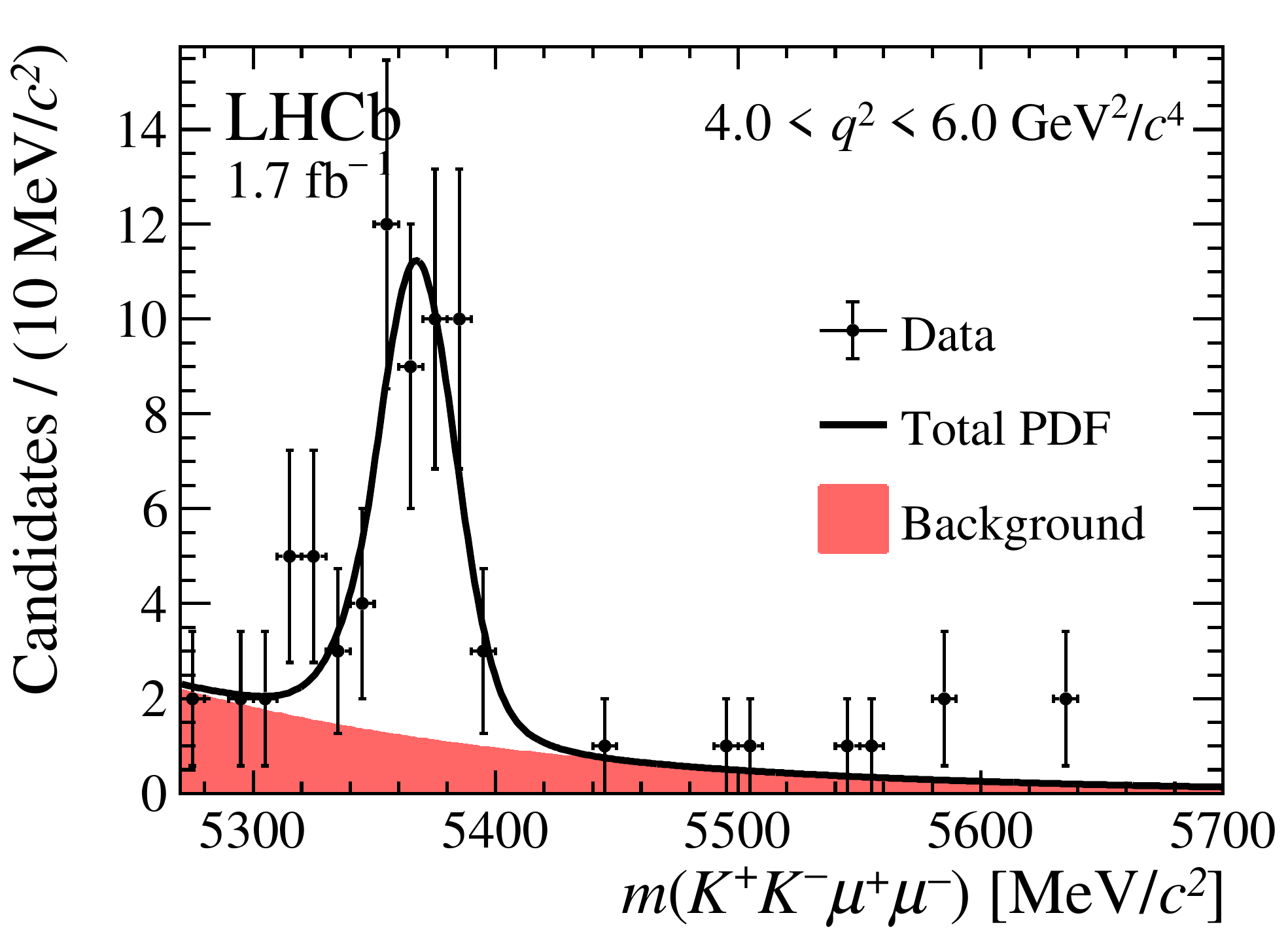}~
    \includegraphics[width=.4\textwidth]{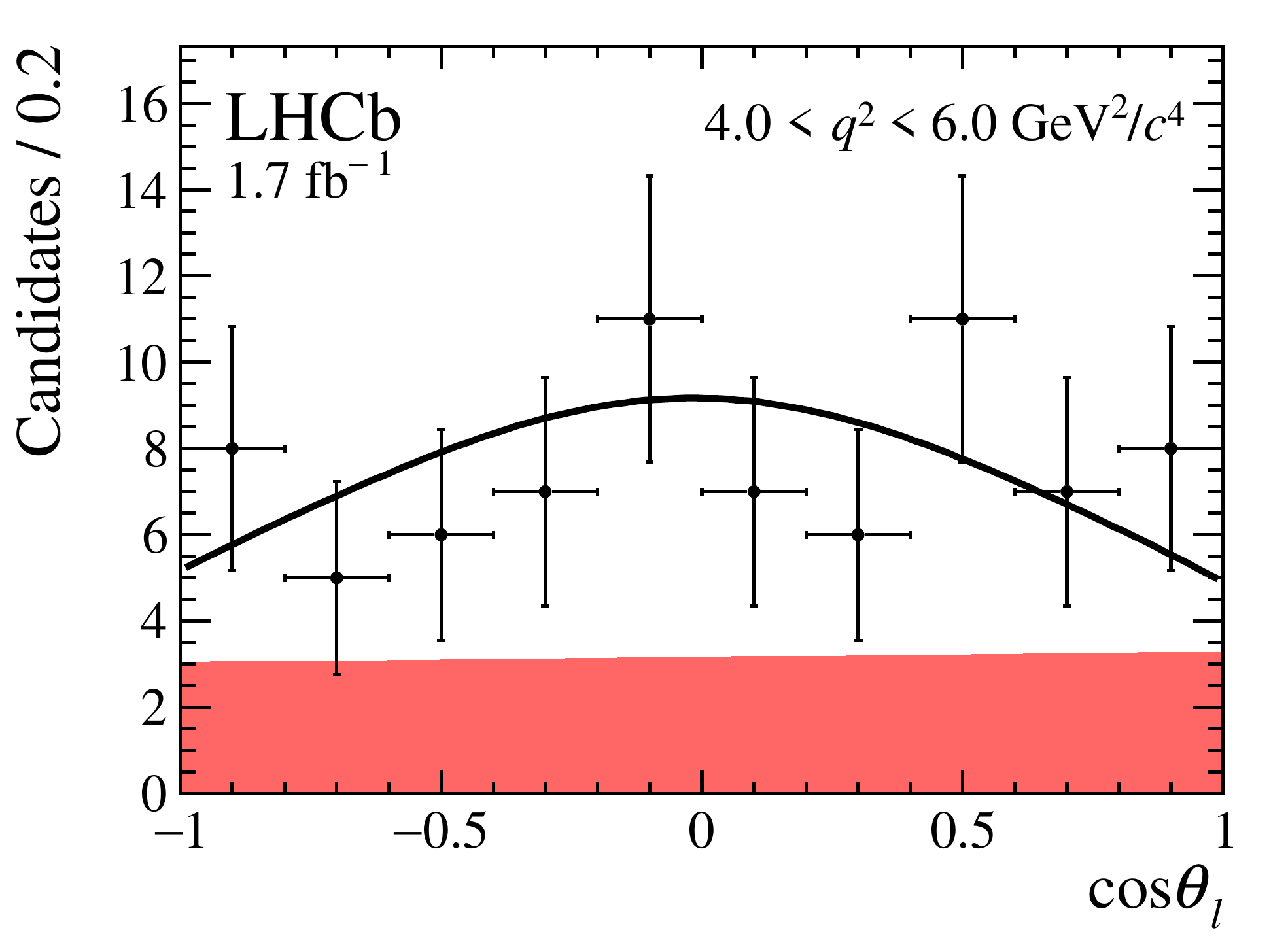}\\
    \includegraphics[width=.4\textwidth]{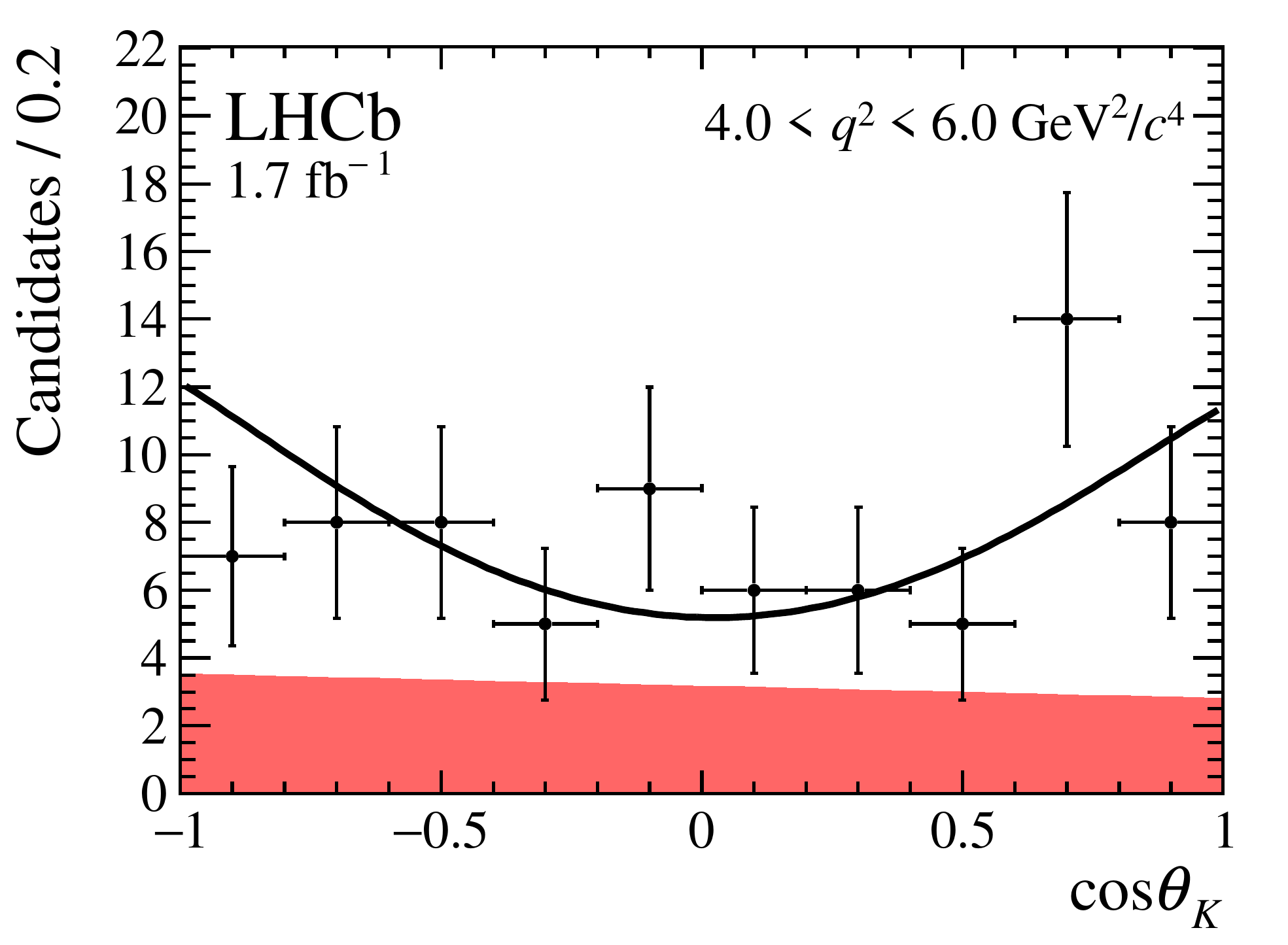}~
    \includegraphics[width=.4\textwidth]{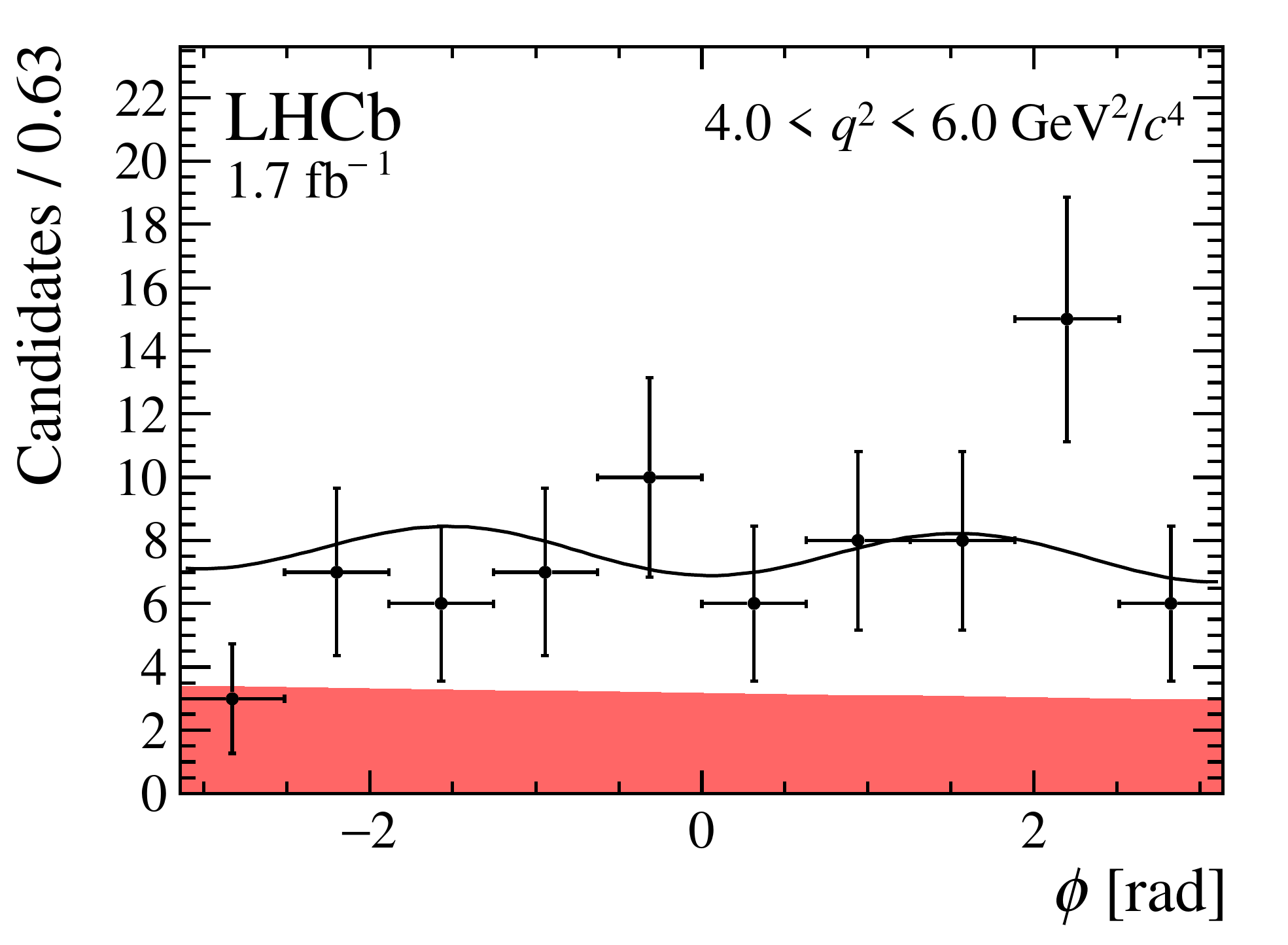}~
    \caption{\label{fig:results_bin3_run2p1} Mass and angular distributions of \BsToPhimm\ candidates in the region \mbox{$4.0<\qsq<6.0\gevgevcccc$} for data taken in 2016. The data are overlaid with the projections of the fitted PDF.}
\end{figure}

\begin{figure}[hb]
    \centering
    \includegraphics[width=.4\textwidth]{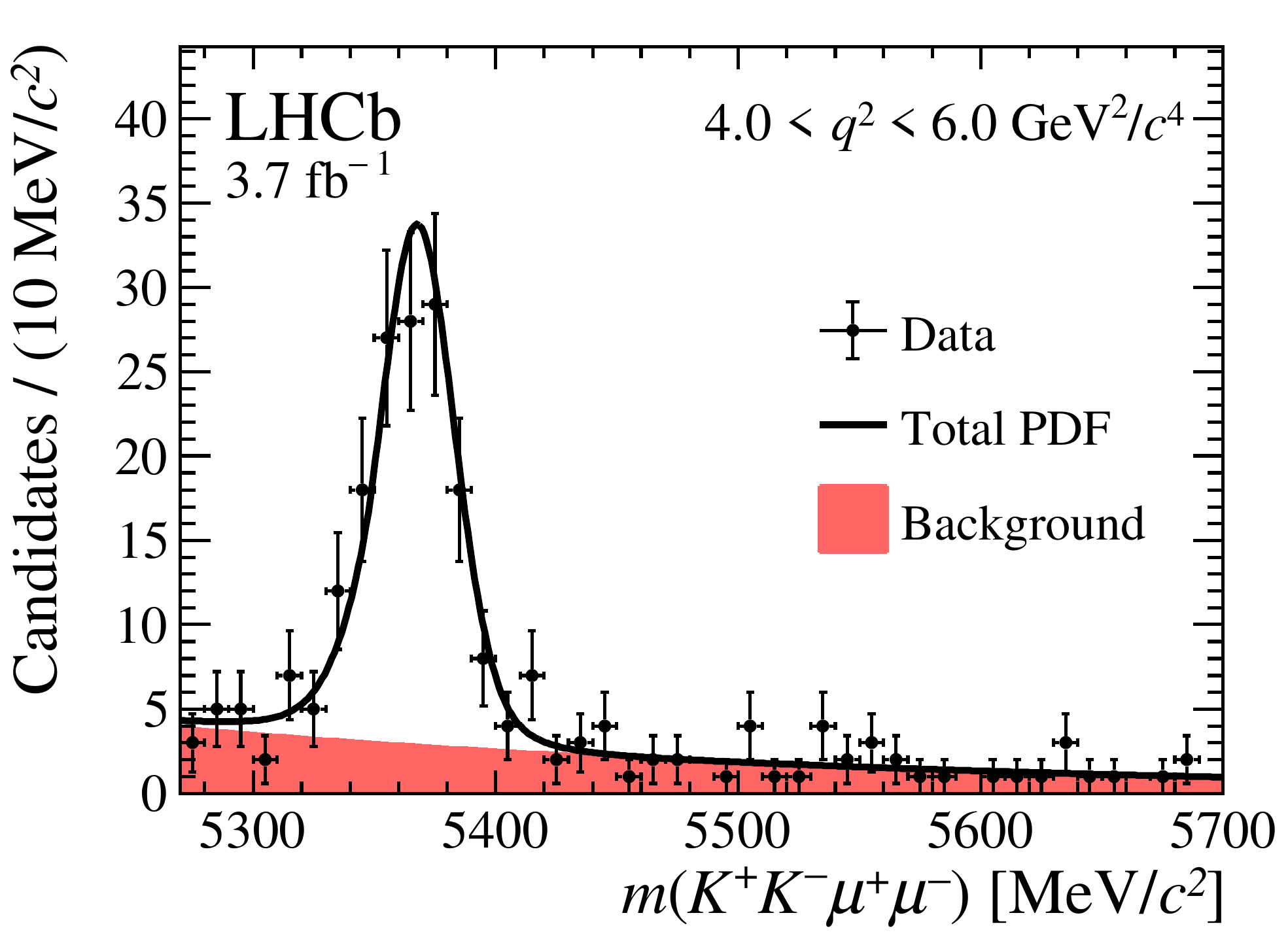}~
    \includegraphics[width=.4\textwidth]{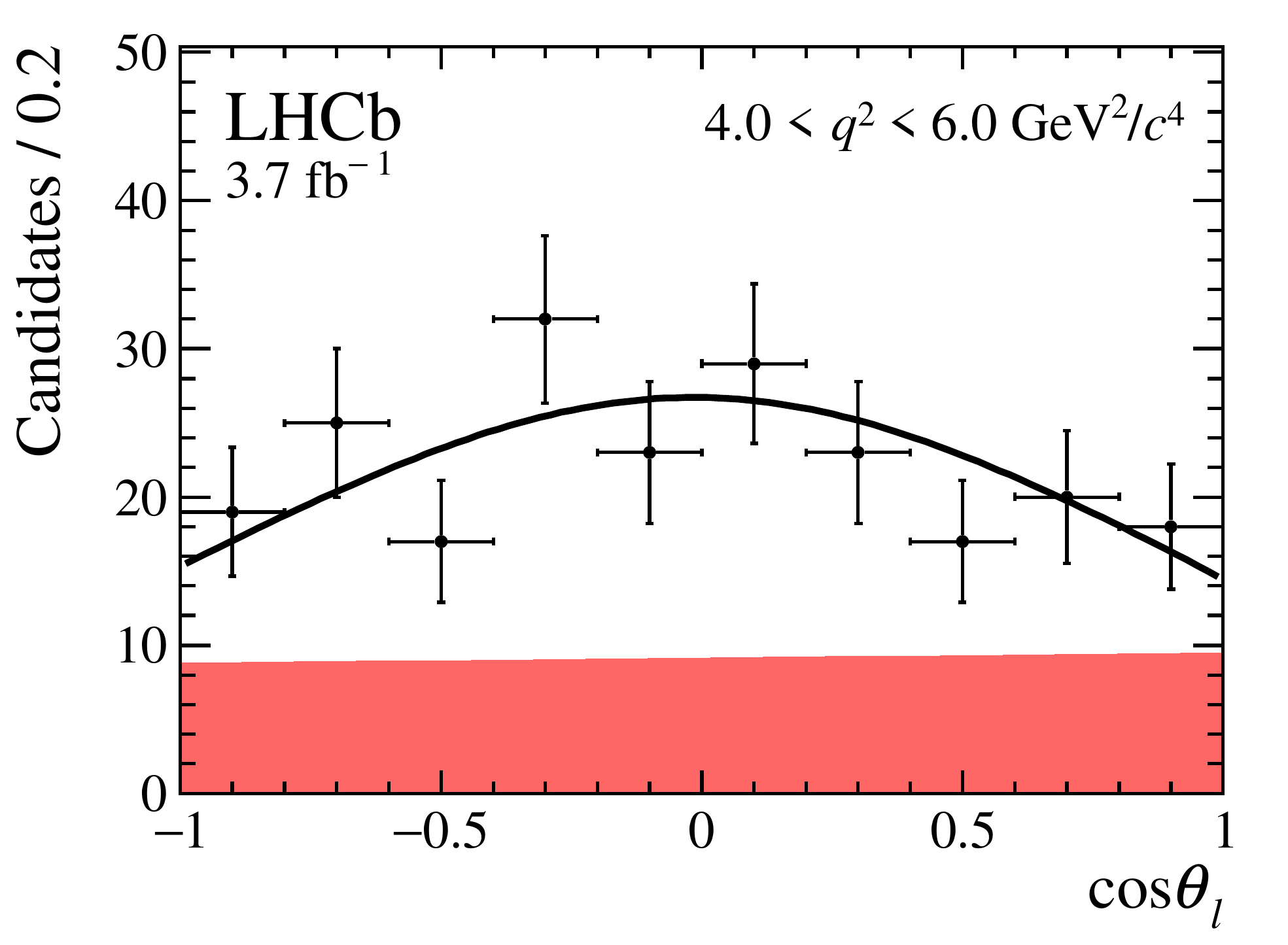}\\
    \includegraphics[width=.4\textwidth]{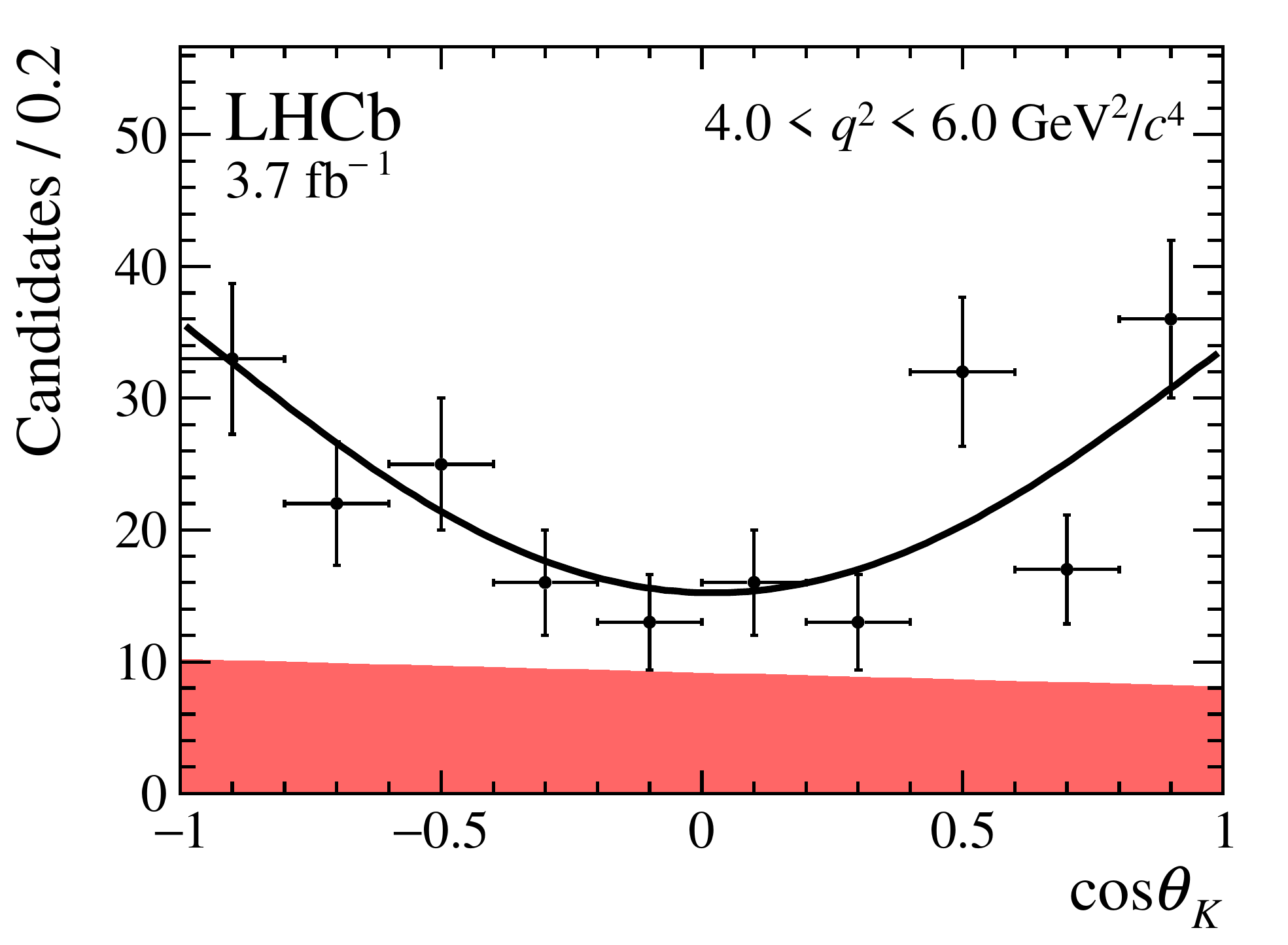}~
    \includegraphics[width=.4\textwidth]{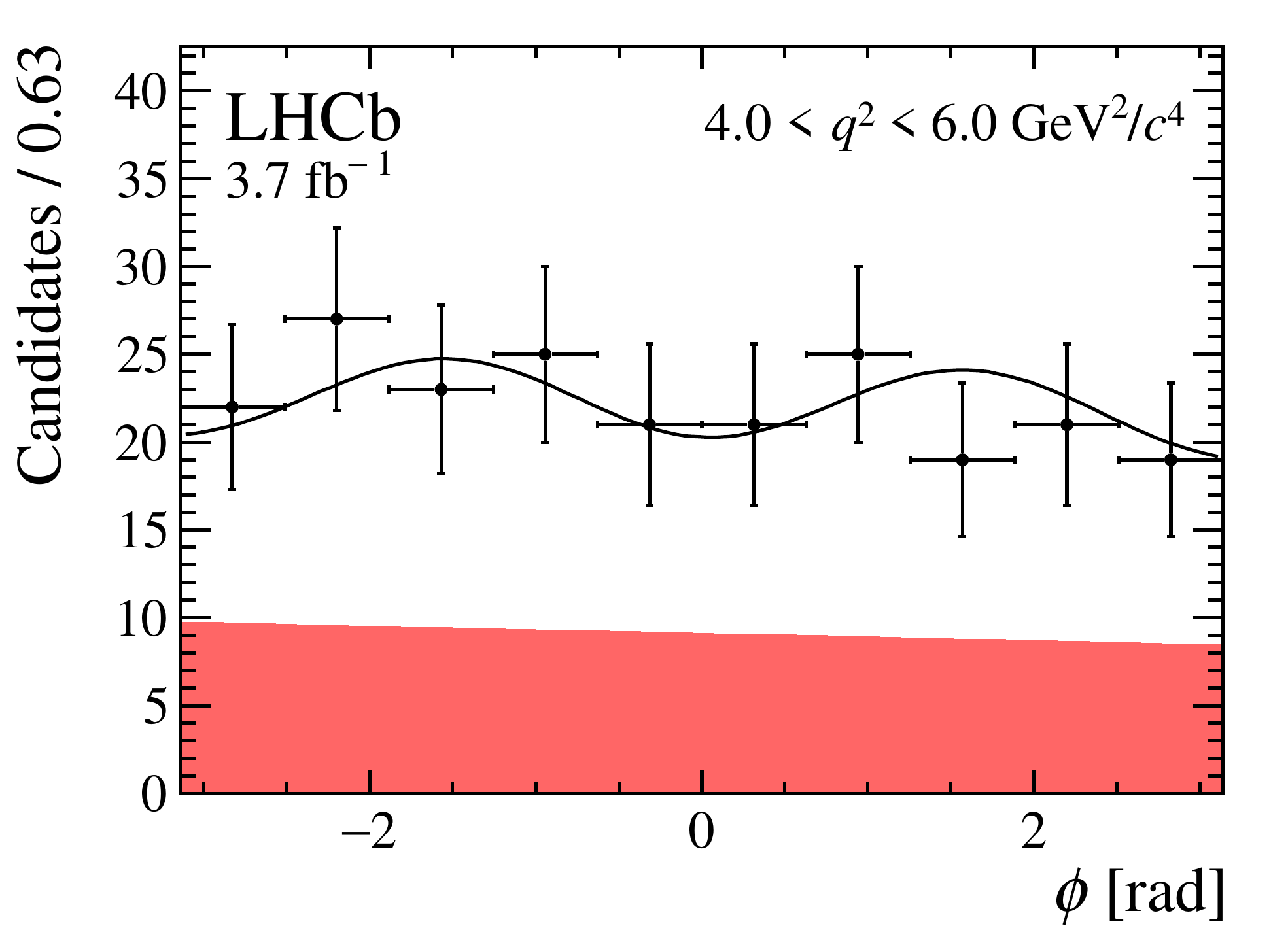}~
    \caption{\label{fig:results_bin3_run2p2} Mass and angular distributions of \BsToPhimm\ candidates in the region \mbox{$4.0<\qsq<6.0\gevgevcccc$} for data taken in 2017--2018. The data are overlaid with the projections of the fitted PDF.}
\end{figure}

\begin{figure}[hb]
    \centering
    \includegraphics[width=.4\textwidth]{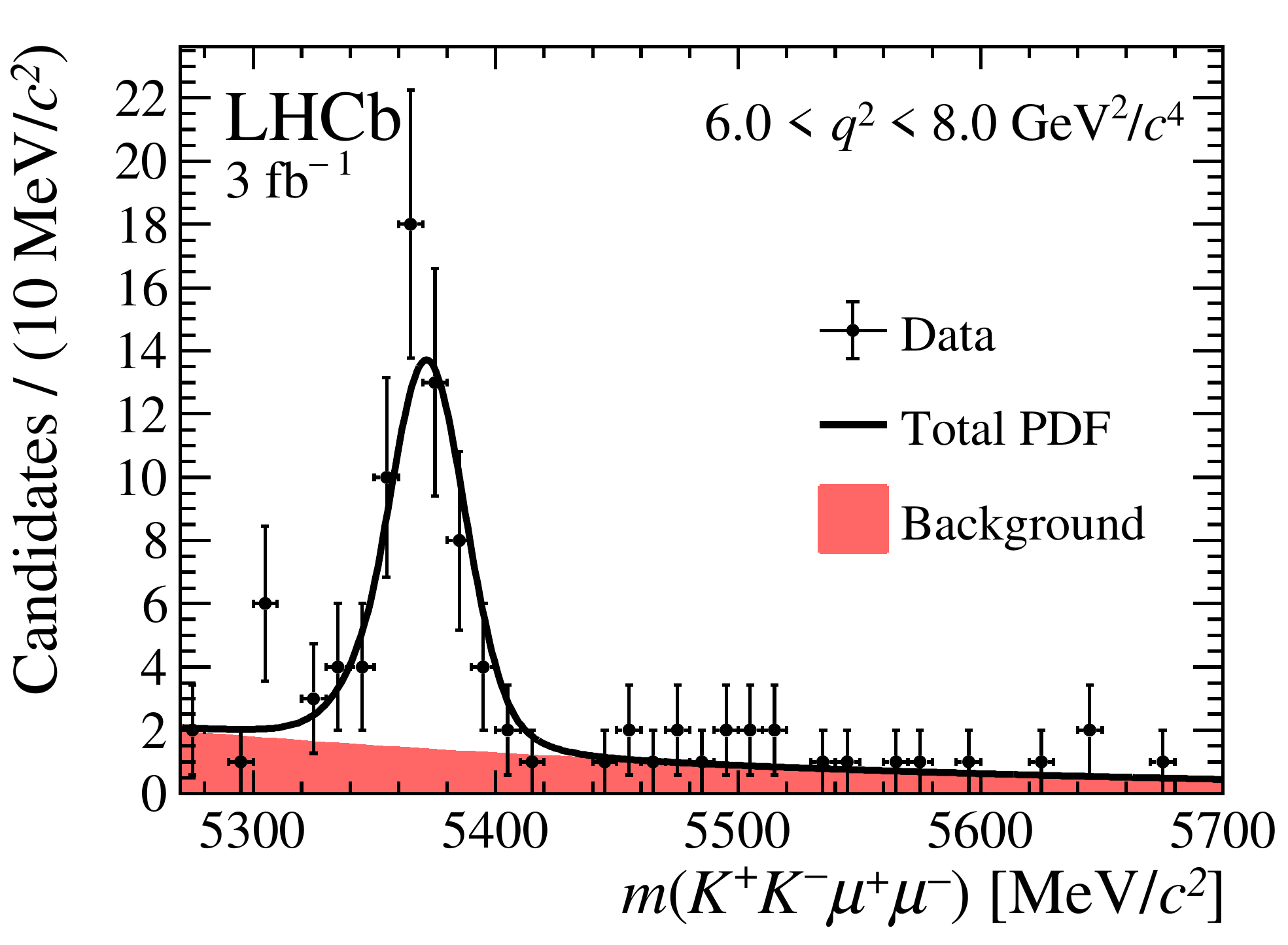}~
    \includegraphics[width=.4\textwidth]{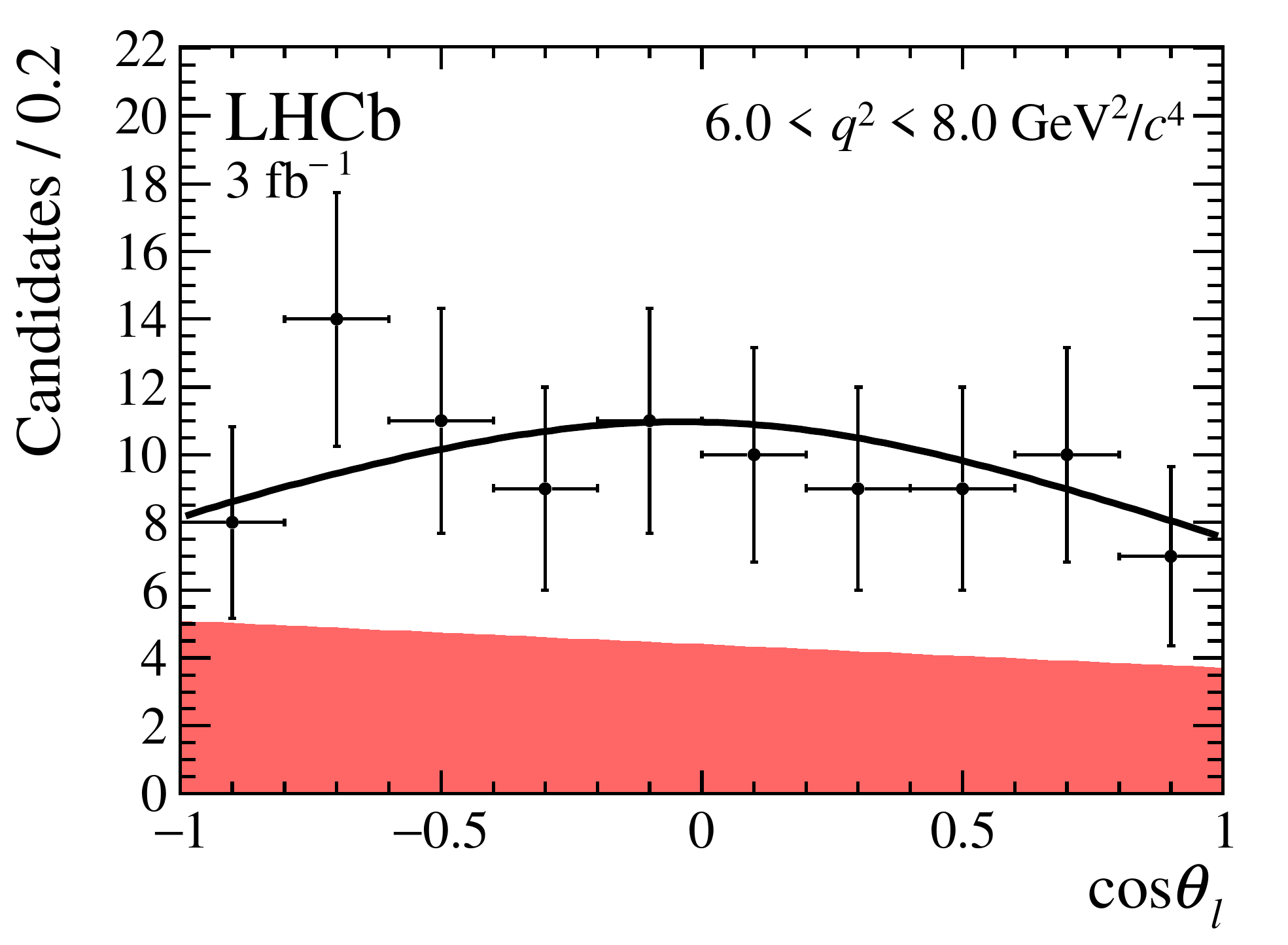}\\
    \includegraphics[width=.4\textwidth]{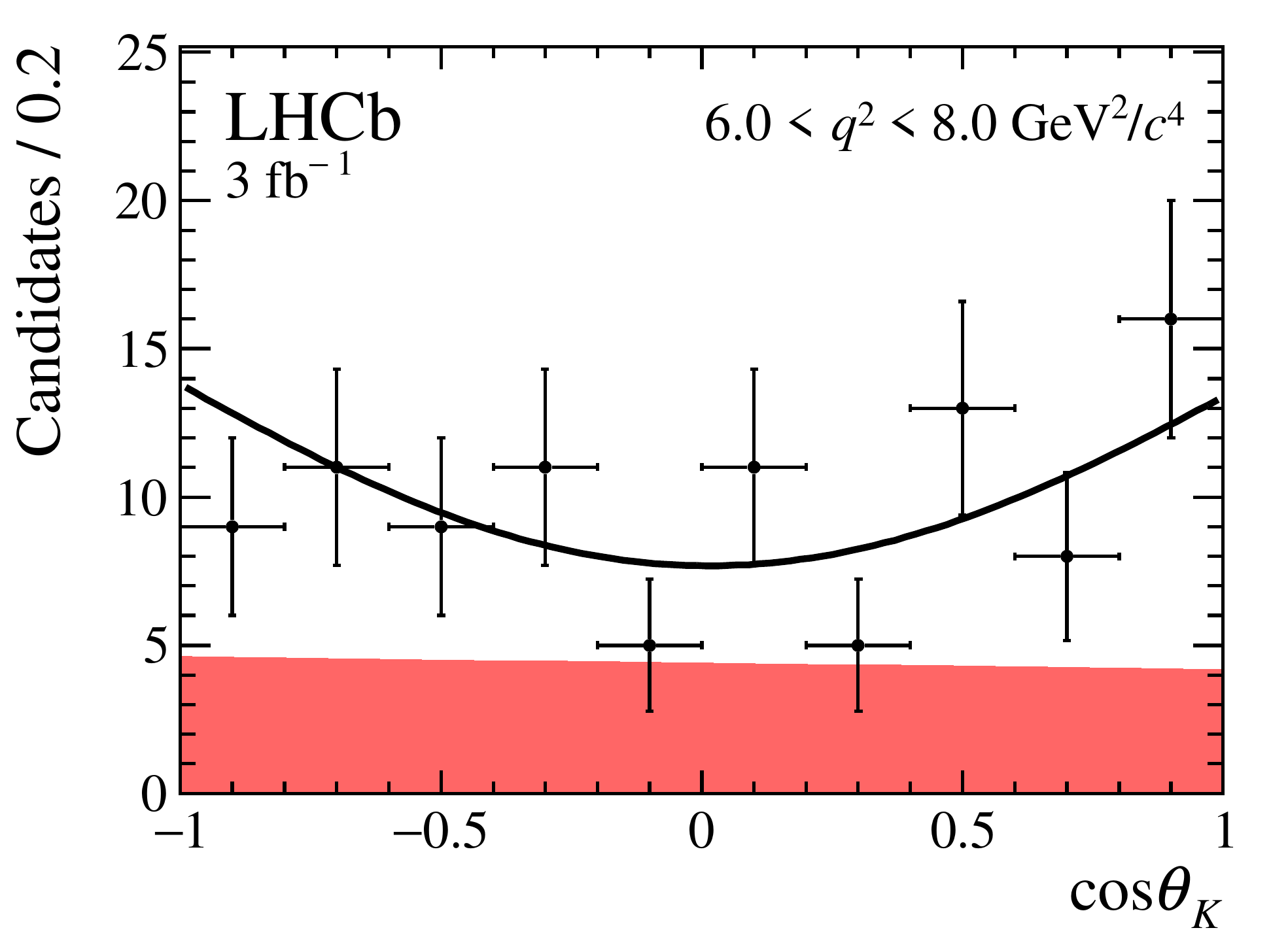}~
    \includegraphics[width=.4\textwidth]{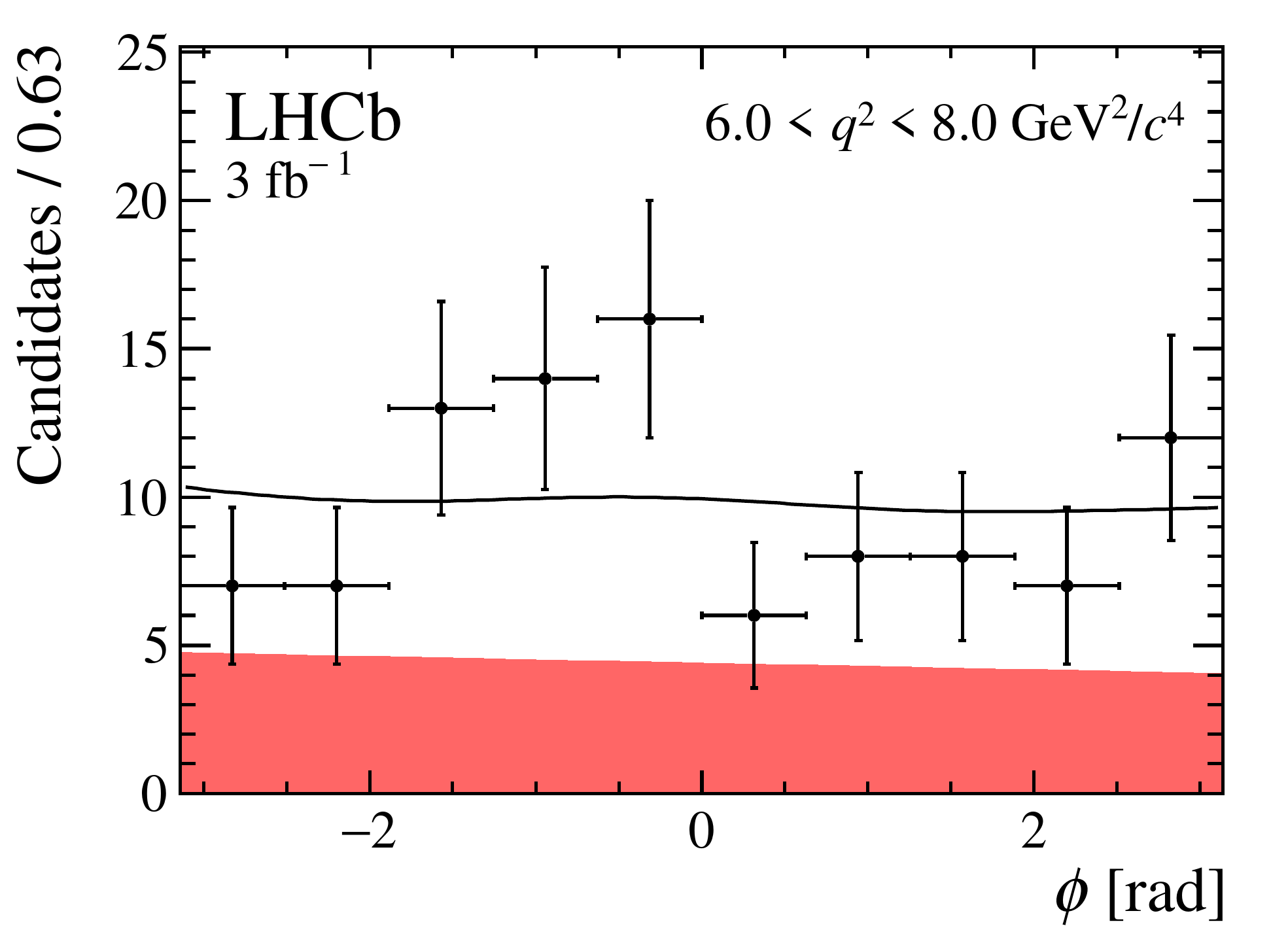}~
    \caption{\label{fig:results_bin4_run1} Mass and angular distributions of \BsToPhimm\ candidates in the region \mbox{$6.0<\qsq<8.0\gevgevcccc$} for data taken in 2011--2012. The data are overlaid with the projections of the fitted PDF.}
\end{figure}

\begin{figure}[hb]
    \centering
    \includegraphics[width=.4\textwidth]{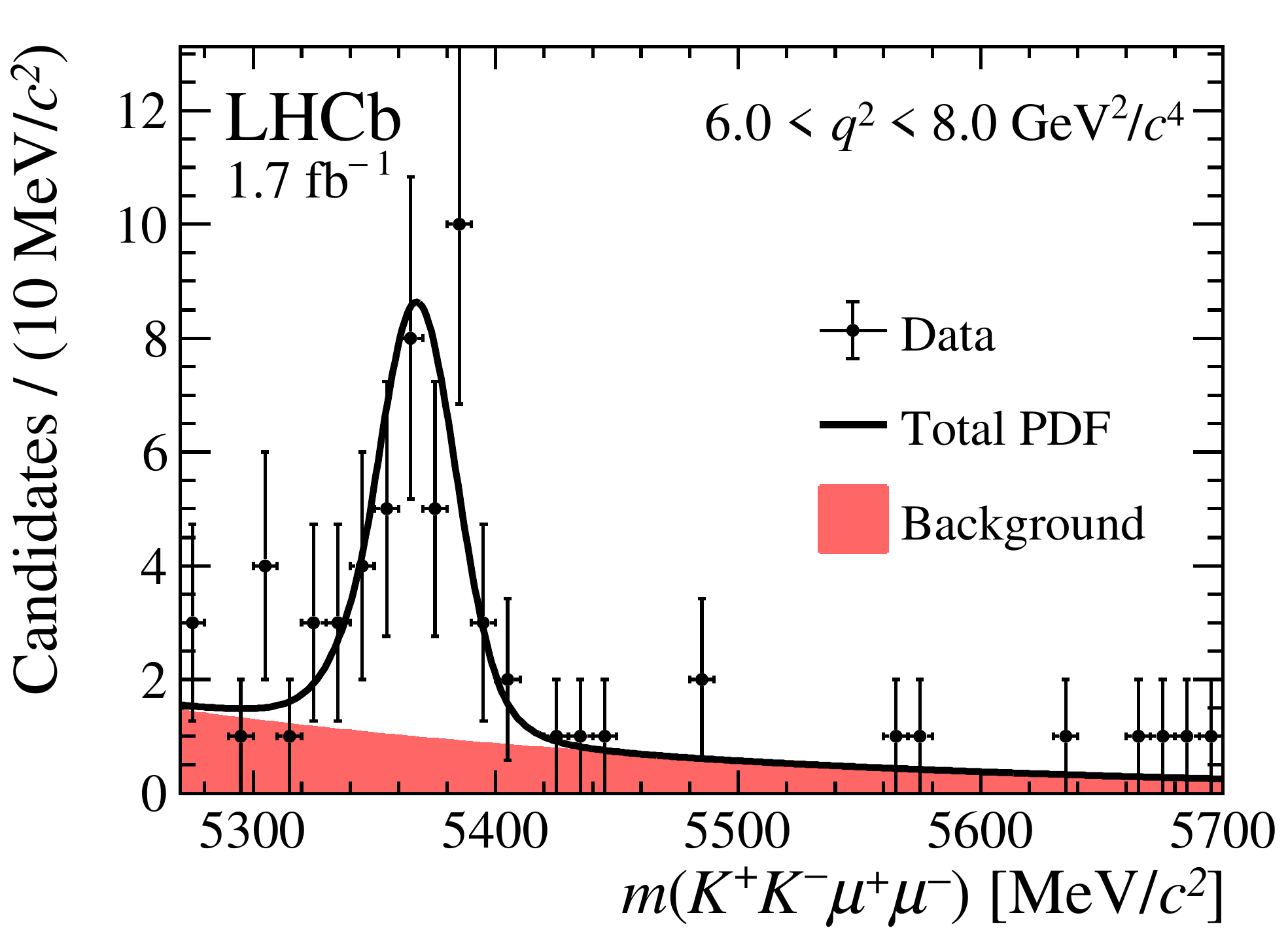}~
    \includegraphics[width=.4\textwidth]{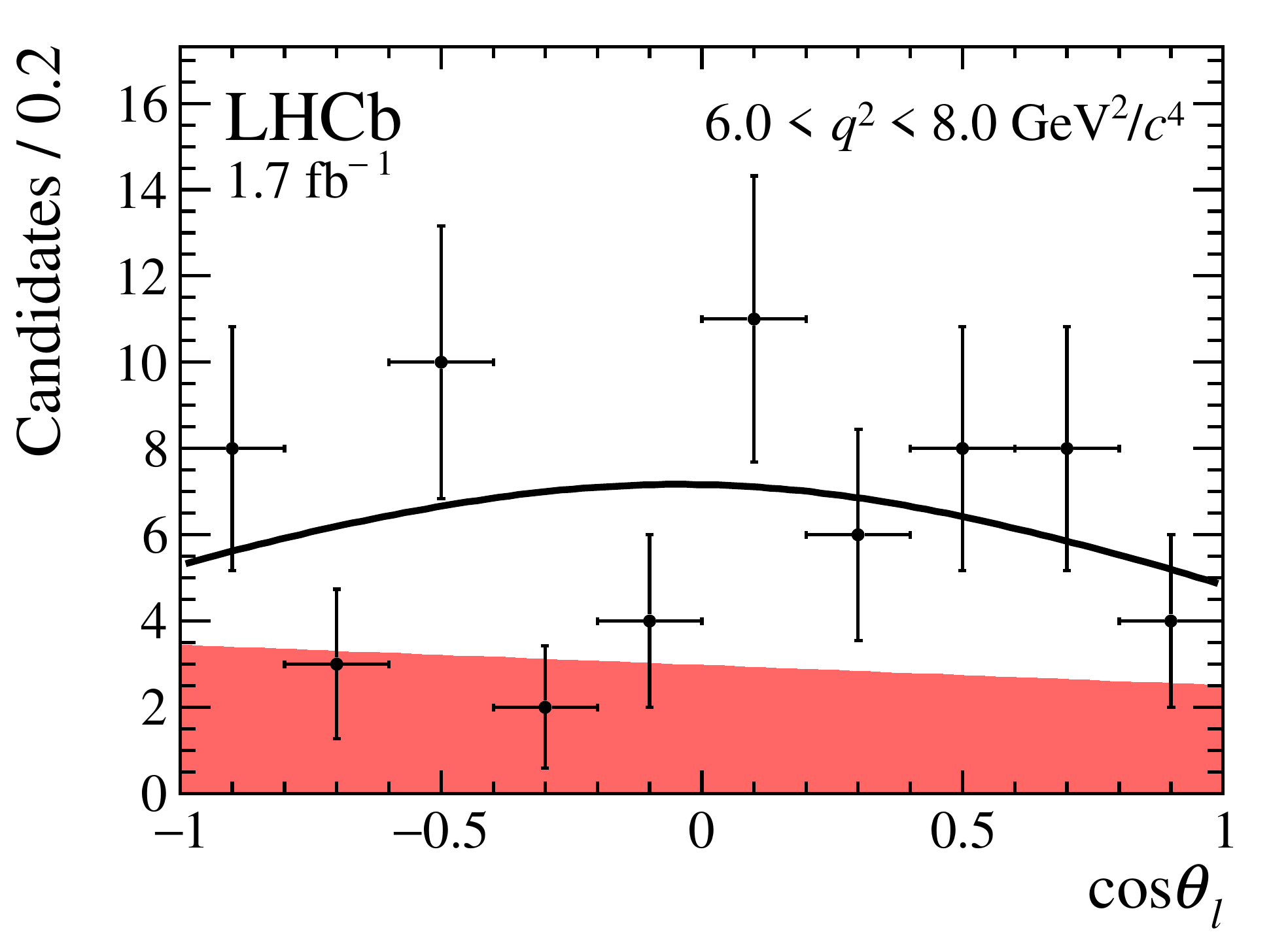}\\
    \includegraphics[width=.4\textwidth]{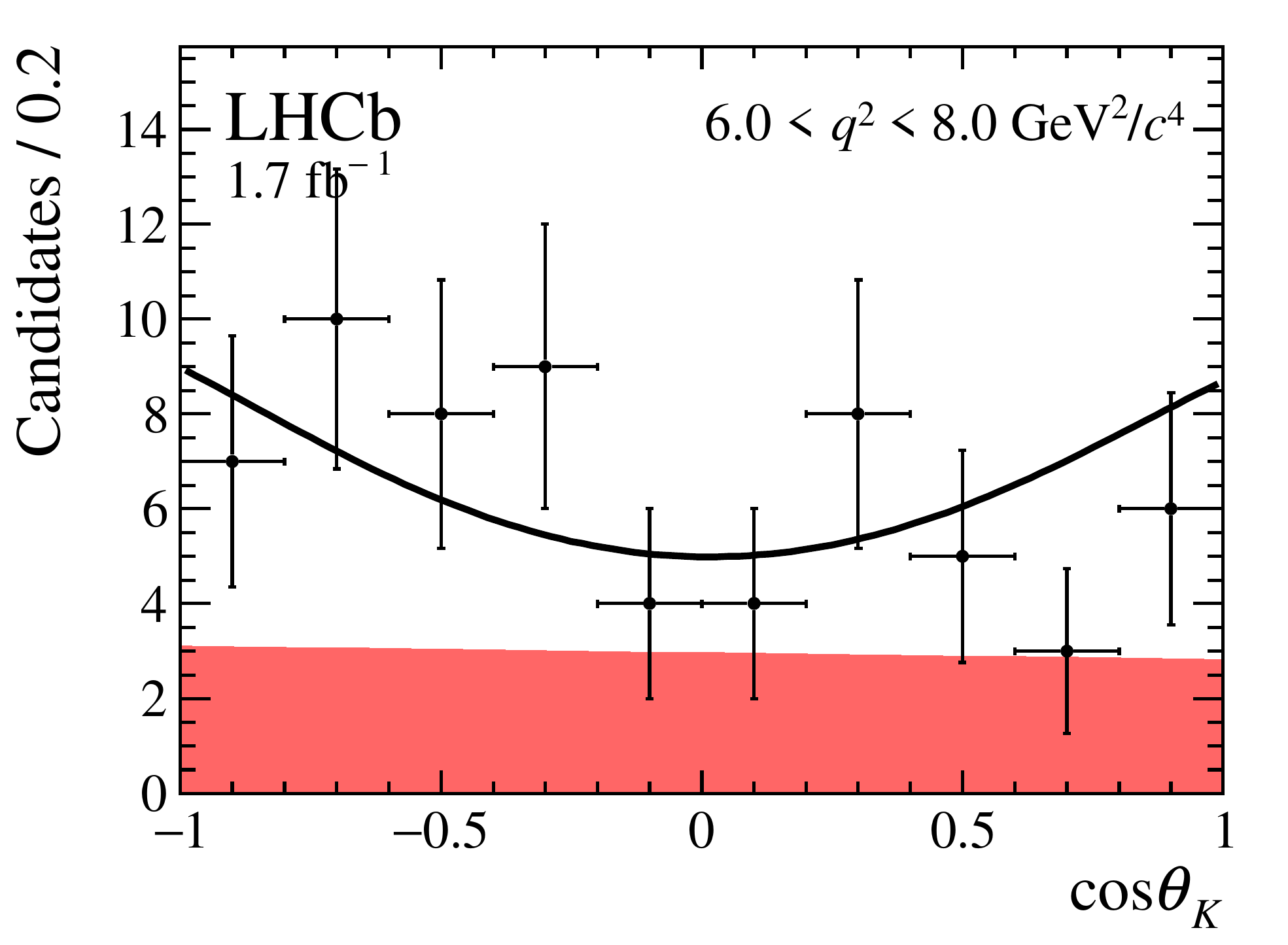}~
    \includegraphics[width=.4\textwidth]{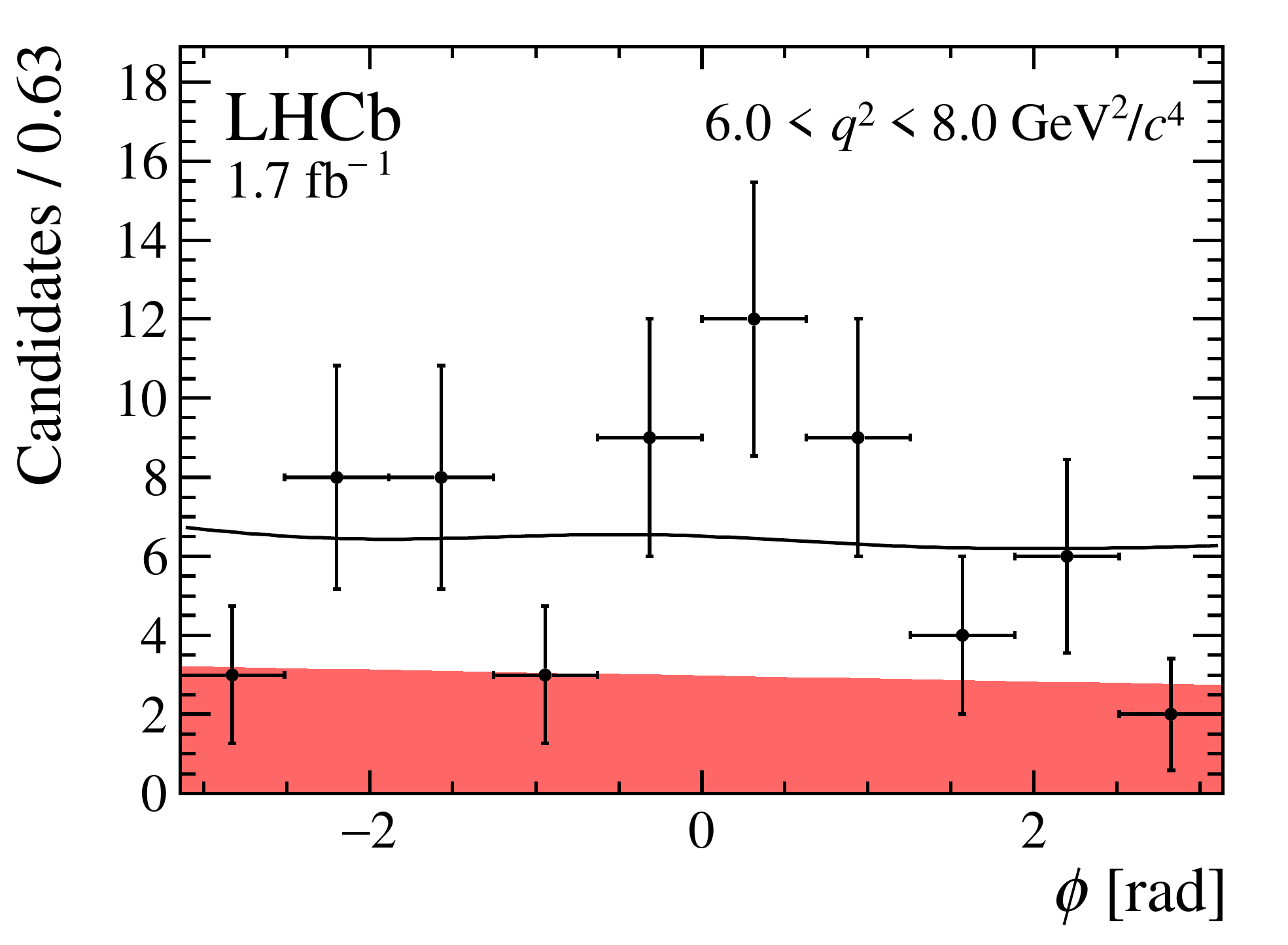}~
    \caption{\label{fig:results_bin4_run2p1} Mass and angular distributions of \BsToPhimm\ candidates in the region \mbox{$6.0<\qsq<8.0\gevgevcccc$} for data taken in 2016. The data are overlaid with the projections of the fitted PDF.}
\end{figure}

\begin{figure}[hb]
    \centering
    \includegraphics[width=.4\textwidth]{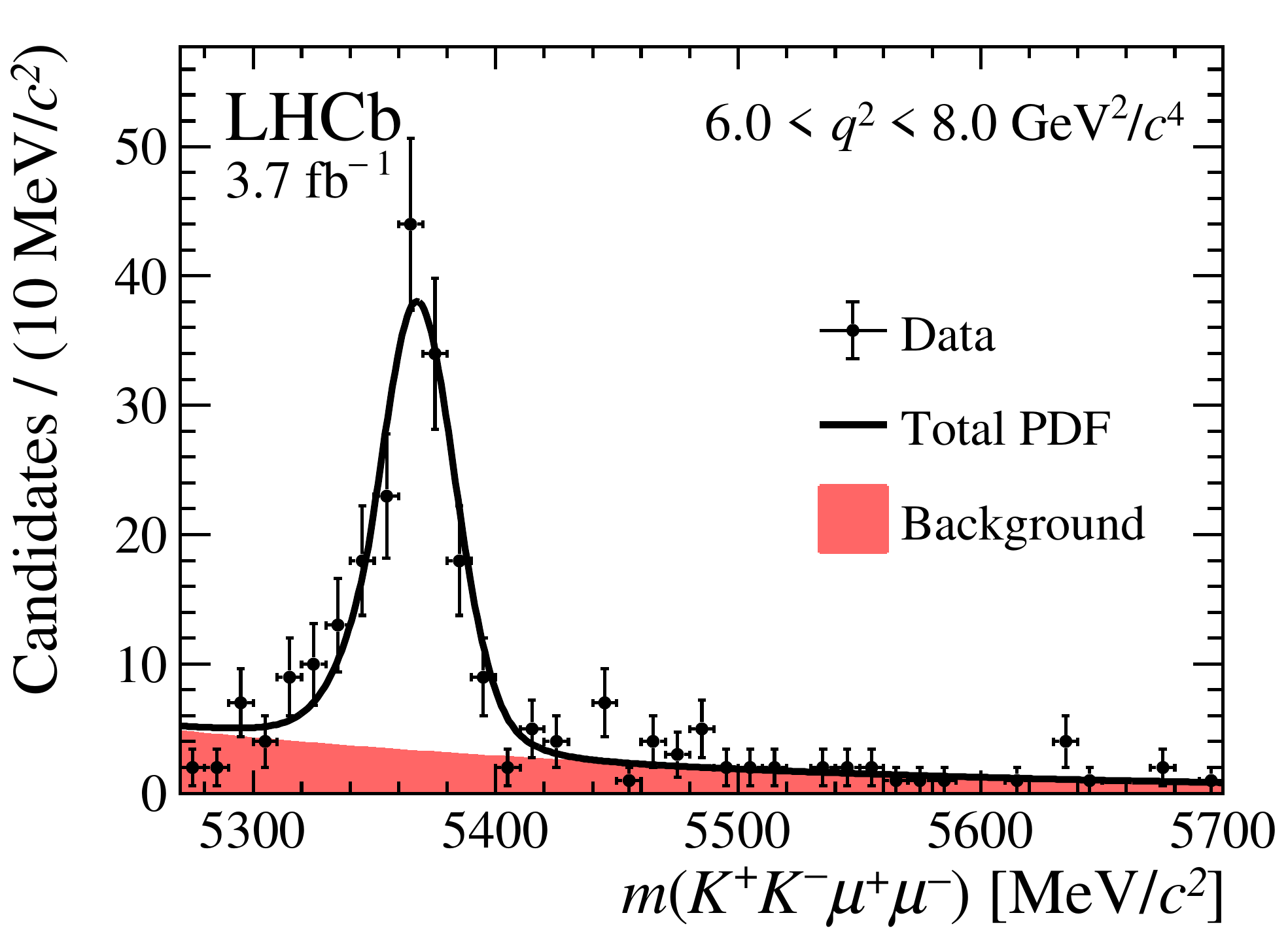}~
    \includegraphics[width=.4\textwidth]{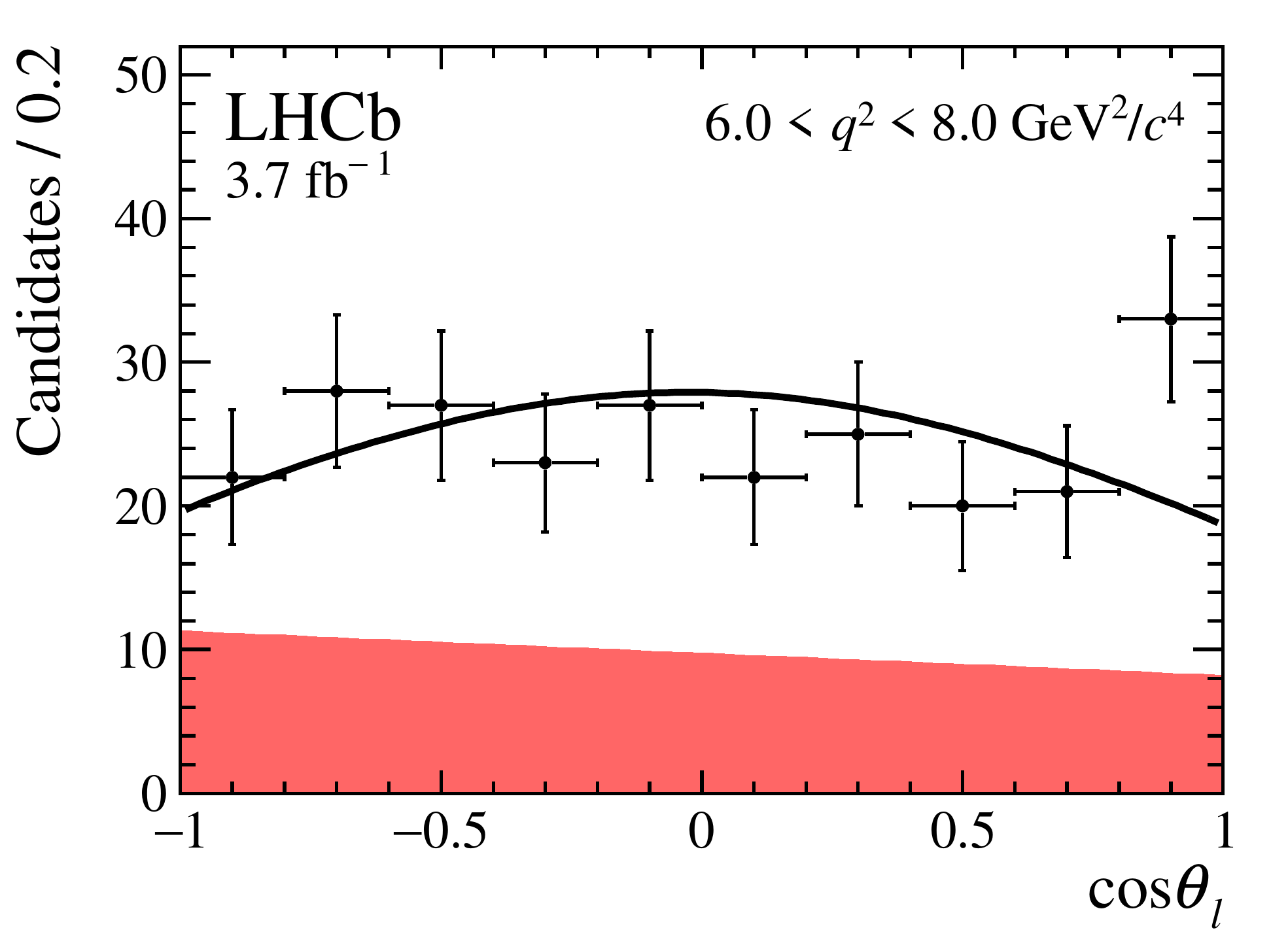}\\
    \includegraphics[width=.4\textwidth]{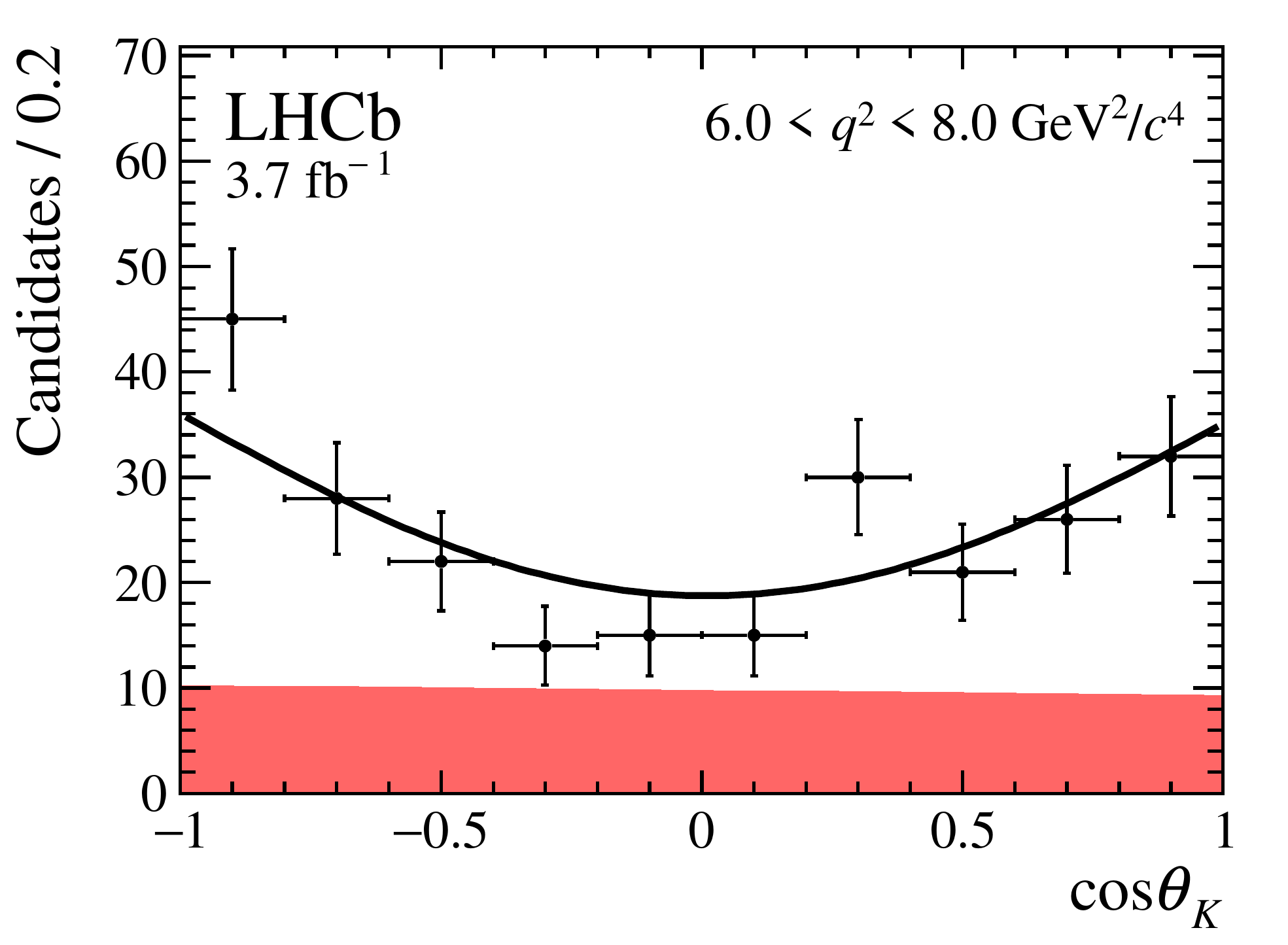}~
    \includegraphics[width=.4\textwidth]{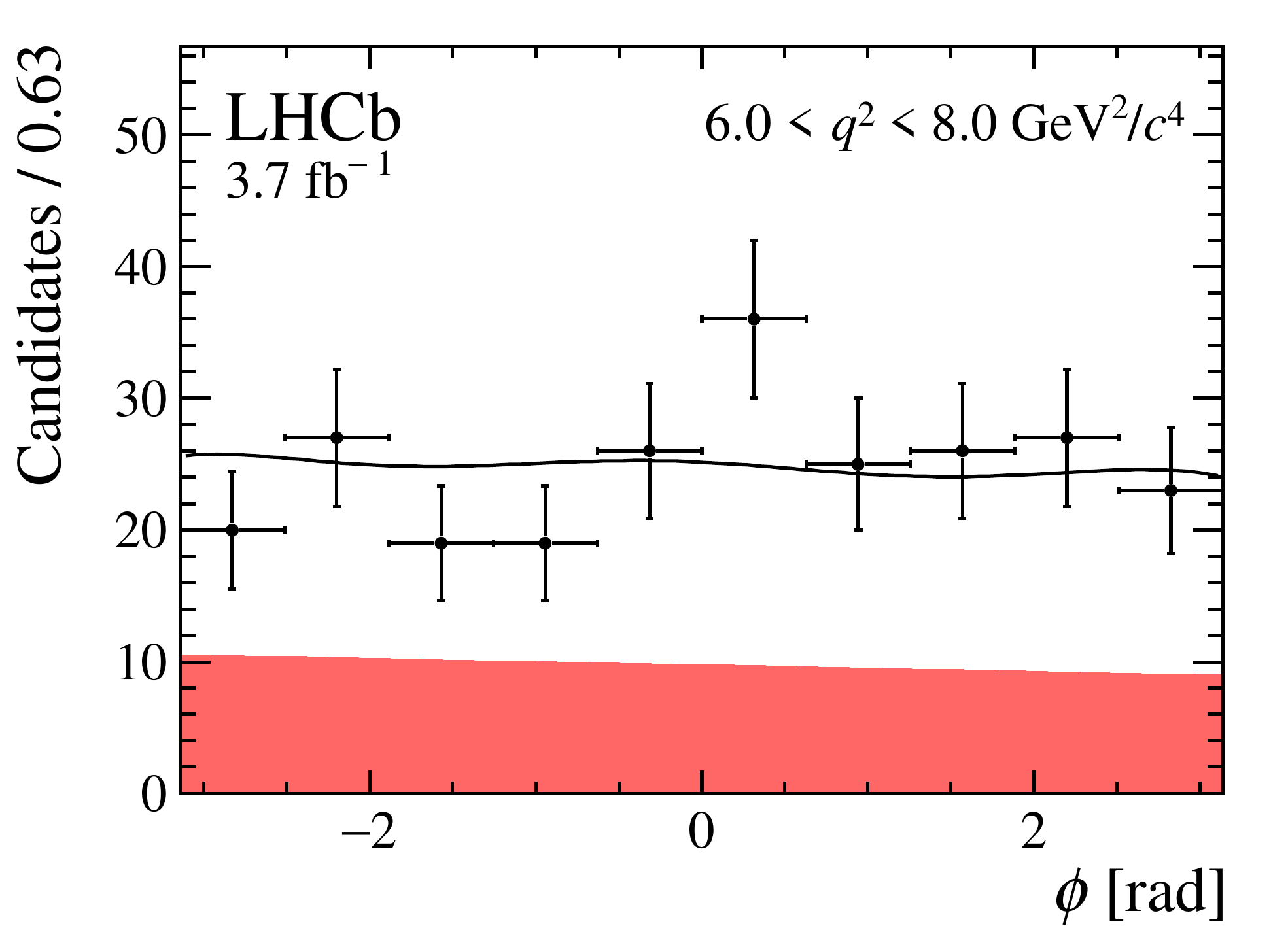}~
    \caption{\label{fig:results_bin4_run2p2} Mass and angular distributions of \BsToPhimm\ candidates in the region \mbox{$6.0<\qsq<8.0\gevgevcccc$} for data taken in 2017--2018. The data are overlaid with the projections of the fitted PDF. }
\end{figure}

\begin{figure}[hb]
    \centering
    \includegraphics[width=.4\textwidth]{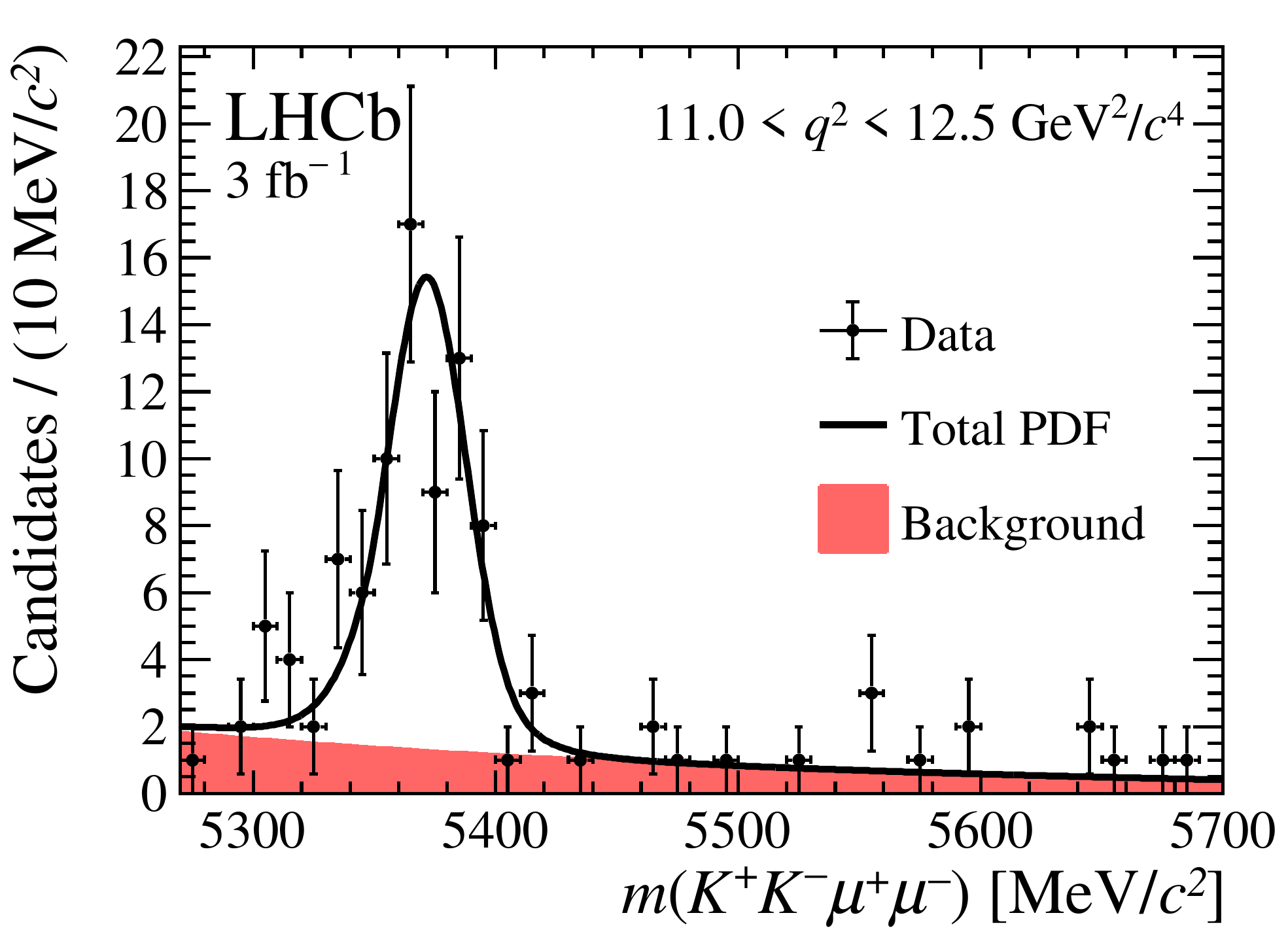}~
    \includegraphics[width=.4\textwidth]{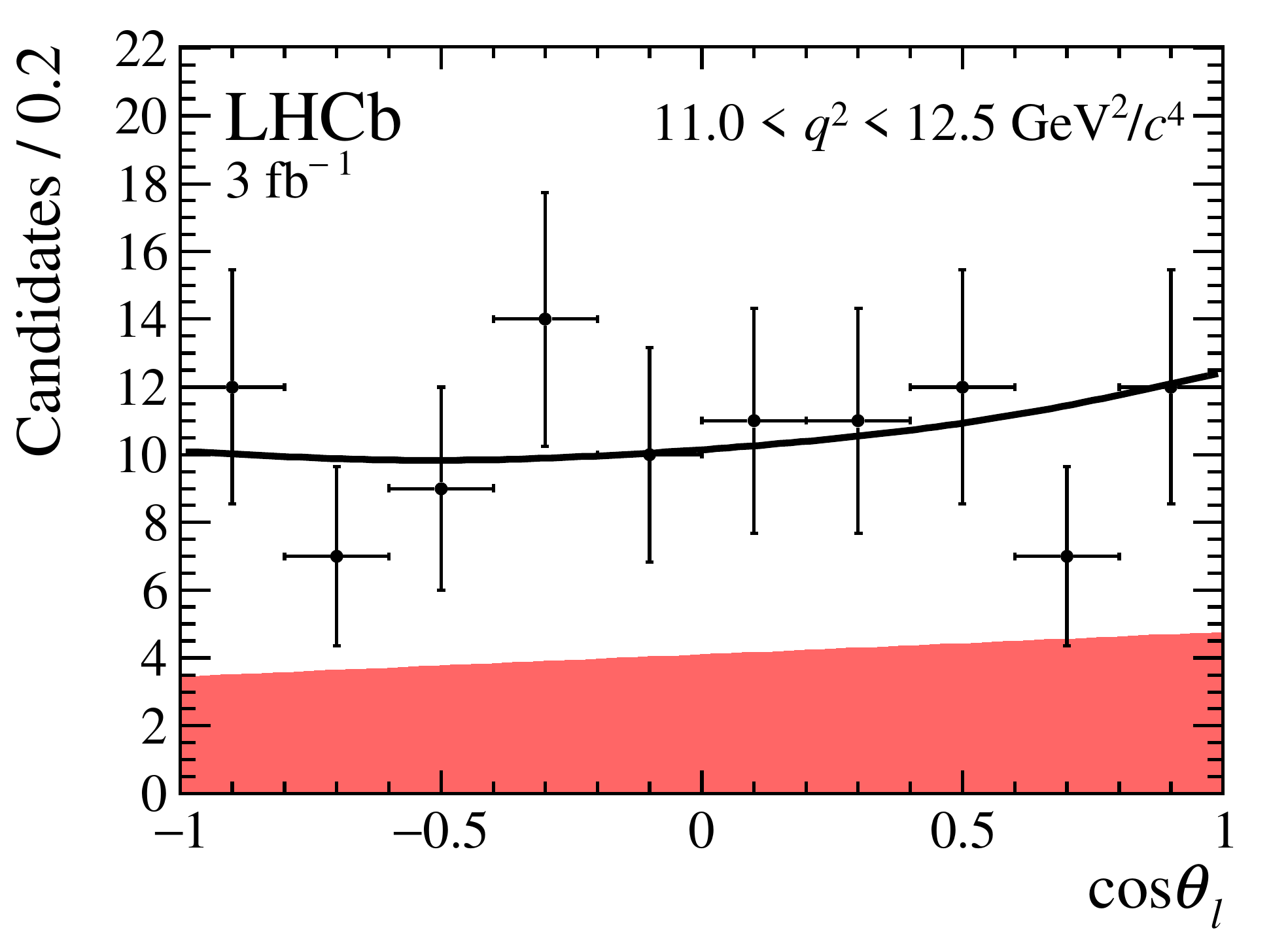}\\
    \includegraphics[width=.4\textwidth]{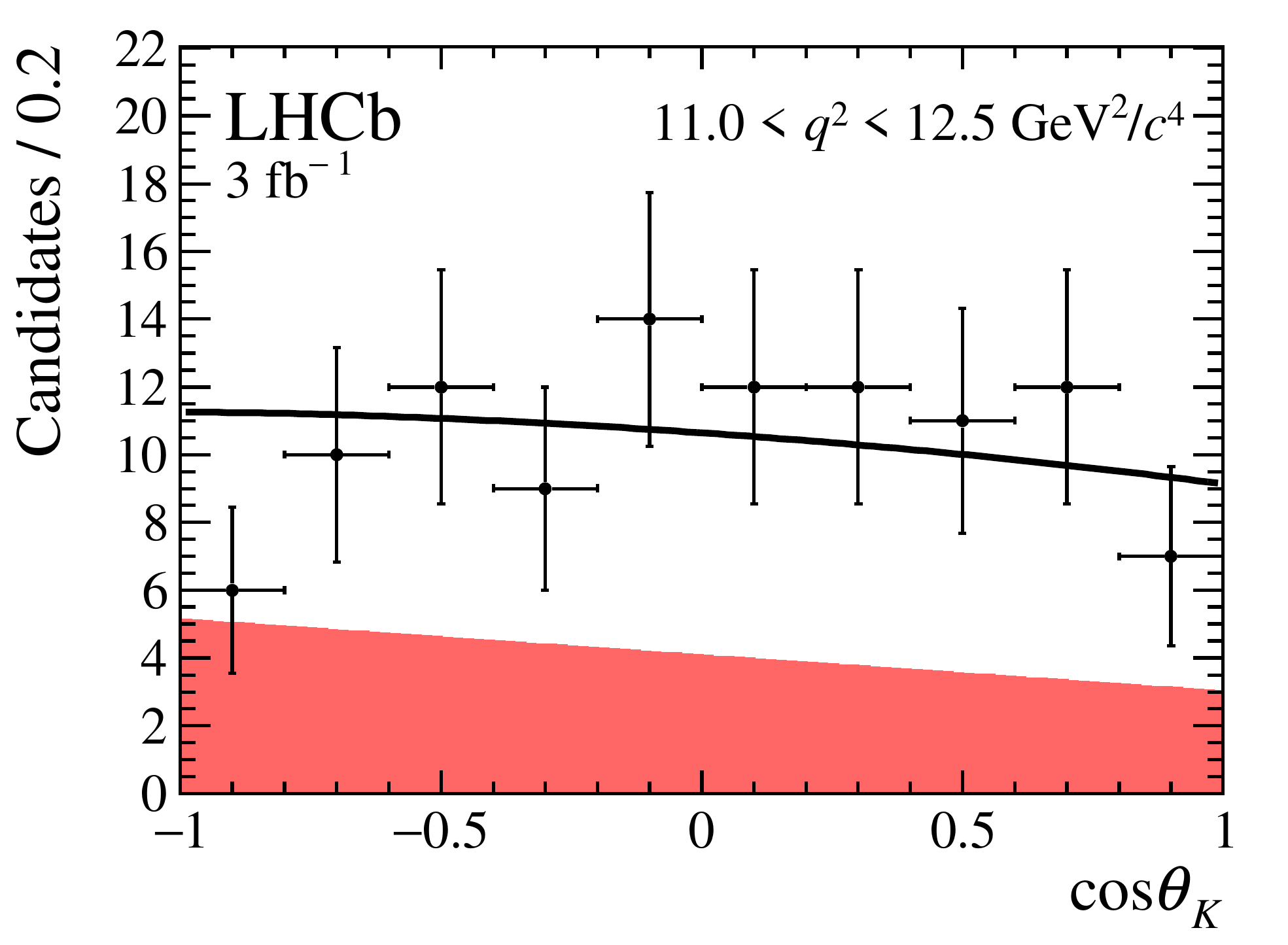}~
    \includegraphics[width=.4\textwidth]{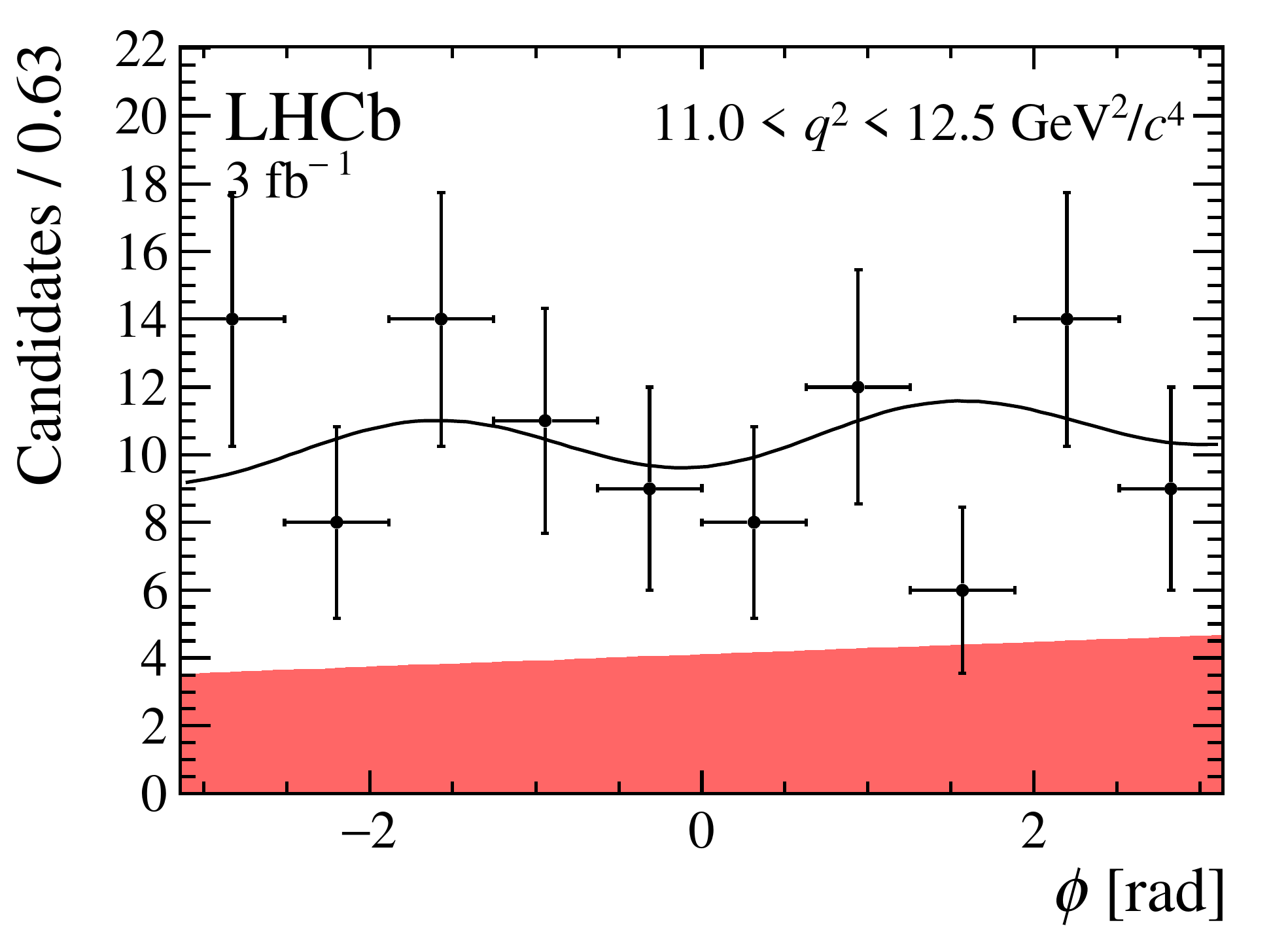}~
    \caption{\label{fig:results_bin5_run1} Mass and angular distributions of \BsToPhimm\ candidates in the region \mbox{$11.0<\qsq<12.5\gevgevcccc$} for data taken in 2011--2012. The data are overlaid with the projections of the fitted PDF.}
\end{figure}

\begin{figure}[hb]
    \centering
    \includegraphics[width=.4\textwidth]{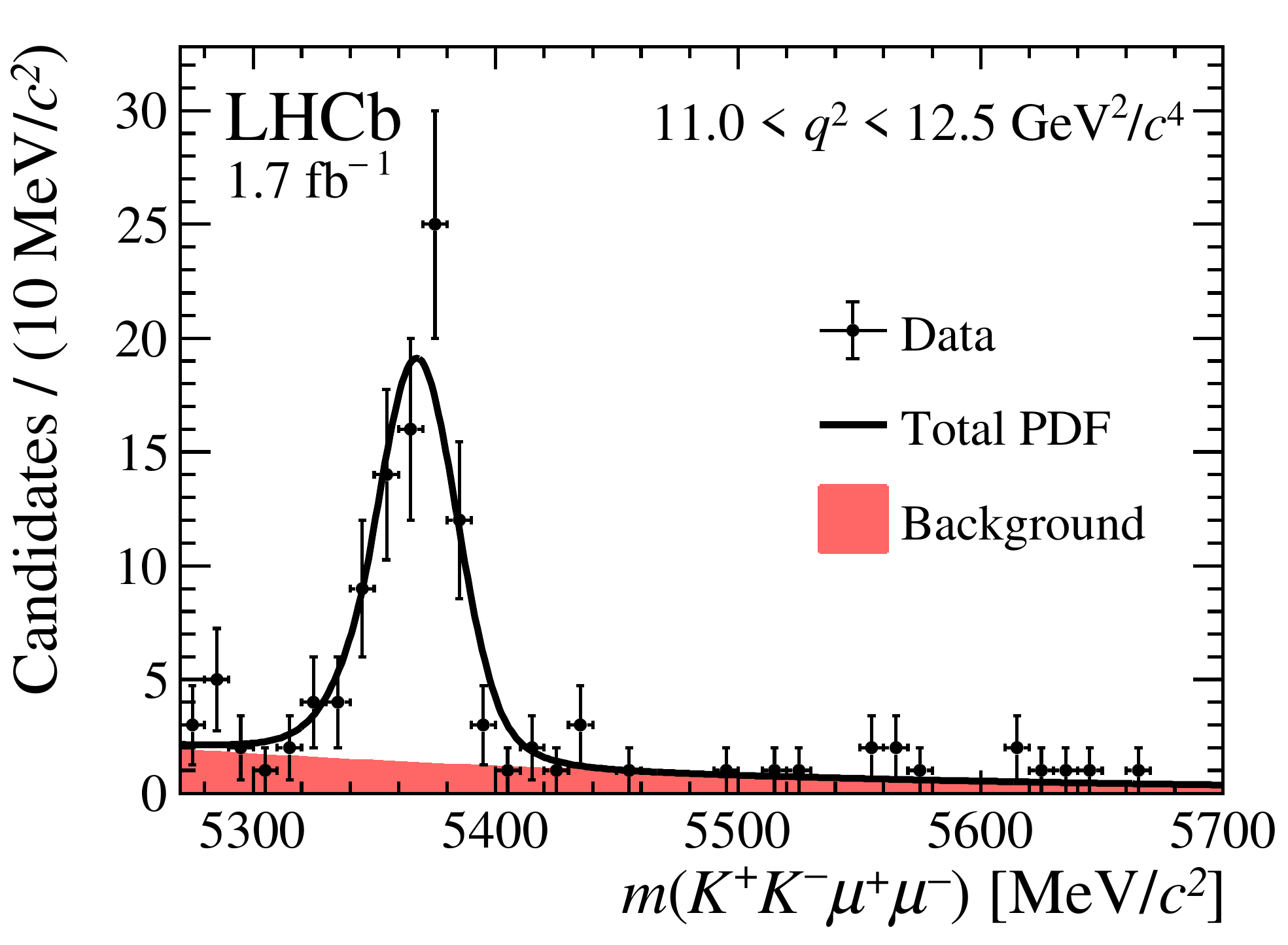}~
    \includegraphics[width=.4\textwidth]{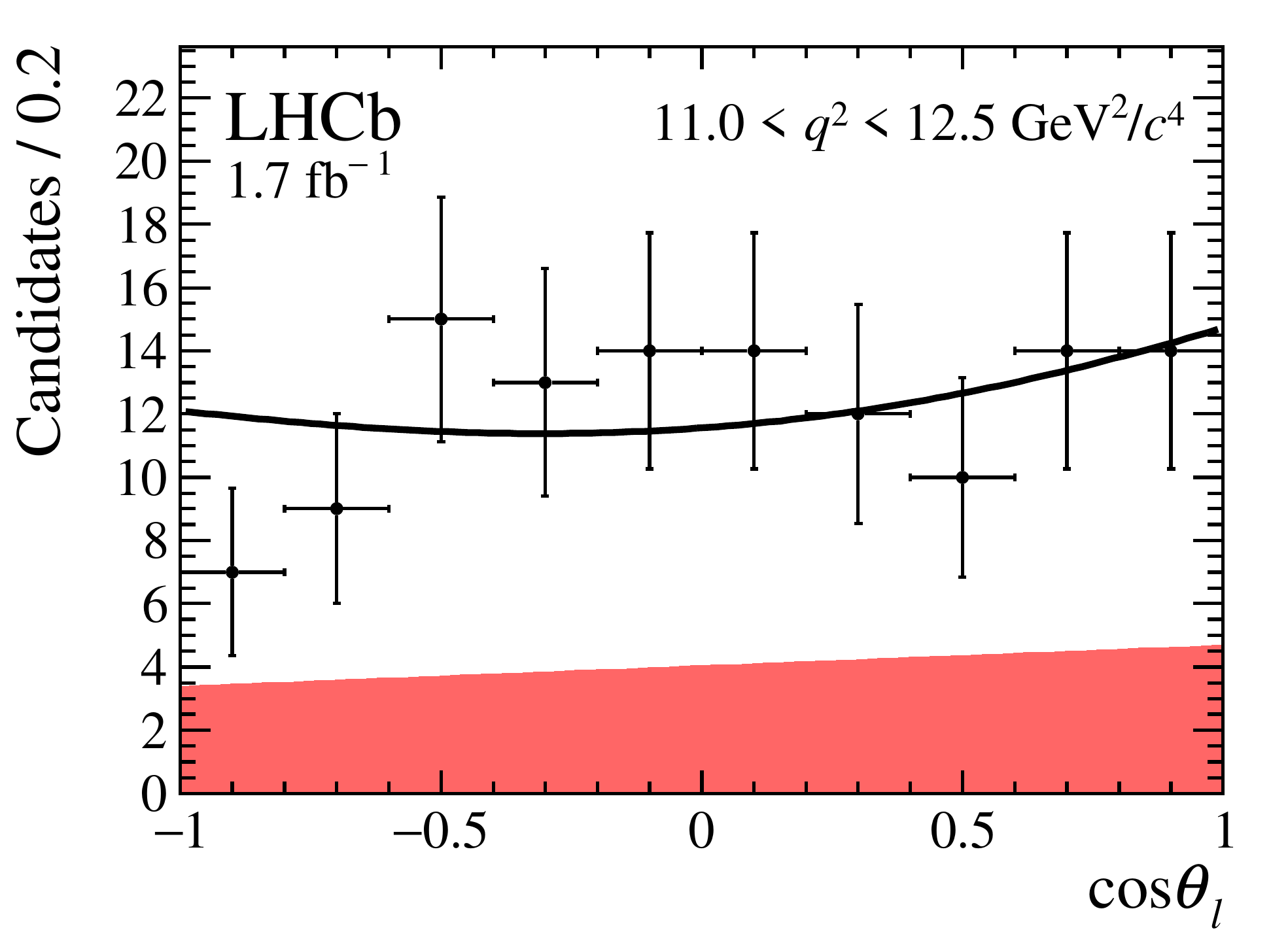}\\
    \includegraphics[width=.4\textwidth]{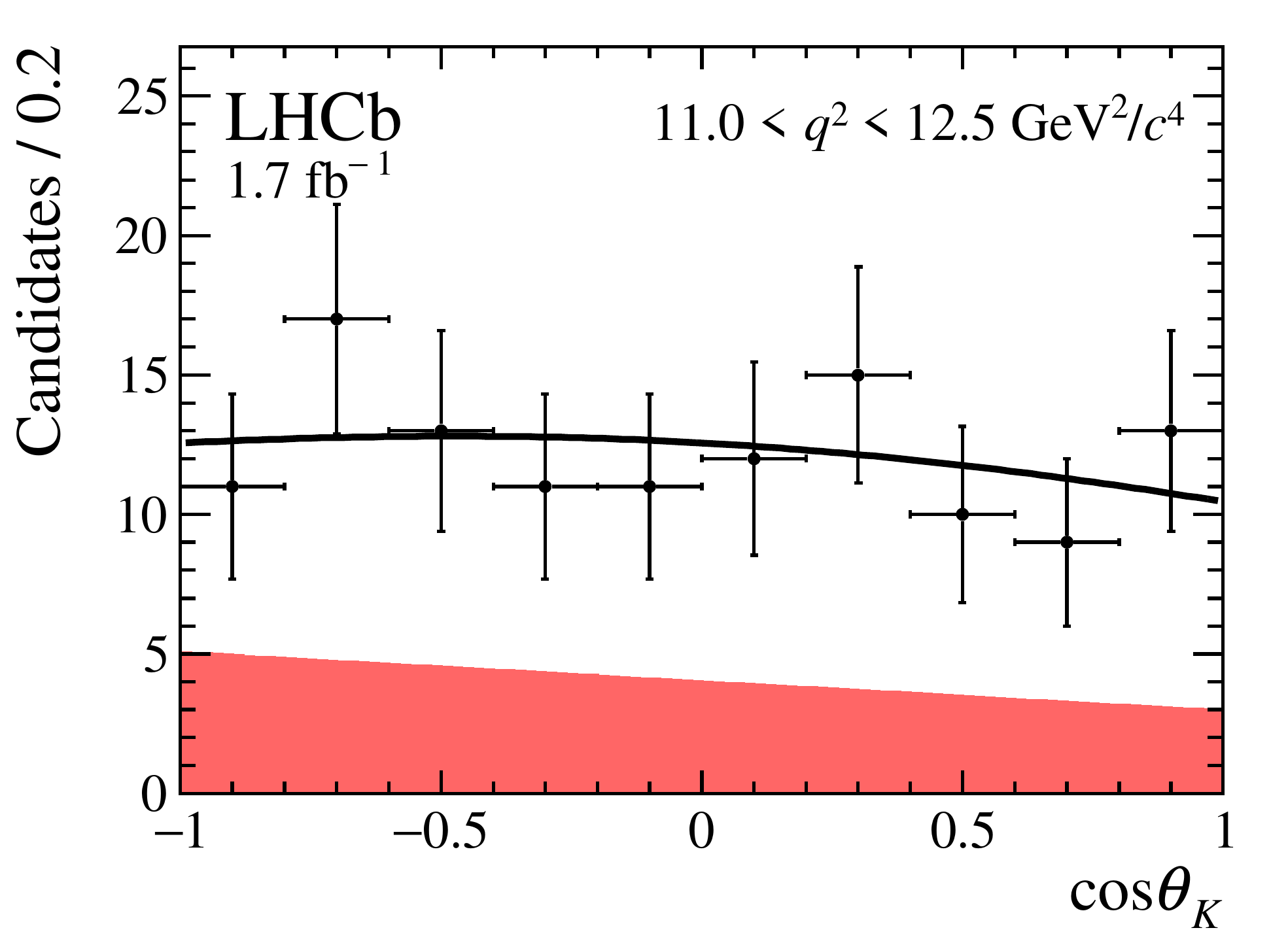}~
    \includegraphics[width=.4\textwidth]{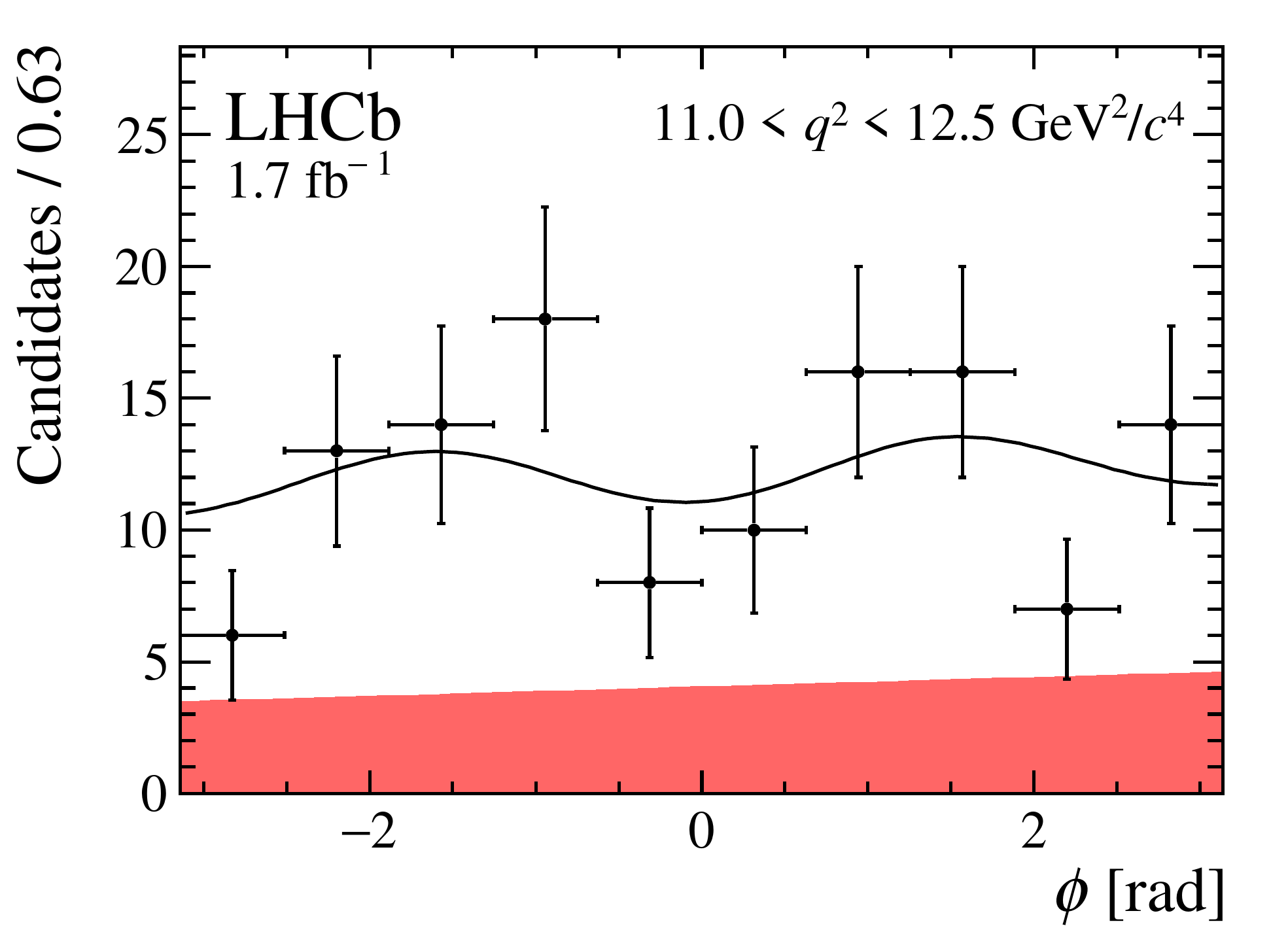}~
    \caption{\label{fig:results_bin5_run2p1} Mass and angular distributions of \BsToPhimm\ candidates in the region \mbox{$11.0<\qsq<12.5\gevgevcccc$} for data taken in 2016. The data are overlaid with the projections of the fitted PDF.}
\end{figure}

\begin{figure}[hb]
    \centering
    \includegraphics[width=.4\textwidth]{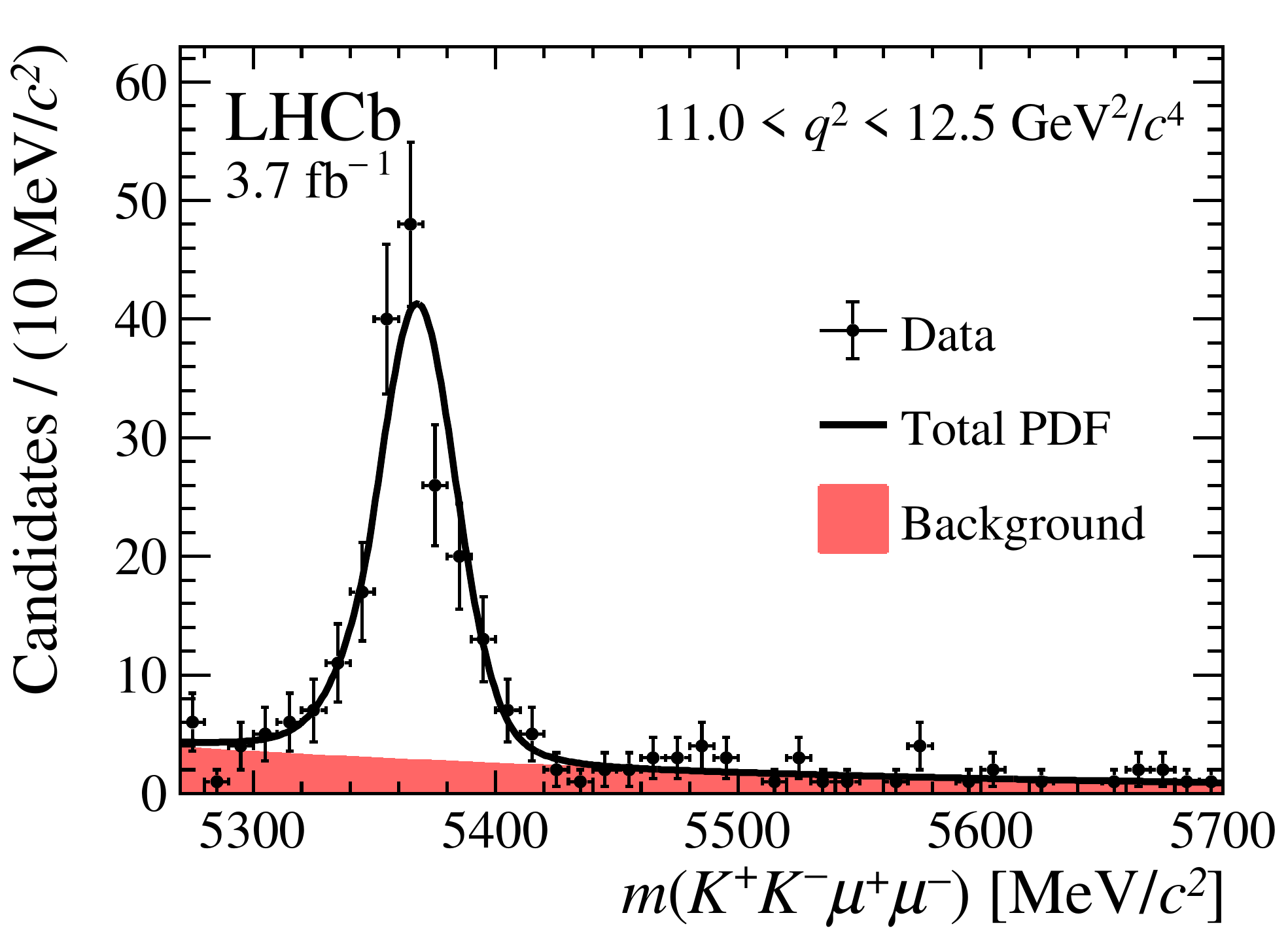}
    \includegraphics[width=.4\textwidth]{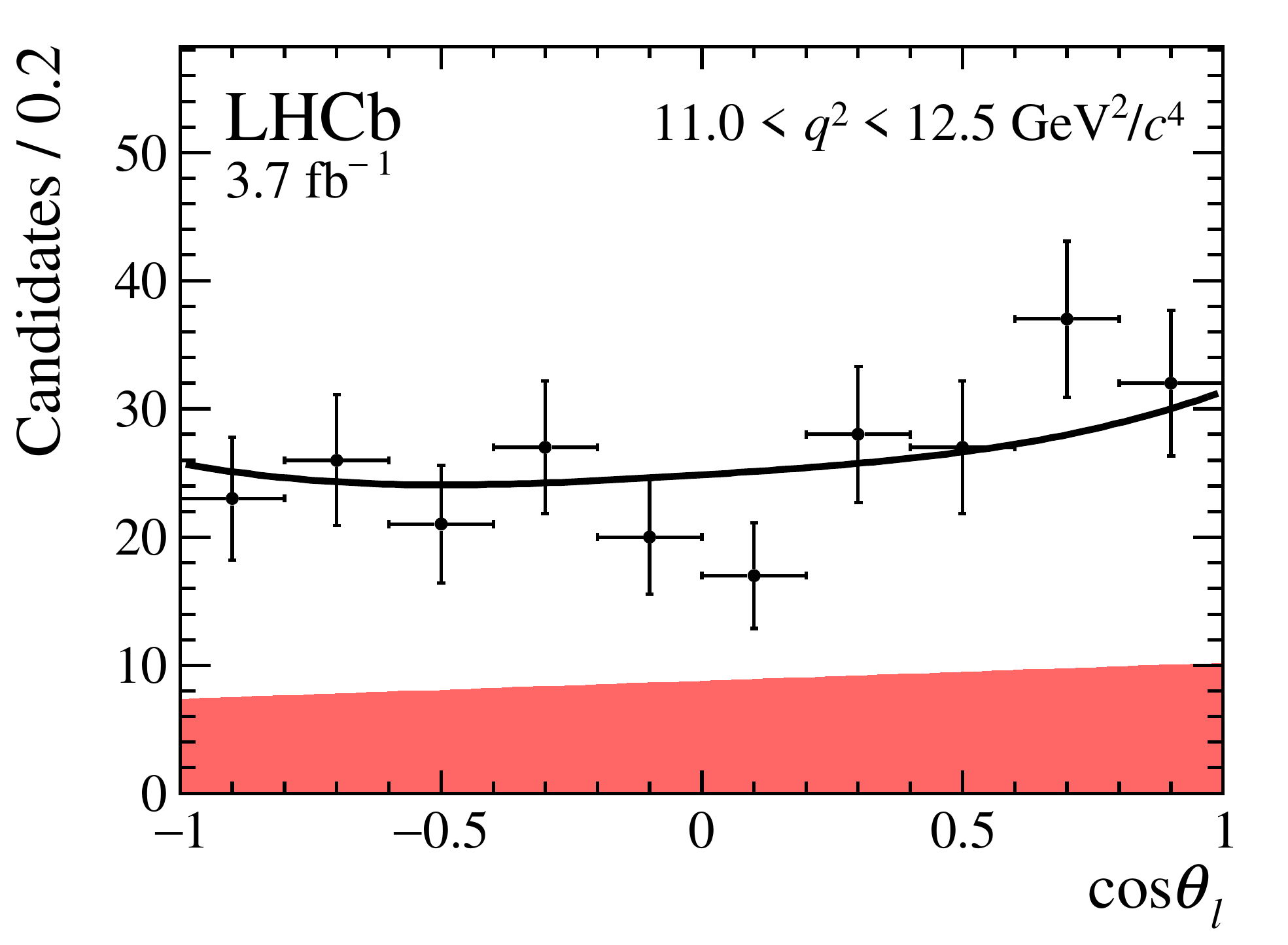}\\
    \includegraphics[width=.4\textwidth]{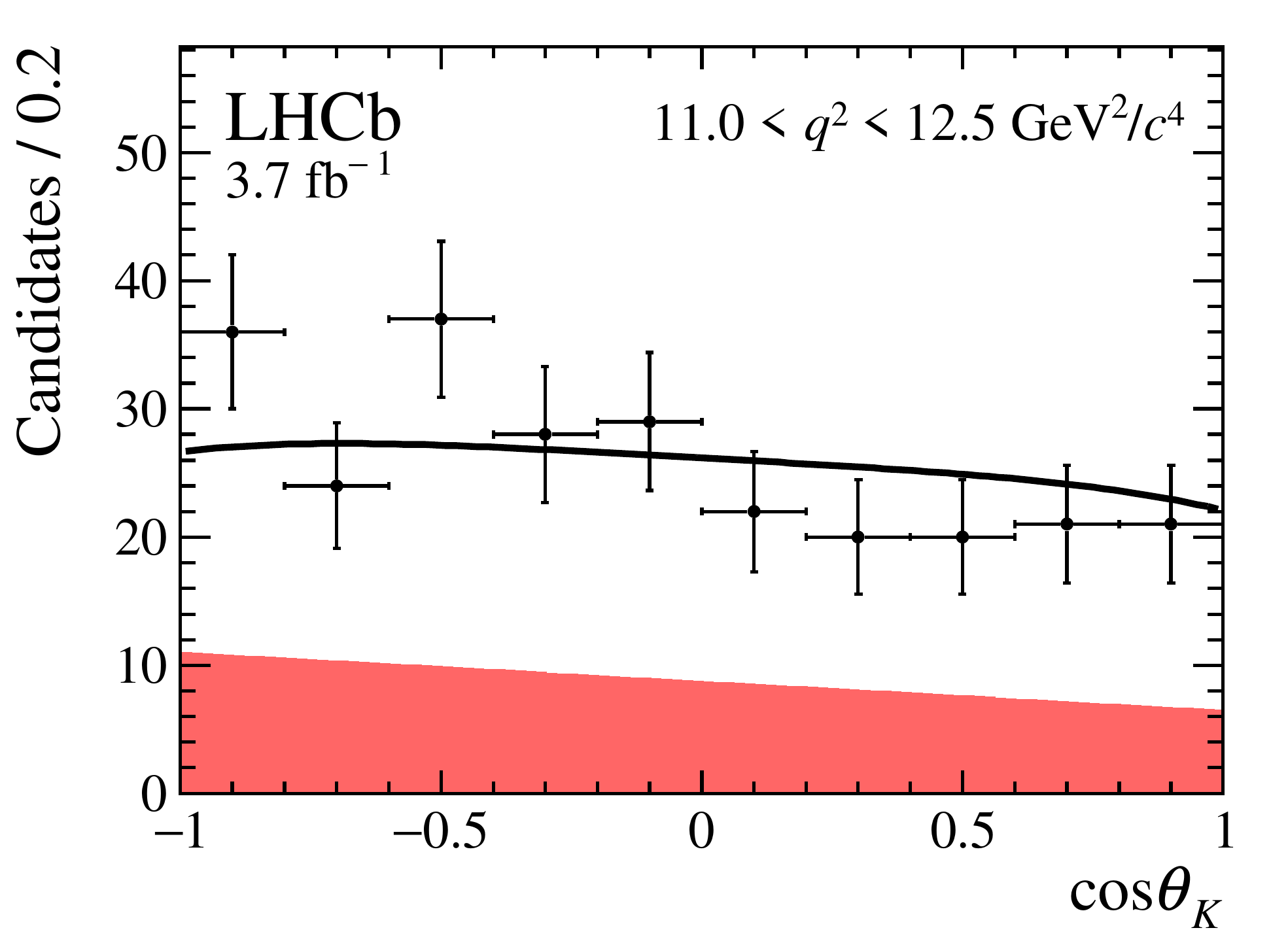}~
    \includegraphics[width=.4\textwidth]{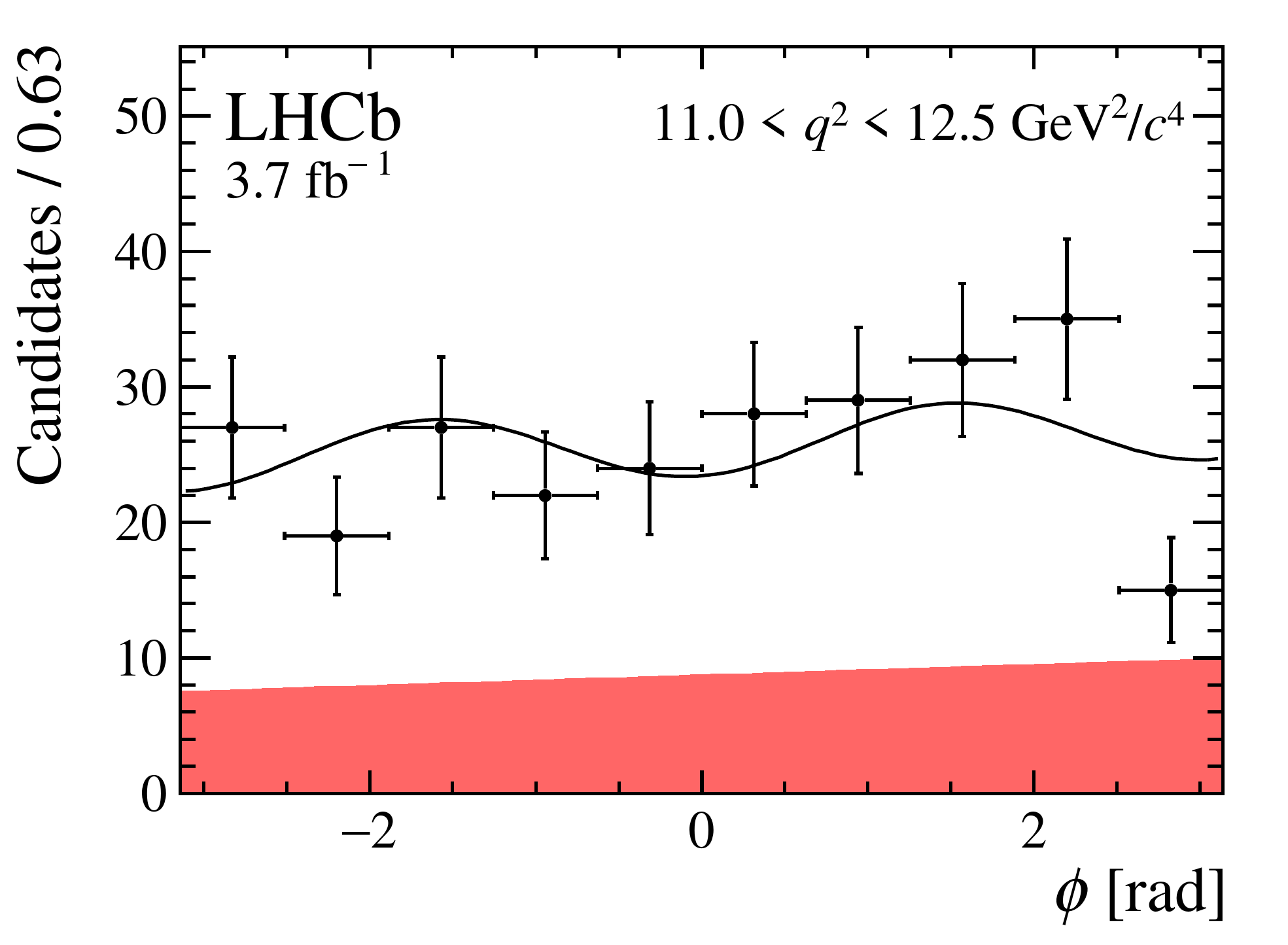}~
    \caption{\label{fig:results_bin5_run2p2} Mass and angular distributions of \BsToPhimm\ candidates in the region \mbox{$11.0<\qsq<12.5\gevgevcccc$} for data taken in 2017--2018. The data are overlaid with the projections of the fitted PDF. }
\end{figure}

\begin{figure}[hb]
    \centering
    \includegraphics[width=.4\textwidth]{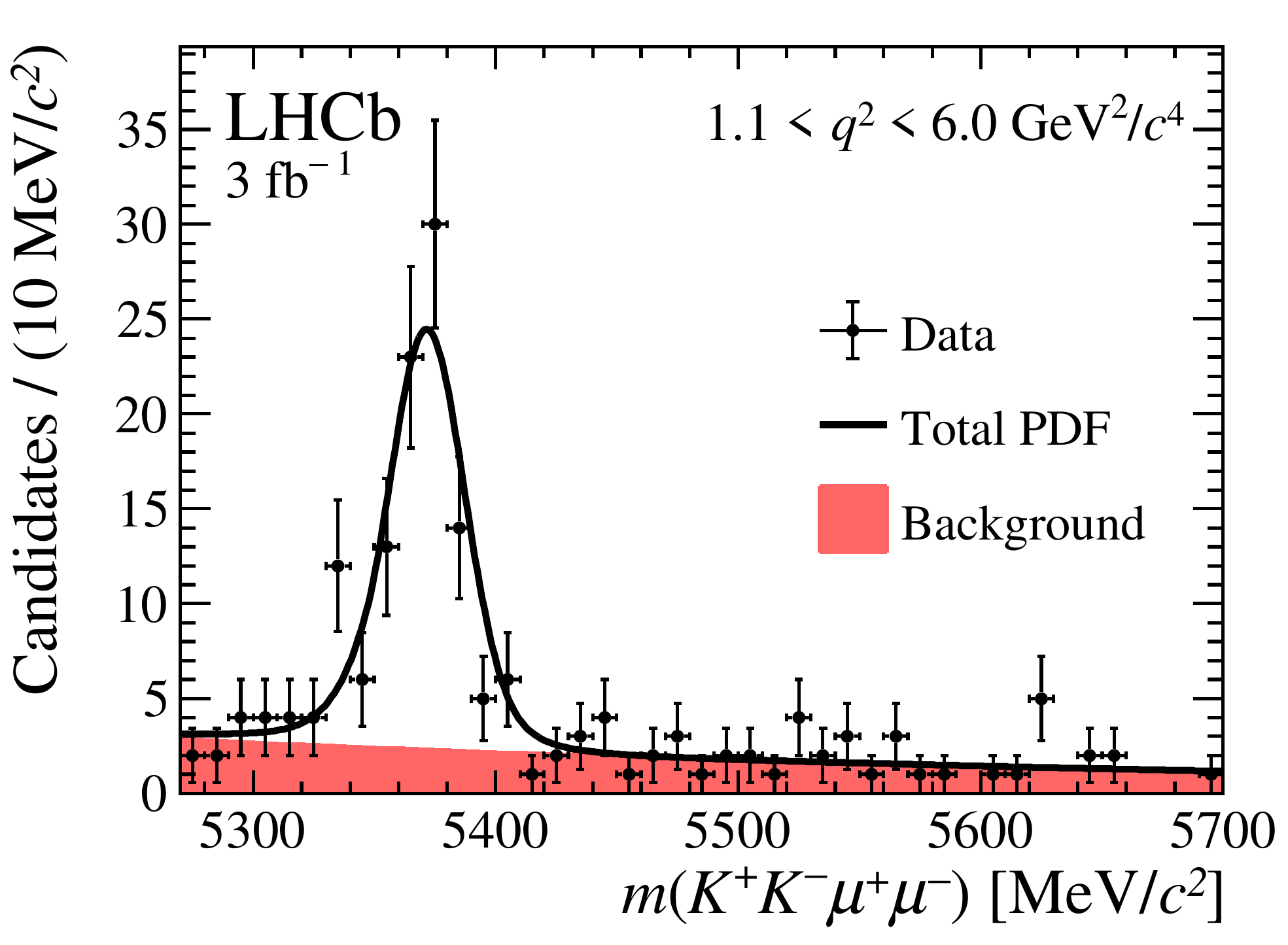}~
    \includegraphics[width=.4\textwidth]{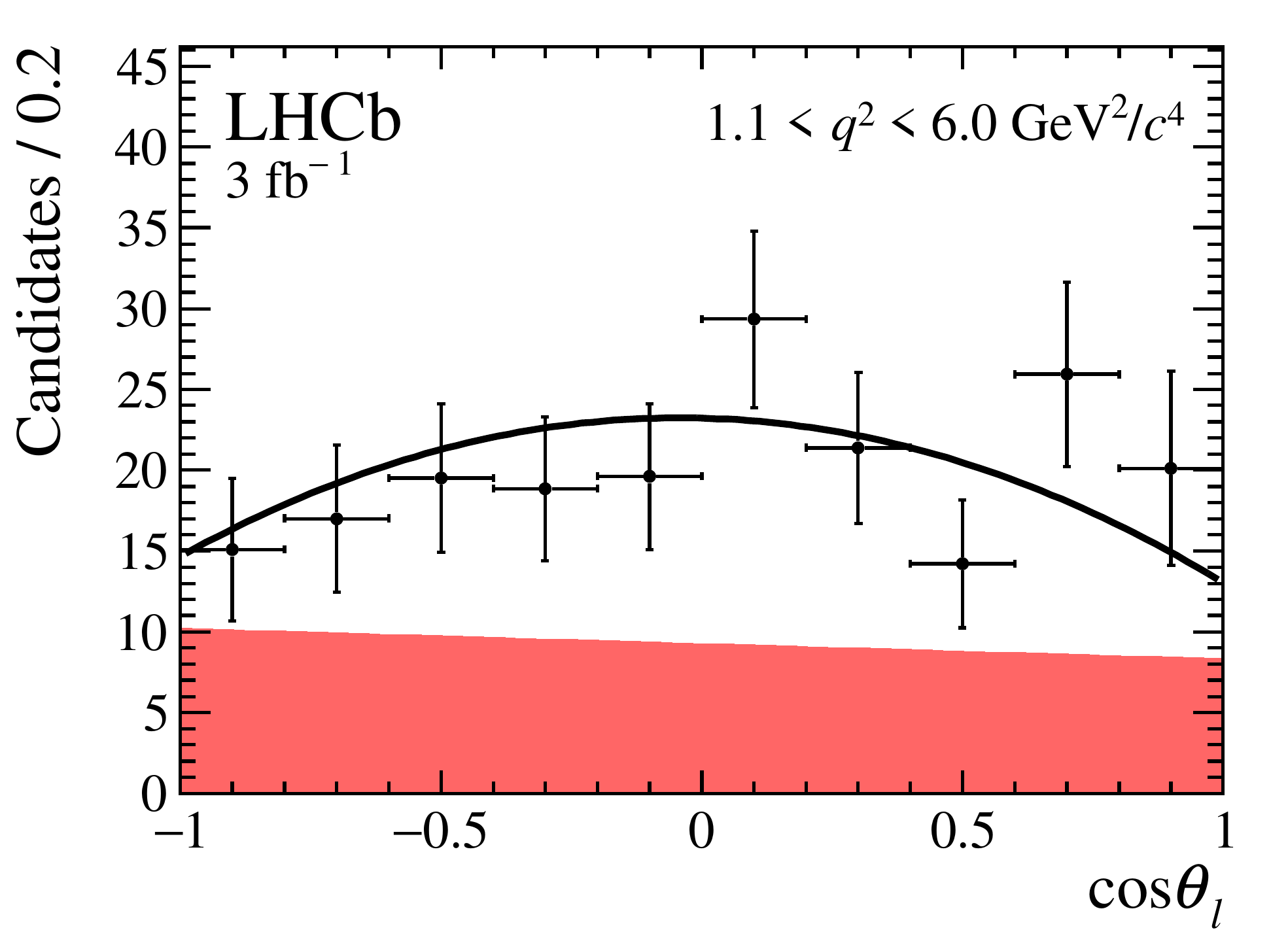}\\
    \includegraphics[width=.4\textwidth]{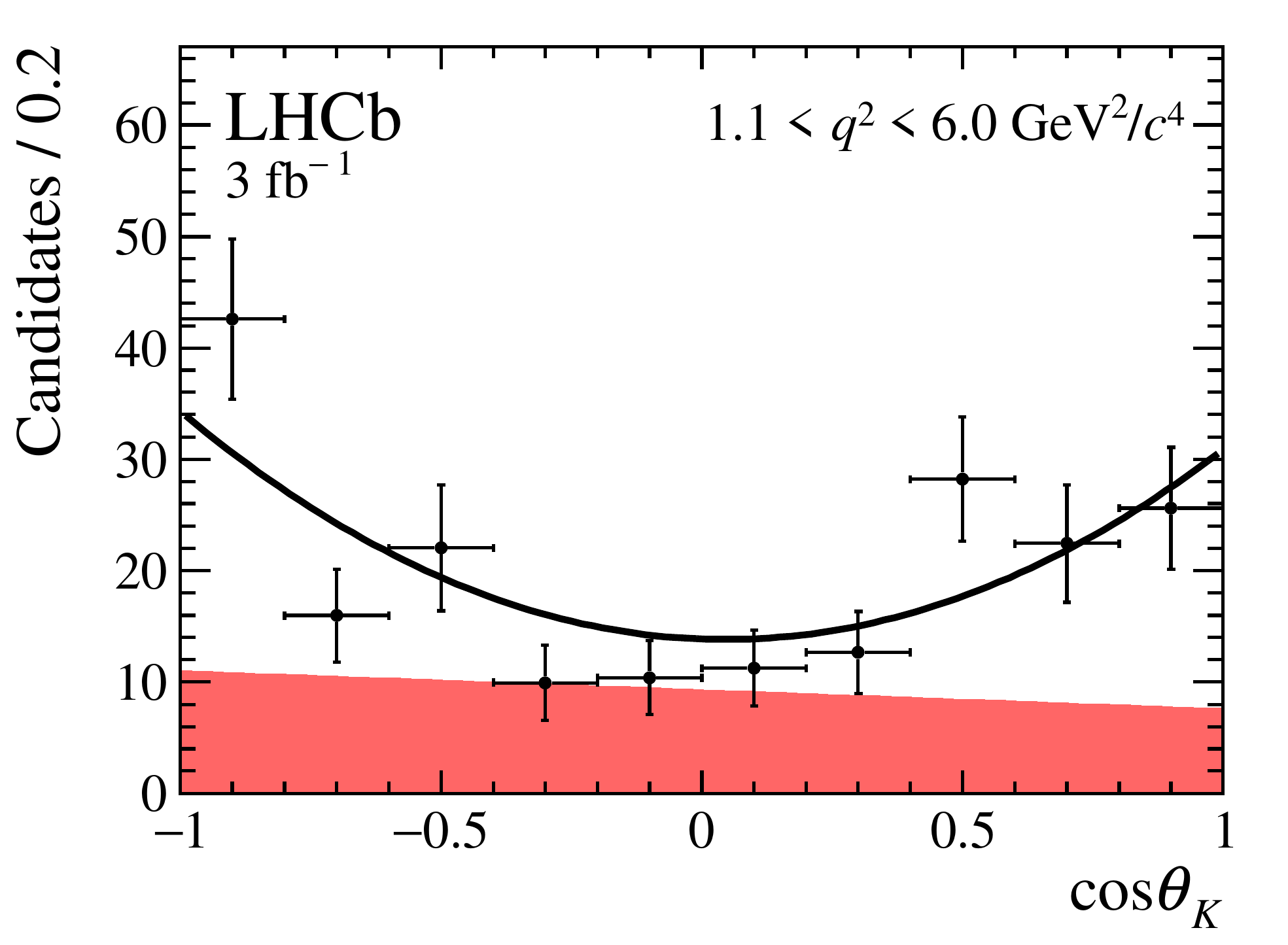}~
    \includegraphics[width=.4\textwidth]{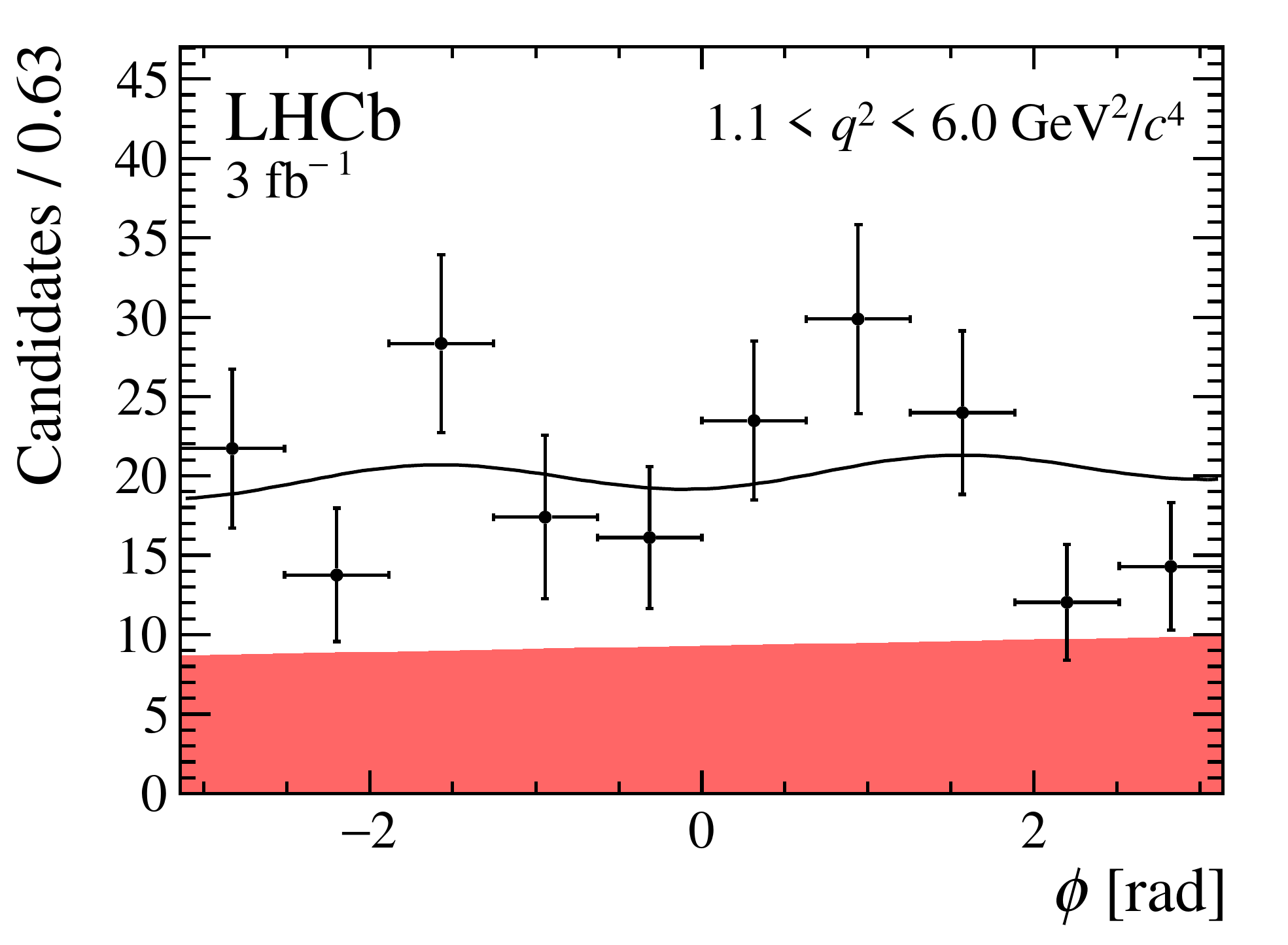}~
    \caption{\label{fig:results_bin6_run1} Mass and angular distributions of \BsToPhimm\ candidates in the region \mbox{$1.1<\qsq<6.0\gevgevcccc$} for data taken in 2011--2012. The data are overlaid with the projections of the fitted PDF. }
\end{figure}

\begin{figure}[hb]
    \centering
    \includegraphics[width=.4\textwidth]{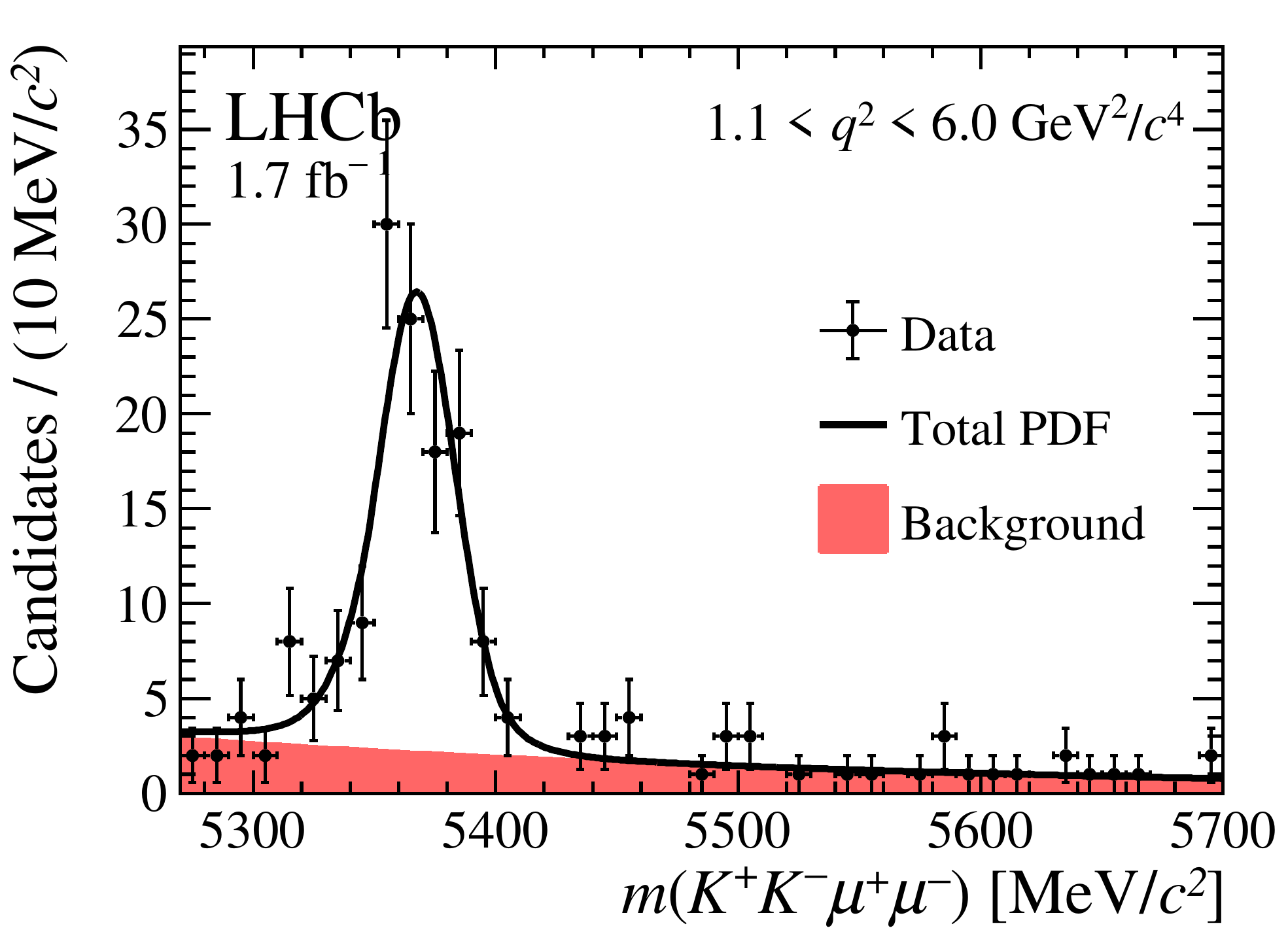}~
    \includegraphics[width=.4\textwidth]{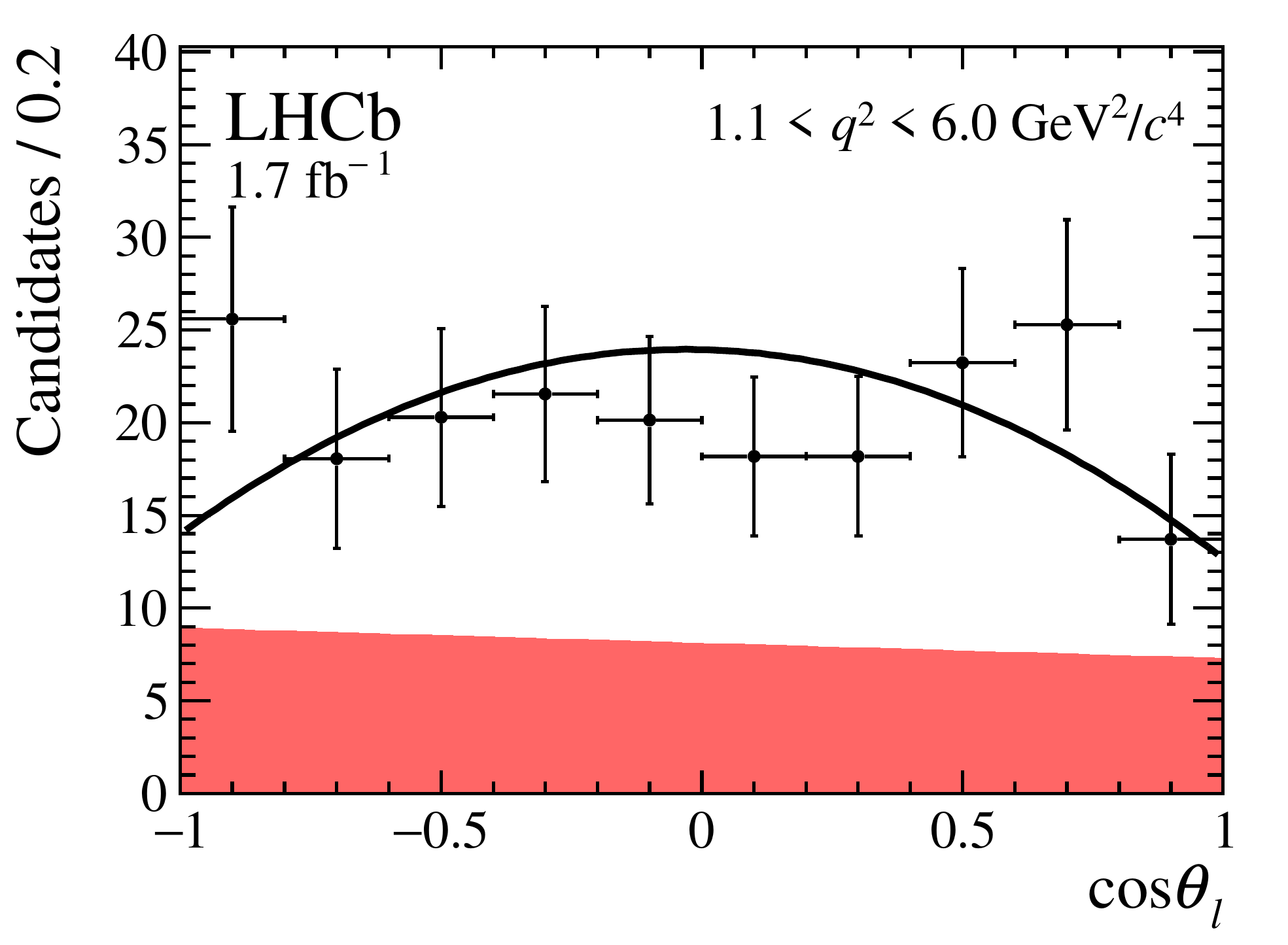}\\
    \includegraphics[width=.4\textwidth]{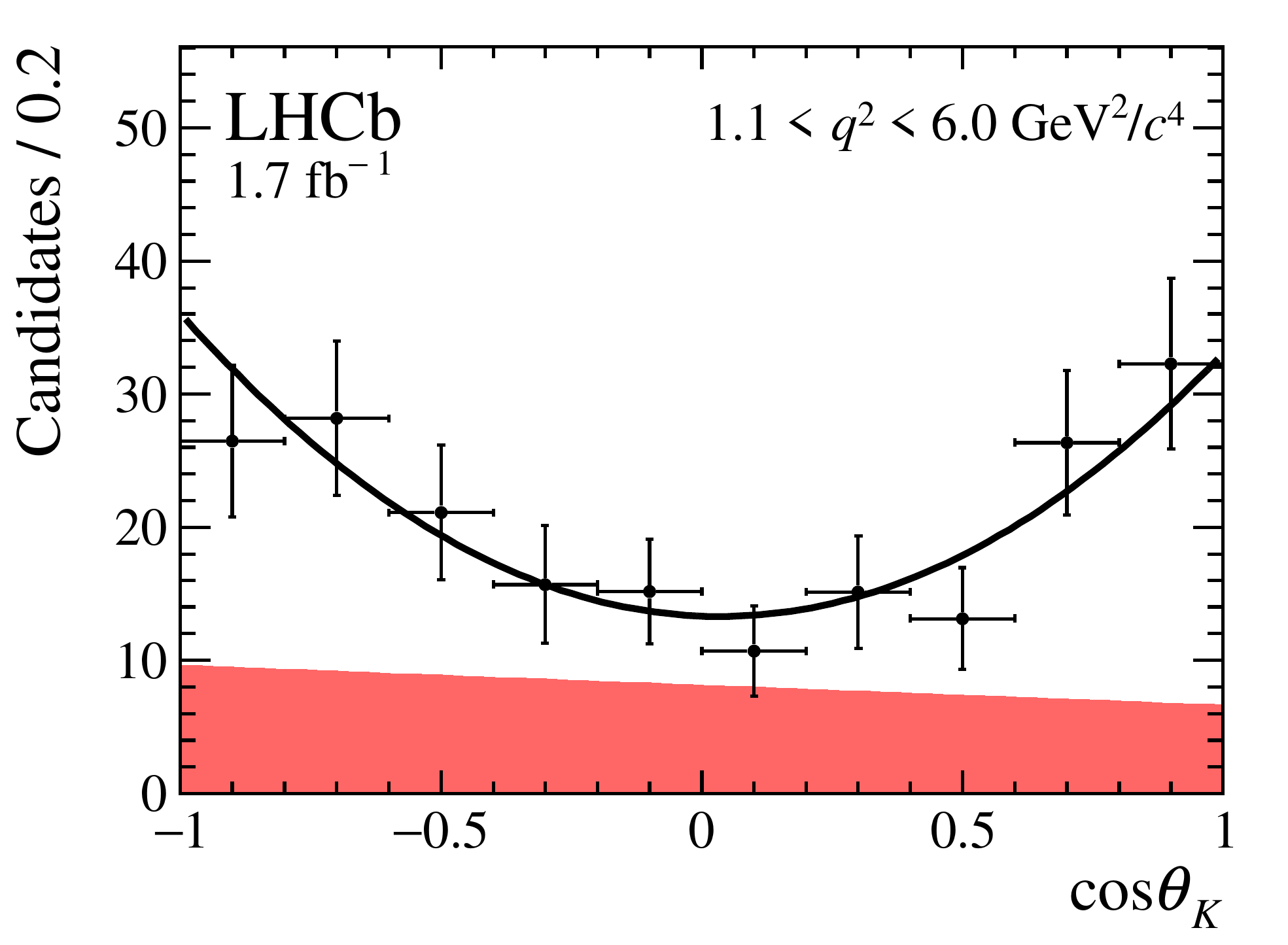}~
    \includegraphics[width=.4\textwidth]{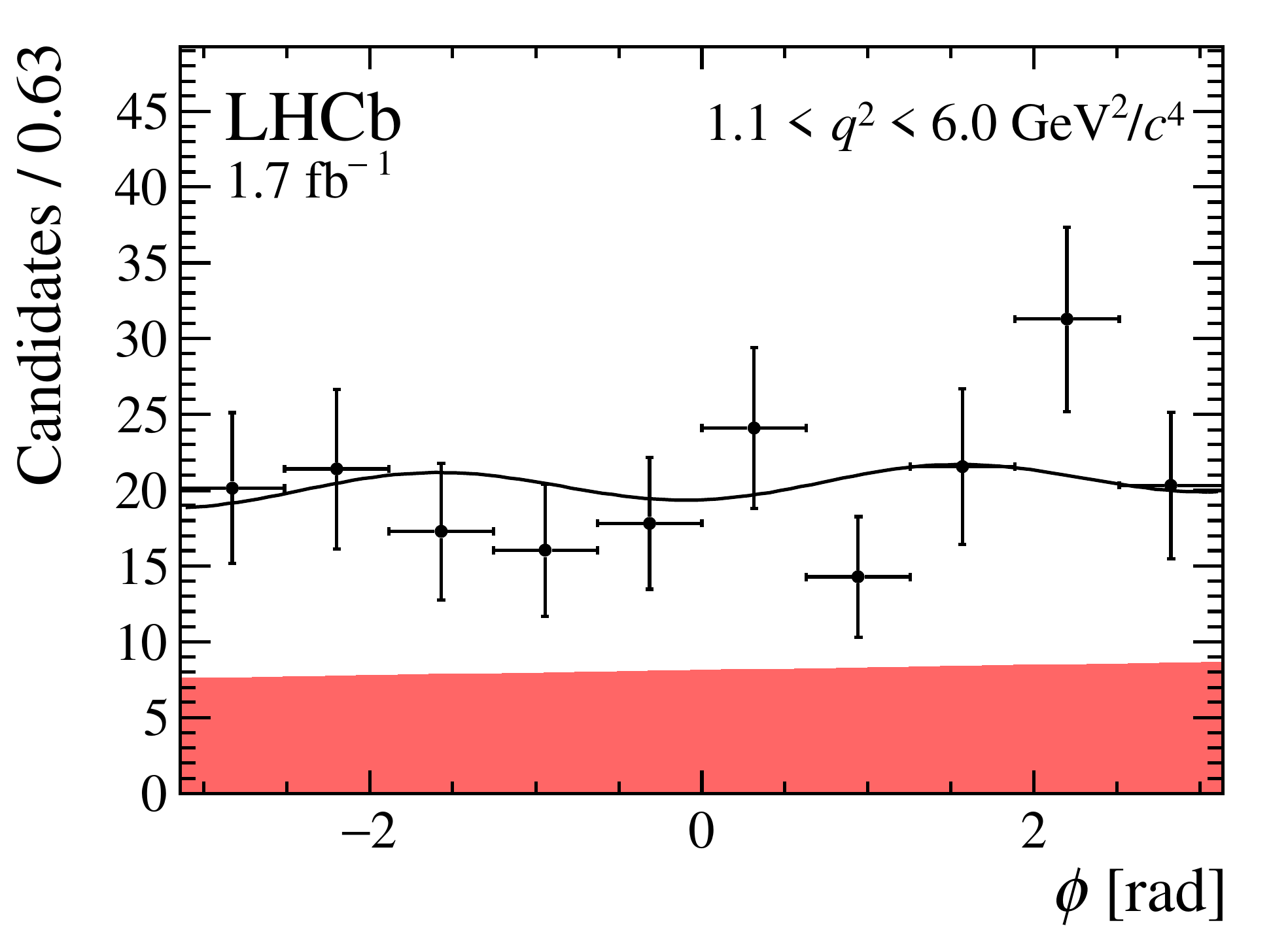}~
    \caption{\label{fig:results_bin6_run2p1} Mass and angular distributions of \BsToPhimm\ candidates in the region \mbox{$1.1<\qsq<6.0\gevgevcccc$} for data taken in 2016. The data are overlaid with the projections of the fitted PDF. }
\end{figure}

\begin{figure}[hb]
    \centering
    \includegraphics[width=.4\textwidth]{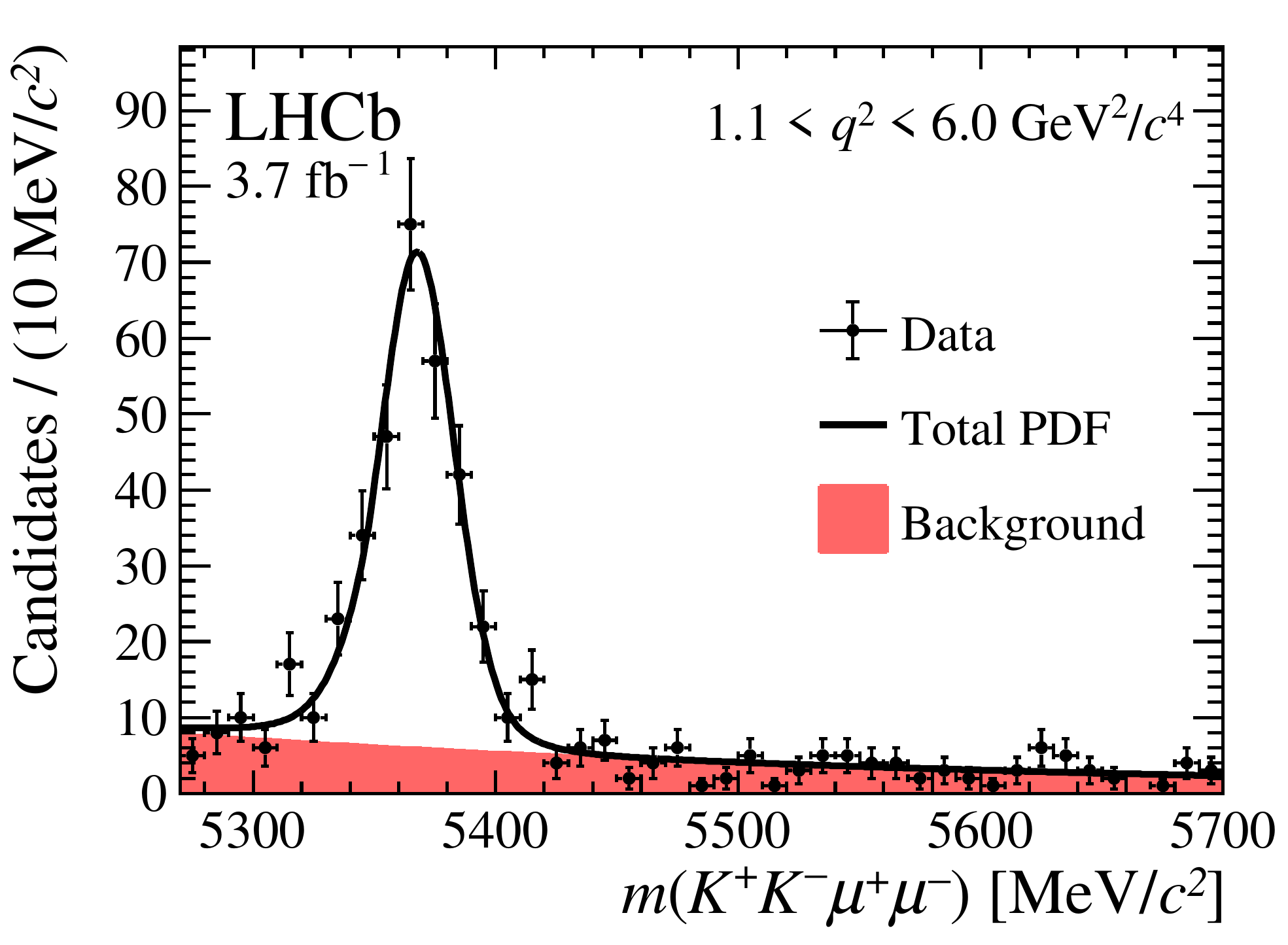}~
    \includegraphics[width=.4\textwidth]{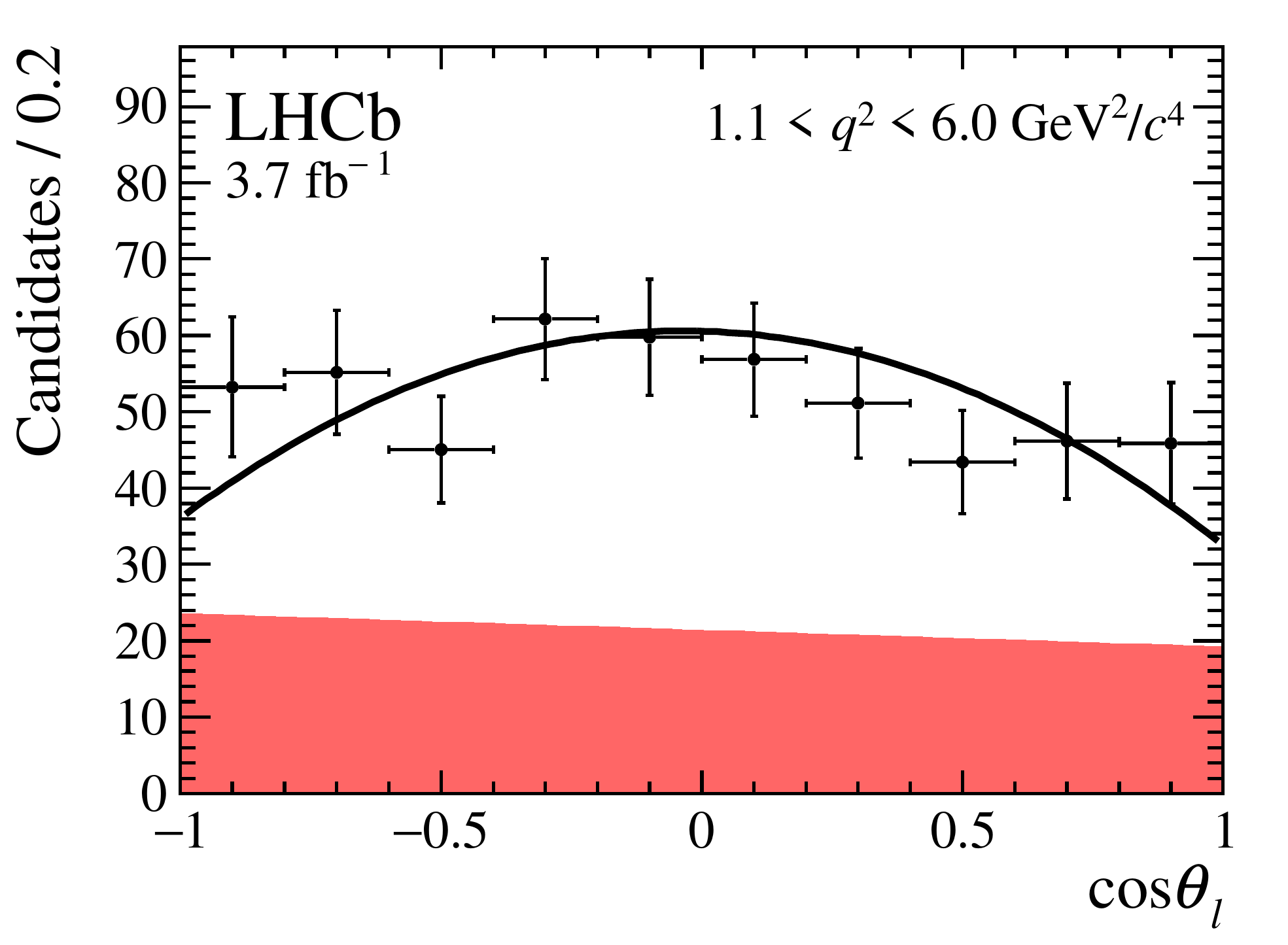}\\
    \includegraphics[width=.4\textwidth]{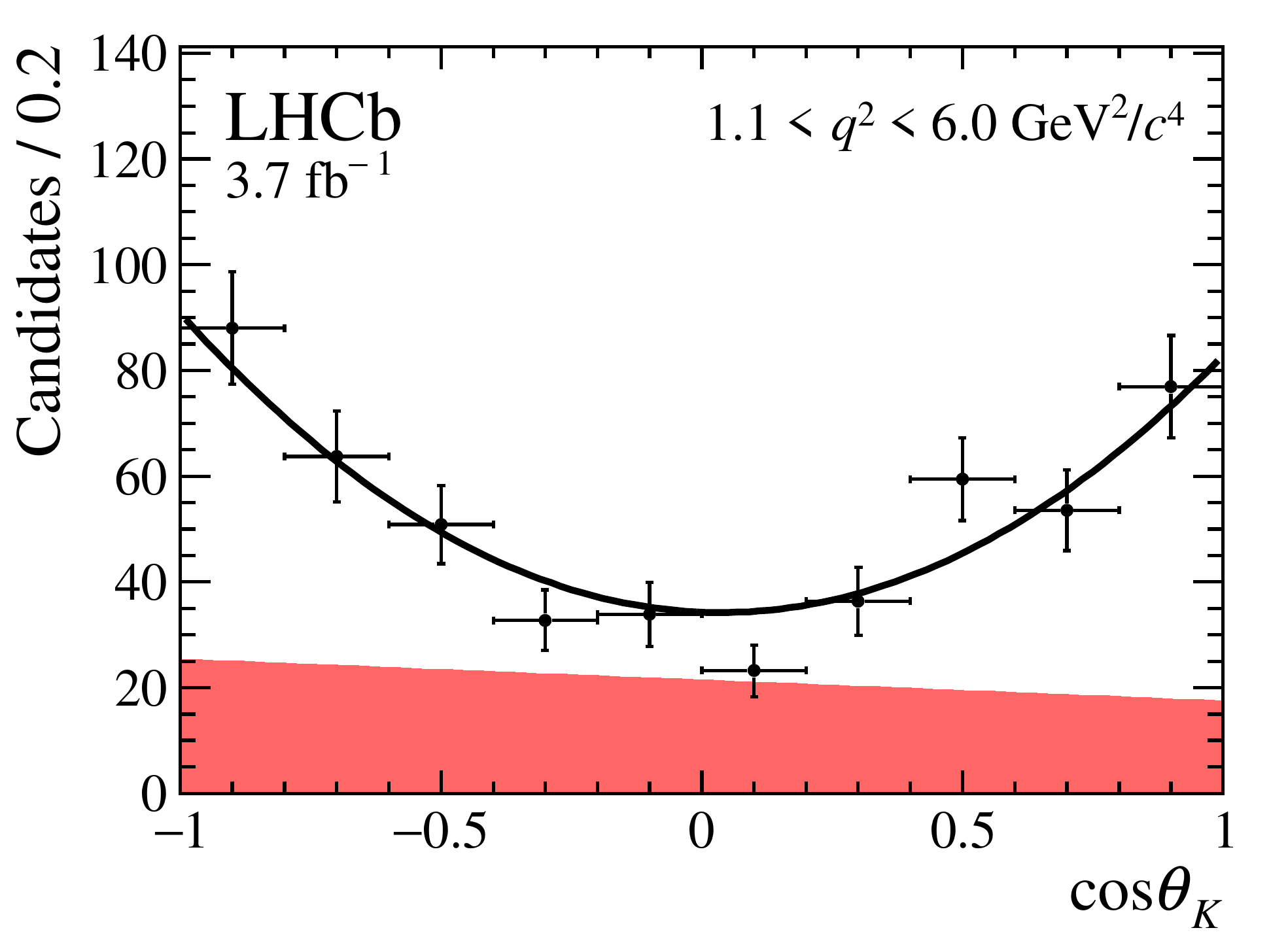}~
    \includegraphics[width=.4\textwidth]{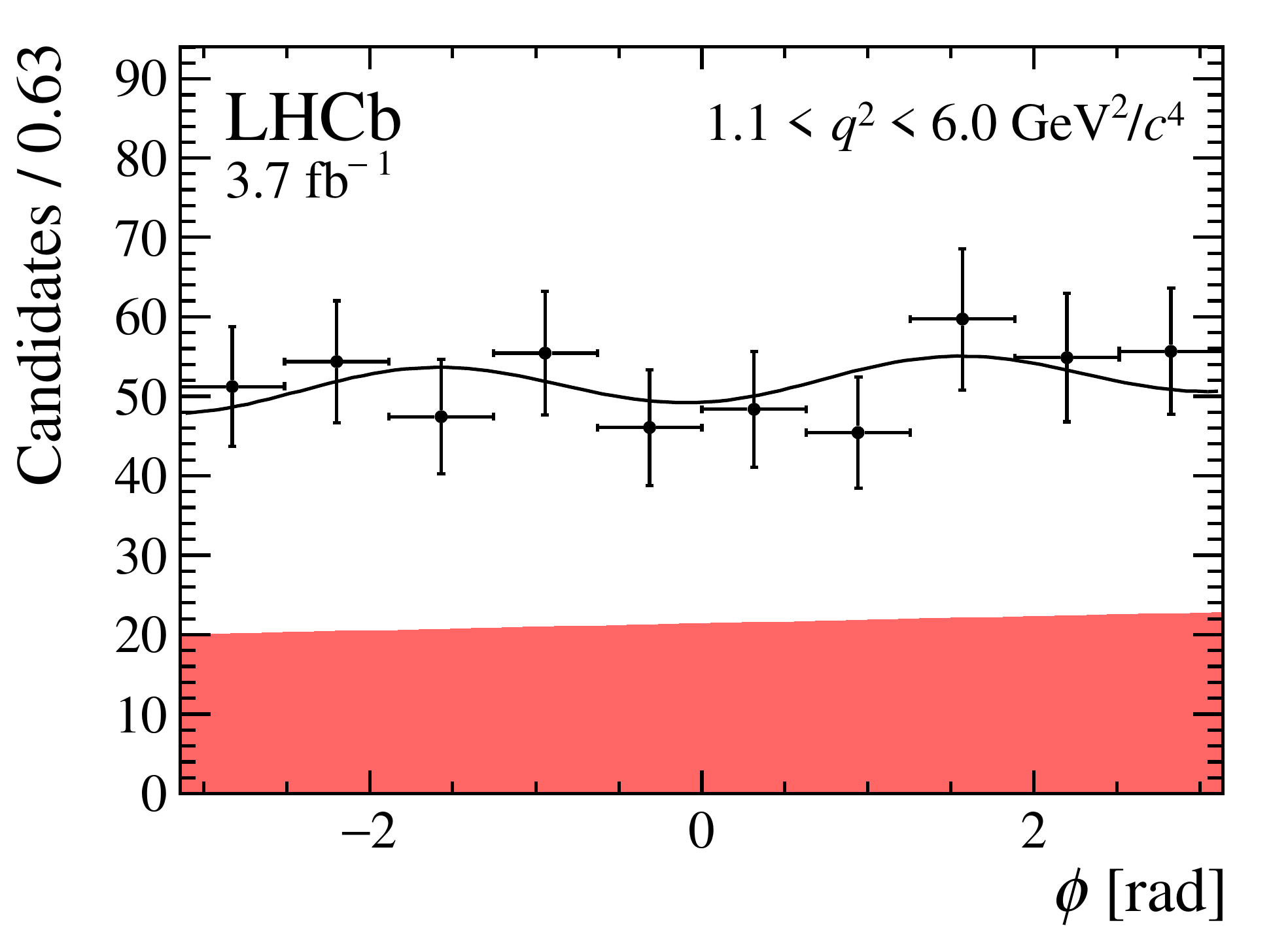}~
    \caption{\label{fig:results_bin6_run2p2} Mass and angular distributions of \BsToPhimm\ candidates in the region \mbox{$1.1<\qsq<6.0\gevgevcccc$} for data taken in 2017--2018. The data are overlaid with the projections of the fitted PDF. }
\end{figure}

\begin{figure}[hb]
    \centering
    \includegraphics[width=.4\textwidth]{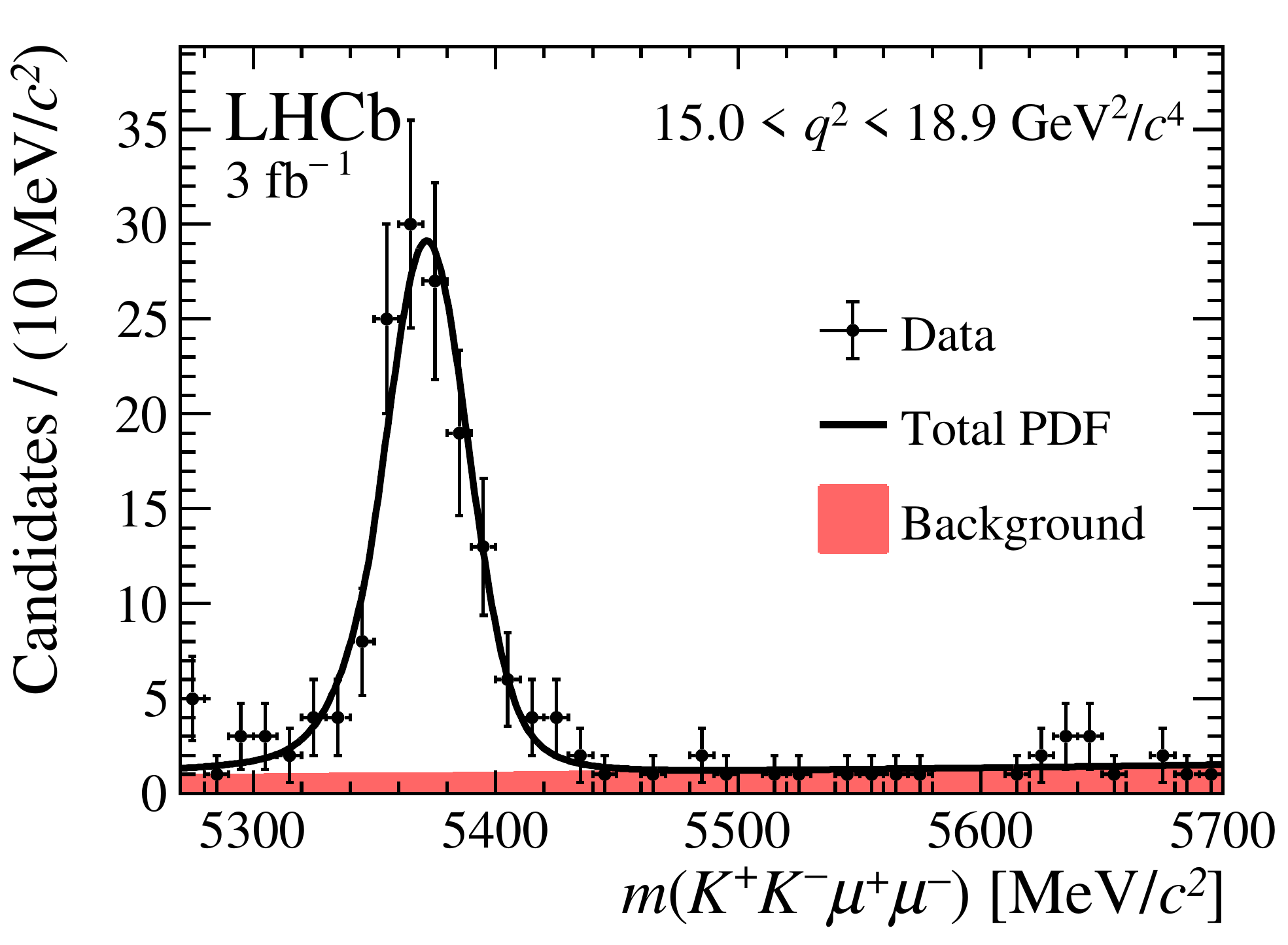}~
    \includegraphics[width=.4\textwidth]{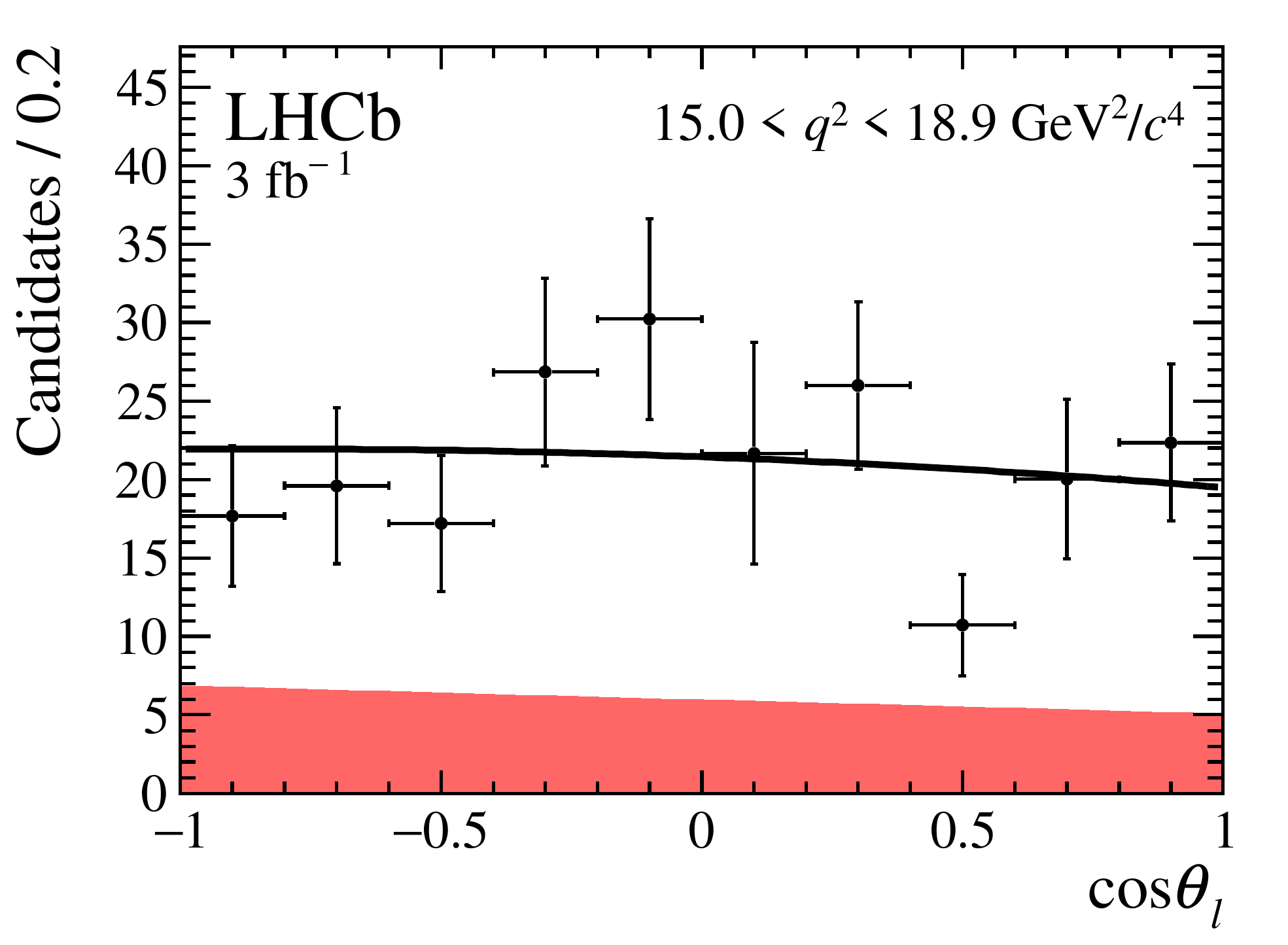}\\
    \includegraphics[width=.4\textwidth]{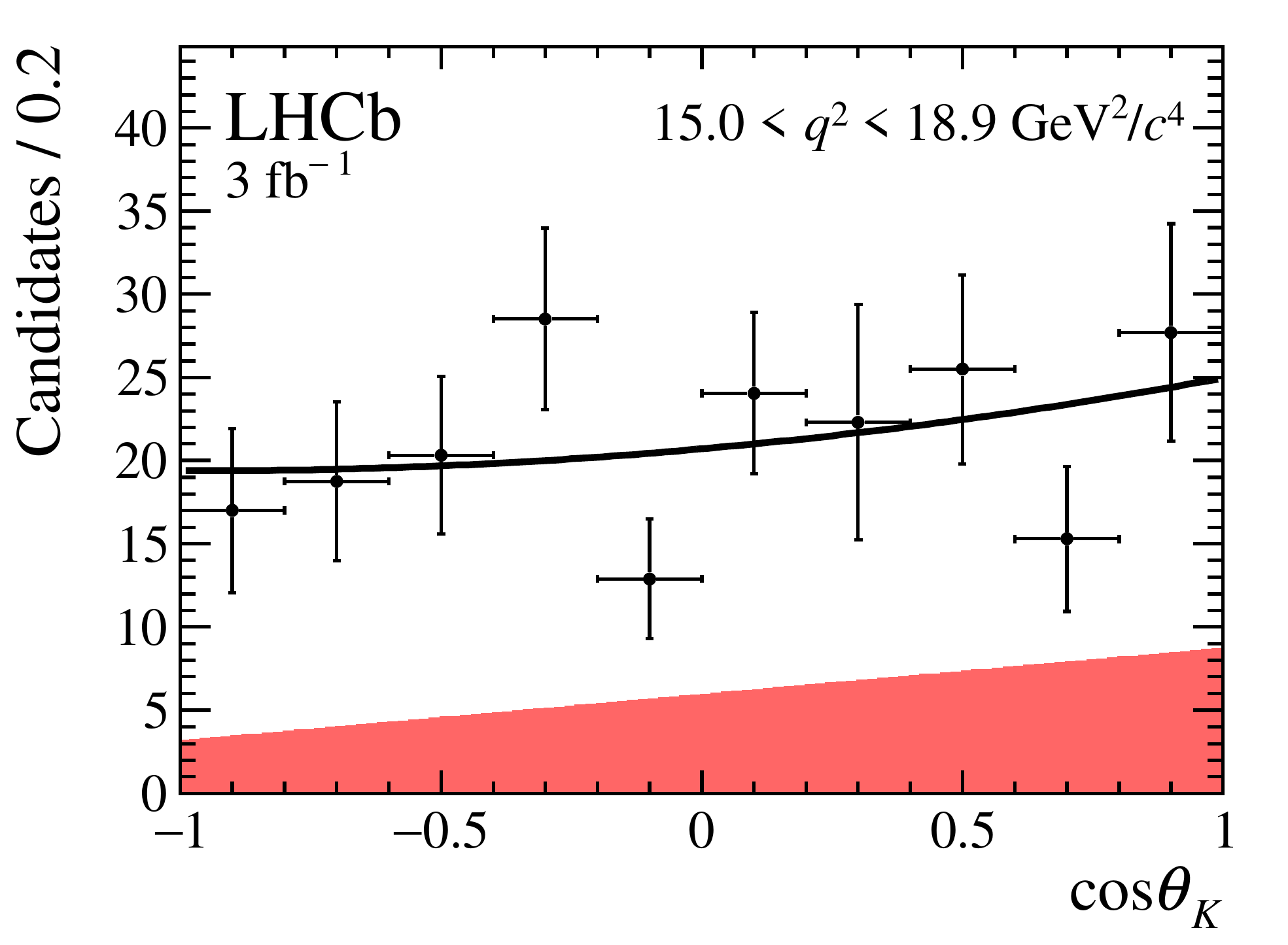}~
    \includegraphics[width=.4\textwidth]{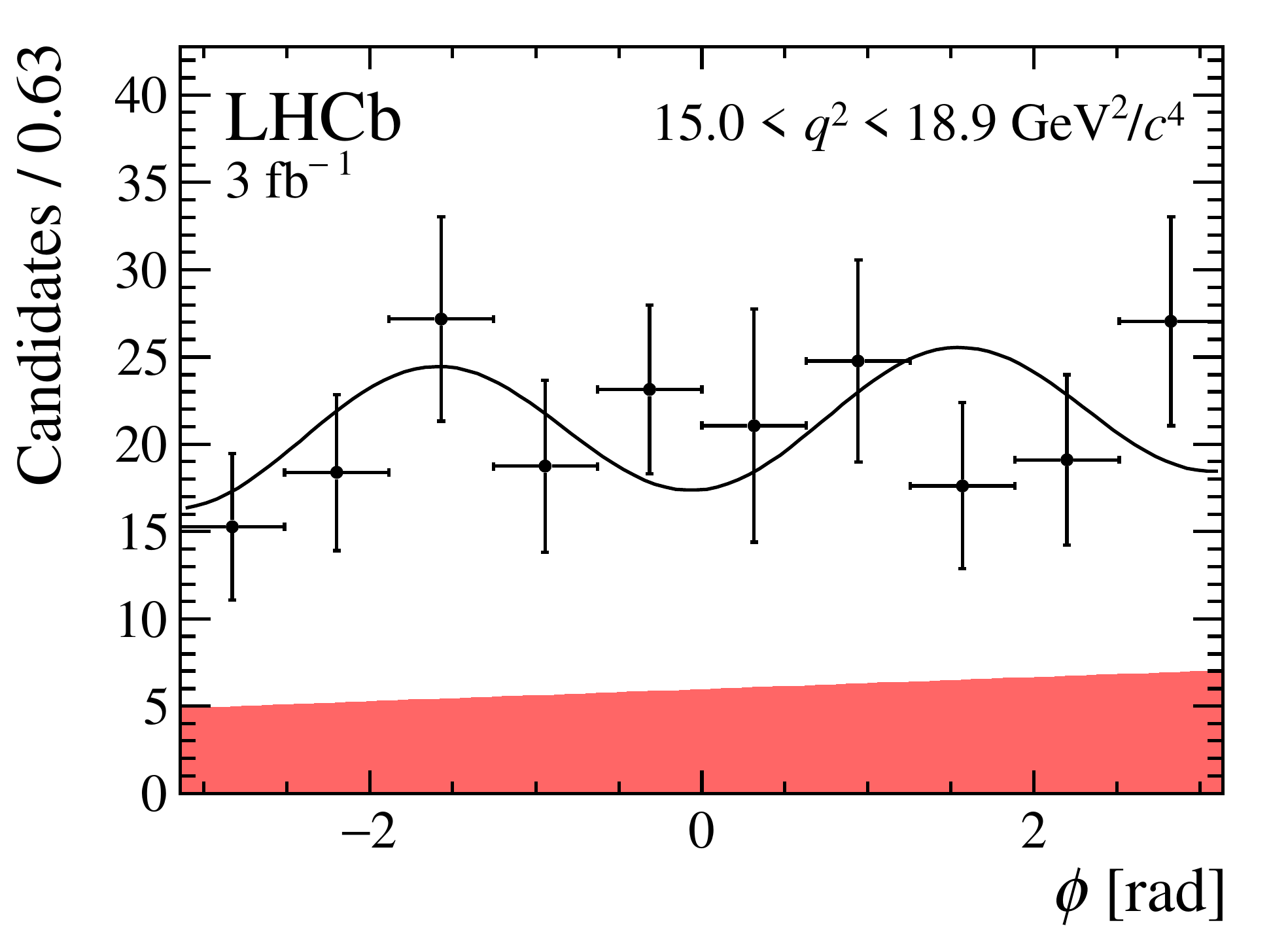}~
    \caption{\label{fig:results_bin7_run1} Mass and angular distributions of \BsToPhimm\ candidates in the region \mbox{$15.0<\qsq<18.9\gevgevcccc$} for data taken in 2011--2012. The data are overlaid with the projections of the fitted PDF. }
\end{figure}

\begin{figure}[hb]
    \centering
    \includegraphics[width=.4\textwidth]{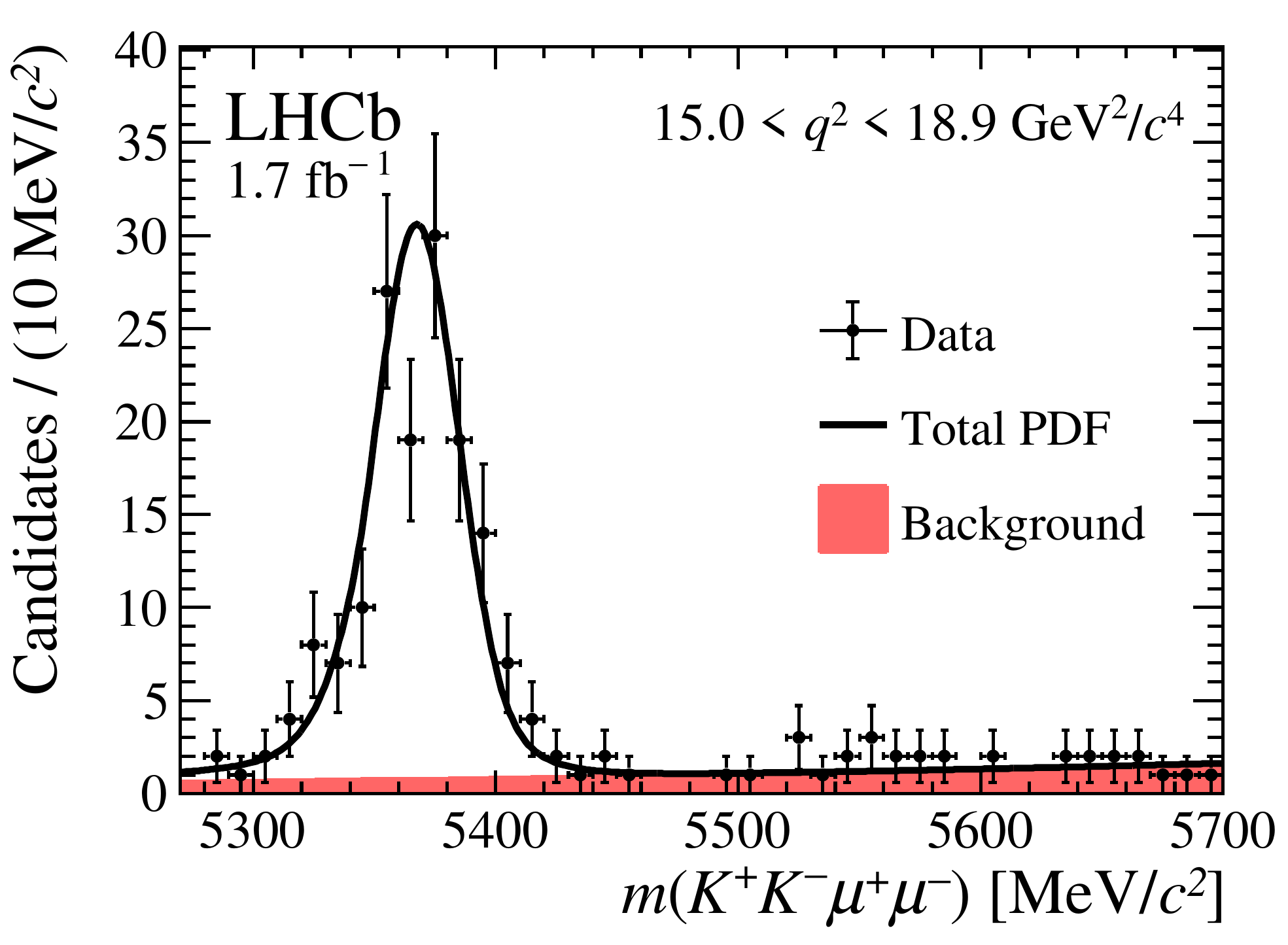}~
    \includegraphics[width=.4\textwidth]{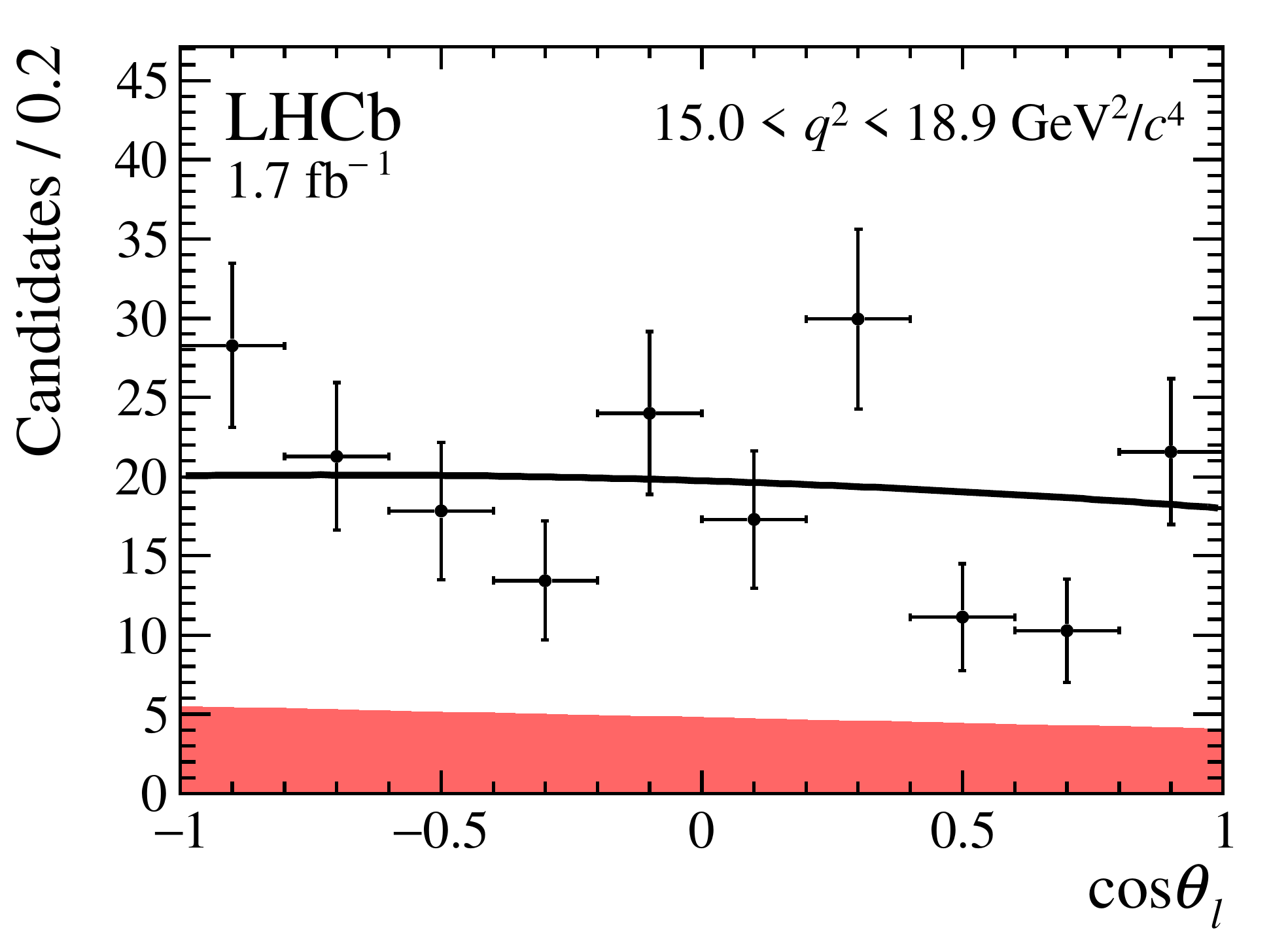}\\
    \includegraphics[width=.4\textwidth]{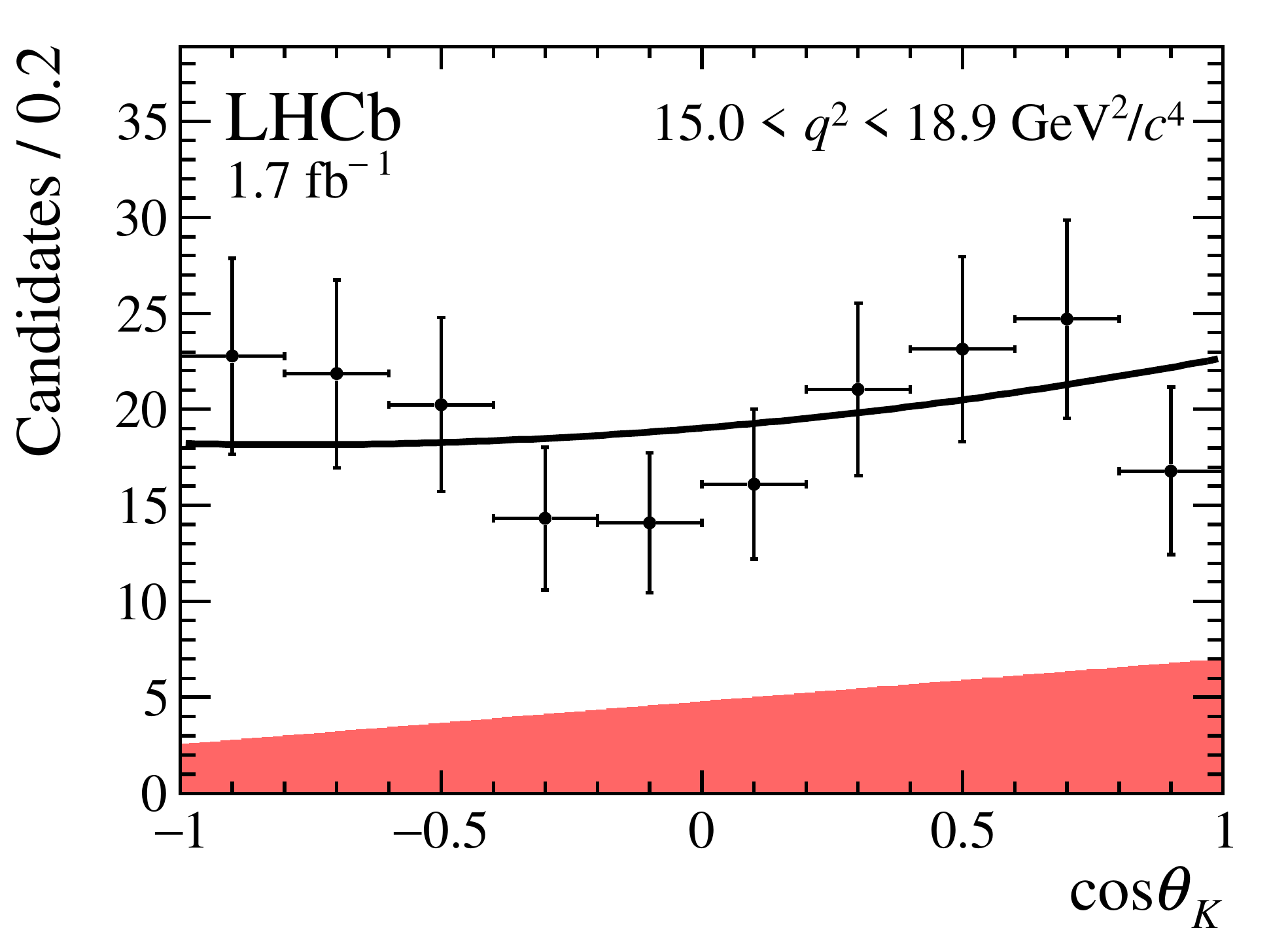}~
    \includegraphics[width=.4\textwidth]{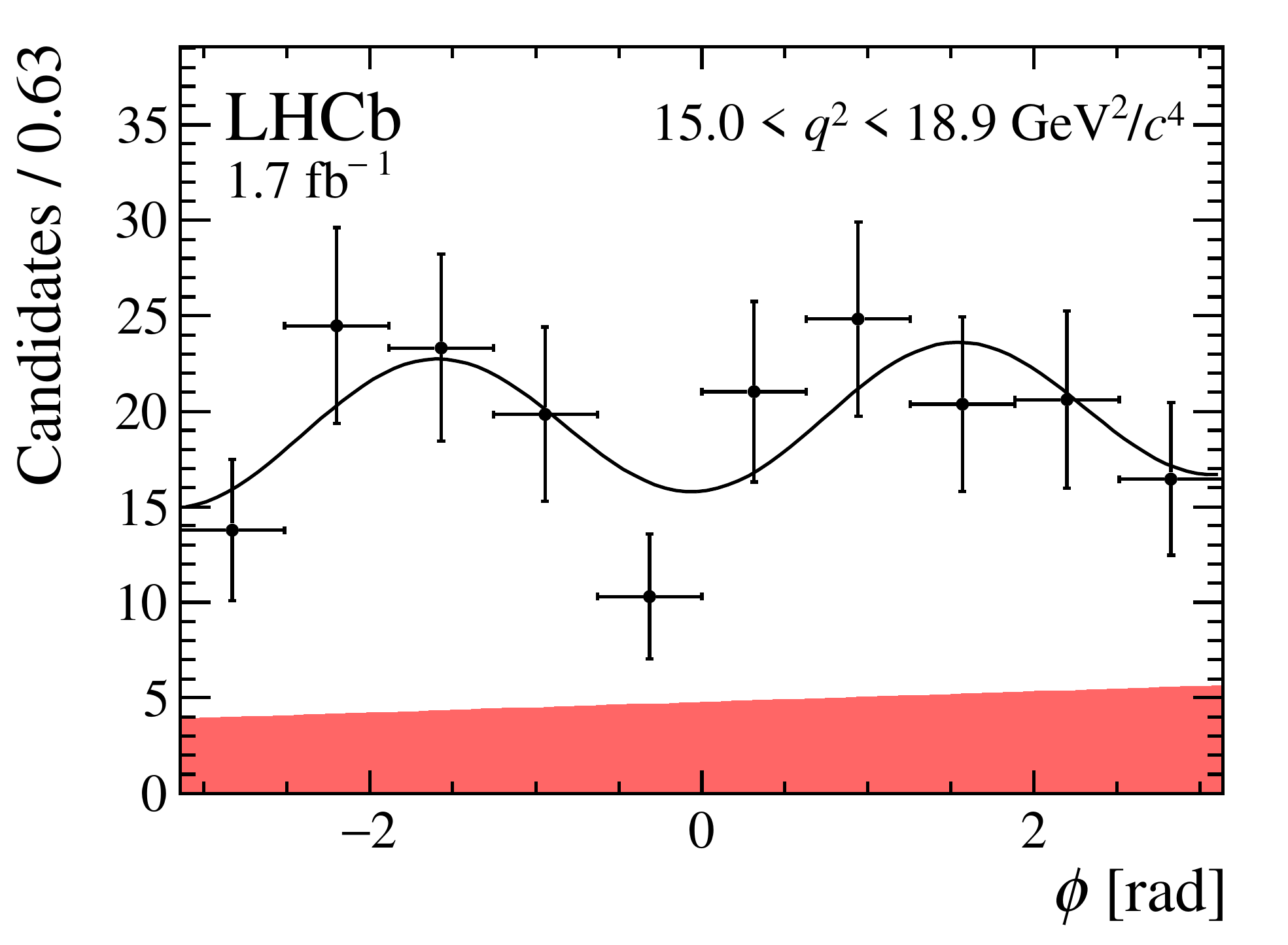}~
    \caption{\label{fig:results_bin7_run2p1} Mass and angular distributions of \BsToPhimm\ candidates in the region \mbox{$15.0<\qsq<18.9\gevgevcccc$} for data taken in 2016. The data are overlaid with the projections of the fitted PDF. }
\end{figure}

\begin{figure}[hb]
    \centering
    \includegraphics[width=.4\textwidth]{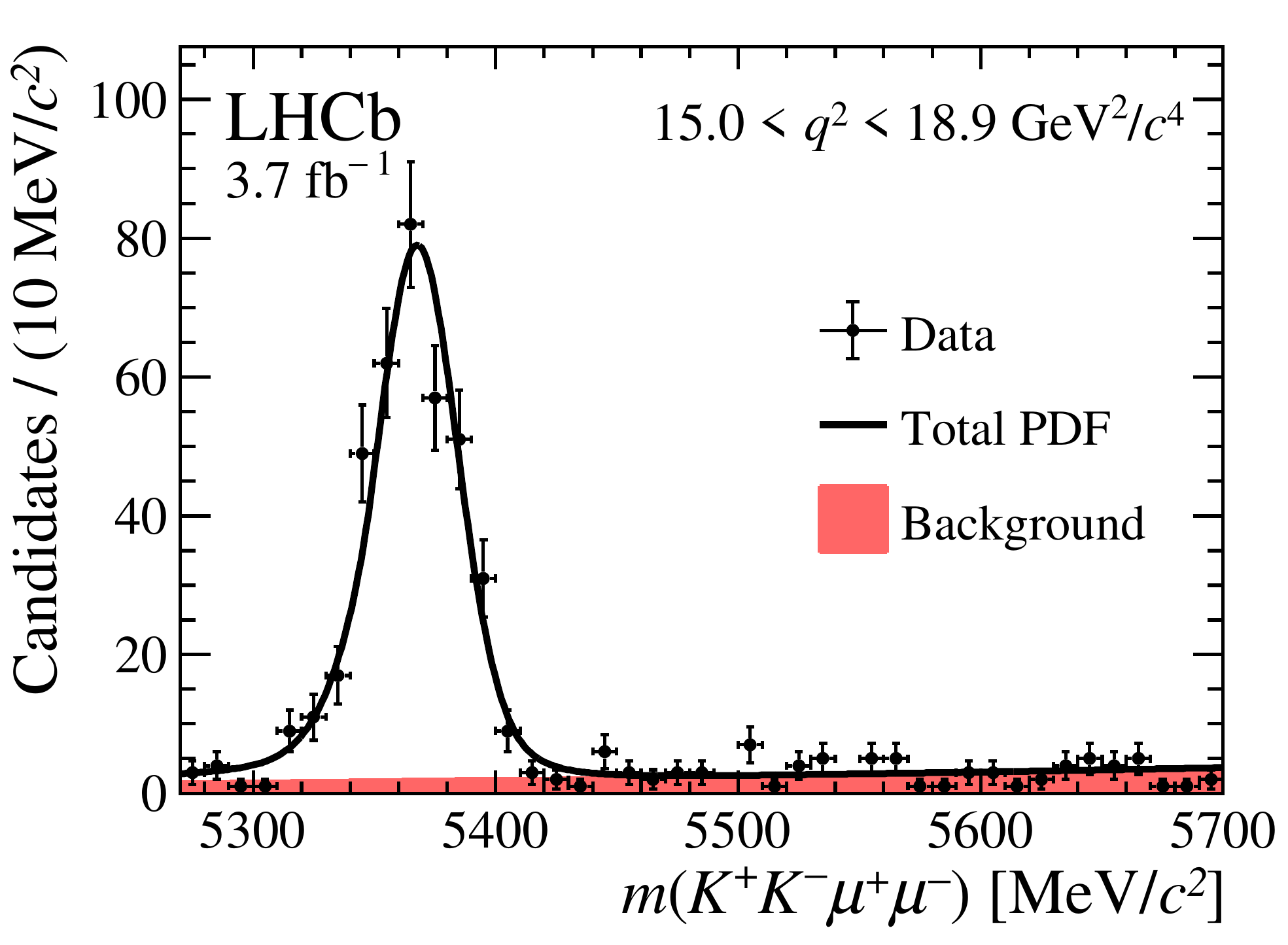}~
    \includegraphics[width=.4\textwidth]{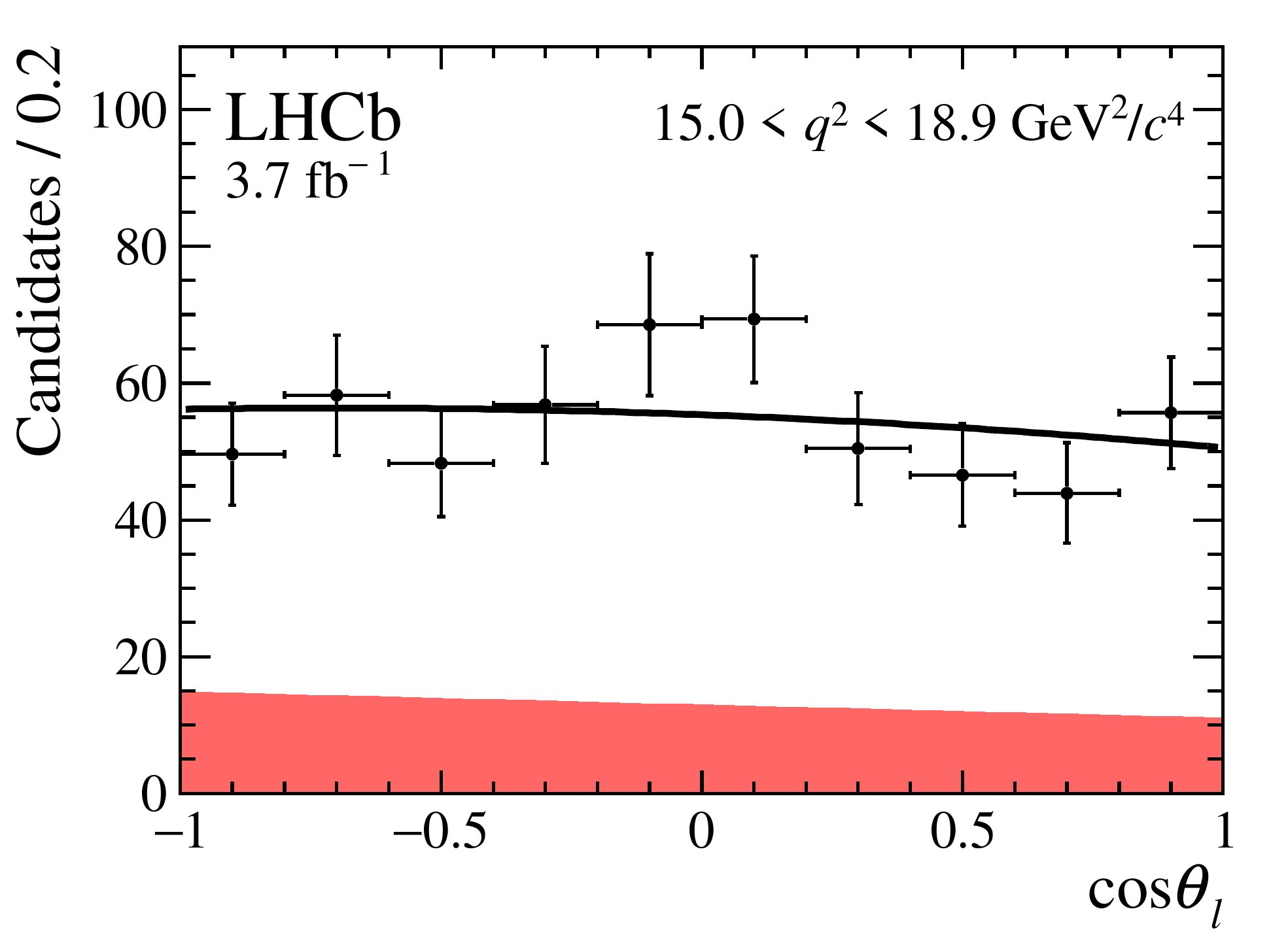}\\
    \includegraphics[width=.4\textwidth]{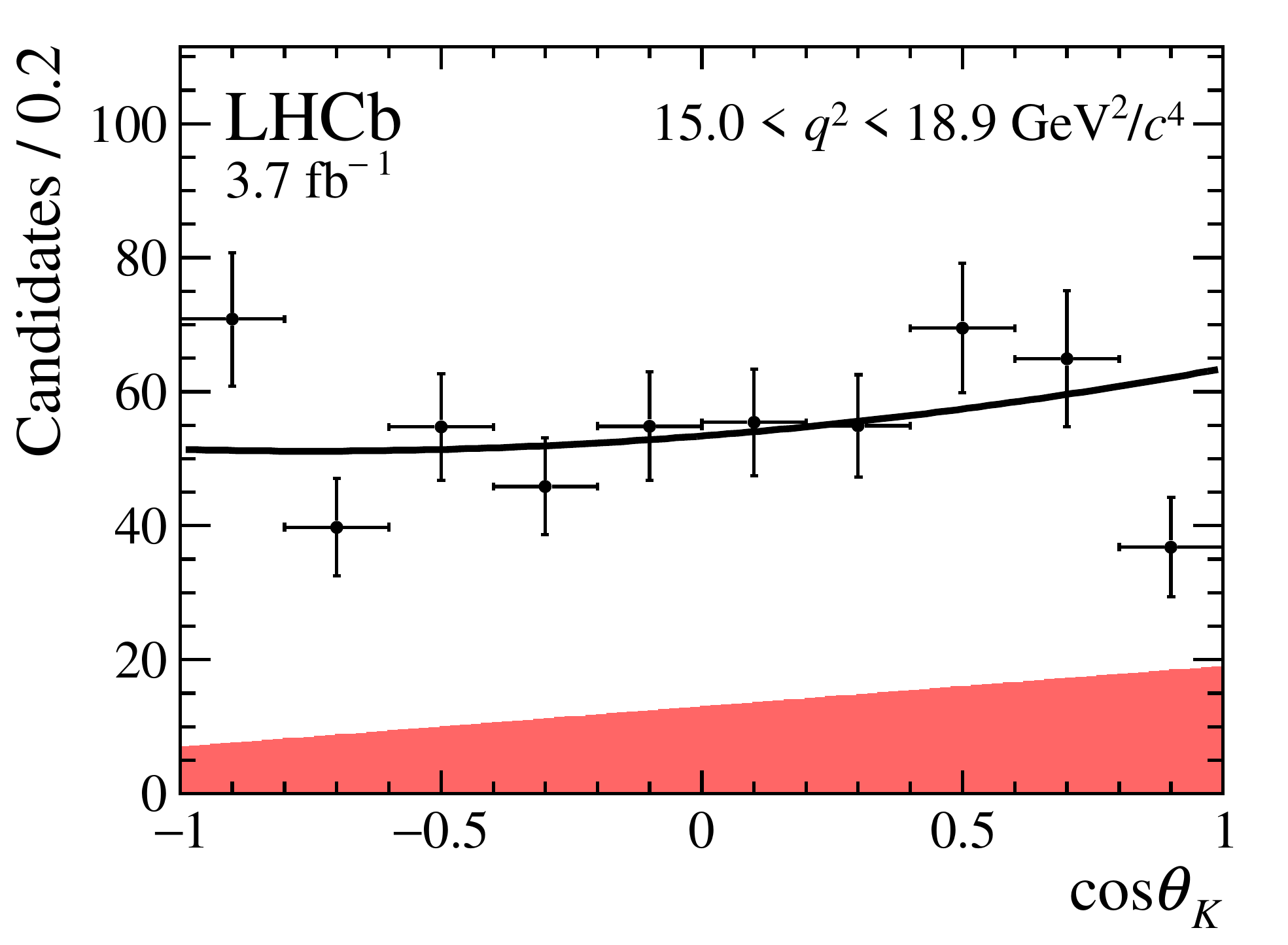}~
    \includegraphics[width=.4\textwidth]{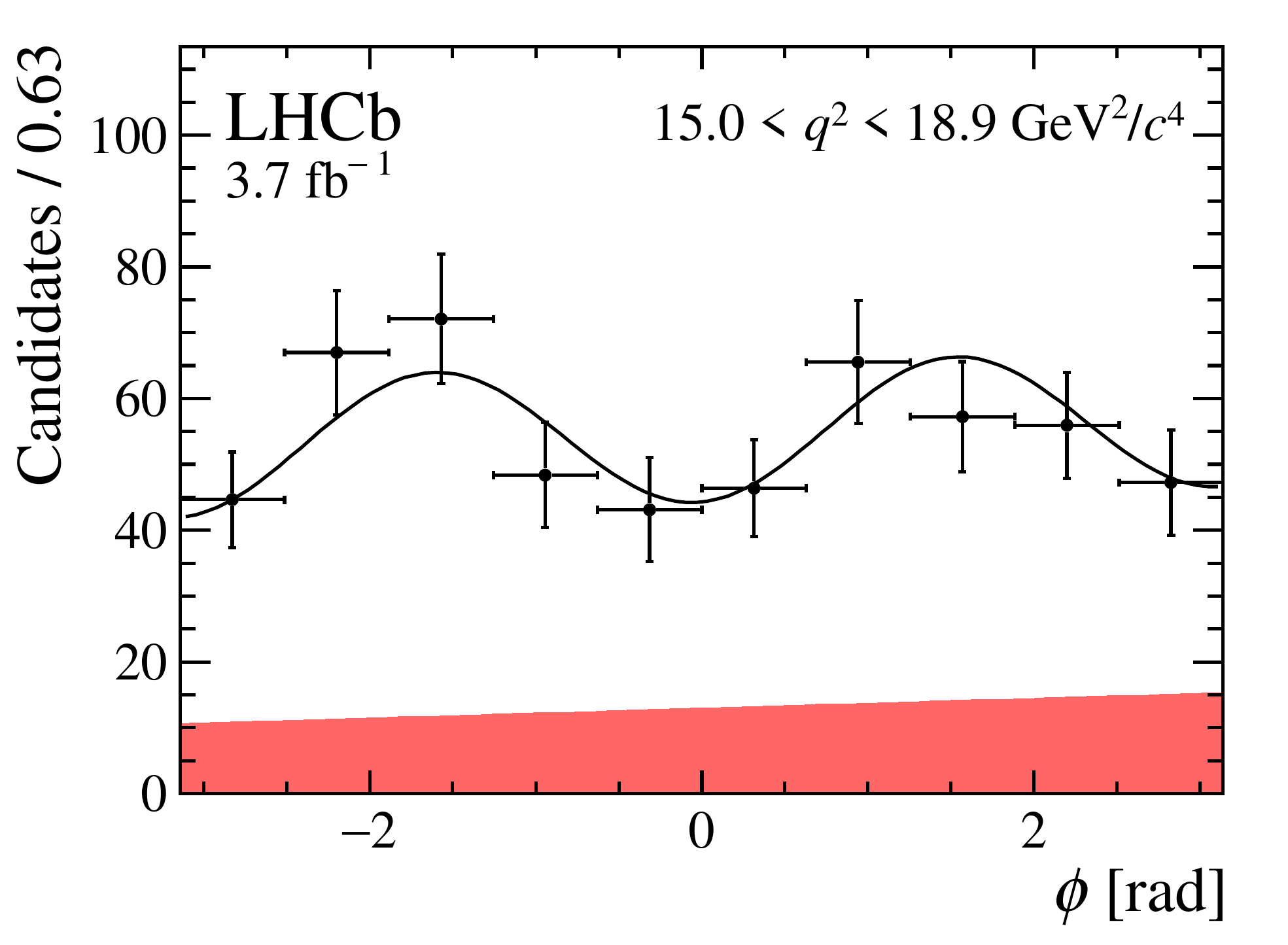}~
    \caption{\label{fig:results_bin7_run2p2} Mass and angular distributions of \BsToPhimm\ candidates in the region \mbox{$15.0<\qsq<18.9\gevgevcccc$} for data taken in 2017--2018. The data are overlaid with the projections of the fitted PDF. }
\end{figure}

The mass distributions for the combined 2011--2012, 2016 and 2017--2018 data sets for each \qsq\ region are shown in Fig.~\ref{fig:results_mass_comb}.
The corresponding angular distributions are shown in Figs.~\ref{fig:results_bin1_comb}--\ref{fig:results_bin7_comb}, for all candidates and the candidates within the signal mass region $\pm 50\mevcc$ around the known \Bs\ mass. The data are overlaid with the projections of the fitted PDF, combined across the data sets.
The projections of the mass and angular distributions, for each data set and \qsq\ region separately, are shown in Figs.~\ref{fig:results_bin1_run1}--\ref{fig:results_bin7_run2p2}. All three data sets used in this analysis are fitted simultaneously.







\clearpage

\section{Correlation matrices}
\label{app:correlations}
The linear correlations obtained from the fit of the angular observables are given in Tables~\ref{tab:correlation1}--\ref{tab:correlation3}.

\begin{table}[hb]
    \centering
    \caption{\label{tab:correlation1} Correlation matrix for the \qsq regions $0.1 < \qsq < 0.98 \gevgevcccc$ and \mbox{$1.1 <\qsq <4.0 \gevgevcccc$}.}
    \begin{tabular}{crrrrrrrr}
    \hline\noalign{\smallskip}
    \multicolumn{9}{c}{Correlation matrix for $ 0.1 < q^2 < 0.98 \gevgevcccc$}\\&\multicolumn{1}{c}{$F_{\rm L}$}&\multicolumn{1}{c}{$S_{3}$}&\multicolumn{1}{c}{$S_{4}$}&\multicolumn{1}{c}{$A_{5}$}&\multicolumn{1}{c}{$A_{\rm FB}^{C\!P}$}&\multicolumn{1}{c}{$S_{7}$}&\multicolumn{1}{c}{$A_{8}$}&\multicolumn{1}{c}{$A_{9}$}\\
    \noalign{\smallskip}\hline\hline
    $F_{\rm L}$&$1.0$0&$-0.03$&$0.06$&$0.10$&$0.02$&$-0.20$&$-0.05$&$-0.04$\\
    $S_{3}$&&$1.00$&$0.11$&$0.00$&$-0.06$&$0.07$&$-0.05$&$0.02$\\
    $S_{4}$&&&$1.00$&$0.03$&$-0.06$&$0.08$&$-0.05$&$-0.01$\\
    $A_{5}$&&&&$1.00$&$0.12$&$-0.07$&$0.06$&$-0.09$\\
    $A_{\rm FB}^{C\!P}$&&&&&$1.00$&$0.02$&$-0.04$&$0.07$\\
    $S_{7}$&&&&&&$1.00$&$0.29$&$-0.03$\\
    $A_{8}$&&&&&&&$1.00$&$0.06$\\
    $A_{9}$&&&&&&&&$1.00$\\
    \hline
    \end{tabular}\\[1cm]
    \begin{tabular}{crrrrrrrr}
    \hline\noalign{\smallskip}
    \multicolumn{9}{c}{Correlation matrix for $ 1.1 < q^2 < 4.0\gevgevcccc$}\\&\multicolumn{1}{c}{$F_{\rm L}$}&\multicolumn{1}{c}{$S_{3}$}&\multicolumn{1}{c}{$S_{4}$}&\multicolumn{1}{c}{$A_{5}$}&\multicolumn{1}{c}{$A_{\rm FB}^{C\!P}$}&\multicolumn{1}{c}{$S_{7}$}&\multicolumn{1}{c}{$A_{8}$}&\multicolumn{1}{c}{$A_{9}$}\\
    \noalign{\smallskip}\hline\hline
    $F_{\rm L}$&$1.00$&$-0.12$&$0.03$&$-0.03$&$-0.07$&$-0.12$&$0.03$&$-0.01$\\
    $S_{3}$&&$1.00$&$-0.07$&$-0.00$&$0.03$&$0.04$&$-0.04$&$0.05$\\
    $S_{4}$&&&$1.00$&$0.02$&$0.09$&$0.01$&$-0.02$&$-0.02$\\
    $A_{5}$&&&&$1.00$&$-0.08$&$0.02$&$0.01$&$0.06$\\
    $A_{\rm FB}^{C\!P}$&&&&&$1.00$&$0.02$&$-0.02$&$0.03$\\
    $S_{7}$&&&&&&$1.00$&$-0.06$&$0.06$\\
    $A_{8}$&&&&&&&$1.00$&$-0.04$\\
    $A_{9}$&&&&&&&&$1.00$\\
    \hline
    \end{tabular}\\[1cm]
\end{table}
\begin{table}
    \centering
    \caption{\label{tab:correlation2} Correlation matrix for the \qsq regions $4.0<\qsq < 6.0\gevgevcccc$, $6.0 < \qsq < 8.0 \gevgevcccc$ and $11.0 < \qsq < 12.5 \gevgevcccc$.}

    \begin{tabular}{crrrrrrrr}
    \hline\noalign{\smallskip}
    \multicolumn{9}{c}{Correlation matrix for $ 4.0 < q^2 < 6.0\gevgevcccc$}\\&\multicolumn{1}{c}{$F_{\rm L}$}&\multicolumn{1}{c}{$S_{3}$}&\multicolumn{1}{c}{$S_{4}$}&\multicolumn{1}{c}{$A_{5}$}&\multicolumn{1}{c}{$A_{\rm FB}^{C\!P}$}&\multicolumn{1}{c}{$S_{7}$}&\multicolumn{1}{c}{$A_{8}$}&\multicolumn{1}{c}{$A_{9}$}\\
    \noalign{\smallskip}\hline\hline
    $F_{\rm L}$&$1.00$&$0.15$&$0.06$&$-0.08$&$0.01$&$-0.01$&$-0.04$&$-0.08$\\
    $S_{3}$&&$1.00$&$-0.04$&$-0.02$&$0.05$&$-0.06$&$-0.04$&$0.22$\\
    $S_{4}$&&&$1.00$&$-0.12$&$-0.02$&$0.05$&$-0.04$&$-0.07$\\
    $A_{5}$&&&&$1.00$&$-0.11$&$-0.06$&$0.04$&$0.05$\\
    $A_{\rm FB}^{C\!P}$&&&&&$1.00$&$-0.01$&$0.10$&$0.01$\\
    $S_{7}$&&&&&&$1.00$&$-0.13$&$0.02$\\
    $A_{8}$&&&&&&&$1.00$&$-0.05$\\
    $A_{9}$&&&&&&&&$1.00$\\
    \hline
    \end{tabular}\\[1cm]

    \begin{tabular}{crrrrrrrr}
    \hline\noalign{\smallskip}
    \multicolumn{9}{c}{Correlation matrix for $ 6.0 < q^2 < 8.0\gevgevcccc$}\\&\multicolumn{1}{c}{$F_{\rm L}$}&\multicolumn{1}{c}{$S_{3}$}&\multicolumn{1}{c}{$S_{4}$}&\multicolumn{1}{c}{$A_{5}$}&\multicolumn{1}{c}{$A_{\rm FB}^{C\!P}$}&\multicolumn{1}{c}{$S_{7}$}&\multicolumn{1}{c}{$A_{8}$}&\multicolumn{1}{c}{$A_{9}$}\\
    \noalign{\smallskip}\hline\hline
    $F_{\rm L}$&$1.00$&$0.03$&$0.07$&$-0.04$&$-0.10$&$0.01$&$-0.03$&$-0.04$\\
    $S_{3}$&&$1.00$&$-0.08$&$-0.02$&$-0.02$&$-0.01$&$0.04$&$-0.05$\\
    $S_{4}$&&&$1.00$&$-0.09$&$0.02$&$-0.05$&$-0.08$&$-0.06$\\
    $A_{5}$&&&&$1.00$&$-0.12$&$-0.05$&$-0.05$&$0.05$\\
    $A_{\rm FB}^{C\!P}$&&&&&$1.00$&$-0.04$&$-0.04$&$-0.01$\\
    $S_{7}$&&&&&&$1.00$&$-0.14$&$0.03$\\
    $A_{8}$&&&&&&&$1.00$&$-0.05$\\
    $A_{9}$&&&&&&&&$1.00$\\
    \hline
    \end{tabular}\\[1cm]
    \begin{tabular}{crrrrrrrr}
    \hline\noalign{\smallskip}
    \multicolumn{9}{c}{Correlation matrix for $ 11.0 < q^2 < 12.5\gevgevcccc$}\\&\multicolumn{1}{c}{$F_{\rm L}$}&\multicolumn{1}{c}{$S_{3}$}&\multicolumn{1}{c}{$S_{4}$}&\multicolumn{1}{c}{$A_{5}$}&\multicolumn{1}{c}{$A_{\rm FB}^{C\!P}$}&\multicolumn{1}{c}{$S_{7}$}&\multicolumn{1}{c}{$A_{8}$}&\multicolumn{1}{c}{$A_{9}$}\\
    \noalign{\smallskip}\hline\hline
    $F_{\rm L}$&$1.00$&$0.09$&$0.02$&$-0.05$&$-0.10$&$-0.03$&$-0.04$&$-0.07$\\
    $S_{3}$&&$1.00$&$-0.14$&$-0.01$&$-0.02$&$0.11$&$-0.06$&$-0.06$\\
    $S_{4}$&&&$1.00$&$-0.04$&$0.18$&$-0.01$&$-0.07$&$-0.05$\\
    $A_{5}$&&&&$1.00$&$-0.23$&$-0.11$&$0.01$&$-0.02$\\
    $A_{\rm FB}^{C\!P}$&&&&&$1.00$&$0.04$&$-0.06$&$0.02$\\
    $S_{7}$&&&&&&$1.00$&$0.06$&$-0.05$\\
    $A_{8}$&&&&&&&$1.00$&$-0.11$\\
    $A_{9}$&&&&&&&&$1.00$\\
    \hline
    \end{tabular}\\[1cm]

\end{table}

\begin{table}
    \centering
    \caption{\label{tab:correlation3} Correlation matrix for the \qsq region $1.1 < \qsq < 6.0 \gevgevcccc$ and \mbox{$15.0 < \qsq < 18.9 \gevgevcccc$}.}
    \begin{tabular}{crrrrrrrr}
    \hline\noalign{\smallskip}
    \multicolumn{9}{c}{Correlation matrix for $ 1.1 < q^2 < 6.0\gevgevcccc$}\\&\multicolumn{1}{c}{$F_{\rm L}$}&\multicolumn{1}{c}{$S_{3}$}&\multicolumn{1}{c}{$S_{4}$}&\multicolumn{1}{c}{$A_{5}$}&\multicolumn{1}{c}{$A_{\rm FB}^{C\!P}$}&\multicolumn{1}{c}{$S_{7}$}&\multicolumn{1}{c}{$A_{8}$}&\multicolumn{1}{c}{$A_{9}$}\\
    \noalign{\smallskip}\hline\hline
    $F_{\rm L}$&$1.00$&$-0.03$&$0.05$&$-0.02$&$-0.07$&$-0.08$&$-0.04$&$-0.03$\\
    $S_{3}$&&$1.00$&$-0.07$&$0.01$&$0.03$&$-0.07$&$-0.07$&$0.10$\\
    $S_{4}$&&&$1.00$&$-0.00$&$-0.00$&$-0.06$&$-0.02$&$-0.01$\\
    $A_{5}$&&&&$1.00$&$-0.07$&$-0.00$&$0.05$&$0.08$\\
    $A_{\rm FB}^{C\!P}$&&&&&$1.00$&$0.01$&$0.05$&$0.07$\\
    $S_{7}$&&&&&&$1.00$&$-0.08$&$0.01$\\
    $A_{8}$&&&&&&&$1.00$&$-0.03$\\
    $A_{9}$&&&&&&&&$1.00$\\
    \hline
    \end{tabular}\\[1cm]
    \begin{tabular}{crrrrrrrr}
    \hline\noalign{\smallskip}
    \multicolumn{9}{c}{Correlation matrix for $ 15.0 < q^2 < 18.9\gevgevcccc$}\\&\multicolumn{1}{c}{$F_{\rm L}$}&\multicolumn{1}{c}{$S_{3}$}&\multicolumn{1}{c}{$S_{4}$}&\multicolumn{1}{c}{$A_{5}$}&\multicolumn{1}{c}{$A_{\rm FB}^{C\!P}$}&\multicolumn{1}{c}{$S_{7}$}&\multicolumn{1}{c}{$A_{8}$}&\multicolumn{1}{c}{$A_{9}$}\\
    \noalign{\smallskip}\hline\hline
    $F_{\rm L}$&$1.00$&$0.20$&$-0.04$&$-0.03$&$0.03$&$0.00$&$-0.03$&$-0.10$\\
    $S_{3}$&&$1.00$&$-0.06$&$0.03$&$-0.11$&$-0.08$&$0.00$&$0.13$\\
    $S_{4}$&&&$1.00$&$-0.13$&$-0.03$&$-0.04$&$0.13$&$0.06$\\
    $A_{5}$&&&&$1.00$&$-0.11$&$0.05$&$-0.08$&$0.05$\\
    $A_{\rm FB}^{C\!P}$&&&&&$1.00$&$0.10$&$-0.00$&$-0.03$\\
    $S_{7}$&&&&&&$1.00$&$0.03$&$0.01$\\
    $A_{8}$&&&&&&&$1.00$&$-0.07$\\
    $A_{9}$&&&&&&&&$1.00$\\
    \hline
    \end{tabular}\\[1cm]
\end{table}

\clearpage
\input{Authorship_LHCb-PAPER-2021-022}

\clearpage

\end{document}

%% file: preamble.tex
\usepackage[top=1in, bottom=1.25in, left=1in, right=1in]{geometry}

\usepackage{microtype}
\usepackage{lineno}  
\usepackage{xspace} 
\usepackage{caption} 

\usepackage{graphicx}  
\usepackage{color}
\usepackage{colortbl}
\graphicspath{{./figs/}} 

\usepackage{amsmath} 
\usepackage{amssymb}
\usepackage{amsfonts}
\usepackage{upgreek} 

\newcommand*\patchAmsMathEnvironmentForLineno[1]{%
\expandafter\let\csname old#1\expandafter\endcsname\csname #1\endcsname
\expandafter\let\csname oldend#1\expandafter\endcsname\csname
end#1\endcsname
 \renewenvironment{#1}%
   {\linenomath\csname old#1\endcsname}%
   {\csname oldend#1\endcsname\endlinenomath}%
}
\newcommand*\patchBothAmsMathEnvironmentsForLineno[1]{%
  \patchAmsMathEnvironmentForLineno{#1}%
  \patchAmsMathEnvironmentForLineno{#1*}%
}
\AtBeginDocument{%
\patchBothAmsMathEnvironmentsForLineno{equation}%
\patchBothAmsMathEnvironmentsForLineno{align}%
\patchBothAmsMathEnvironmentsForLineno{flalign}%
\patchBothAmsMathEnvironmentsForLineno{alignat}%
\patchBothAmsMathEnvironmentsForLineno{gather}%
\patchBothAmsMathEnvironmentsForLineno{multline}%
\patchBothAmsMathEnvironmentsForLineno{eqnarray}%
}

\usepackage{hyperxmp}

\usepackage[pdftex,
            pdfauthor={\paperauthors},
            pdftitle={\paperasciititle},
            pdfkeywords={\paperkeywords},
            pdfcopyright={Copyright (C) \papercopyright},
            pdflicenseurl={\paperlicenceurl}]{hyperref}

\usepackage[colorinlistoftodos,textsize=scriptsize]{todonotes}

\usepackage[bottom,flushmargin,hang,multiple]{footmisc}

\usepackage[all]{hypcap} 

\input{lhcb-symbols-def} 

\usepackage{cite} 
\usepackage{mciteplus}

\usepackage{tikz}
\usetikzlibrary{decorations.pathmorphing}
\usetikzlibrary{decorations.markings}
\usepackage{subfig}

%% file: lhcb-symbols-def.tex
\usepackage{xspace} 
\usepackage{upgreek}

\def\lhcb   {\mbox{LHCb}\xspace}

\def\MagUp {\mbox{\em Mag\kern -0.05em Up}\xspace}

\ifthenelse{\boolean{uprightparticles}}%
{

 \def\Pmu         {\ensuremath{\upmu}\xspace}

 \def\Ppi         {\ensuremath{\uppi}\xspace}

 \def\Ppsi        {\ensuremath{\uppsi}\xspace}

 \def\PDelta      {\ensuremath{\Delta}\xspace}                 
 \def\PXi         {\ensuremath{\Xi}\xspace}                 
 \def\PLambda     {\ensuremath{\Lambda}\xspace}                 
 \def\PSigma      {\ensuremath{\Sigma}\xspace}                 
 \def\POmega      {\ensuremath{\Omega}\xspace}                 
 \def\PUpsilon    {\ensuremath{\Upsilon}\xspace}

 \def\PB      {\ensuremath{\mathrm{B}}\xspace}                 
                  
 \def\PD      {\ensuremath{\mathrm{D}}\xspace}

 \def\PJ      {\ensuremath{\mathrm{J}}\xspace}                 
 \def\PK      {\ensuremath{\mathrm{K}}\xspace}

 \def\Pb      {\ensuremath{\mathrm{b}}\xspace}                 
 \def\Pc      {\ensuremath{\mathrm{c}}\xspace}

 \def\Pi      {\ensuremath{\mathrm{i}}\xspace}

 \def\Pp      {\ensuremath{\mathrm{p}}\xspace}

 \def\Ps      {\ensuremath{\mathrm{s}}\xspace}

 \def\thebaroffset{0.0em}
}
{

 \def\Pmu         {\ensuremath{\mu}\xspace}

 \def\Ppi         {\ensuremath{\pi}\xspace}

 \def\Ppsi        {\ensuremath{\psi}\xspace}                 
                  
 \mathchardef\PDelta="7101
 \mathchardef\PXi="7104
 \mathchardef\PLambda="7103
 \mathchardef\PSigma="7106
 \mathchardef\POmega="710A
 \mathchardef\PUpsilon="7107
                  
 \def\PB      {\ensuremath{B}\xspace}                 
                  
 \def\PD      {\ensuremath{D}\xspace}

 \def\PJ      {\ensuremath{J}\xspace}                 
 \def\PK      {\ensuremath{K}\xspace}

 \def\Pb      {\ensuremath{b}\xspace}                 
 \def\Pc      {\ensuremath{c}\xspace}

 \def\Pi      {\ensuremath{i}\xspace}

 \def\Pp      {\ensuremath{p}\xspace}

 \def\Ps      {\ensuremath{s}\xspace}

 \def\thebaroffset{0.18em}
}
\newcommand{\offsetoverline}[2][\thebaroffset]{\kern #1\overline{\kern -#1 #2}}%

\makeatletter
\ifcase \@ptsize \relax
  \newcommand{\miniscule}{\@setfontsize\miniscule{4}{5}}
\or
  \newcommand{\miniscule}{\@setfontsize\miniscule{5}{6}}
\or
  \newcommand{\miniscule}{\@setfontsize\miniscule{5}{6}}
\fi
\makeatother

\DeclareRobustCommand{\optbar}[1]{\shortstack{{\miniscule (\rule[.5ex]{1.25em}{.18mm})}
  \\ [-.7ex] $#1$}}

\def\mup        {{\ensuremath{\Pmu^+}}\xspace}
\def\mun        {{\ensuremath{\Pmu^-}}\xspace} 
\def\mupm       {{\ensuremath{\Pmu^\pm}}\xspace} 
 
\def\mumu       {{\ensuremath{\Pmu^+\Pmu^-}}\xspace}

\def\squark    {{\ensuremath{\Ps}}\xspace}

\def\cquark    {{\ensuremath{\Pc}}\xspace}

\def\bquark    {{\ensuremath{\Pb}}\xspace}

\def\pion   {{\ensuremath{\Ppi}}\xspace}

\def\pim    {{\ensuremath{\pion^-}}\xspace}

\def\kaon    {{\ensuremath{\PK}}\xspace}

\def\KorKbar {\kern \thebaroffset\optbar{\kern -\thebaroffset \PK}{}\xspace}

\def\Kp      {{\ensuremath{\kaon^+}}\xspace}
\def\Km      {{\ensuremath{\kaon^-}}\xspace}

\def\Kmp     {{\ensuremath{\kaon^\mp}}\xspace}

\def\Kstarz  {{\ensuremath{\kaon^{*0}}}\xspace}

\def\D       {{\ensuremath{\PD}}\xspace}

\def\DorDbar {\kern \thebaroffset\optbar{\kern -\thebaroffset \PD}\xspace}

\def\Dp      {{\ensuremath{\D^+}}\xspace}
\def\Dm      {{\ensuremath{\D^-}}\xspace}

\def\DpDm    {\ensuremath{\Dp {\kern -0.16em \Dm}}\xspace}

\def\B       {{\ensuremath{\PB}}\xspace}
\def\Bbar    {{\ensuremath{\offsetoverline{\PB}}}\xspace}

\def\BorBbar {\kern \thebaroffset\optbar{\kern -\thebaroffset \PB}\xspace}

\def\Bd      {{\ensuremath{\B^0}}\xspace}

\def\BdorBdbar {\kern \thebaroffset\optbar{\kern -\thebaroffset \Bd}\xspace}

\def\Bs      {{\ensuremath{\B^0_\squark}}\xspace}
\def\Bsb     {{\ensuremath{\Bbar{}^0_\squark}}\xspace}
\def\BsorBsbar {\kern \thebaroffset\optbar{\kern -\thebaroffset \Bs}\xspace}

\def\jpsi     {{\ensuremath{{\PJ\mskip -3mu/\mskip -2mu\Ppsi}}}\xspace}
\def\psitwos  {{\ensuremath{\Ppsi{(2S)}}}\xspace}

\def\Y#1S{\ensuremath{\PUpsilon{(#1S)}}\xspace}

\def\proton      {{\ensuremath{\Pp}}\xspace}

\def\Lz          {{\ensuremath{\PLambda}}\xspace}

\def\LorLbar     {\kern \thebaroffset\optbar{\kern -\thebaroffset \PLambda}\xspace}

\def\Lb           {{\ensuremath{\Lz^0_\bquark}}\xspace}

\newcommand{\decay}[2]{\ensuremath{#1\!\to #2}\xspace} 

\def\to                 {\ensuremath{\rightarrow}\xspace}

\def\qsq       {{\ensuremath{q^2}}\xspace}

\def\CP                {{\ensuremath{C\!P}}\xspace}

\def\BdToKstmm    {\decay{\Bd}{\Kstarz\mup\mun}}

\def\BsToJPsiPhi  {\decay{\Bs}{\jpsi\phi}}

\def\BdToKstmm    {\decay{\Bd}{\Kstarz\mup\mun}}

\def\BsToPhimm    {\decay{\Bs}{\phi\mup\mun}}

\def\LbTopKmm      {\decay{\Lb}{\proton\Km\mumu}}
\def\bsll     {\decay{\bquark}{\squark \ell^+ \ell^-}}

\def\AFBCP      {\ensuremath{A_{\mathrm{FB}}^{\CP}}\xspace}
\def\FL       {\ensuremath{F_{\mathrm{L}}}\xspace}
\def\thetal   {{\ensuremath{\theta_{l}}}\xspace}
\def\thetak   {{\ensuremath{\theta_{K}}}\xspace}
\def\AT#1     {\ensuremath{A_{\mathrm{T}}^{#1}}\xspace}           

\def\ctl       {\ensuremath{\cos{\theta_\ell}}\xspace}
\def\ctk       {\ensuremath{\cos{\theta_K}}\xspace}

\def\C#1      {\ensuremath{\mathcal{C}_{#1}}\xspace}                       
\def\Cp#1     {\ensuremath{\mathcal{C}_{#1}^{'}}\xspace}                    
\def\Ceff#1   {\ensuremath{\mathcal{C}_{#1}^{\mathrm{(eff)}}}\xspace}        
\def\Cpeff#1  {\ensuremath{\mathcal{C}_{#1}^{'\mathrm{(eff)}}}\xspace}       
\def\Ope#1    {\ensuremath{\mathcal{O}_{#1}}\xspace}                       
\def\Opep#1   {\ensuremath{\mathcal{O}_{#1}^{'}}\xspace}                    

\newcommand{\nospaceunit}[1]{\ensuremath{\text{#1}}}       
\newcommand{\aunit}[1]{\ensuremath{\text{\,#1}}}       

\newcommand{\tev}{\aunit{Te\kern -0.1em V}\xspace}
\newcommand{\gev}{\aunit{Ge\kern -0.1em V}\xspace}
\newcommand{\mev}{\aunit{Me\kern -0.1em V}\xspace}
\newcommand{\kev}{\aunit{ke\kern -0.1em V}\xspace}
\newcommand{\ev}{\aunit{e\kern -0.1em V}\xspace}
 
\newcommand{\mevc}{\ensuremath{\aunit{Me\kern -0.1em V\!/}c}\xspace}
\newcommand{\gevc}{\ensuremath{\aunit{Ge\kern -0.1em V\!/}c}\xspace}
\newcommand{\mevcc}{\ensuremath{\aunit{Me\kern -0.1em V\!/}c^2}\xspace}
\newcommand{\gevcc}{\ensuremath{\aunit{Ge\kern -0.1em V\!/}c^2}\xspace}
\newcommand{\gevgevcccc}{\ensuremath{\gev^2\!/c^4}\xspace} 

\def\mum  {\ensuremath{\,\upmu\nospaceunit{m}}\xspace}

\def\fb   {\ensuremath{\aunit{fb}}\xspace}
\def\invfb   {\ensuremath{\fb^{-1}}\xspace}

\newcommand{\chisq}{\ensuremath{\chi^2}\xspace}

\newcommand{\chisqip}{\ensuremath{\chi^2_{\text{IP}}}\xspace}

\def\deriv {\ensuremath{\mathrm{d}}}

\def\gsim{{~\raise.15em\hbox{$>$}\kern-.85em
          \lower.35em\hbox{$\sim$}~}\xspace}
\def\lsim{{~\raise.15em\hbox{$<$}\kern-.85em
          \lower.35em\hbox{$\sim$}~}\xspace}

\def\pt         {\ensuremath{p_{\mathrm{T}}}\xspace}

\def\ptot       {\ensuremath{p}\xspace}

\def\evtgen     {\mbox{\textsc{EvtGen}}\xspace}

\def\geant      {\mbox{\textsc{Geant4}}\xspace}

\def\photos     {\mbox{\textsc{Photos}}\xspace}

\def\pythia     {\mbox{\textsc{Pythia}}\xspace}

\def\tell1  {TELL1\xspace}
\def\ukl1   {UKL1\xspace}

\newcommand{\ie}{\mbox{\itshape i.e.}\xspace}



%% file: title-LHCb-PAPER.tex
\begin{titlepage}
\pagenumbering{roman}

\vspace*{-1.5cm}
\centerline{\large EUROPEAN ORGANIZATION FOR NUCLEAR RESEARCH (CERN)}
\vspace*{1.5cm}
\noindent
\begin{tabular*}{\linewidth}{lc@{\extracolsep{\fill}}r@{\extracolsep{0pt}}}
\ifthenelse{\boolean{pdflatex}}
{\vspace*{-1.5cm}\mbox{\!\!\!\includegraphics[width=.14\textwidth]{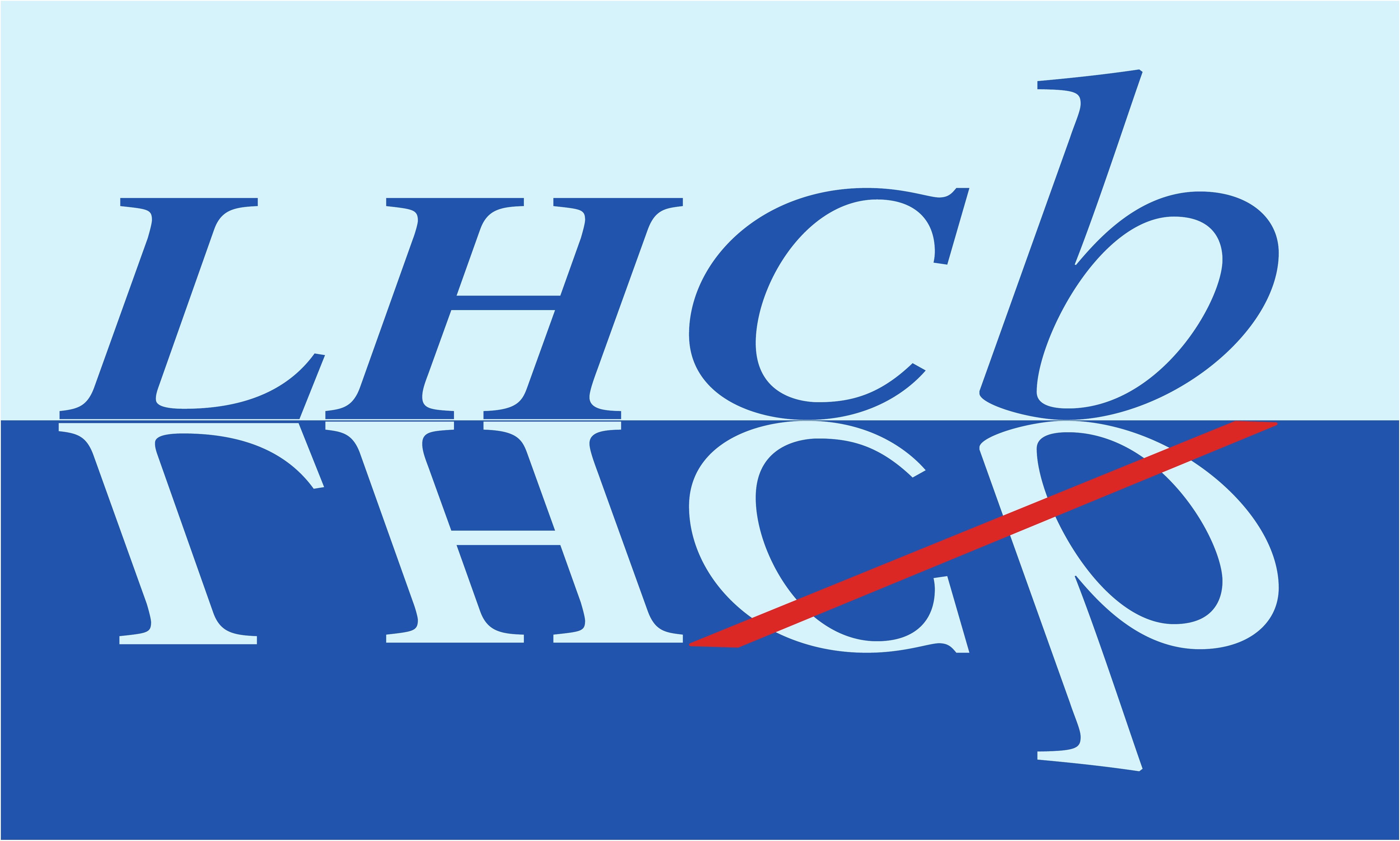}} & &}%
{\vspace*{-1.2cm}\mbox{\!\!\!\includegraphics[width=.12\textwidth]{figs/lhcb-logo.eps}} & &}%
\\
 & & CERN-EP-2021-138 \\
 & & LHCb-PAPER-2021-022 \\
 & & November 10, 2021 \\ 

 & & \\
\end{tabular*}

\vspace*{4.0cm}

{\normalfont\bfseries\boldmath\huge
\begin{center}
  \papertitle
\end{center}
}

\vspace*{2.0cm}

\begin{center}
\paperauthors\footnote{Authors are listed at the end of this paper.}
\end{center}

\vspace{\fill}

\begin{abstract}
  \noindent
An angular analysis of the rare decay \mbox{$B^0_s\rightarrow\phi\mu^+\mu^-$} is presented, using proton-proton collision data collected by the LHCb experiment at centre-of-mass energies of $7$, $8$ and $13~\rm{Te\kern -0.1em V}$, corresponding to an integrated luminosity of 8.4~$\rm{fb}^{-1}$.
The observables describing the angular distributions of the decay \mbox{$B^0_s\rightarrow\phi\mu^+\mu^-$} are determined in regions of $q^2$, the square of the dimuon invariant mass.
The results are consistent with Standard Model predictions.
\end{abstract}

\vspace*{2.0cm}

\begin{center}
  Published in JHEP 11 (2021) 043
\end{center}

\vspace{\fill}

{\footnotesize
\centerline{\copyright~\papercopyright. \href{\paperlicenceurl}{\paperlicence}.}}
\vspace*{2mm}

\end{titlepage}

\newpage
\setcounter{page}{2}
\mbox{~}

%% file: acknowledgements.tex
\section*{Acknowledgements}
%
%
\noindent We express our gratitude to our colleagues in the CERN
accelerator departments for the excellent performance of the LHC. We
thank the technical and administrative staff at the LHCb
institutes.
We acknowledge support from CERN and from the national agencies:
CAPES, CNPq, FAPERJ and FINEP (Brazil); 
MOST and NSFC (China); 
CNRS/IN2P3 (France); 
BMBF, DFG and MPG (Germany); 
INFN (Italy); 
NWO (Netherlands); 
MNiSW and NCN (Poland); 
MEN/IFA (Romania); 
MSHE (Russia); 
MICINN (Spain); 
SNSF and SER (Switzerland); 
NASU (Ukraine); 
STFC (United Kingdom); 
DOE NP and NSF (USA).
We acknowledge the computing resources that are provided by CERN, IN2P3
(France), KIT and DESY (Germany), INFN (Italy), SURF (Netherlands),
PIC (Spain), GridPP (United Kingdom), RRCKI and Yandex
LLC (Russia), CSCS (Switzerland), IFIN-HH (Romania), CBPF (Brazil),
PL-GRID (Poland) and NERSC (USA).
We are indebted to the communities behind the multiple open-source
software packages on which we depend.
Individual groups or members have received support from
ARC and ARDC (Australia);
AvH Foundation (Germany);
EPLANET, Marie Sk\l{}odowska-Curie Actions and ERC (European Union);
A*MIDEX, ANR, IPhU and Labex P2IO, and R\'{e}gion Auvergne-Rh\^{o}ne-Alpes (France);
Key Research Program of Frontier Sciences of CAS, CAS PIFI, CAS CCEPP, 
Fundamental Research Funds for the Central Universities, 
and Sci. \& Tech. Program of Guangzhou (China);
RFBR, RSF and Yandex LLC (Russia);
GVA, XuntaGal and GENCAT (Spain);
the Leverhulme Trust, the Royal Society
 and UKRI (United Kingdom).

%% file: Authorship_LHCb-PAPER-2021-022.tex
\centerline
{\large\bf LHCb collaboration}
\begin
{flushleft}
\small
R.~Aaij$^{32}$,
A.S.W.~Abdelmotteleb$^{56}$,
C.~Abell{\'a}n~Beteta$^{50}$,
T.~Ackernley$^{60}$,
B.~Adeva$^{46}$,
M.~Adinolfi$^{54}$,
H.~Afsharnia$^{9}$,
C.~Agapopoulou$^{13}$,
C.A.~Aidala$^{86}$,
S.~Aiola$^{25}$
% ...

{\footnotesize \it

% ...
$^{9}$Universit{\'e} Clermont Auvergne, CNRS/IN2P3, LPC, Clermont-Ferrand, France\\
% ...
$^{13}$LPNHE, Sorbonne Universit{\'e}, Paris Diderot Sorbonne Paris Cit{\'e}, CNRS/IN2P3, Paris, France\\
% ...
$^{25}$INFN Sezione di Milano, Milano, Italy\\
% ...
$^{32}$Nikhef National Institute for Subatomic Physics, Amsterdam, Netherlands\\
% ...
$^{46}$Instituto Galego de F{\'\i}sica de Altas Enerx{\'\i}as (IGFAE), Universidade de Santiago de Compostela, Santiago de Compostela, Spain\\
% ...
$^{50}$Physik-Institut, Universit{\"a}t Z{\"u}rich, Z{\"u}rich, Switzerland\\
% ...
$^{54}$H.H. Wills Physics Laboratory, University of Bristol, Bristol, United Kingdom\\
% ...
$^{56}$Department of Physics, University of Warwick, Coventry, United Kingdom\\
% ...
$^{60}$Oliver Lodge Laboratory, University of Liverpool, Liverpool, United Kingdom\\
% ...
$^{68}$Syracuse University, Syracuse, NY, United States\\
% ...
$^{86}$University of Michigan, Ann Arbor, United States, associated to $^{68}$\\
\bigskip
% ...
\medskip
}
\end{flushleft}

%% file: main.bbl
\ifx\mcitethebibliography\mciteundefinedmacro
\PackageError{LHCb.bst}{mciteplus.sty has not been loaded}
{This bibstyle requires the use of the mciteplus package.}\fi
\providecommand{\href}[2]{#2}
\begin{mcitethebibliography}{10}
\mciteSetBstSublistMode{n}
\mciteSetBstMaxWidthForm{subitem}{\alph{mcitesubitemcount})}
\mciteSetBstSublistLabelBeginEnd{\mcitemaxwidthsubitemform\space}
{\relax}{\relax}

\bibitem{LHCb-PAPER-2013-017}
LHCb collaboration, R.~Aaij {\em et~al.},
  \ifthenelse{\boolean{articletitles}}{\emph{{Differential branching fraction
  and angular analysis of the decay \mbox{\decay{\Bs}{\phiz\mumu}}}},
  }{}\href{https://doi.org/10.1007/JHEP07(2013)084}{JHEP \textbf{07} (2013)
  084}, \href{http://arxiv.org/abs/1305.2168}{{\normalfont\ttfamily
  arXiv:1305.2168}}\relax
\mciteBstWouldAddEndPuncttrue
\mciteSetBstMidEndSepPunct{\mcitedefaultmidpunct}
{\mcitedefaultendpunct}{\mcitedefaultseppunct}\relax
\EndOfBibitem
\bibitem{LHCb-PAPER-2014-006}
LHCb collaboration, R.~Aaij {\em et~al.},
  \ifthenelse{\boolean{articletitles}}{\emph{{Differential branching fractions
  and isospin asymmetries of \mbox{\decay{\B}{K^{(*)}\mumu}} decays}},
  }{}\href{https://doi.org/10.1007/JHEP06(2014)133}{JHEP \textbf{06} (2014)
  133}, \href{http://arxiv.org/abs/1403.8044}{{\normalfont\ttfamily
  arXiv:1403.8044}}\relax
\mciteBstWouldAddEndPuncttrue
\mciteSetBstMidEndSepPunct{\mcitedefaultmidpunct}
{\mcitedefaultendpunct}{\mcitedefaultseppunct}\relax
\EndOfBibitem
\bibitem{LHCb-PAPER-2015-009}
LHCb collaboration, R.~Aaij {\em et~al.},
  \ifthenelse{\boolean{articletitles}}{\emph{{Differential branching fraction
  and angular analysis of \mbox{\decay{\Lb}{\Lz\mumu}} decays}},
  }{}\href{https://doi.org/10.1007/JHEP06(2015)115}{JHEP \textbf{06} (2015)
  115}, Erratum \href{https://doi.org/10.1007/JHEP09(2018)145}{ibid.\
  \textbf{09} (2018) 145},
  \href{http://arxiv.org/abs/1503.07138}{{\normalfont\ttfamily
  arXiv:1503.07138}}\relax
\mciteBstWouldAddEndPuncttrue
\mciteSetBstMidEndSepPunct{\mcitedefaultmidpunct}
{\mcitedefaultendpunct}{\mcitedefaultseppunct}\relax
\EndOfBibitem
\bibitem{LHCb-PAPER-2015-023}
LHCb collaboration, R.~Aaij {\em et~al.},
  \ifthenelse{\boolean{articletitles}}{\emph{{Angular analysis and differential
  branching fraction of the decay \mbox{\decay{\Bs}{\phi\mumu}}}},
  }{}\href{https://doi.org/10.1007/JHEP09(2015)179}{JHEP \textbf{09} (2015)
  179}, \href{http://arxiv.org/abs/1506.08777}{{\normalfont\ttfamily
  arXiv:1506.08777}}\relax
\mciteBstWouldAddEndPuncttrue
\mciteSetBstMidEndSepPunct{\mcitedefaultmidpunct}
{\mcitedefaultendpunct}{\mcitedefaultseppunct}\relax
\EndOfBibitem
\bibitem{LHCb-PAPER-2016-012}
LHCb collaboration, R.~Aaij {\em et~al.},
  \ifthenelse{\boolean{articletitles}}{\emph{{Measurements of the S-wave
  fraction in \mbox{\decay{\Bz}{\Kp\pim\mumu}} decays and the
  \mbox{\decay{\Bz}{\Kstar(892)^0\mumu}} differential branching fraction}},
  }{}\href{https://doi.org/10.1007/JHEP11(2016)047}{JHEP \textbf{11} (2016)
  047}, Erratum \href{https://doi.org/10.1007/JHEP04(2017)142}{ibid.\
  \textbf{04} (2017) 142},
  \href{http://arxiv.org/abs/1606.04731}{{\normalfont\ttfamily
  arXiv:1606.04731}}\relax
\mciteBstWouldAddEndPuncttrue
\mciteSetBstMidEndSepPunct{\mcitedefaultmidpunct}
{\mcitedefaultendpunct}{\mcitedefaultseppunct}\relax
\EndOfBibitem
\bibitem{LHCb-PAPER-2021-014}
LHCb collaboration, R.~Aaij {\em et~al.},
  \ifthenelse{\boolean{articletitles}}{\emph{{Branching fraction measurements
  of the rare $B^0_s \to \phi \mu^+\mu^-$ and $B^0_s \to f_2^\prime(1525)
  \mu^+\mu^-$ decays}},
  }{}\href{http://arxiv.org/abs/2105.14007}{{\normalfont\ttfamily
  arXiv:2105.14007}}, {submitted to PRL}\relax
\mciteBstWouldAddEndPuncttrue
\mciteSetBstMidEndSepPunct{\mcitedefaultmidpunct}
{\mcitedefaultendpunct}{\mcitedefaultseppunct}\relax
\EndOfBibitem
\bibitem{LHCb-PAPER-2015-051}
LHCb collaboration, R.~Aaij {\em et~al.},
  \ifthenelse{\boolean{articletitles}}{\emph{{Angular analysis of the
  \mbox{\decay{\Bz}{\Kstarz\mumu}} decay using $3\invfb$ of integrated
  luminosity}}, }{}\href{https://doi.org/10.1007/JHEP02(2016)104}{JHEP
  \textbf{02} (2016) 104},
  \href{http://arxiv.org/abs/1512.04442}{{\normalfont\ttfamily
  arXiv:1512.04442}}\relax
\mciteBstWouldAddEndPuncttrue
\mciteSetBstMidEndSepPunct{\mcitedefaultmidpunct}
{\mcitedefaultendpunct}{\mcitedefaultseppunct}\relax
\EndOfBibitem
\bibitem{LHCb-PAPER-2020-002}
LHCb collaboration, R.~Aaij {\em et~al.},
  \ifthenelse{\boolean{articletitles}}{\emph{{Measurement of \CP-averaged
  observables in the \mbox{\decay{\Bz}{\Kstarz\mumu}} decay}},
  }{}\href{https://doi.org/10.1103/PhysRevLett.125.011802}{Phys.\ Rev.\ Lett.\
  \textbf{125} (2020) 011802},
  \href{http://arxiv.org/abs/2003.04831}{{\normalfont\ttfamily
  arXiv:2003.04831}}\relax
\mciteBstWouldAddEndPuncttrue
\mciteSetBstMidEndSepPunct{\mcitedefaultmidpunct}
{\mcitedefaultendpunct}{\mcitedefaultseppunct}\relax
\EndOfBibitem
\bibitem{LHCb-PAPER-2020-041}
LHCb collaboration, R.~Aaij {\em et~al.},
  \ifthenelse{\boolean{articletitles}}{\emph{{Angular analysis of the $B^{+}\to
  K^{\ast+}\mu^+\mu^-$ decay}},
  }{}\href{https://doi.org/10.1103/PhysRevLett.126.161802}{Phys.\ Rev.\ Lett.\
  \textbf{126} (2021) 161802},
  \href{http://arxiv.org/abs/2012.13241}{{\normalfont\ttfamily
  arXiv:2012.13241}}\relax
\mciteBstWouldAddEndPuncttrue
\mciteSetBstMidEndSepPunct{\mcitedefaultmidpunct}
{\mcitedefaultendpunct}{\mcitedefaultseppunct}\relax
\EndOfBibitem
\bibitem{Aaboud:2018krd}
ATLAS collaboration, M.~Aaboud {\em et~al.},
  \ifthenelse{\boolean{articletitles}}{\emph{{Angular analysis of $B^0_d
  \rightarrow K^{*}\mu^+\mu^-$ decays in $pp$ collisions at $\sqrt{s}= 8$ TeV
  with the ATLAS detector}},
  }{}\href{https://doi.org/10.1007/JHEP10(2018)047}{JHEP \textbf{10} (2018)
  047}, \href{http://arxiv.org/abs/1805.04000}{{\normalfont\ttfamily
  arXiv:1805.04000}}\relax
\mciteBstWouldAddEndPuncttrue
\mciteSetBstMidEndSepPunct{\mcitedefaultmidpunct}
{\mcitedefaultendpunct}{\mcitedefaultseppunct}\relax
\EndOfBibitem
\bibitem{Khachatryan:2015isa}
CMS collaboration, V.~Khachatryan {\em et~al.},
  \ifthenelse{\boolean{articletitles}}{\emph{{Angular analysis of the decay
  \mbox{$B^0 \to K^{*0} \mu^+ \mu^-$} from pp collisions at $\sqrt s = 8$
  TeV}}, }{}\href{https://doi.org/10.1016/j.physletb.2015.12.020}{Phys.\ Lett.\
   \textbf{B753} (2016) 424},
  \href{http://arxiv.org/abs/1507.08126}{{\normalfont\ttfamily
  arXiv:1507.08126}}\relax
\mciteBstWouldAddEndPuncttrue
\mciteSetBstMidEndSepPunct{\mcitedefaultmidpunct}
{\mcitedefaultendpunct}{\mcitedefaultseppunct}\relax
\EndOfBibitem
\bibitem{Sirunyan:2017dhj}
CMS collaboration, A.~M. Sirunyan {\em et~al.},
  \ifthenelse{\boolean{articletitles}}{\emph{{Measurement of angular parameters
  from the decay ${B}^0 \to {K}^{*0} \mu^+ \mu^-$ in proton-proton collisions
  at $\sqrt{s} = $ 8 TeV}},
  }{}\href{https://doi.org/10.1016/j.physletb.2018.04.030}{Phys.\ Lett.\
  \textbf{B781} (2018) 517},
  \href{http://arxiv.org/abs/1710.02846}{{\normalfont\ttfamily
  arXiv:1710.02846}}\relax
\mciteBstWouldAddEndPuncttrue
\mciteSetBstMidEndSepPunct{\mcitedefaultmidpunct}
{\mcitedefaultendpunct}{\mcitedefaultseppunct}\relax
\EndOfBibitem
\bibitem{Wehle:2016yoi}
Belle collaboration, S.~Wehle {\em et~al.},
  \ifthenelse{\boolean{articletitles}}{\emph{{Lepton-Flavor-Dependent Angular
  Analysis of $B\to K^\ast \ell^+\ell^-$}},
  }{}\href{https://doi.org/10.1103/PhysRevLett.118.111801}{Phys.\ Rev.\ Lett.\
  \textbf{118} (2017) 111801},
  \href{http://arxiv.org/abs/1612.05014}{{\normalfont\ttfamily
  arXiv:1612.05014}}\relax
\mciteBstWouldAddEndPuncttrue
\mciteSetBstMidEndSepPunct{\mcitedefaultmidpunct}
{\mcitedefaultendpunct}{\mcitedefaultseppunct}\relax
\EndOfBibitem
\bibitem{LHCb-PAPER-2014-024}
LHCb collaboration, R.~Aaij {\em et~al.},
  \ifthenelse{\boolean{articletitles}}{\emph{{Test of lepton universality using
  \mbox{\decay{\Bp}{\Kp\ellell}} decays}},
  }{}\href{https://doi.org/10.1103/PhysRevLett.113.151601}{Phys.\ Rev.\ Lett.\
  \textbf{113} (2014) 151601},
  \href{http://arxiv.org/abs/1406.6482}{{\normalfont\ttfamily
  arXiv:1406.6482}}\relax
\mciteBstWouldAddEndPuncttrue
\mciteSetBstMidEndSepPunct{\mcitedefaultmidpunct}
{\mcitedefaultendpunct}{\mcitedefaultseppunct}\relax
\EndOfBibitem
\bibitem{LHCb-PAPER-2017-013}
LHCb collaboration, R.~Aaij {\em et~al.},
  \ifthenelse{\boolean{articletitles}}{\emph{{Test of lepton universality with
  \mbox{\decay{\Bz}{\Kstarz\ellell}} decays}},
  }{}\href{https://doi.org/10.1007/JHEP08(2017)055}{JHEP \textbf{08} (2017)
  055}, \href{http://arxiv.org/abs/1705.05802}{{\normalfont\ttfamily
  arXiv:1705.05802}}\relax
\mciteBstWouldAddEndPuncttrue
\mciteSetBstMidEndSepPunct{\mcitedefaultmidpunct}
{\mcitedefaultendpunct}{\mcitedefaultseppunct}\relax
\EndOfBibitem
\bibitem{LHCb-PAPER-2019-009}
LHCb collaboration, R.~Aaij {\em et~al.},
  \ifthenelse{\boolean{articletitles}}{\emph{{Search for lepton-universality
  violation in \mbox{\decay{\Bp}{\Kp\ellell}} decays}},
  }{}\href{https://doi.org/10.1103/PhysRevLett.122.191801}{Phys.\ Rev.\ Lett.\
  \textbf{122} (2019) 191801},
  \href{http://arxiv.org/abs/1903.09252}{{\normalfont\ttfamily
  arXiv:1903.09252}}\relax
\mciteBstWouldAddEndPuncttrue
\mciteSetBstMidEndSepPunct{\mcitedefaultmidpunct}
{\mcitedefaultendpunct}{\mcitedefaultseppunct}\relax
\EndOfBibitem
\bibitem{LHCb-PAPER-2019-040}
LHCb collaboration, R.~Aaij {\em et~al.},
  \ifthenelse{\boolean{articletitles}}{\emph{{Test of lepton universality using
  \mbox{\decay{\Lb}{p\Km\ellell}} decays}},
  }{}\href{https://doi.org/10.1007/JHEP05(2020)040}{JHEP \textbf{05} (2020)
  040}, \href{http://arxiv.org/abs/1912.08139}{{\normalfont\ttfamily
  arXiv:1912.08139}}\relax
\mciteBstWouldAddEndPuncttrue
\mciteSetBstMidEndSepPunct{\mcitedefaultmidpunct}
{\mcitedefaultendpunct}{\mcitedefaultseppunct}\relax
\EndOfBibitem
\bibitem{LHCb-PAPER-2021-004}
LHCb collaboration, R.~Aaij {\em et~al.},
  \ifthenelse{\boolean{articletitles}}{\emph{{Test of lepton universality in
  beauty-quark decays}},
  }{}\href{http://arxiv.org/abs/2103.11769}{{\normalfont\ttfamily
  arXiv:2103.11769}}, {submitted to Nature Physics}\relax
\mciteBstWouldAddEndPuncttrue
\mciteSetBstMidEndSepPunct{\mcitedefaultmidpunct}
{\mcitedefaultendpunct}{\mcitedefaultseppunct}\relax
\EndOfBibitem
\bibitem{Lees:2012tva}
BaBar collaboration, J.~P. Lees {\em et~al.},
  \ifthenelse{\boolean{articletitles}}{\emph{{Measurement of Branching
  Fractions and Rate Asymmetries in the Rare Decays $B \to K^{(*)} l^+ l^-$}},
  }{}\href{https://doi.org/10.1103/PhysRevD.86.032012}{Phys.\ Rev.\
  \textbf{D86} (2012) 032012},
  \href{http://arxiv.org/abs/1204.3933}{{\normalfont\ttfamily
  arXiv:1204.3933}}\relax
\mciteBstWouldAddEndPuncttrue
\mciteSetBstMidEndSepPunct{\mcitedefaultmidpunct}
{\mcitedefaultendpunct}{\mcitedefaultseppunct}\relax
\EndOfBibitem
\bibitem{Abdesselam:2019lab}
Belle collaboration, S.~Choudhury {\em et~al.},
  \ifthenelse{\boolean{articletitles}}{\emph{{Test of lepton flavor
  universality and search for lepton flavor violation in $B \rightarrow K\ell
  \ell$ decays}}, }{}\href{https://doi.org/10.1007/JHEP03(2021)105}{JHEP
  \textbf{03} (2021) 105},
  \href{http://arxiv.org/abs/1908.01848}{{\normalfont\ttfamily
  arXiv:1908.01848}}\relax
\mciteBstWouldAddEndPuncttrue
\mciteSetBstMidEndSepPunct{\mcitedefaultmidpunct}
{\mcitedefaultendpunct}{\mcitedefaultseppunct}\relax
\EndOfBibitem
\bibitem{Abdesselam:2019wac}
Belle collaboration, A.~Abdesselam {\em et~al.},
  \ifthenelse{\boolean{articletitles}}{\emph{{Test of Lepton-Flavor
  Universality in ${B\to K^\ast\ell^+\ell^-}$ Decays at Belle}},
  }{}\href{https://doi.org/10.1103/PhysRevLett.126.161801}{Phys.\ Rev.\ Lett.\
  \textbf{126} (2021) 161801},
  \href{http://arxiv.org/abs/1904.02440}{{\normalfont\ttfamily
  arXiv:1904.02440}}\relax
\mciteBstWouldAddEndPuncttrue
\mciteSetBstMidEndSepPunct{\mcitedefaultmidpunct}
{\mcitedefaultendpunct}{\mcitedefaultseppunct}\relax
\EndOfBibitem
\bibitem{Altmannshofer:2014rta}
W.~Altmannshofer and D.~M. Straub,
  \ifthenelse{\boolean{articletitles}}{\emph{{New physics in $b\rightarrow s$
  transitions after LHC run 1}},
  }{}\href{https://doi.org/10.1140/epjc/s10052-015-3602-7}{Eur.\ Phys.\ J.\
  \textbf{C75} (2015) 382},
  \href{http://arxiv.org/abs/1411.3161}{{\normalfont\ttfamily
  arXiv:1411.3161}}\relax
\mciteBstWouldAddEndPuncttrue
\mciteSetBstMidEndSepPunct{\mcitedefaultmidpunct}
{\mcitedefaultendpunct}{\mcitedefaultseppunct}\relax
\EndOfBibitem
\bibitem{Straub:2015ica}
A.~Bharucha, D.~M. Straub, and R.~Zwicky,
  \ifthenelse{\boolean{articletitles}}{\emph{{$B\to V\ell^+\ell^-$ in the
  Standard Model from light-cone sum rules}},
  }{}\href{https://doi.org/10.1007/JHEP08(2016)098}{JHEP \textbf{08} (2016)
  098}, \href{http://arxiv.org/abs/1503.05534}{{\normalfont\ttfamily
  arXiv:1503.05534}}\relax
\mciteBstWouldAddEndPuncttrue
\mciteSetBstMidEndSepPunct{\mcitedefaultmidpunct}
{\mcitedefaultendpunct}{\mcitedefaultseppunct}\relax
\EndOfBibitem
\bibitem{Straub:2018kue}
D.~M. Straub, \ifthenelse{\boolean{articletitles}}{\emph{{flavio: a Python
  package for flavour and precision phenomenology in the Standard Model and
  beyond}}, }{}\href{http://arxiv.org/abs/1810.08132}{{\normalfont\ttfamily
  arXiv:1810.08132}}\relax
\mciteBstWouldAddEndPuncttrue
\mciteSetBstMidEndSepPunct{\mcitedefaultmidpunct}
{\mcitedefaultendpunct}{\mcitedefaultseppunct}\relax
\EndOfBibitem
\bibitem{Horgan:2013pva}
R.~R. Horgan, Z.~Liu, S.~Meinel, and M.~Wingate,
  \ifthenelse{\boolean{articletitles}}{\emph{{Calculation of $B^0 \to K^{*0}
  \mu^+ \mu^-$ and $B_s^0 \to \phi \mu^+ \mu^-$ observables using form factors
  from lattice QCD}},
  }{}\href{https://doi.org/10.1103/PhysRevLett.112.212003}{Phys.\ Rev.\ Lett.\
  \textbf{112} (2014) 212003},
  \href{http://arxiv.org/abs/1310.3887}{{\normalfont\ttfamily
  arXiv:1310.3887}}\relax
\mciteBstWouldAddEndPuncttrue
\mciteSetBstMidEndSepPunct{\mcitedefaultmidpunct}
{\mcitedefaultendpunct}{\mcitedefaultseppunct}\relax
\EndOfBibitem
\bibitem{Horgan:2015vla}
R.~R. Horgan, Z.~Liu, S.~Meinel, and M.~Wingate,
  \ifthenelse{\boolean{articletitles}}{\emph{{Rare $B$ decays using lattice QCD
  form factors}}, }{}\href{https://doi.org/10.22323/1.214.0372}{PoS
  \textbf{LATTICE2014} (2015) 372},
  \href{http://arxiv.org/abs/1501.00367}{{\normalfont\ttfamily
  arXiv:1501.00367}}\relax
\mciteBstWouldAddEndPuncttrue
\mciteSetBstMidEndSepPunct{\mcitedefaultmidpunct}
{\mcitedefaultendpunct}{\mcitedefaultseppunct}\relax
\EndOfBibitem
\bibitem{LHCB-PAPER-2016-031}
LHCb collaboration, R.~Aaij {\em et~al.},
  \ifthenelse{\boolean{articletitles}}{\emph{{Measurement of the \bquark-quark
  production cross-section in 7 and 13$\tev$ $\proton\proton$ collisions}},
  }{}\href{https://doi.org/10.1103/PhysRevLett.118.052002}{Phys.\ Rev.\ Lett.\
  \textbf{118} (2017) 052002}, Erratum
  \href{https://doi.org/10.1103/PhysRevLett.119.169901}{ibid.\   \textbf{119}
  (2017) 169901}, \href{http://arxiv.org/abs/1612.05140}{{\normalfont\ttfamily
  arXiv:1612.05140}}\relax
\mciteBstWouldAddEndPuncttrue
\mciteSetBstMidEndSepPunct{\mcitedefaultmidpunct}
{\mcitedefaultendpunct}{\mcitedefaultseppunct}\relax
\EndOfBibitem
\bibitem{LHCb-PAPER-2017-037}
LHCb collaboration, R.~Aaij {\em et~al.},
  \ifthenelse{\boolean{articletitles}}{\emph{{Measurement of the \Bpm
  production cross-section in \proton\proton collisions at $\sqs = $7 and
  13\tev}}, }{}\href{https://doi.org/10.1007/JHEP12(2017)026}{JHEP \textbf{12}
  (2017) 026}, \href{http://arxiv.org/abs/1710.04921}{{\normalfont\ttfamily
  arXiv:1710.04921}}\relax
\mciteBstWouldAddEndPuncttrue
\mciteSetBstMidEndSepPunct{\mcitedefaultmidpunct}
{\mcitedefaultendpunct}{\mcitedefaultseppunct}\relax
\EndOfBibitem
\bibitem{Descotes-Genon:2015hea}
S.~Descotes-Genon and J.~Virto,
  \ifthenelse{\boolean{articletitles}}{\emph{{Time dependence in $B \to
  V\ell\ell$ decays}}, }{}\href{https://doi.org/10.1007/JHEP04(2015)045}{JHEP
  \textbf{04} (2015) 045}, Erratum
  \href{https://doi.org/10.1007/JHEP07(2015)049}{ibid.\   \textbf{07} (2015)
  049}, \href{http://arxiv.org/abs/1502.05509}{{\normalfont\ttfamily
  arXiv:1502.05509}}\relax
\mciteBstWouldAddEndPuncttrue
\mciteSetBstMidEndSepPunct{\mcitedefaultmidpunct}
{\mcitedefaultendpunct}{\mcitedefaultseppunct}\relax
\EndOfBibitem
\bibitem{Altmannshofer:2008dz}
W.~Altmannshofer {\em et~al.},
  \ifthenelse{\boolean{articletitles}}{\emph{{Symmetries and Asymmetries of $B
  \to K^{*} \mu^{+} \mu^{-}$ Decays in the Standard Model and Beyond}},
  }{}\href{https://doi.org/10.1088/1126-6708/2009/01/019}{JHEP \textbf{01}
  (2009) 019}, \href{http://arxiv.org/abs/0811.1214}{{\normalfont\ttfamily
  arXiv:0811.1214}}\relax
\mciteBstWouldAddEndPuncttrue
\mciteSetBstMidEndSepPunct{\mcitedefaultmidpunct}
{\mcitedefaultendpunct}{\mcitedefaultseppunct}\relax
\EndOfBibitem
\bibitem{Bobeth:2008ij}
C.~Bobeth, G.~Hiller, and G.~Piranishvili,
  \ifthenelse{\boolean{articletitles}}{\emph{{CP Asymmetries in $\bar{B} \to
  \bar{K}^* (\to \bar{K} \pi) \bar{\ell} \ell$ and Untagged $\bar{B}_s$, $B_s
  \to \phi (\to K^{+} K^-) \bar{\ell} \ell$ Decays at NLO}},
  }{}\href{https://doi.org/10.1088/1126-6708/2008/07/106}{JHEP \textbf{07}
  (2008) 106}, \href{http://arxiv.org/abs/0805.2525}{{\normalfont\ttfamily
  arXiv:0805.2525}}\relax
\mciteBstWouldAddEndPuncttrue
\mciteSetBstMidEndSepPunct{\mcitedefaultmidpunct}
{\mcitedefaultendpunct}{\mcitedefaultseppunct}\relax
\EndOfBibitem
\bibitem{LHCb-DP-2008-001}
LHCb collaboration, A.~A. Alves~Jr.\ {\em et~al.},
  \ifthenelse{\boolean{articletitles}}{\emph{{The \lhcb detector at the LHC}},
  }{}\href{https://doi.org/10.1088/1748-0221/3/08/S08005}{JINST \textbf{3}
  (2008) S08005}\relax
\mciteBstWouldAddEndPuncttrue
\mciteSetBstMidEndSepPunct{\mcitedefaultmidpunct}
{\mcitedefaultendpunct}{\mcitedefaultseppunct}\relax
\EndOfBibitem
\bibitem{LHCb-DP-2014-002}
LHCb collaboration, R.~Aaij {\em et~al.},
  \ifthenelse{\boolean{articletitles}}{\emph{{LHCb detector performance}},
  }{}\href{https://doi.org/10.1142/S0217751X15300227}{Int.\ J.\ Mod.\ Phys.\
  \textbf{A30} (2015) 1530022},
  \href{http://arxiv.org/abs/1412.6352}{{\normalfont\ttfamily
  arXiv:1412.6352}}\relax
\mciteBstWouldAddEndPuncttrue
\mciteSetBstMidEndSepPunct{\mcitedefaultmidpunct}
{\mcitedefaultendpunct}{\mcitedefaultseppunct}\relax
\EndOfBibitem
\bibitem{LHCb-DP-2014-001}
R.~Aaij {\em et~al.}, \ifthenelse{\boolean{articletitles}}{\emph{{Performance
  of the LHCb Vertex Locator}},
  }{}\href{https://doi.org/10.1088/1748-0221/9/09/P09007}{JINST \textbf{9}
  (2014) P09007}, \href{http://arxiv.org/abs/1405.7808}{{\normalfont\ttfamily
  arXiv:1405.7808}}\relax
\mciteBstWouldAddEndPuncttrue
\mciteSetBstMidEndSepPunct{\mcitedefaultmidpunct}
{\mcitedefaultendpunct}{\mcitedefaultseppunct}\relax
\EndOfBibitem
\bibitem{LHCb-DP-2013-003}
R.~Arink {\em et~al.}, \ifthenelse{\boolean{articletitles}}{\emph{{Performance
  of the LHCb Outer Tracker}},
  }{}\href{https://doi.org/10.1088/1748-0221/9/01/P01002}{JINST \textbf{9}
  (2014) P01002}, \href{http://arxiv.org/abs/1311.3893}{{\normalfont\ttfamily
  arXiv:1311.3893}}\relax
\mciteBstWouldAddEndPuncttrue
\mciteSetBstMidEndSepPunct{\mcitedefaultmidpunct}
{\mcitedefaultendpunct}{\mcitedefaultseppunct}\relax
\EndOfBibitem
\bibitem{LHCb-DP-2017-001}
P.~d'Argent {\em et~al.}, \ifthenelse{\boolean{articletitles}}{\emph{{Improved
  performance of the LHCb Outer Tracker in LHC Run 2}},
  }{}\href{https://doi.org/10.1088/1748-0221/12/11/P11016}{JINST \textbf{12}
  (2017) P11016}, \href{http://arxiv.org/abs/1708.00819}{{\normalfont\ttfamily
  arXiv:1708.00819}}\relax
\mciteBstWouldAddEndPuncttrue
\mciteSetBstMidEndSepPunct{\mcitedefaultmidpunct}
{\mcitedefaultendpunct}{\mcitedefaultseppunct}\relax
\EndOfBibitem
\bibitem{LHCb-DP-2012-003}
M.~Adinolfi {\em et~al.},
  \ifthenelse{\boolean{articletitles}}{\emph{{Performance of the \lhcb RICH
  detector at the LHC}},
  }{}\href{https://doi.org/10.1140/epjc/s10052-013-2431-9}{Eur.\ Phys.\ J.\
  \textbf{C73} (2013) 2431},
  \href{http://arxiv.org/abs/1211.6759}{{\normalfont\ttfamily
  arXiv:1211.6759}}\relax
\mciteBstWouldAddEndPuncttrue
\mciteSetBstMidEndSepPunct{\mcitedefaultmidpunct}
{\mcitedefaultendpunct}{\mcitedefaultseppunct}\relax
\EndOfBibitem
\bibitem{LHCb-DP-2012-002}
A.~A. Alves~Jr.\ {\em et~al.},
  \ifthenelse{\boolean{articletitles}}{\emph{{Performance of the LHCb muon
  system}}, }{}\href{https://doi.org/10.1088/1748-0221/8/02/P02022}{JINST
  \textbf{8} (2013) P02022},
  \href{http://arxiv.org/abs/1211.1346}{{\normalfont\ttfamily
  arXiv:1211.1346}}\relax
\mciteBstWouldAddEndPuncttrue
\mciteSetBstMidEndSepPunct{\mcitedefaultmidpunct}
{\mcitedefaultendpunct}{\mcitedefaultseppunct}\relax
\EndOfBibitem
\bibitem{LHCb-DP-2012-004}
R.~Aaij {\em et~al.}, \ifthenelse{\boolean{articletitles}}{\emph{{The \lhcb
  trigger and its performance in 2011}},
  }{}\href{https://doi.org/10.1088/1748-0221/8/04/P04022}{JINST \textbf{8}
  (2013) P04022}, \href{http://arxiv.org/abs/1211.3055}{{\normalfont\ttfamily
  arXiv:1211.3055}}\relax
\mciteBstWouldAddEndPuncttrue
\mciteSetBstMidEndSepPunct{\mcitedefaultmidpunct}
{\mcitedefaultendpunct}{\mcitedefaultseppunct}\relax
\EndOfBibitem
\bibitem{Sjostrand:2007gs}
T.~Sj\"{o}strand, S.~Mrenna, and P.~Skands,
  \ifthenelse{\boolean{articletitles}}{\emph{{A brief introduction to PYTHIA
  8.1}}, }{}\href{https://doi.org/10.1016/j.cpc.2008.01.036}{Comput.\ Phys.\
  Commun.\  \textbf{178} (2008) 852},
  \href{http://arxiv.org/abs/0710.3820}{{\normalfont\ttfamily
  arXiv:0710.3820}}\relax
\mciteBstWouldAddEndPuncttrue
\mciteSetBstMidEndSepPunct{\mcitedefaultmidpunct}
{\mcitedefaultendpunct}{\mcitedefaultseppunct}\relax
\EndOfBibitem
\bibitem{Sjostrand:2006za}
T.~Sj\"{o}strand, S.~Mrenna, and P.~Skands,
  \ifthenelse{\boolean{articletitles}}{\emph{{PYTHIA 6.4 physics and manual}},
  }{}\href{https://doi.org/10.1088/1126-6708/2006/05/026}{JHEP \textbf{05}
  (2006) 026}, \href{http://arxiv.org/abs/hep-ph/0603175}{{\normalfont\ttfamily
  arXiv:hep-ph/0603175}}\relax
\mciteBstWouldAddEndPuncttrue
\mciteSetBstMidEndSepPunct{\mcitedefaultmidpunct}
{\mcitedefaultendpunct}{\mcitedefaultseppunct}\relax
\EndOfBibitem
\bibitem{LHCb-PROC-2010-056}
I.~Belyaev {\em et~al.}, \ifthenelse{\boolean{articletitles}}{\emph{{Handling
  of the generation of primary events in Gauss, the LHCb simulation
  framework}}, }{}\href{https://doi.org/10.1088/1742-6596/331/3/032047}{J.\
  Phys.\ Conf.\ Ser.\  \textbf{331} (2011) 032047}\relax
\mciteBstWouldAddEndPuncttrue
\mciteSetBstMidEndSepPunct{\mcitedefaultmidpunct}
{\mcitedefaultendpunct}{\mcitedefaultseppunct}\relax
\EndOfBibitem
\bibitem{Lange:2001uf}
D.~J. Lange, \ifthenelse{\boolean{articletitles}}{\emph{{The EvtGen particle
  decay simulation package}},
  }{}\href{https://doi.org/10.1016/S0168-9002(01)00089-4}{Nucl.\ Instrum.\
  Meth.\  \textbf{A462} (2001) 152}\relax
\mciteBstWouldAddEndPuncttrue
\mciteSetBstMidEndSepPunct{\mcitedefaultmidpunct}
{\mcitedefaultendpunct}{\mcitedefaultseppunct}\relax
\EndOfBibitem
\bibitem{davidson2015photos}
N.~Davidson, T.~Przedzinski, and Z.~Was,
  \ifthenelse{\boolean{articletitles}}{\emph{{PHOTOS interface in C++:
  Technical and physics documentation}},
  }{}\href{https://doi.org/https://doi.org/10.1016/j.cpc.2015.09.013}{Comp.\
  Phys.\ Comm.\  \textbf{199} (2016) 86},
  \href{http://arxiv.org/abs/1011.0937}{{\normalfont\ttfamily
  arXiv:1011.0937}}\relax
\mciteBstWouldAddEndPuncttrue
\mciteSetBstMidEndSepPunct{\mcitedefaultmidpunct}
{\mcitedefaultendpunct}{\mcitedefaultseppunct}\relax
\EndOfBibitem
\bibitem{Allison:2006ve}
Geant4 collaboration, J.~Allison {\em et~al.},
  \ifthenelse{\boolean{articletitles}}{\emph{{Geant4 developments and
  applications}}, }{}\href{https://doi.org/10.1109/TNS.2006.869826}{IEEE
  Trans.\ Nucl.\ Sci.\  \textbf{53} (2006) 270}\relax
\mciteBstWouldAddEndPuncttrue
\mciteSetBstMidEndSepPunct{\mcitedefaultmidpunct}
{\mcitedefaultendpunct}{\mcitedefaultseppunct}\relax
\EndOfBibitem
\bibitem{Agostinelli:2002hh}
Geant4 collaboration, S.~Agostinelli {\em et~al.},
  \ifthenelse{\boolean{articletitles}}{\emph{{Geant4: A simulation toolkit}},
  }{}\href{https://doi.org/10.1016/S0168-9002(03)01368-8}{Nucl.\ Instrum.\
  Meth.\  \textbf{A506} (2003) 250}\relax
\mciteBstWouldAddEndPuncttrue
\mciteSetBstMidEndSepPunct{\mcitedefaultmidpunct}
{\mcitedefaultendpunct}{\mcitedefaultseppunct}\relax
\EndOfBibitem
\bibitem{LHCb-PROC-2011-006}
M.~Clemencic {\em et~al.}, \ifthenelse{\boolean{articletitles}}{\emph{{The
  \lhcb simulation application, Gauss: Design, evolution and experience}},
  }{}\href{https://doi.org/10.1088/1742-6596/331/3/032023}{J.\ Phys.\ Conf.\
  Ser.\  \textbf{331} (2011) 032023}\relax
\mciteBstWouldAddEndPuncttrue
\mciteSetBstMidEndSepPunct{\mcitedefaultmidpunct}
{\mcitedefaultendpunct}{\mcitedefaultseppunct}\relax
\EndOfBibitem
\bibitem{PDG2020}
Particle Data Group, P.~A. Zyla {\em et~al.},
  \ifthenelse{\boolean{articletitles}}{\emph{{\href{http://pdg.lbl.gov/}{Review
  of particle physics}}}, }{}\href{https://doi.org/10.1093/ptep/ptaa104}{Prog.\
  Theor.\ Exp.\ Phys.\  \textbf{2020} (2020) 083C01}\relax
\mciteBstWouldAddEndPuncttrue
\mciteSetBstMidEndSepPunct{\mcitedefaultmidpunct}
{\mcitedefaultendpunct}{\mcitedefaultseppunct}\relax
\EndOfBibitem
\bibitem{Breiman}
L.~Breiman, J.~H. Friedman, R.~A. Olshen, and C.~J. Stone, {\em Classification
  and regression trees}, Wadsworth international group, Belmont, California,
  USA, 1984\relax
\mciteBstWouldAddEndPuncttrue
\mciteSetBstMidEndSepPunct{\mcitedefaultmidpunct}
{\mcitedefaultendpunct}{\mcitedefaultseppunct}\relax
\EndOfBibitem
\bibitem{AdaBoost}
Y.~Freund and R.~E. Schapire, \ifthenelse{\boolean{articletitles}}{\emph{A
  decision-theoretic generalization of on-line learning and an application to
  boosting}, }{}\href{https://doi.org/10.1006/jcss.1997.1504}{J.\ Comput.\
  Syst.\ Sci.\  \textbf{55} (1997) 119}\relax
\mciteBstWouldAddEndPuncttrue
\mciteSetBstMidEndSepPunct{\mcitedefaultmidpunct}
{\mcitedefaultendpunct}{\mcitedefaultseppunct}\relax
\EndOfBibitem
\bibitem{Hocker:2007ht}
H.~Voss, A.~Hoecker, J.~Stelzer, and F.~Tegenfeldt,
  \ifthenelse{\boolean{articletitles}}{\emph{{TMVA - Toolkit for Multivariate
  Data Analysis with ROOT}}, }{}\href{https://doi.org/10.22323/1.050.0040}{PoS
  \textbf{ACAT} (2007) 040}\relax
\mciteBstWouldAddEndPuncttrue
\mciteSetBstMidEndSepPunct{\mcitedefaultmidpunct}
{\mcitedefaultendpunct}{\mcitedefaultseppunct}\relax
\EndOfBibitem
\bibitem{TMVA4}
A.~Hoecker {\em et~al.}, \ifthenelse{\boolean{articletitles}}{\emph{{TMVA 4 ---
  Toolkit for Multivariate Data Analysis with ROOT. Users Guide.}},
  }{}\href{http://arxiv.org/abs/physics/0703039}{{\normalfont\ttfamily
  arXiv:physics/0703039}}\relax
\mciteBstWouldAddEndPuncttrue
\mciteSetBstMidEndSepPunct{\mcitedefaultmidpunct}
{\mcitedefaultendpunct}{\mcitedefaultseppunct}\relax
\EndOfBibitem
\bibitem{Blum:1999:BHB:307400.307439}
A.~Blum, A.~Kalai, and J.~Langford,
  \ifthenelse{\boolean{articletitles}}{\emph{Beating the hold-out: Bounds for
  k-fold and progressive cross-validation}, }{} in {\em Proceedings of the
  Twelfth Annual Conference on Computational Learning Theory},
  \href{https://doi.org/10.1145/307400.307439}{ COLT '99, (New York, NY, USA),
  203--208, ACM, 1999}\relax
\mciteBstWouldAddEndPuncttrue
\mciteSetBstMidEndSepPunct{\mcitedefaultmidpunct}
{\mcitedefaultendpunct}{\mcitedefaultseppunct}\relax
\EndOfBibitem
\bibitem{Pivk:2004ty}
M.~Pivk and F.~R. Le~Diberder,
  \ifthenelse{\boolean{articletitles}}{\emph{{sPlot: A statistical tool to
  unfold data distributions}},
  }{}\href{https://doi.org/10.1016/j.nima.2005.08.106}{Nucl.\ Instrum.\ Meth.\
  \textbf{A555} (2005) 356},
  \href{http://arxiv.org/abs/physics/0402083}{{\normalfont\ttfamily
  arXiv:physics/0402083}}\relax
\mciteBstWouldAddEndPuncttrue
\mciteSetBstMidEndSepPunct{\mcitedefaultmidpunct}
{\mcitedefaultendpunct}{\mcitedefaultseppunct}\relax
\EndOfBibitem
\bibitem{Skwarnicki:1986xj}
T.~Skwarnicki, {\em {A study of the radiative cascade transitions between the
  Upsilon-prime and Upsilon resonances}}, PhD thesis, Institute of Nuclear
  Physics, Krakow, 1986,
  {\href{http://inspirehep.net/record/230779/}{DESY-F31-86-02}}\relax
\mciteBstWouldAddEndPuncttrue
\mciteSetBstMidEndSepPunct{\mcitedefaultmidpunct}
{\mcitedefaultendpunct}{\mcitedefaultseppunct}\relax
\EndOfBibitem
\bibitem{LHCb-PAPER-2012-040}
LHCb collaboration, R.~Aaij {\em et~al.},
  \ifthenelse{\boolean{articletitles}}{\emph{{Amplitude analysis and branching
  fraction measurement of \mbox{\decay{\Bsb}{\jpsi\Kp\Km}}}},
  }{}\href{https://doi.org/10.1103/PhysRevD.87.072004}{Phys.\ Rev.\
  \textbf{D87} (2013) 072004},
  \href{http://arxiv.org/abs/1302.1213}{{\normalfont\ttfamily
  arXiv:1302.1213}}\relax
\mciteBstWouldAddEndPuncttrue
\mciteSetBstMidEndSepPunct{\mcitedefaultmidpunct}
{\mcitedefaultendpunct}{\mcitedefaultseppunct}\relax
\EndOfBibitem
\bibitem{LHCb-PAPER-2014-059}
LHCb collaboration, R.~Aaij {\em et~al.},
  \ifthenelse{\boolean{articletitles}}{\emph{{Precision measurement of \CP
  violation in \mbox{\decay{\Bs}{\jpsi\Kp\Km}} decays}},
  }{}\href{https://doi.org/10.1103/PhysRevLett.114.041801}{Phys.\ Rev.\ Lett.\
  \textbf{114} (2015) 041801},
  \href{http://arxiv.org/abs/1411.3104}{{\normalfont\ttfamily
  arXiv:1411.3104}}\relax
\mciteBstWouldAddEndPuncttrue
\mciteSetBstMidEndSepPunct{\mcitedefaultmidpunct}
{\mcitedefaultendpunct}{\mcitedefaultseppunct}\relax
\EndOfBibitem
\bibitem{LHCb-PAPER-2019-013}
LHCb collaboration, R.~Aaij {\em et~al.},
  \ifthenelse{\boolean{articletitles}}{\emph{{Updated measurement of
  time-dependent \CP-violating observables in \mbox{\decay{\Bs}{\jpsi \Kp\Km}}
  decays}}, }{}\href{https://doi.org/10.1140/epjc/s10052-019-7159-8}{Eur.\
  Phys.\ J.\  \textbf{C79} (2019) 706}, Erratum
  \href{https://doi.org/10.1140/epjc/s10052-020-7875-0}{ibid.\   \textbf{C80}
  (2020) 601}, \href{http://arxiv.org/abs/1906.08356}{{\normalfont\ttfamily
  arXiv:1906.08356}}\relax
\mciteBstWouldAddEndPuncttrue
\mciteSetBstMidEndSepPunct{\mcitedefaultmidpunct}
{\mcitedefaultendpunct}{\mcitedefaultseppunct}\relax
\EndOfBibitem
\bibitem{Ball:2004rg}
P.~Ball and R.~Zwicky, \ifthenelse{\boolean{articletitles}}{\emph{{$B_{d,s} \to
  \rho, \omega, K^*, \phi$ decay form-factors from light-cone sum rules
  revisited}}, }{}\href{https://doi.org/10.1103/PhysRevD.71.014029}{Phys.\
  Rev.\  \textbf{D71} (2005) 014029},
  \href{http://arxiv.org/abs/hep-ph/0412079}{{\normalfont\ttfamily
  arXiv:hep-ph/0412079}}\relax
\mciteBstWouldAddEndPuncttrue
\mciteSetBstMidEndSepPunct{\mcitedefaultmidpunct}
{\mcitedefaultendpunct}{\mcitedefaultseppunct}\relax
\EndOfBibitem
\bibitem{PDG2014}
Particle Data Group, K.~A. Olive {\em et~al.},
  \ifthenelse{\boolean{articletitles}}{\emph{{\href{http://pdg.lbl.gov/}{Review
  of particle physics}}},
  }{}\href{https://doi.org/10.1088/1674-1137/38/9/090001}{Chin.\ Phys.\
  \textbf{C38} (2014) 090001}\relax
\mciteBstWouldAddEndPuncttrue
\mciteSetBstMidEndSepPunct{\mcitedefaultmidpunct}
{\mcitedefaultendpunct}{\mcitedefaultseppunct}\relax
\EndOfBibitem
\end{mcitethebibliography}
